\documentclass[11pt, letterpaper]{thesis}
\usepackage[lmargin=1in,rmargin=1in,tmargin=1in,bmargin=1in]{geometry}

\usepackage[utf8]{inputenc}
\usepackage[english]{babel}
\addto\captionsenglish{}
\usepackage{verbatim}
\usepackage[dvipsnames]{xcolor}
\usepackage{indentfirst}
\usepackage{graphicx}
\usepackage{comment}
\usepackage{csquotes}
\usepackage{float}
\usepackage{enumitem}
\usepackage[toc,acronym]{glossaries}
\usepackage{url}
\usepackage{hyperref}
\hypersetup{
	colorlinks   = true,
	citecolor    = blue,
	linkcolor    = blue,
	urlcolor     = blue,
}

\usepackage{longtable}
\usepackage{tabularx}
\usepackage{multirow}
\usepackage{array}
\usepackage{makecell}
\usepackage{caption}
\usepackage{subcaption}
\usepackage{hhline}
\usepackage{pifont}
\usepackage{pdflscape}
\usepackage{booktabs}
\usepackage[nottoc]{tocbibind}
\usepackage[tocskip=6pt]{parskip}
\usepackage{setspace}
\usepackage{adjustbox}

\usepackage{amsmath,amssymb,amsfonts}

\usepackage{listings}
\lstdefinestyle{mystyle}{
    basicstyle=\ttfamily\footnotesize,
    breakatwhitespace=false,
    breaklines=true,
    captionpos=b,
    keepspaces=true,
    numbers=none,
    numbersep=5pt,
    showspaces=false,
    showstringspaces=false,
    showtabs=false,
    tabsize=2
}
\newcommand{\blackparrot}{BlackParrot}
\newcommand{\bedrock}{BedRock}
\newcommand{\bpbedrock}{BP-BedRock}

\newcolumntype{x}[1]{>{\centering\arraybackslash\hspace{0pt}}p{#1}}
\newcolumntype{C}[1]{>{\centering\arraybackslash\hspace{0pt}}p{#1}}
\newcolumntype{R}[1]{>{\raggedleft\arraybackslash\hspace{0pt}}p{#1}}
\newcolumntype{L}[1]{>{\raggedright\arraybackslash\hspace{0pt}}p{#1}}
\newcommand{\ra}[1]{\renewcommand{\arraystretch}{#1}}
\newcommand{\tss}[1]{\textsuperscript{#1}}

\usepackage[backend=biber,
            style=ieee,
			maxbibnames=99,
            maxcitenames=99,
			hyperref=true,
			backref=false,
            url=false,
			sorting=nyt]{biblatex}
\addto\extrasenglish{}
\addto\extrasenglish{}
\addto\extrasenglish{}
\addto\extrasenglish{}
\addto\extrasenglish{}
\addto\extrasenglish{}
\addto\extrasenglish{}

\makeglossaries
\newacronym{lce}{LCE}{Local Cache Engine}

\newacronym{cce}{CCE}{Cache Coherence Engine}

\newacronym{bp}{BP}{BlackParrot}

\newacronym{bpbedrock}{BP-BedRock}{BlackParrot-BedRock}

\newacronym{M}{M}{Modified (cache coherence state)}

\newacronym{O}{O}{Owned (cache coherence state)}

\newacronym{E}{E}{Exclusive (cache coherence state)}

\newacronym{S}{S}{Shared (cache coherence state)}

\newacronym{I}{I}{Invalid (cache coherence state)}

\newacronym{F}{F}{Forward (cache coherence state)}

\newacronym{ucode}{ucode}{Microcode}

\newacronym{fsm}{FSM}{Finite State Machine}

\newacronym{isa}{ISA}{Instruction Set Architecture}

\newacronym{io}{I/O}{Input / Output}

\newacronym{ram}{RAM}{Random-Access Memory}

\newacronym{dram}{DRAM}{Dynamic Random-Access Memory}

\newacronym{sram}{SRAM}{Static Random-Access Memory}

\newacronym{vipt}{VIPT}{Virtually-Indexed, Physically-Tagged}

\newacronym{pipt}{PIPT}{Physically-Indexed, Physically-Tagged}

\newacronym{va}{VA}{Virtual Address}

\newacronym{pa}{PA}{Physical Address}

\newacronym{swmr}{SWMR}{Single-Writer, Multiple-Reader}

\newacronym{llc}{LLC}{Last-Level Cache}

\newacronym{noc}{NoC}{Network-on-Chip}

\newacronym{tlp}{TLP}{Thread-Level Parallelism}

\newacronym{ilp}{ILP}{Instruction-Level Parallelism}

\newacronym{pc}{PC}{Program Counter}

\newacronym{gpr}{GPR}{General Purpose Register}

\newacronym{ai}{AI}{Artificial Intelligence}

\newacronym{ml}{ML}{Machine Learning}

\newacronym{ppa}{PPA}{Power, Performance, and Area}

\newacronym{fpga}{FPGA}{Field Programmable Gate Array}

\newacronym{asic}{ASIC}{Application-Specific Integrated Circuit}

\newacronym{fpu}{FPU}{Floating-Point Unit}

\newacronym{dsa}{DSA}{Domain-Specific Accelerator}

\newacronym{risc}{RISC}{Reduced Instruction Set Computer}

\newacronym{cisc}{CISC}{Complex Instruction Set Computer}

\newacronym{hls}{HLS}{High-Level Synthesis}

\newacronym{dma}{DMA}{Direct Memory Access}

\newacronym{soc}{SoC}{System-on-Chip}

\newacronym{eda}{EDA}{Electronic Design Automation}

\begin{document}

%%% General desired structure:
% Title
% Copyright
% Abstract
% TOC (w/ copyright, abstract, LOF, LOT, Glossary, and Acronyms)
% List of Figures
% List of Tables
% Acronyms
% Intro, ...
% References

% turn off page numbering
\pagestyle{empty}
\pagenumbering{gobble}

%% Front Matter

%\Title{The BlackParrot-BedRock Cache Coherence System}
\Title{The Open-Source BlackParrot-BedRock Cache Coherence System}
\Author{Mark Unruh Wyse}
\Year{2025}
\Program{Paul G. Allen School of Computer Science \& Engineering}
\Chair{Mark Oskin}{}{Paul G. Allen School of Computer Science \& Engineering}
\Signature{Michael Taylor}
\Signature{Ajay Joshi}
%\Signature{Patrick Boyle, GSR}

\titlepage
\clearpage
\copyrightpage
\clearpage

\abstract{%
Hardware-based cache coherence is employed nearly universally in modern shared-memory multicore processors due to its power, performance, and area efficiency. During the development of shared-memory processors, researchers proposed programmable controllers to support both cache coherence and message passing designs. Despite the flexibility afforded by programmability for post-design protocol changes or runtime selection of communication paradigms, such systems were never widely adopted. Hardware-based fixed-function designs provided excellent performance and area efficiency, as well as an easy to use shared-memory programming model that alleviated programmers of explicit cache and data-movement management. Combined with the rapid advancement of integrated circuit design technologies, single-chip shared-memory multicore processors with hardware-based cache coherence became the dominant design.

However, in the decades since the first shared-memory multicore processors emerged, the computing landscape has changed dramatically. Transistor density and power scaling have slowed or collapsed while the diversity and application of computing systems has increased significantly. It is no longer clear whether design decisions adopted by and retained from early multiprocessor designs are correct today.

This dissertation revisits the topic of programmable cache coherence engines in the context of modern shared-memory multicore processors. First, the open-source \bedrock{} cache coherence protocol is described. \bedrock{} employs the canonical MOESIF coherence states and reduces implementation burden by eliminating transient coherence states from the protocol. The protocol's design complexity, concurrency, and verification effort are analyzed and compared to a canonical directory-based invalidate coherence protocol. Second, the architecture and microarchitecture of three separate cache coherence directories implementing the \bedrock{} protocol within the \blackparrot{} 64-bit RISC-V multicore processor, collectively called \blackparrot{}-\bedrock{} (\bpbedrock{}), are described. A fixed-function coherence directory engine implementation provides a baseline design for performance and area comparisons. A microcode-programmable coherence directory implementation demonstrates the feasibility of implementing a programmable coherence engine capable of maintaining sufficient protocol processing performance. A hybrid fixed-function and programmable coherence directory blends the protocol processing performance of the fixed-function design with the programmable flexibility of the microcode-programmable design. Commentary and analysis are provided to illuminate the practical architectural and microarchitectural design and implementation challenges of cache coherence systems, both with and without programmability. All three designs are available open-source, providing researchers with an easy-to-use platform for further investigation. Collectively, the \bedrock{} coherence protocol and its three \bpbedrock{} implementations demonstrate the feasibility and challenges of including programmable logic within the coherence system of modern shared-memory multicore processors, paving the way for future research into the application- and system-level benefits of programmable coherence engines.
}

\clearpage

% reset page style
\pagestyle{plain}
\pagenumbering{roman}
\setcounter{page}{1}

\tableofcontents
\listoffigures
\listoftables
\printglossary[type=\acronymtype,nonumberlist]
\glsaddall

\chapter*{Acknowledgements}

Completing a Ph.D. is only achievable through a combination of individual and communal effort. My time at the Paul G. Allen School of Computer Science \& Engineering has been no different, and I am fortunate to have been supported in a myriad of ways by a large community of people - advisors, lab mates, students, teaching assistants, research assistants, employers, managers, friends, and family.

First, I thank my advisor Mark Oskin for his unwavering guidance, encouragement, and support throughout my studies. Little could I have imagined, over a decade ago as an undergraduate student in one of his courses, that I would spend so many years working with him. Mark taught me that successful research is a combination of technical effort and human compassion; without either one it is not possible to be successful. I also extend sincere thanks to the other faculty from whom I have received guidance and support. Michael Taylor provided many years of research funding and guidance and always reminded me to focus on the details. Ajay Joshi was a valued collaborator on BlackParrot and graciously agreed to serve on my dissertation committee. Luis Ceze advised my M.S. studies, welcomed me with open arms to the computer architecture group at UW CSE, and brought an unending supply of enthusiasm and excitement about our shared field of study.

I am grateful to have been part of the Allen School community during my time at the University of Washington. The faculty, staff, and students of the Allen School work everyday to create a truly special community and environment for learning and research. The collective efforts of this community have preserved its best qualities, even as it grew from a humble department into a full-fledged school. I express special thanks to Elise deGoede Dorough, who has shepherded countless graduate students, including myself, from our first orientation through hooding. I also extend special thanks to the members of the Allen School's computer science education community, in particular Justin Hsia, Ruth Anderson, and Brett Wortzman, for encouraging and supporting my passion for teaching and learning. I thank all of the students I have had a privilege of instructing in lecture halls, study sessions, and labs - I learned far more from you than you did from me.

The students of Sampa and BSG, who I am proud to call my friends and colleagues, made my experience during graduate school much more enjoyable, even during the most difficult times. I am grateful for all of the shared meals, beers, outings, commiseration, laughs, memories, and research discussions. I had the privilege of working with Thierry Moreau on my first real research project and thank him for his enthusiasm and early guidance. I extend deep thanks to Dan Petrisko, Farzam Gilani, Paul Gao, Yuan-Mao Chueh, Dai Cheol Jung, and Scott Davidson for their insights and contributions to the work described in this dissertation and on the \blackparrot{} project. I am grateful to James Bornholt, Meghan Cowan, Emily Furst, Brandon Holt, Vincent Lee, Ming Liu, Liang Luo, Amrita Mazumdar, Carlo del Mundo, Brandon Myers, Jacob Nelson, Adrian Sampson, Gus Smith, Luis Vega, Max Willsey, and Eddie Yan for more great memories than can be listed.

My Ph.D. studies have been book-ended by employment opportunities with the Research and Advanced Development group at Advanced Micro Devices, Inc. To my colleagues, supervisors, and managers, past and present, thank you for exposing me to industrial research and for the flexibility to conclude my studies while employed.

% Family
I extend heartfelt gratitude to my family for their continual love and support throughout my studies. To my parents, thank you for providing me with the opportunity and privilege to attend college and encouraging me to always reach higher and keep going, even when the path was not clear. Brian and Bekah, thank you for taking me in when I moved back to Seattle, encouraging my culinary exploits, and more game nights than I can count.

% Erin
Lastly, I express my deepest gratitude to my spouse and partner-in-life, Dr.~Erin McAuliffe. I feel truly fortunate to have found and married someone who shares my love for learning, has compassion for others, believes in the good of public service, and works every day to create a world that is fair, just, and egalitarian. I will also forever be indebted to Erin for her support, encouragement, and love. Whether we were in the same city or thousands of miles apart, Erin always made me her priority and always put us first. I am especially grateful to have shared the experience of doctoral studies with Erin, and to have a partner who deeply and fully understands the commitment, desire, effort, and sacrifice required to complete a Ph.D., regardless of the field of study. Thank you for standing by my side, through all of the ups and downs, achievements and challenges. I am also particularly grateful for the much needed variety that Erin has added to my life, and for tolerating my quirks, hobbies, and love of video games. Erin has encouraged my cooking experiments and tolerated my accumulation of far too many kitchen gadgets, shared in more board game nights than I can count, led me on countless adventures, both near and far, and given me space to relax with a long video game session when it was desperately needed. I never imagined that I would meet someone who truly understands me as well as she does. I also have Erin to thank for adding our two loving and entertaining cats, Cali and Gemini, to our family; my quality of life has certainly improved with them around the house. Finally, I owe an incalculable debt to Erin for her selflessness and sacrifices as I worked on my research, especially in the closing months while I completed this dissertation. Erin, I would never have reached this point without you. Thank you with all my heart.

\clearpage
\pagestyle{empty}
\cleardoublepage
\pagestyle{plain}
\clearpage
\pagenumbering{arabic}

%% Dissertation Contents
\chapter{Introduction}
\label{chap:introduction}

Computer processor architecture has historically been driven by three primary factors: power, performance, and area (PPA). However, as computers proliferate through every aspect of life, the programmability, security, and adaptability of computer systems have also become first-class design considerations. Across varied domains, shared-memory multicore processors emerged as the dominant processor architecture underlying nearly all general-purpose computer systems in use today. Furthermore, these processor designs are employed with increasing frequency in domains traditionally reliant on less complex processors, rapidly overtaking microcontroller-based systems.

Regardless of domain, the architecture of shared-memory multicore processors has largely remained constant since the earliest multicore processors were introduced commercially about two decades ago. This dissertation revisits one of the design choices employed nearly universally in shared-memory multicore processors: the use of hardware-based cache coherence mechanisms. In the rest of this chapter, \autoref{sec:introduction-motivation} motivates revisiting the design choice of hardware-based cache coherence, \autoref{sec:introduction-thesis-overview} describes the primary questions addressed by this research, \autoref{sec:introduction-outline} outlines the contents of this dissertation, and \autoref{sec:introduction-pubs} briefly lists published materials derived from the research described in this dissertation.

\section{Motivation}
\label{sec:introduction-motivation}

As computer systems spread into all areas of modern life, the performance and adaptability of computer processors becomes increasingly important. The rapid adoption and application of computing systems in diverse domains ranging from mobile consumer devices to datacenters, supercomputers, industrial systems, healthcare, and autonomous vehicles combined with the explosive growth of edge computing, artificial intelligence (AI), and machine learning (ML) has brought about a period of extreme demand for computer processors. Across domains and applications, the processor architecture of choice for the majority of computer systems remains shared-memory multicore processors, whether as the primary computational element or as the host processor for powerful domain-specific accelerators. However, as new system demands emerge, driven by emerging applications and the novel use of computing systems in new environments and domains, it is not entirely clear whether existing systems and processor architectures will be capable of meeting those demands. The research described in this dissertation is motivated by a confluence of factors driving change in computing systems, including rapidly changing application and domain demands, the rise of open-source computer architectures and hardware implementations, and the breakdown of fundamental technology scaling laws that have drastically altered computer architecture's design scaling assumptions.

First, emerging applications and the application of computing systems across an ever growing set of domains is driving the need for computer processors to become both more adaptable and specialized. Artificial intelligence and machine learning applications demand novel computational hardware capable of implementing the key algorithms with greater power and performance efficiency than general purpose processors can provide. At the system level, this places additional demands on the flexibility and adaptability of existing processor architectures that are now taking on the role of host processors to domain-specific accelerators. The emergence of computing systems across a wide variety of domains such as healthcare, industrial, automotive, and edge computing further expands the set of features and functionality that computing systems must provide, and that traditional processor architectures must accommodate, integrate, or interface with. The research in this dissertation investigates how to improve the adaptability and flexibility of the cache coherence subsystem found in modern shared-memory multicore processors.

Second, the rapid growth of open-source hardware and processor architectures, driven largely by the introduction of the RISC-V Instruction Set Architecture (ISA) \cite{asanovic_riscv}, has democratized computer processor design. No longer limited to using closed source ISAs, computer architects are able to leverage the RISC-V ISA to design and implement novel processors for any domain. The RISC-V ISA provides an extensible architecture, allowing architects to introduce domain-specific instructions or interfaces directly into the architecture in a structured manner, preserving comparability with general-purpose software infrastructure and tools while enabling application-specific, high-performance functionality. A key contribution of this dissertation is the design and implementation of a fully open-source RISC-V shared-memory multicore processor that researchers and computer designers can use to build systems of the future.

Third, over the past two decades the fundamental technology scaling laws relied upon by computer architects have stalled or broken down. Moore's Law \cite{moore_1965, moore_1998}, which effectively provided architects with an increasing number of transistors to implement ever complex designs has stalled and slowed significantly. Dennard scaling \cite{dennard_1974}, which stated that the power consumed by transistors decreased as they became smaller, has failed, resulting in decreased power- and energy-efficiency in processor designs. Collectively, the breakdown of these laws drove computer processor designers to shift from creating faster single-core designs to constructing single-chip multicore processors and from focusing solely on general-purpose CPU architectures to leveraging domain-specific accelerator processors and system heterogeneity. However, since the shift to multicore processors occurred, their design has largely remain unchanged.

Prior to the introduction of single-chip shared-memory multicore processors, computer architects developed multi-processor systems comprising multiple processor chips connected via communication networks and having access to shared global memory. System architects explored the use of both message passing and shared memory mechanisms to create large scale systems capable of executing parallel and concurrent programs. Shared-memory systems began relying on cache coherence mechanisms, and some early systems investigated the use of programmability to support both message passing and shared memory architectures on the same system, based on application needs. However, most system architectures relied on either message passing or shared memory with shared-memory systems being implemented primarily on top of hardware-based cache coherence mechanisms. When the shift from single-core to multicore processors occurred in the early 2000s, these newly developed shared-memory multicore processors adopted the use of hardware-based cache coherence mechanisms underneath the shared-memory system.

Today, shared-memory multicore processors rely nearly universally on fixed-function hardware-based cache coherence systems. These systems demonstrate strong protocol processing performance and the use of cache coherence to provide shared memory in multicore processors is unlikely to change in the foreseeable future \cite{martin_why_2012}. However, hardware-based cache coherence systems are inflexible and unable to rapidly adapt to emerging application- or system-specific demands. As shared-memory multicore processors, and computing systems in general, are employed in a growing set of domains, the ability to adapt to domain-specific needs becomes more important. While programmability within the cache coherence and shared-memory systems was explored in the past, these investigations occurred when the computing and technology landscape was significantly different than the one that exists today. Therefore, this dissertation revisits programmability within the cache coherence system of modern shared-memory single-chip multicore processors.

\section{Thesis Overview}
\label{sec:introduction-thesis-overview}

This dissertation hypothesizes that programmability can be be introduced to the cache coherence system without introducing significant performance or area overheads in modern shared-memory multicore processors. Given the significant changes in the computing landscape since the topic was last investigated in depth, the widespread deployment of computing systems throughout all aspects of society and industry, and emerging trends demanding more flexible and adaptable systems, it is an apt time to revisit programmable coherence engines. Addressing this hypothesis, the research described herein investigates the design, architecture, implementation, and trade-offs of a programmable coherence directory controller, using a bottom-up, architecture-first approach guided by the following questions:

\begin{enumerate}[label=\textbf{Q\arabic*},ref=Q\arabic*]
    \item \label{Q1} What is an appropriate cache coherence protocol for a small- to medium-scale shared-memory multicore processor with efficient in-order processor cores and private data and instruction caches?
    \item \label{Q2} What are the key architectural design decisions required for a programmable coherence directory controller to be performance and area competitive with a hardware-based fixed-function controller?
    \item \label{Q3} Is it possible to design a coherence controller that achieves the flexibility benefits of a programmable controller and the power, performance, and area benefits of a fixed-function controller?
\end{enumerate}

Investigating these three questions led to the development of a complete cache coherence protocol and system implemented within the \blackparrot{} 64-bit RISC-V shared-memory multicore processor. \autoref{chap:bedrock} addresses \ref{Q1}, describing the \bedrock{} coherence protocol, which is a practical, easy to implement coherence protocol for small- to medium-scale shared-memory multicore processors. \bedrock{} emphasizes reducing protocol complexity over maximizing request concurrency and utilizes a space-efficient directory organization. \autoref{chap:bp-bedrock} addresses \ref{Q2} with \bpbedrock{}, a \blackparrot{}-based implementation of the \bedrock{} coherence protocol. \bpbedrock{} provides two implementations of the cache coherence directory: one that is fixed-function hardware and another that is microcode programmable. Applying lessons learned from prior work and new innovations, \bpbedrock{} shows that it is possible to build a programmable coherence engine that is performance competitive with a fixed-function design, with only minimal area overhead. These investigations lead to \ref{Q3}, which is address in \autoref{chap:hybrid}. A hybrid fixed-function and programmable coherence engine design combines the best aspects of the fixed-function (performance) and programmable (flexibility) coherence engines. Collectively, the research described in this dissertation demonstrates the feasibility of cache coherence engines that include programmability and provides researchers with a fully-open source architecture and implementation for use in future research.

\section{Dissertation Outline}
\label{sec:introduction-outline}

The rest of this dissertation is organized as follows.~\autoref{chap:background} contains background discussion on cache coherence, open-source hardware development, and the \blackparrot{} RISC-V processor design.~\autoref{chap:bedrock} describes the design of the \bedrock{} cache coherence protocol.~\autoref{chap:bp-bedrock} details the design and implementation of \bedrock{} within a multicore \blackparrot{} processor design, which is called \blackparrot{-}\bedrock{} (\bpbedrock{}). \bpbedrock{} provides two implementations of the coherence directory: one constructed from fixed-function hardware and another that is microcode programmable.~\autoref{chap:hybrid} explores the integration of the programmable and fixed-function coherence directories into a unified, hybrid coherence engine capable of both efficiently executing the coherence protocol and providing useful programmability to system programmers to implement system-specific functionality.~\autoref{chap:related} describes relevant related work in the areas of cache coherence protocol design, approaches to implementing cache coherence systems, programmability within the coherence system, and open-source hardware and multicore RISC-V processor design. Lastly,~\autoref{chap:conclusion} concludes the dissertation and provides a brief discussion of potential future work.
\section{Published Materials}
\label{sec:introduction-pubs}

The following list provides publications related to this dissertation. Some parts of this dissertation have been published as peer-reviewed, pre-print, or technical papers.

\begin{itemize}
    \item \fullcite{petrisko2020}
    \item \fullcite{bedrock}
    \item \fullcite{wyse2022}
\end{itemize}
\chapter{Background}
\label{chap:background}

In this chapter, brief overviews of shared-memory multicore processor architecture, cache coherence, and the \blackparrot{} multicore processor are provided.~\autoref{sec:background-smp} presents an architectural model of a shared-memory multicore processor that is used as a reference architecture throughout the rest of this dissertation.~\autoref{sec:background-cache-coherence} describes the fundamental problem of cache coherence, the importance of cache coherence protocols, and the relationship between cache coherence and memory consistency.~\autoref{sec:background-blackparrot} concludes the chapter with a description of the \blackparrot{} processor, an open-source processor that can be configured as a shared-memory multicore, within which \bpbedrock{} is implemented.

\section{Shared-Memory Multicore Processors}
\label{sec:background-smp}

\begin{figure}[t]
	\centering
	\includegraphics[width=0.7\linewidth]{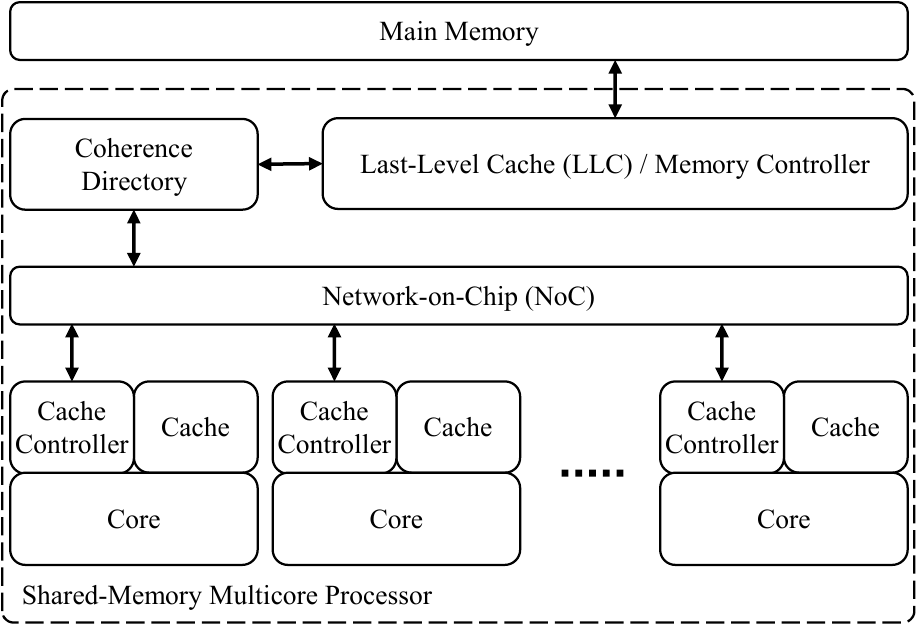}
	\caption{Canonical Shared-Memory Multicore Processor Architecture}
	\label{fig:background-system}
\end{figure}

Driven by the breakdown of transistor density, power, and manufacturing scaling in the early 2000s, computer architects shifted focus from designing increasingly larger and more complex single-core processors to designing tightly-integrated single-chip shared-memory multicore processors. These processors shifted design focus from maximizing Instruction-Level Parallelsim (ILP) and single-core clock frequency to exploiting Thread-Level Parallelism (TLP) via parallel and concurrent algorithms.

A high-level diagram of a canonical shared-memory multicore processor with a directory-based cache coherence system is shown in \autoref{fig:background-system}\footnote{Other shared-memory multicore processor organizations are possible and employed in real systems. The canonical design chosen illustrates the design concepts while remaining representative of small- to medium-scale multicore processor designs.}. This type of multicore comprises two or more processing cores with private caches interconnected by a Network-on-Chip (NoC), which is further connected to the last-level cache (LLC) and main memory by way of the cache coherence directory. As with single-core designs, each core in a multicore processor is capable of performing load, store, or atomic read-modify-write operations to any physical address in the system and has one or more levels of private hardware data caches. The private cache attached to each core services memory requests issued by the core by forwarding those requests to the cache coherence directory. Every level of cache in the system serves to reduce memory access latency and increase effective memory bandwidth of the shared main memory.

Multicore processors allow programmers to exploit Thread-Level Parallelism (TLP) through concurrent execution of cooperating threads within a single program address space. Multicores enable efficient execution of concurrent algorithms and cooperating or synchronizing tasks, however they rely on the assumption that every cooperating thread executing on the individual cores always sees the \emph{correct} value when loading any memory location, regardless of how many cores may be accessing the location. However, since multiple hardware threads of execution may exist and execute concurrently, with each processor core having private data caches to improve execution performance and memory access latency, it becomes possible for multiple copies of an arbitrary piece of memory data to exist within the entire system. For example, if two threads of execution running on independent cores within the multicore both access a shared memory location, each core will fetch and store a local copy of that memory location in its hardware cache to reduce the access latency of any successive read or write operations. A situation may arise where both cores then write all or a portion of the bits associated with the memory location or a group of consecutive memory locations that have been cached locally by many caches. Therefore, a mechanism and ordering rules are required to guarantee that any particular read or write of memory data returns the correct value to a thread of execution, accounting for any and all local hardware caching and updates that any given core and thread of execution make to the location.

\section{Cache Coherence}
\label{sec:background-cache-coherence}

Coordinating the read and write access of memory data across multiple processor cores and their private caches is a complex problem that multicore processor architects have been working on for decades. More concretely, defining the semantics of a multicore processor's memory system and the visibility of load and store operations can be described using the two cooperating mechanisms of \emph{memory consistency} and \emph{cache coherence}.

"Memory consistency is a precise, architecturally visible definition of shared memory correctness" \cite{nagarajan_primer_2020}. It defines the semantics for visibility and ordering of memory operations, as observed at the level of programs, instructions, and threads of execution for the shared memory system. Memory consistency defines the allowable ordering of operations across all accesses and all memory locations in the system. A detailed discussion of memory consistency is beyond the scope of this dissertation.

Cache coherence, on the other hand, defines the ordering and outcomes of read and write operations made by more than one processing element to a single memory location in a system employing private data caches that hold local copies of memory data. In many systems, cache coherence defines protocols and mechanisms upon which a full memory consistency model is constructed. Typically, the memory consistency model is visible to the programmer through the hardware-software interface of the machine's Instruction Set Architecture (ISA) while the cache coherence protocol and mechanisms remain completely invisible to programmer's and software. The rest of this section describes the fundamental problem of cache coherence using a two invariant model that is sufficient to fully define the allowable access properties and data value changes for a memory location.

\subsection{The Cache Coherence Problem}

\begin{figure}[t]
	\centering
	\begin{subfigure}[b]{0.3\textwidth}
	    \centering
	    \includegraphics[width=\textwidth]{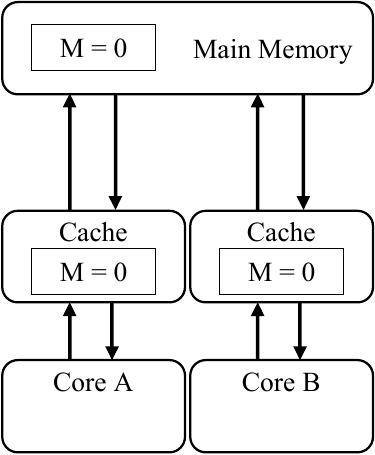}
	    \caption{}
	    \label{fig:coherence-0}
	\end{subfigure}
    \hfill
    \begin{subfigure}[b]{0.3\textwidth}
	    \centering
	    \includegraphics[width=\textwidth]{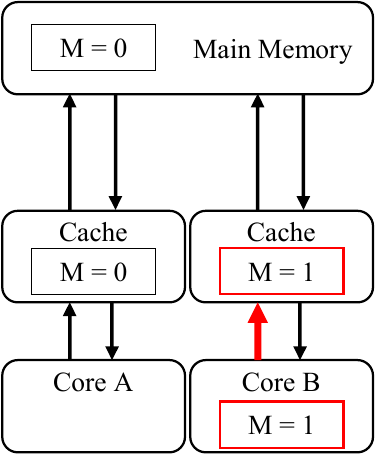}
	    \caption{}
	    \label{fig:coherence-1}
	\end{subfigure}
    \hfill
    \begin{subfigure}[b]{0.3\textwidth}
	    \centering
	    \includegraphics[width=\textwidth]{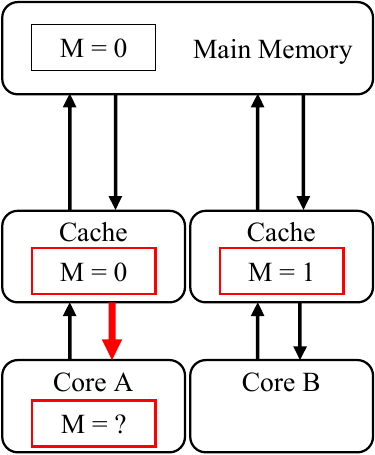}
	    \caption{}
	    \label{fig:coherence-2}
	\end{subfigure}
	\caption{The Cache Coherence Problem}
	\label{fig:coherence-problem}
\end{figure}

The cache coherence problem, as alluded to above, occurs when multiple private copies of a single memory location can exist within a system at the same time. In a shared-memory multicore processor with private local caches per core this is easily possible. \autoref{fig:coherence-problem} depicts the cache coherence problem. As above, consider two independent cores, core A and core B, executing programs that access a single shared memory location M. Initially, assume M has a value of 0 and both core A and core B have loaded location M, creating a copy of the location in each core's private data cache, as shown in \autoref{fig:coherence-0}. Caches reduce memory access latency by storing copies of memory data closer to the processor core, and if a memory location is cached then a load or store operation only needs to manipulate the value of the location in the cache storage rather than incurring the high latency cost of a round trip access to memory. Next, assume core B modifies its cached copy of location M to have a value of 1, as shown in \autoref{fig:coherence-1}. Since core B's cache has a copy of location M, the store operation from the core hits in the cache and updates only the copy of location M in core B's cache. Lastly, as shown in \autoref{fig:coherence-2}, core A executes a load operation for location M. When this occurs, the question is what value should core A receive in response to its load operation?

This example illustrates the fundamental problem of incoherence. Whenever multiple agents, in this case caches, can access a shared memory, there may exist multiple copies of memory locations stored in different parts of the system, and those copies can easily be manipulated such that the stored value is no longer the same across all copies. In the example above, determining the value returned by core A's load of location M requires understanding how changes to the value of a shared memory location propagate from cache to cache or between memory and caches. Intuitively, a programmer likely expects that the load from core A will return a value of 1, which is the most recently written value to the location M, even though that write was performed by core B. Managing incoherence and solving the cache coherence problem requires defining the semantics, ordering, and observed values for load and store operations to a single shared memory location that can be cached in multiple locations throughout the system.

\subsection{Cache Coherence Defined}

Defining \emph{cache coherence} is an important step when implementing a shared-memory multicore processor. Informally, "a memory scheme is coherent if the value returned on a LOAD instruction is always the value given by the latest STORE instruction with the same address"~\cite{censier_new_1978}. This definition from Censier and Feautrier, while informal, is easy to understand for most programmers and architects. However, it does not precisely define the allowable ordering or observable values of a memory location as multiple agents (cores, caches) operate on the location.

More formally, cache coherence can be defined using two invariants that specify the allowable read and write access permissions any agent has for a memory location and how the observable value of the memory location changes as access permissions change \cite{nagarajan_primer_2020}.

\subsubsection{Single Writer Multiple Reader Invariant}

\begin{figure}[t]
	\centering
	\begin{subfigure}[b]{0.3\textwidth}
	    \centering
	    \includegraphics[width=\textwidth]{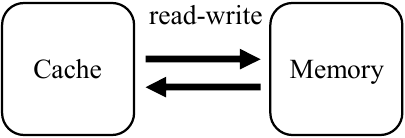}
	    \caption{Single Writer}
	    \label{fig:background-swmr-rw}
	\end{subfigure}
    \hspace{1cm}
    \begin{subfigure}[b]{0.5\textwidth}
	    \centering
	    \includegraphics[width=\textwidth]{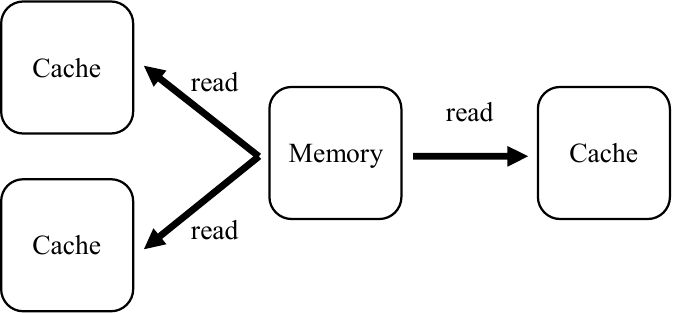}
	    \caption{Multiple Readers}
	    \label{fig:background-swmr-ro}
	\end{subfigure}
	\caption{Cache Coherence Single Writer Multiple Reader (SWMR) Invariant}
	\label{fig:background-swmr}
\end{figure}

The \emph{Single Writer Multiple Reader (SWMR)} Invariant, depicted in \autoref{fig:background-swmr}, states that at any given time for a particular memory location either exactly one agent may have read and write permissions for the location or one or more agents may have read-only permissions for the location. This rule ensures that at any given time during execution, there is always at most one agent that is allowed to modify the value of a particular memory location. Further, that modification can only occur at times when no other agent can access the location. In practice, this rule implies that if a core and its cache have write permissions for a memory location then that memory location will be cached only in that core's cache and no other cached copies of the data will exist in the system being managed by the coherence protocol. Likewise, memory locations that are cached in more than one private cache are guaranteed to be in a read-only state and the value of every copy of the location is guaranteed to be identical.

\subsubsection{Data Value Invariant}

\begin{figure}[t]
	\centering
	\includegraphics[width=0.9\linewidth]{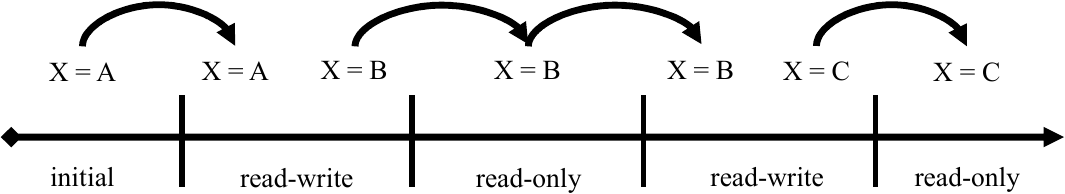}
	\caption{Cache Coherence Data Value Invariant}
	\label{fig:background-data-value-invariant}
\end{figure}

The \emph{Data Value} Invariant, depicted in \autoref{fig:background-data-value-invariant} defines how a memory location's data value changes are observed in relation to read-only and read-write epochs during execution. The SWMR invariant effectively defines two types of epochs, one that is read-only (Multiple Readers) and one that is read-write (Single Writer). The data value invariant then says that starting from some initial system state, the observable value of a memory location at the start of any epoch is equivalent to the value of that memory location at the end of its last read-write epoch. Consecutive read-write epochs are possible when two agents perform writes to the memory location one after the other. The transfer of write-permissions from the first agent to the second demarcates the end of the first agent's read-write epoch and the start of the second agent's read-write epoch.

Informally, the data value invariant simply says that the value of a memory location cannot invisibly change between epochs. Further, there is a precise, serialized ordering of epochs that is observed during execution. The realized ordering of epochs depends on the actual timing of execution across the multiple threads of execution and processor cores, so while there may be many possible epoch orderings, every allowable ordering obeys the data value invariant and there is a well-defined transfer of values between epochs.

\section{\blackparrot{}}
\label{sec:background-blackparrot}

\blackparrot{}~\cite{petrisko2020} is an open-source, industrial-strength 64-bit RISC-V multicore processor that aims to become the default open-source Linux-capable RISC-V multicore used by the world. \blackparrot{} features a modular, tiled design, and a \blackparrot{} multicore processor instance is composed by selecting the appropriate number of each type of available tile and connecting them with appropriate networks. \blackparrot{} multicore processors implement the \bedrock{} cache coherence protocol (\autoref{chap:bedrock})~\cite{bedrock} to maintain cache coherence throughout the cacheable memory regions defined by in system. In this section, a brief overview of \blackparrot{} is provided, describing the types of \blackparrot{} tiles, the interconnection networks used between them, the \blackparrot{} address space, and the cache engine interface used between coherent caches and cache controllers. This section's overview focuses on the design aspects relevant to the \bpbedrock{} coherence system implementation. Additional details on the \blackparrot{} design can be found in the documentation and code at~\url{https://github.com/black-parrot/black-parrot}\footnote{The open-source \blackparrot{} code repository maintains the most up-to-date implementation of the processor. In the event of differences between this document and the published code, the code takes precedence.}.

\begin{figure}[t]
	\centering
	\includegraphics[width=0.7\linewidth]{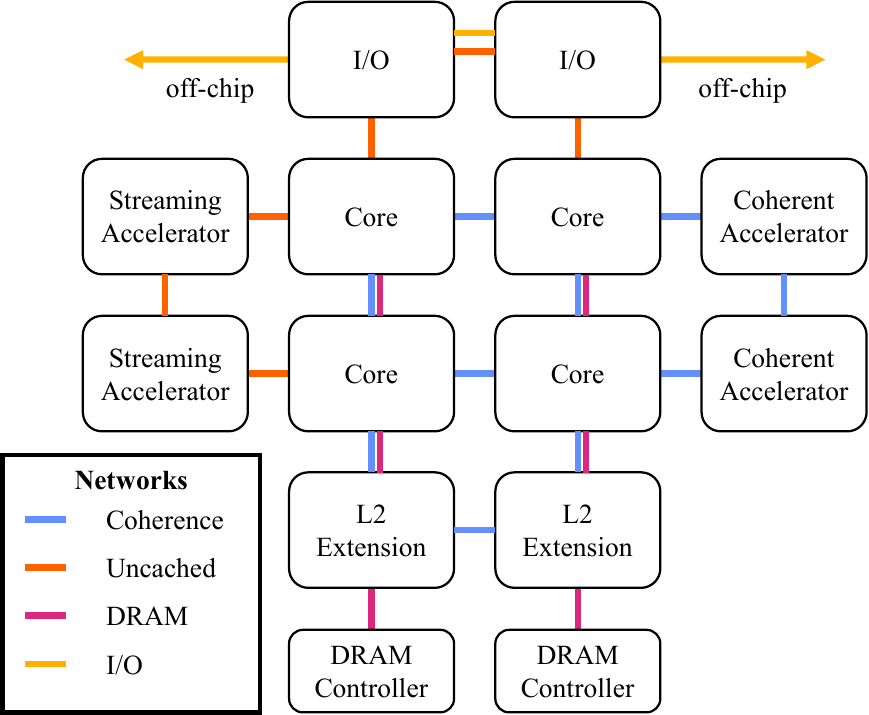}
	\caption{\blackparrot{} Tiled Multicore Processor}
	\label{fig:blackparrot-multicore}
\end{figure}

\autoref{fig:blackparrot-multicore} depicts a canonical \blackparrot{} multicore processor design. At the heart of the multicore is an array of \blackparrot{} core tiles, called the core complex. The number of core tiles in the core complex is configurable and supports most common organizations. The remaining tile types surround the core complex in 1-dimensional complexes. Along the North side of the core complex are the I/O tiles, which connect off-chip to the East and West. Accelerator tiles are arranged along the West and East of the core complex. To the South of the core complex are the memory controllers and optional L2 extension tiles.

\subsection{\blackparrot{} Multicore Tiles}
\label{sec:background-blackparrot-mc}

\begin{figure}[ht]
	\centering
	\begin{subfigure}[b]{0.3\textwidth}
	    \centering
	    \includegraphics[width=\textwidth]{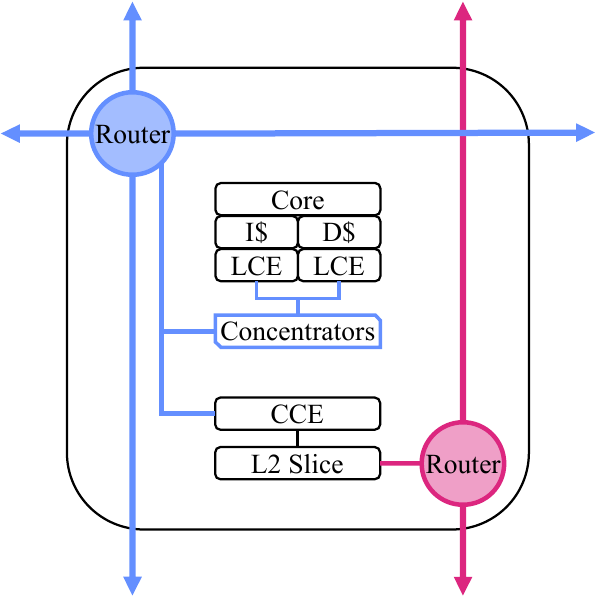}
	    \caption{\blackparrot{} Core Tile}
	    \label{fig:blackparrot-core-tile}
	\end{subfigure}
	\hfill
	\begin{subfigure}[b]{0.3\textwidth}
	    \centering
	    \includegraphics[width=\textwidth]{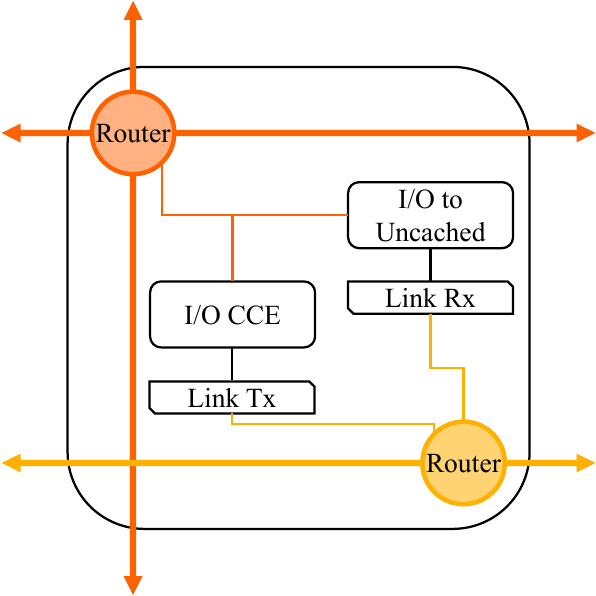}
	    \caption{\blackparrot{} I/O Tile}
	    \label{fig:blackparrot-io-tile}
	\end{subfigure}
	\hfill
	\begin{subfigure}[b]{0.3\textwidth}
	    \centering
	    \includegraphics[width=\textwidth]{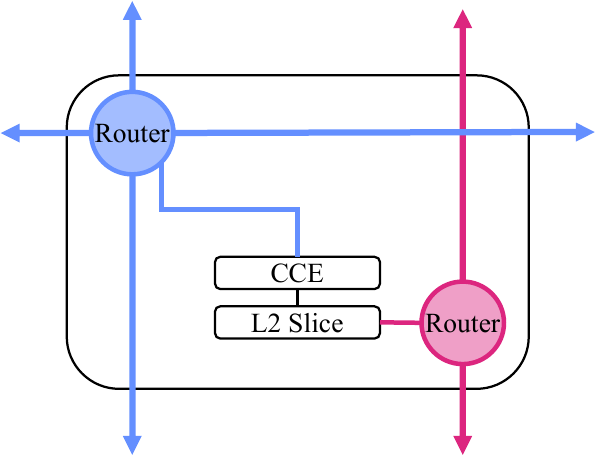}
	    \caption{\blackparrot{} L2 Tile}
	    \label{fig:blackparrot-l2ext-tile}
	\end{subfigure}
	\caption{\blackparrot{} Multicore Tiles}
	\label{fig:blackparrot-tiles}
\end{figure}

\autoref{fig:blackparrot-tiles} provides a detailed view of the core, I/O, and L2 extension tiles. There are four different networks that connect the various tiles: Coherence, Uncached, DRAM, and I/O. The type of network required for each tile is determined by the functionality of the tile and the type of memory operations it handles. Note that the coloring of the networks in \autoref{fig:blackparrot-tiles} matches that of \autoref{fig:blackparrot-multicore}.

\subsubsection{Coherence and Uncached Networks}

The Coherence network implements the full \bedrock{} coherence protocol and its Request, Command, Fill, and Response networks. These networks are 2-D mesh networks with point-to-point ordering and Y-X dimension-ordered routing. Each of the four \bedrock{} coherence protocol networks is carried on a separate physical network. The width of the networks is parameterizable, with 64-, 128-, and 256-bits being common widths. This network runs between the core, L2 extension, and coherent accelerator tiles.

The Uncached network is a subset-extension of the Coherence network comprising only the Request and Command \bedrock{} networks from the Coherence network mesh that are extended out to the  I/O and streaming accelerator tiles. These tiles only support uncached load and store operations, and therefore do not utilize the Fill or Response \bedrock{} networks.

\subsubsection{DRAM Network}

The DRAM network supports both cacheable and uncacheable memory load and store operations. It is constructed as a set of vertical physical networks that run through the core complex columns and then South through the optional L2 extension tile and into the memory controller. The width of all DRAM networks is parameterizable, with 64-, 128-, and 256-bits being common widths. Memory accesses are only issued by the coherence directories (CCEs) found in the core and L2 extension tiles. A memory access request may originate in a core, accelerator, or off-chip via an I/O tile, and requests are routed through the Coherence and Uncached networks to a CCE, which then issues the operation to memory over the DRAM network.

\subsubsection{I/O Network}

The I/O network is a horizontal network that connects a \blackparrot{} multicore to the external world. It supports uncached operations. The I/O network is bi-directional and supports both outbound and inbound paired command/response channels. Typically, each channel is implemented as its own physical network. The width of all I/O network channels is parameterizable, with 64-, 128-, or 256-bits being common widths.

\subsubsection{Core Tile}

\autoref{fig:blackparrot-core-tile} shows the contents of a \blackparrot{} core tile. Each tile comprises a single \blackparrot{} processor core with private L1 instruction and data caches, cache controllers (LCE) attached to each L1 cache, a coherence directory (CCE), a slice of L2 the distributed L2 cache, and connections to the Coherence and Memory networks. The Coherence network is a 2-D network that spans the core complex and reaches out to coherent accelerator and L2 extension tiles. The Memory network is a 1-D vertical network that connects each column of core tiles to a memory controller at the South border of the multicore.

The \blackparrot{} L1 caches maintain cache block state and data using three distinct memories, called the data, tag, and stat memories. The data memory stores the cache block data. The tag memory stores the address tags and coherence state of for every cache block. The stat memory maintains replacement metadata, such as pseudo-LRU bits, for every cache set and dirty bits for every cache block. The L1 caches have parameterizable cache block width, associativity, sets, and fill widths. The fill width is a multiple of the cache's internal SRAM bank width and determines the width of writes to the cache's data memory. If the fill width is smaller than the cache block width, writing a complete cache block into the data memory takes multiple cycles.

\blackparrot{}'s L1 caches are virtually-indexed, physically-tagged (VIPT). Caches may be 2-, 4-, or 8-way associative and have a block size of 64, 128, 256, or 512 bits. The private L1 caches rely on banked SRAMs to store data. The width of each bank must be at least 64-bits and is computed as the cache block width divided by associativity. The cache fill width, or the width that data can be supplied to the cache must be a multiple of the cache bank width. \blackparrot{} implements the RISC-V Sv39 virtual-memory system with 4 KiB memory pages. There are therefore 12-bits available for the L1 cache block byte offset and cache set index bits, and it is typically assumed that exactly 12-bits are used for these two fields. The number of cache sets and block size are closely related and determined by the formula: $\log_{2}(4 KiB) = \log_{2}(sets) + \log_{2}(block size)$. The number of cache sets must be a power-of-two. The default \blackparrot{} L1 cache organization is 64-sets, 8-way associative, 512-bit blocks, with a total capacity of 32 KiB.

The \blackparrot{} L1 caches are blocking and support one request at a time for all cacheable requests and uncacheable loads. The L1 caches are non-blocking for uncacheable stores, and the number of outstanding uncacheable stores is limited by a request credit counter in the LCE. The L1 cache is capable of executing atomic read-modify-write operations on 32- or 64-bit data words that are cached in a valid block with read-write permissions. The L1 cache may also issue atomic read-modify-write operations to the LCE, which are treated similar to uncacheable requests by the LCE.

The Coherence network instantiates each of the four \bedrock{} coherence networks: Request, Command, Fill, and Response. Each of the four coherence networks is carried on a separate physical network. As explained in the \bedrock{} coherence protocol specification~\cite{bedrock}, only the Fill network is bi-directional. This means that the Request, Command, and Response network input and output links connect either to the CCE or the LCE concentrators. The concentrators are used to to multiplex the coherence network connections to the two LCEs. The CCE connects directly to the Request input, Command output, and Response input coherence networks. The Fill network connects to a bi-directional concentrator since the LCEs must both send and receive Fill messages.

The L2 cache is non-inclusive of the L1 caches, and logically acts as a memory-side DRAM buffer. The coherence directory (CCE) manages a subset of physical memory, and the L2 cache only stores blocks from the same subset or slice of physical memory. An address hash or swizzle is used to efficiently utilize the entire L2 cache regardless of memory access patterns. The L2 cache slice connects to the Memory network that runs vertically through the core complex to access DRAM.

\subsubsection{I/O Tile}

\autoref{fig:blackparrot-io-tile} details the \blackparrot{} I/O tile. Each I/O tile acts as a bridge between the I/O network that runs East and West at the top of the multicore to provide off-chip connections with a 2-D Uncached network that interfaces with the core complex. The Uncached network is simply a subset of the Coherence network that only supports uncached requests. In \blackparrot{}, all of the I/O address space is considered uncached, however it is possible for an off-chip device to perform uncached access to \blackparrot{}'s DRAM address space. In this situation, software is responsible for maintaining coherence between the off-chip and DRAM domains. In practice, off-chip I/O access to cacheable DRAM addresses is used to load software or data for execution into \blackparrot{}'s DRAM prior to the processor cores beginning execution.

Each I/O tile includes an I/O CCE module that converts uncacheable request messages issued by an LCE over the \bedrock{} coherence/uncached network into \bedrock{} I/O network commands, and then converts the I/O network responses to uncacheable commands that are sent back to the requesting LCE. There is also an external I/O to \bedrock{} network converter that accepts uncacheable memory commands from off-chip sources, converts them to uncacheable requests on the \bedrock{} coherence/uncached network, and then receives the uncachable commands satisfying the requests and converts them back to I/O network response messages for the off-chip device.

\subsubsection{L2 Extension Tile}

If a row of L2 extension tiles are present, the Coherence and Memory networks are extended from the core complex to the L2 tiles.~\autoref{fig:blackparrot-l2ext-tile} depicts the contents of the L2 extension tiles. Each tile contains a slice of the distributed L2 cache and a coherence directory (CCE) that manages that slice of physical memory, in the same manner that the L2 and CCE operate in a core tile. The L2 extension tiles are simplified core tiles that omit the \blackparrot{} core, private caches, LCEs, and concentrators. Additionally, there is no need to route the \bedrock{} Fill coherence network since this network is only used to carry messages between LCEs, and does not connect to the CCE.

\subsubsection{Accelerator Tiles}

Accelerator tiles may be streaming or coherent. A streaming tile is capable of performing only uncached memory operations and typically contains an accelerator IP block and a simplified LCE-like module that connects with limited functionality to the \bedrock{} Request and Command coherence networks. A coherent accelerator tile typically contains an accelerator IP, a private L1 cache, and an LCE that participates in the full \bedrock{} coherence protocol.

\subsection{\blackparrot{} Address Spaces}

\begin{figure}[t]
	\centering
	\includegraphics[width=0.9\linewidth]{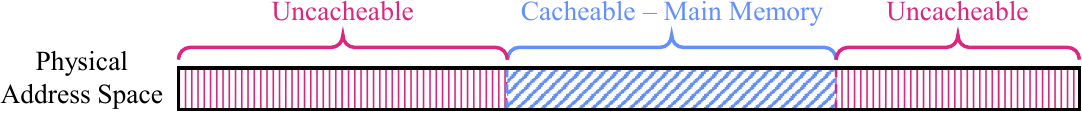}
	\caption{\blackparrot{} Physical Address Space}
	\label{fig:blackparrot-address-space}
\end{figure}

\blackparrot{} divides its address space into cacheable and uncacheable regions.~\autoref{fig:blackparrot-address-space} depicts this address space at a high-level. The address space is divided into three sections: uncacheable local memory, cacheable DRAM, and uncacheable global memory. The default physical address width in \blackparrot{} is 40 bits.

\subsubsection{Uncacheable Local Memory}
Uncacheable local memory is a 2 GiB region of memory, starting at address 0, that contains mappings for devices found on \blackparrot{} tiles. These include bootroms, configuration links, and the Core Local Interrupt Controller (CLINT). This region supports only uncacheable accesses and is not managed by the \bedrock{} coherence protocol. The \blackparrot{} Platform Guide found in the \blackparrot{} documentation provides a more detailed breakdown of the local memory region.

\subsubsection{Cacheable Global Memory}
Cacheable global memory is a 2 GiB region of of memory, starting at address \lstinline{0x00_8000_0000}, that maps to a cacheable portion of DRAM. This region of memory is managed using the \bedrock{} cache coherence protocol, and is striped by cache line across the L2 cache slices and coherence directories of the core and L2 extension tiles. The region supports both cacheable and uncacheable accesses, and all accesses are serialized at the coherence directories. Performing an uncacheable access to the region results in the targeted cache block being recalled and written back, if dirty, from all caches that contain a valid copy of the block, prior to the request being issued to memory.

\subsubsection{Uncacheable Global Memory}
Uncacheable global memory occupies the remainder of the address space, beginning at address \lstinline{0x01_0000_0000}. These addresses may map to uncacheable DRAM or off-chip memory.

\subsection{\blackparrot{} Cache Engine Interface}
\label{sec:background-blackparrot-cache-engine-if}

\begin{figure}[t]
	\centering
	\begin{subfigure}[b]{0.45\textwidth}
	    \centering
	    \includegraphics[width=\textwidth]{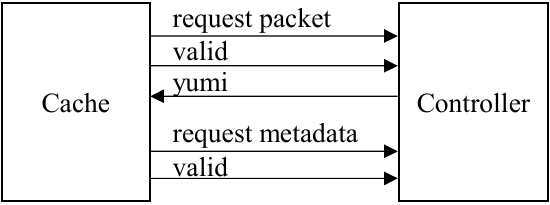}
	    \caption{Request Interface}
	    \label{fig:blackparrot-cache-engine-req-if}
	\end{subfigure}
	\hfill
	\begin{subfigure}[b]{0.45\textwidth}
	    \centering
	    \includegraphics[width=\textwidth]{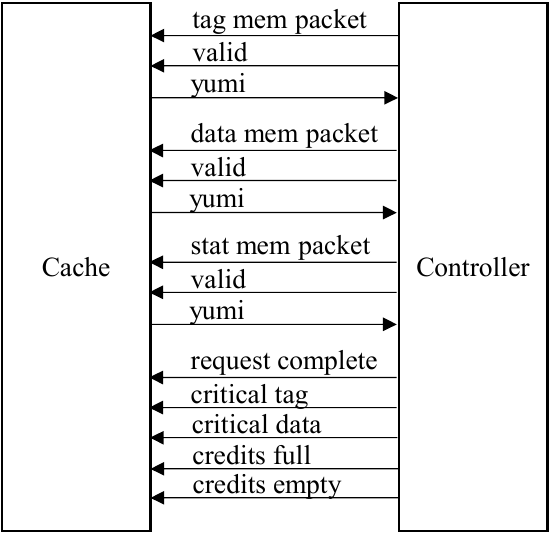}
	    \caption{Fill Interface}
	    \label{fig:blackparrot-cache-engine-fill-if}
	\end{subfigure}
	\caption{\blackparrot{} Cache Engine Interface}
	\label{fig:blackparrot-cache-engine-if}
\end{figure}

\blackparrot{} employs a flexible cache engine interface that connects each cache to an attached cache engine or controller. \autoref{fig:blackparrot-cache-engine-if} shows the cache engine interface found in \blackparrot{}. The interface comprises an cache to controller request interface and a controller to cache fill interface. The cache engines are responsible for servicing misses, invalidations, and coherence transactions. The interface is latency insensitive and supports both coherent and non-coherent caches. In a \bpbedrock{} multicore, the cache engine interface connects the private, coherent L1 caches of each core with a dedicated Local Cache Engine (LCE). The LCE services all cache requests and maintains cache coherence for the attached cache by participating in the \bpbedrock{} coherence system.~\autoref{sec:bp-bedrock-lce} provides a detailed overview of the \bpbedrock{} LCE implementation.

\subsubsection{Cache Request and Metadata Interface}

The cache engine request interface, shown in \autoref{fig:blackparrot-cache-engine-req-if}, carries cache miss requests and associated metadata from the cache to the controller. In \bpbedrock{}, the cache can issue cacheable and uncacheable load and store miss requests, and atomic read-modify-write requests. A request includes the operation type, address, size, data for uncacheable stores and atomic operations, and a hit bit to indicate if the target cache block is cached in a valid state by the cache. The request packet utilizes a valid-then-yumi handshake\footnote{valid-then-yumi is also called valid-then-ready in \cite{taylor2018}}. The cache presents a request packet and raises the \lstinline{valid} signal when it has a new cache request for the controller. The controller consumes the packet by raising the \lstinline{yumi} signal, which depends on the \lstinline{valid} input signal.

Request metadata is provided to the controller using a valid-only handshake in the cycle following the request packet handshake. The controller must be ready to accept the metadata information in the cycle following its consumption of the request packet. The request metadata includes whether the cache block including the request address is dirty and, depending on context, a way ID to use as either the replacement way or the way of a cache block hit for the request address.

\subsubsection{Cache Fill Interface}

\autoref{fig:blackparrot-cache-engine-fill-if} shows the cache fill interface that allows the controller to read and write the cache's data, stat, and tag memories while servicing cache requests. The cache fill interface also includes request completion and credit signals to help with flow control and critical word first behavior.

In general, a cache request may result in many transactions on the fill interface as the controller reads and writes the cache's memories. The cache fill interface comprises three valid-then-yumi interfaces to access the data, stat, and tag memories, and a set of request completion signals. Each of the cache's three memories has a separate, independently operating, interface, which enables flexibility in the cache engine implementation. Each of these interfaces supports read and write operations so the controller can examine and update the current state of the cache.

Three request completion signals are used by the controller to indicate when important phases of a cache request transaction complete. A request complete signal is raised to indicate that the request has been fully completed. The critical tag and data signals support critical word first behavior for cache misses. Each is raised for a single cycle when the critical data word and its associated tag are written to the cache's data and tag memories, respectively. Two credit signals are provided by the cache controller to help with flow control and hazard detection.

\chapter{\bedrock{}}
\label{chap:bedrock}

The \bedrock{} Cache Coherence Protocol defines a family of cache coherence protocols and the system components required to implement a specific coherence protocol. The \bedrock{} protocols are directory-based invalidate protocols using the standard MOESIF coherence states. Protocol variants are defined for the MI, MSI, MESI, MOSI, MOESI, MESIF, and MOESIF state subsets. \bedrock{} relies on a complete coherence directory to precisely track the coherence state of every cache block managed by the coherence system. The coherence directory is the point of serialization for all coherence transactions, and coherence is enforced using the \emph{Single-Writer, Multiple-Reader (SWMR) Invariant} and \emph{Data-Value Invariant}. A \bedrock{} coherence system is constructed from three components: cache controllers (Local Cache Engines), coherence directories (Cache Coherence Engines), and coherence networks. 

The canonical \bedrock{} protocol presented in this chapter is well-suited for small to medium size shared-memory multicore processors. An initial implementation of \bedrock{} within the \blackparrot{} 64-bit RISC-V multicore processor~\cite{petrisko2020} is described in depth in \autoref{chap:bp-bedrock}. Although \bedrock{} has been influenced by the design needs and implementation practicalities of \blackparrot{}, this chapter presents \bedrock{} agnostic to system implementation decisions.

This chapter's description of \bedrock{} assumes readers are familiar with the basics of cache coherence protocols. Nagarajan et al. provide an excellent overview~\cite{nagarajan_primer_2020} for those unfamiliar with the topic; Chapters 2 and 8 cover the basics of cache coherence and directory-based coherence protocols, respectively, and are highly relevant to the following presentation of \bedrock{}. To the extent possible, this chapter adopts the terminology from~\cite{nagarajan_primer_2020} to remain consistent with the majority of published cache coherence literature.

The rest of this chapter is organized as follows.~\autoref{sec:bedrock-system} describes the \bedrock{} coherence system components.~\autoref{sec:bedrock-protocol} presents the family of \bedrock{} cache coherence protocols using both high-level overviews and detailed tabular specifications.~\autoref{sec:bedrock-uc-amo} discusses the incorporation of uncached and atomic accesses within the \bedrock{} protocol.~\autoref{sec:bedrock-verification} describes verification of the MESI variant of \bedrock{} using the CMurphi~\cite{cmurphi} model checker software.~\autoref{sec:bedrock-protocol-analysis} compares \bedrock{} to a canonical directory-based coherence protocol. Additionally, \autoref{app:protocol-lce}, \autoref{app:protocol-cce}, \autoref{app:state-tables-lce}, and \autoref{app:state-tables-cce} provide full listings of the coherence protocol and state transition tables for all defined subsets of \bedrock{}.

\section{System Components} \label{sec:bedrock-system}

\begin{figure}[t]
	\centering
	\includegraphics[width=0.80\linewidth]{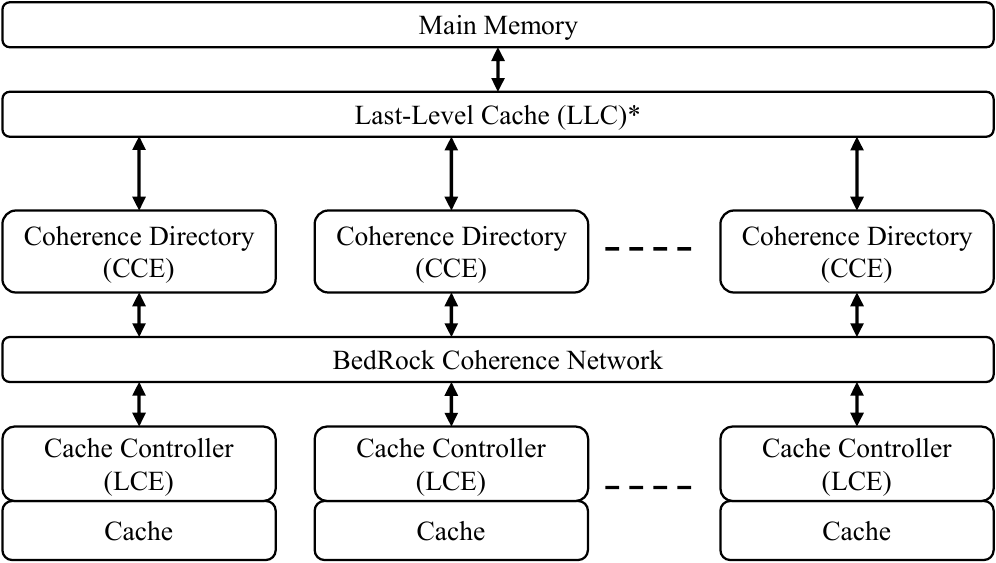}
	\caption{Canonical \bedrock{} Coherence System Organization}
	\label{fig:bedrock-generic}
\end{figure}

\bedrock{} defines both a coherence protocol and the system components required to implement the protocol. This section introduces the three major system components in a \bedrock{} system: cache controllers, coherence directories, and the coherence networks.

\autoref{fig:bedrock-generic} depicts a canonical \bedrock{} coherence system. Each cache controller (LCE) manages a single cache participating in the coherence protocol. Each coherence directory manages a disjoint subset of the physical address space and contains coherence directory storage to track all cached blocks from that subset. The controllers and directories are connected via the \bedrock{} coherence network. The coherence directories also connect to main memory, with an optional (indicated by an asterisk) memory-side, non-inclusive Last-Level Cache (LLC) between the directories and the main memory. The LLC does not participate in the cache coherence protocol, is logically considered to be part of main memory, acting as a memory bandwidth amplifier for the higher-level caches. \bedrock{} places no constraints on the organization of the LLC, but its implementation must provide a block-based access interface consistent with the cache block size of the \bedrock{} system.

\subsection{\bedrock{} Coherence Networks} \label{sec:bedrock-system-network}

\begin{figure}[t]
	\centering
	\includegraphics[width=0.75\linewidth]{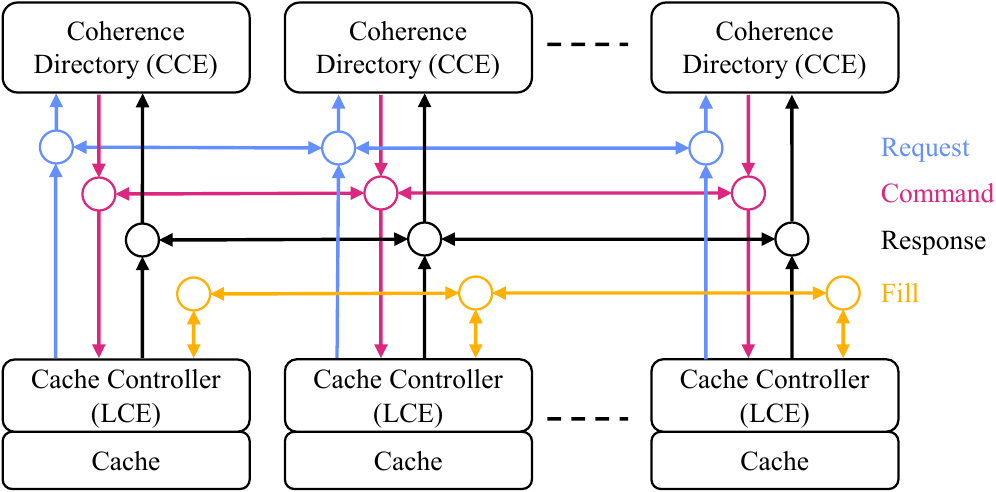}
	\caption{\bedrock{} Coherence Networks}
	\label{fig:bedrock-networks}
\end{figure}

The \bedrock{} Coherence Networks carry coherence protocol messages between the cache controllers and coherence directories. The \bedrock{} coherence protocol comprises four distinct coherence networks to carry Request, Command, Fill, and Response messages.~\autoref{fig:bedrock-networks} depicts the \bedrock{} coherence networks and their connections to the cache controllers and coherence directories. The implementation of these networks is system specific, and independent of the \bedrock{} coherence protocol. The protocol requires each network to provide error-free message delivery and that all networks operate independently from each other. Each network may be a physical or virtual network and totally ordered, point-to-point ordered, or unordered, provided the preceding requirements are satisfied.

\subsubsection{Request Network}

The Request network carries messages from a cache controller to a coherence directory. Coherence requests are initiated when a cache miss occurs due to the cache having insufficient permissions to complete the requested operation. This includes attempting to write to a read-only block and attempting to read or write a block that is not cached (i.e., the cache has no permissions for the block). The Request network may fill up and exert backpressure on the cache controllers. Backpressure may temporarily prevent a cache controller from issuing new requests, but new requests will eventually send as the coherence directory drains and processes older requests from the network. The network has no ordering constraints. All requests are eventually serialized by the network and arrive as a single request stream at each coherence directory.

\subsubsection{Command Network}

The Command network carries coherence commands from the coherence directory to the cache controllers. The Command network has no ordering constraints and only requires that commands are processed in a timely manner after arriving at the cache controller. A cache controller may not delay processing a command in order to send a new coherence request.

\subsubsection{Fill Network}

The Fill network carries cache to cache data transfers between the cache controllers. The Fill network has no ordering constraints and only requires that messages are processed in a timely manner after arriving at the cache controller. A cache controller may not delay processing a fill message in order to send a new coherence request.

\subsubsection{Response Network}

The Response network carries messages from the cache controller to the coherence directory in response to commands issued by the directory or forwarded from another cache controller. Responses include invalidation acknowledgements, writeback responses with or without data, and coherence transaction acknowledgements. The Response network has no ordering constraints. Each response message sent by a cache controller is in response to a single command or fill network message. The coherence directory must process responses in a timely manner to prevent deadlock in the coherence system. The directory must also prioritize processing responses over processing new requests or issuing additional commands.

\subsubsection{Network Priority}

The coherence networks described above are ordered in priority, from highest to lowest, as follows:

\begin{enumerate}
    \item Response
    \item Fill
    \item Command
    \item Request
\end{enumerate}

When a cache controller or directory receives a message it may only cause a message of higher priority to send. Requests may cause the directory to send Commands, and Commands cause the cache controllers to send Fills or Responses. Fills cause the cache controllers to send Responses.

The cache controller and coherence directory must favor processing higher priority messages over lower priority messages to avoid deadlocking the protocol. Enforcing a priority ordering of the coherence networks helps guarantee deadlock-free operation of the protocol and is commonly used by many other protocols. Readers are referred to Sections 8.2.3 and 9.3 in \cite{nagarajan_primer_2020} for more information on deadlock avoidance.

\subsection{Cache Controller - Local Cache Engine (LCE)} \label{sec:protocl-system-lce}

The cache controller in \bedrock{} is called a Local Cache Engine (LCE) and manages coherence transactions for a single cache. The associated cache may be a private or shared cache, but is assumed to be a write-back cache that is inclusive of any higher-level caches in its hierarchy. The cache controller interfaces with its associated cache and with the \bedrock{} coherence network. It may be tightly integrated into the cache pipeline or it may be more loosely coupled and interact with the cache over a well-defined interface that allows the controller to read and write cache block metadata and data.

Each cache controller manages coherence transactions for its associated cache. It issues new requests when a cache miss occurs and responds to coherence commands that arrive on the Command network. The associated cache is only allowed to access a block using the block's current permissions and may not change the permissions on a block unless directed by the coherence directory. Any operation that requires a change in permissions results in the controller issuing a new request. This includes cache block invalidations, which are detected and initiated by the coherence directory while processing coherence requests. A cache controller may have multiple cache requests outstanding at any given time as long as each request is associated with a unique cache block, however controllers should not issue multiple requests for the same cache block. The maximum number of outstanding requests per controller is a property of the specific \bedrock{} implementation.

The cache controller must respond in a timely manner to coherence commands. Most coherence commands generate a single response message while a few messages generate no response. Cache to cache transfer commands may generate one or both of a single fill message to another controller and a single response to the directory, depending on the specific command. The cache controller must not stall command or fill message processing in order to issue a new coherence request. The \bedrock{} system and protocol allow the request network to fill and block new requests from sending, but the command and fill networks must be processed as they arrive independent of the request network status.

\subsection{Coherence Directory - Cache Coherence Engine (CCE)} \label{sec:bedrock-system-cce}

The \bedrock{} coherence directory is called the Cache Coherence Engine (CCE) and is responsible for maintaining coherence for a disjoint subset of the physical address space. A \bedrock{} system may have one or more coherence directories. If multiple coherence directories exist, management of the address space is divided evenly among all coherence engines with the physical address space striped across directories at the cache set granularity. All cache blocks that map to the same cache set in a cache controller are managed by exactly one coherence directory. The \bedrock{} coherence directory must be a complete directory that can precisely track the coherence state of all cache blocks under its management. All state transitions within the coherence protocol, including the eviction and replacement of cache blocks at the cache controllers, are controlled by the directory.

The coherence directory must process response messages in a timely manner to prevent deadlock in the coherence protocol. The directory may stall additional requests and apply back-pressure on the request network while it processes the current request. The directory is also able to issue memory commands to fetch cache blocks or perform writebacks. The directory must either process memory responses as they arrive or provide sufficient buffering between the memory command and response channels to avoid stalling when issuing memory commands. A canonical system that processes requests in-order, as described below, must either block and wait for a memory response after each memory command or provide buffering for two memory command and response pairs per directory. The two memory operations that may be required for each coherence request are for cache block eviction writeback and either an additional writeback or a memory fetch.

\subsubsection{Abstract CCE Request Processing Flow}

\begin{figure}[t]
	\centering
	\includegraphics[width=0.95\linewidth]{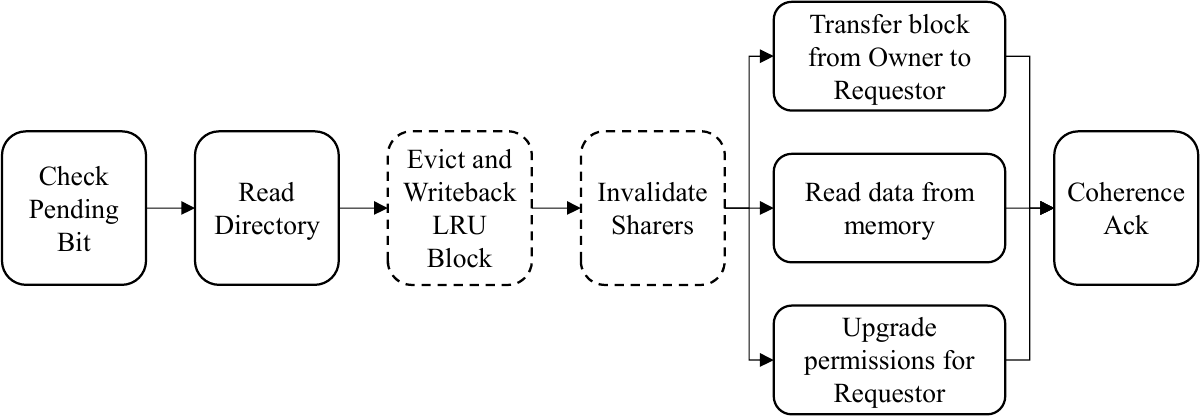}
	\caption{Canonical \bedrock{} Coherence Directory Request Processing Flow}
	\label{fig:bedrock-cce-flow-generic}
\end{figure}

The coherence directory processes coherence requests as they arrive. Each request results in one or more coherence commands being sent to the cache controllers. Request processing concludes when a coherence acknowledgment message is received at the directory from the controller that initiated the request. \autoref{fig:bedrock-cce-flow-generic} depicts the canonical, high-level request processing flow for the coherence directory. As requests arrive, the directory is read to determine the current coherence state of the requested cache block. If an eviction is required to make room for the newly request block a writeback command is issued to the requesting LCE. The writeback response is forwarded to memory if the cache responds with dirty cache block data. Next, any other caches with the block in the shared state are invalidated, as required by the specific request type. Then, the directory either initiates a memory read for the block, commands a cache to cache transfer from the current block owner to the requester (with a possible writeback to update memory), or responds with upgraded permissions if the requesting cache already has a copy of the target block. Finally, the directory waits for a coherence acknowledgment message to complete the transaction.

A simple directory implementation may execute the processing flow serially, stalling to wait for responses from every issued cache controller or memory command. Optimized coherence engine implementations may introduce concurrency both within a single request and across multiple requests. For example, an advanced implementation of the coherence directory may include logic to process responses in parallel to issuing commands, thereby avoiding unnecessary serialization in the processing flow. Regardless of implementation details, all requests are serialized relative to one another at the directory.

\subsection{Uncacheable and Atomic Accesses}

Any practical system must also be capable of processing uncacheable and atomic accesses to both cacheable and uncacheable memory. However, how these requests are handled and whether coherence is enforced for them is implementation specific. In general, atomic operations to cacheable memory should occur coherently while uncacheable operations to cacheable memory may or may not be coherent, depending on their use within the system. \autoref{sec:bedrock-uc-amo} discusses one approach to keeping both uncacheable and atomic accesses to cacheable memory coherent within an implementation of the \bedrock{} protocol and system.

\subsection{System Assumptions} \label{sec:bedrock-system-assumptions}

To facilitate discussion of the \bedrock{} protocol's function, the following assumptions are made, unless explicitly stated otherwise, for the remainder of this chapter's description of the protocol.

\begin{enumerate}
    \item The coherence networks require no specific ordering properties and may be completely unordered.
    \item The coherence networks guarantee that messages are delivered error-free.
    \item Each coherence network operates independently of the other networks.
    \item Each cache controller manages a single cache that is inclusive of all higher-level caches in its hierarchy, if any exist.
    \item Each cache block is managed by a single coherence directory and all cache blocks that map to the same cache set are managed by the same coherence directory.
\end{enumerate}

\section{Coherence Protocol}
\label{sec:bedrock-protocol}

\bedrock{}'s cache coherence protocol is a four-phase, directory-based, invalidate protocol featuring the common MOESIF coherence protocol states. \bedrock{} was designed assuming a full-duplicate tag directory organization\footnote{Any directory organization that provides complete knowledge of the system's coherence state could be used with \bedrock{}.}. The coherence protocol functions similarly to a standard directory protocol~\cite{nagarajan_primer_2020}, however the coherence directory has complete control over all coherence state transitions in the protocol. In \bedrock{}, the cache controllers may only use a cache block with its current permissions. Any change to permissions, including invalidation, must be requested from and directed by the coherence directory. \bedrock{}'s other major difference from canonical directory protocols is its use of four transaction phases and four coherence networks, including a dedicated network for cache to cache data transfers.

\subsection{Address Space Properties}
\label{sec:protocol-addresses}

\begin{figure}[t]
	\centering
	\includegraphics[width=0.90\linewidth]{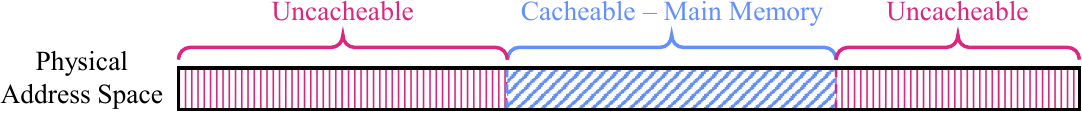}
	\caption{Canonical Address Space Layout and Cacheability Properties}
	\label{fig:address-space}
\end{figure}

\bedrock{} defines a coherence protocol that is enforced for the cacheable region of the physical address space. \autoref{fig:address-space} depicts a canonical address space, with a single cacheable region backed by main memory (e.g., DRAM) and multiple uncacheable regions. Typically, the cacheable region of physical memory consists of the system's installed DRAM address space (or a subset thereof). Without loss of generality, the rest of this chapter assumes this system model and that cacheable memory accesses are allowed only to cacheable memory. Extending the cacheable address space to cover a differently sized memory region or multiple regions requires the coherence directory to have knowledge about these ranges (or to trust the cache controllers to only issue cacheable access requests for cacheable memory). Uncacheable accesses are allowed to any physical address, and the handling of uncacheable accesses to cacheable memory is implementation specific. The implications of handling uncacheable accesses in \bedrock{} is discussed in \autoref{sec:bedrock-uc-amo}.

\subsection{Coherence Protocol States}
\label{sec:protocol-states}

\begin{table}[t]\centering
\ra{1.3}
\begin{tabular}{@{}C{0.08\linewidth}L{0.1\linewidth}C{0.08\linewidth}C{0.08\linewidth}C{0.08\linewidth}C{0.1\linewidth}C{0.08\linewidth}C{0.1\linewidth}@{}}\toprule
State & Name & Valid & Dirty & Owned & Not-Excl & RW & Encoding\\ 
\midrule
I & Invalid & \ding{55} & \ding{55} & \ding{55} & \ding{55} & \textbf{\textendash} & 000\\
S & Shared & \ding{51} & \ding{55} & \ding{55} & \ding{51} & R & 001\\
E & Exclusive & \ding{51} & \ding{55} & \ding{51} & \ding{55} & RW & 010\\
F & Forward & \ding{51} & \ding{55} & \ding{51}  & \ding{51} & R & 011\\
M & Modified & \ding{51} & \ding{51} & \ding{51}  & \ding{55} & RW & 110\\
O & Owned & \ding{51} & \ding{51} & \ding{51}  & \ding{51} & R & 111\\
\bottomrule
\end{tabular}
\caption{\bedrock{} Coherence State Properties}
\label{table:moesif-properties}
\end{table}

Every cache block in the \bedrock{} protocol exists in a stable coherence state: \emph{Invalid (I)}, \emph{Shared (S)}, \emph{Exclusive (E)}, \emph{Modified (M)}, \emph{Owned (O)}, or \emph{Forward(F)}. Adopting the terminology of \cite{nagarajan_primer_2020}, each coherence state is described using four well-defined properties: \textbf{validity}, \textbf{dirtiness}, \textbf{exclusivity}, and \textbf{ownership}. \autoref{table:moesif-properties} summarizes the mapping of properties to coherence states for the \bedrock{} coherence protocol with exclusivity encoded as its negation (not-exclusive indicates the block may be shared and cached in one or more caches) and validity as the logical OR of the remaining three properties. A \ding{55} indicates the property is false, \ding{51} indicates the property is true. The \textbf{RW} column indicates if a block is read-only (\textbf{R}) or read-write (\textbf{RW}). A cache block is writable if the not-exclusive property is false (i.e., a single cache has ownership and write permissions) and the state is Valid. The \textbf{Encoding} column is a direct, three-bit encoding of the coherence state properties \{\textit{dirty, owned, not-exclusive}\}, usable by hardware implementations.

\subsection{Protocol Messages}
\label{sec:protocol-messages}

As described above, the \bedrock{} coherence protocol relies on the Request, Command, and Response networks to carry coherence protocol messages between the cache controllers and coherence directories. This section describes the messages carried on each of the coherence networks in detail. The messages for each network are presented in a separate table. Each table lists the message name, its abbreviation as used throughout the remainder of this document, and a description of the message's functionality.

\begin{table}[t]\centering
\ra{1.3}
\begin{tabular}{@{}L{0.2\linewidth}@{}L{0.18\linewidth}@{}L{0.62\linewidth}@{}}
\toprule
Message & Abbreviation & Description\\
\midrule
Read Miss & ReqRd & Request cache block after a load miss\\
Read Miss (Non-Exclusive) & ReqRd-NE & Request cache block after a load miss with hint to not provide block in E state\\
Write Miss & ReqWr & Request cache block after a store miss\\
\bottomrule
\end{tabular}
\caption{\bedrock{} Request Network Messages}
\label{table:bedrock-requests}
\end{table}

\subsubsection{Request Network}

\autoref{table:bedrock-requests} lists the Request Network message types that a cache controller may send to the coherence directory. Read requests are issued when the cache encounters a cache miss on a load operation, and write requests are issued when the cache encounters a cache miss on a store operation. Read requests may encode an optional Non-Exclusive hint to inform the coherence directory that there is no benefit in providing the cache block in the Exclusive coherence state instead of the standard Shared coherence state. This hint is useful when a cache knows that it will never need write permissions for a block and allows the directory to issue the block with read-only permissions instead of read-write permissions. Instruction caches typically issue non-exclusive read requests because the cache does not have the ability to perform writes and modify instruction memory. \bedrock{} does not call a non-exclusive read request an instruction fetch request because this type of request may be issued by any cache controller and for either data or instruction memory locations.

A cache controller is not allowed to re-issue a coherence request until the existing request has been processed by the directory and resolved at the cache controller. A write request for a specific cache block may be issued after a read request for the same block, but multiple write or read requests for the same block from the same controller are not allowed.

Readers familiar with directory-based cache coherence may notice that \bedrock{} does not contain a separate Upgrade request message. Upgrades exist in some protocols to indicate that write permissions are needed for a cache block that the cache controller currently has cached with read-only permissions. \bedrock{} omits this message type because it introduces a race into the coherence protocol between the cache issuing the upgrade and any other cache issuing an upgrade or write request for the same block. Instead, \bedrock{} requires the LCE to issue a Write request to acquire write permissions for a block cached with read-only permissions. Section 8.8.1 of \cite{nagarajan_primer_2020} explains this type of coherence protocol race in detail.

\subsubsection{Command Network}

\begin{table}[t]\centering
\ra{1.3}
\begin{tabular}{@{}L{0.2\linewidth}@{}L{0.18\linewidth}@{}L{0.62\linewidth}@{}}
\toprule
Message & Abbreviation & Description\\
\midrule
Invalidate & INV & Invalidate cache block specified by address \\
Data & DATA & Provide data and coherence state for block specified by address and wake up cache to complete request \\
Set State & ST & Modify state of cache block specified by address \\
Set State \& Wakeup & STW & Modify state of cache block specified by address and wake up cache to complete request \\
Writeback & WB & Command cache to writeback block specified by address \\
Transfer & TR & Command cache to transfer cache block specified by address to another cache \\
Set State \& Writeback & ST-WB & Modify coherence state and then command a writeback of cache block specified by address \\
Set State \& Transfer & ST-TR & Modify coherence state and then command cache to send cache block specified by address to another cache \\
Set State \& Transfer \& Writeback & ST-TR-WB & Modify coherence state, command cache to send cache block specified by address to another cache, and then command a writeback of cache block specified by address \\
\bottomrule
\end{tabular}
\caption{\bedrock{} Command Network Messages}
\label{table:bedrock-commands}
\end{table}

\autoref{table:bedrock-commands} lists the Command Network messages that the coherence directory may send to the cache controllers. Command messages are used to modify the state of cache blocks currently cached at a cache controller, provide new cache blocks to a cache, and evict blocks from a cache when a replacement is required to make room for a newly requested block. Every command generates a response of some form from the cache controller.

Command messages are divided into a group of base commands and compound commands. The base commands include Invalidate (INV), Data (DATA), Set State (ST), Set State \& Wakeup (STW), and Transfer (TR). Each of these commands instructs the LCE to perform a single indivisible operation. Invalidate orders the LCE to set the specified cache block's state to Invalid (I), thereby revoking access permissions of the block from the LCE and its cache. A Data (DATA) command provides an LCE with cache block data and coherence state in response to a coherence request. Set State (ST) and Set State \& Wakeup (STW) modify the coherence state of the specified cache block. STW also indicates that the coherence request is resolved and execution can resume, which is used to implement permission upgrades. Transfer (TR) commands direct an LCE to initiate a cache-to-cache transfer of the specified block by sending a message on the Fill network.

The Set State \& Writeback (ST-WB), Set State \& Transfer (ST-TR), and Set State \& Transfer \& Writeback (ST-TR-WB) messages are compound messages constructed from the Set State (ST), Transfer (TR), and Writeback (WB) command primitives. Semantically, the compound messages perform the indicated primitives in the order listed. That is, a ST-TR-WB causes the cache controller to perform a set state operation, followed by a transfer, and lastly a writeback. These operations must happen "atomically" at the cache controller, in that no other operation or command should occur between any component of the compound command. Abstractly, the compound messages are similar to atomic read-modify-write operations or transactions in that the actions taken must be all-or-none, although there is no possibility that the operations will not happen in the coherence protocol.

\bedrock{} utilizes compound messages to reduce network traffic and enhance coherence system performance. They are sent as a single message across the network, and the main benefit of these compound messages is that it simplifies protocol correctness when using unordered networks. On an unordered network it would be possible for a sequence of ST, TR, and WB commands to arrive out of order at the cache controller, requiring protocol redesign or transient states to handle these race conditions. Races make the protocol significantly more complex and is in conflict with the design goal of simplifying the coherence system. The cache controllers must process coherence commands, including compound compounds, as atomic operations. This guarantees that the coherence state of every block at the cache controller is only visible in one of the stable MOESIF states at all times. In practice this means that coherence commands need to be serialized with cache accesses such that a cache access cannot see a block's metadata or data in different states during its lifetime.

\begin{table}[t]\centering
\ra{1.3}
\begin{tabular}{@{}L{0.2\linewidth}@{}L{0.18\linewidth}@{}L{0.62\linewidth}@{}}
\toprule
Message & Abbreviation & Description\\
\midrule
Data & DATA & Provide data and coherence state for block specified by address and wake up cache to complete request \\
\bottomrule
\end{tabular}
\caption{\bedrock{} Fill Network Messages}
\label{table:bedrock-fills}
\end{table}

\subsubsection{Fill Network}

\autoref{table:bedrock-fills} lists the Fill Network messages that a cache controller may send to another cache controller. Currently, the Fill network carries only Data (DATA) messages that send a full cache block and its associated state and tag to the destination controller. The state and tag portions of the message are provided by the coherence directory in an arriving Transfer command. Unlike the other coherence networks, the cache controller must support both sending and receiving messages on the Fill network.

The cache controllers must process coherence fill messages as atomic operations. This guarantees that the coherence state of every block at the cache controller is only visible in one of the stable MOESIF states at all times. In practice this means that updates to the cache data and metadata from fill messages must be serialized with cache accesses such that a cache access cannot see a block's metadata or data in different states during its lifetime.

\begin{table}[t]\centering
\ra{1.3}
\begin{tabular}{@{}L{0.2\linewidth}@{}L{0.18\linewidth}@{}L{0.62\linewidth}@{}}
\toprule
Message & Abbreviation & Description\\
\midrule
Invalidate Ack & InvAck & Response to an Invalidate command\\
Coherence Ack & CohAck & Response to finish coherence transaction\\
Writeback & DirtyWB & Response to Writeback command with cache block data\\
Null Writeback & NullWB & Response to Writeback command with no data\\
\bottomrule
\end{tabular}
\caption{\bedrock{} Response Network Messages}
\label{table:bedrock-responses}
\end{table}

\subsubsection{Response Network}

\autoref{table:bedrock-responses} lists the Response Network messages that the cache controller may send to the coherence directory. Responses triggered by the receipt of a Command message. STW and DATA commands result in a Coherence Acknowledgment (CohAck) being sent back to the directory. Invalidate (INV) commands trigger an Invalidate Ack (InvAck) to the directory while Writeback (WB) commands trigger either Writeback (DirtyWB) or Null Writeback (NullWB) responses. Only DirtyWB messages carry data on the Response Network. The Response network is the highest priority network. Messages on the response network receive priority over command and request messages for processing by the coherence directory, and response messages do not cause any other messages to be sent.

\subsection{Coherence Transactions and Tracking Coherence State}
\label{sec:protocol-transactions}

The \bedrock{} coherence protocol relies on the concept of a \textbf{coherence transaction} to define the concurrency behavior of the protocol. A coherence transaction encompasses the total duration of a coherence request, beginning with the cache controller issuing a Request message to the directory and ending when the coherence directory receives the Coherence Acknowledgment (CohAck) response from the cache controller. At the cache controller, a coherence transaction begins when it issues a coherence request to the coherence directory. The coherence transaction completes when the controller receives either a STW command or a DATA message (on the Command or Fill network) and issues a CohAck response to the coherence directory. At the coherence directory, a coherence transaction begins when it starts processing a new coherence request message. The coherence transaction completes when the directory receives a CohAck response message from the cache controller that initiated the coherence request.

\subsection{Protocol Assumptions}
\label{sec:protocol-assumptions}

The \bedrock{} protocol makes the following assumptions that must be enforced to ensure correct operation of the coherence protocol.

\begin{enumerate}
    \item The coherence directory controls all coherence state transitions, with the single exception that a cache controller may silently upgrade a block from Exclusive (E) to Modified (M) on a write operation.
    \item The cache controllers must not issue a duplicate coherence request until the original outstanding request is resolved.
    \item The cache controllers must process all coherence commands atomically.
\end{enumerate}

\subsection{Coherence Protocol Tables}
\label{sec:protocol-tables}

This section presents \bedrock{}'s MOESIF cache coherence protocol in tabular form for both the cache controller and coherence directory~\cite{sorin_specifying_2002}. The tables use a notation of "Action/Next State" to describe the behavior of the controllers. Given the current coherence state for a cache block, as indicated at the start of the row, an entry in the table describes the action taken by the controller and the next coherence state of the block at the controller for the event indicated by the column header. If no action is required, a --- is written in place of the action. Blank entries indicate that the event listed in the column header cannot occur for the state given at the start of the row. Additional tables for the other protocol variants in the MOESIF family can be found in the \autoref{app:protocol-lce} and \autoref{app:protocol-cce}. 

\subsubsection{Cache Controller Protocol Table}

\autoref{table:bedrock-protocol-lce-moesif} presents the cache controller protocol table for the \bedrock{} MOESIF protocol. A cache controller may experience Cache Action and Coherence Message events. Cache Actions are cacheable load and store operations. Coherence Messages are the arrival of a \bedrock{} Command message.

Each "Action/State" entry indicates a message sent in response to the command and the next state of the target block at the cache controller. The Action may be a response message sent to the directory or a command messages sent to another cache controller (as directed by the arriving command from the directory). Some entries in the table have a next coherence state of X that indicates the next coherence state is not known \emph{a priori} by the cache controller. In these situations, the arriving command provides the correct coherence state, which is applied to the block by the cache controller. The specific next state is determined by the coherence directory.

\subsubsection{Coherence Directory Protocol Table}

\autoref{table:bedrock-protocol-cce-moesif} presents the coherence directory protocol table for the \bedrock{} MOESIF protocol. The coherence directory experiences two types of coherence events: new Coherence Requests from the cache controllers and cache block replacements (a Directory Action) that are initiated by the coherence directory while processing a coherence request. Replacements occur when a cache block in the target cache set of a coherence request must be evicted to make room for the newly requested cache block.

Each "Action/Next State" entry provides the messages sent by the directory to cache controllers to complete processing of the request and the next state of the target block at the coherence directory. The coherence state superscripts attached to some messages in the table indicate a coherence state associated with the message or compound message component. For example, a ST\tss{F}-TR\tss{S}-WB message instructs a cache controller to set the target cache block's state to F, forward that same block to another controller with the S state, and lastly send a writeback to the coherence directory.

Replacements are only required for blocks that may be dirty (cached in the E, M or O states). Cache blocks in the S or F state are clean and do not require a separate invalidation or writeback message prior to the new cache block arriving and overwriting the existing block. The block being replaced may be freely used by the cache until it is overwritten or the invalidation and writeback occurs. Dirty blocks must be either invalidated or downgraded to a read-only state to prevent a write racing with the completion of the coherence request. Typically, dirty blocks are invalidated to conceptually align with the PUT procedure found in canonical directory-based coherence protocols.

% clear page for protocol table
\clearpage

\begin{table}[t]\centering
\footnotesize
\begin{adjustbox}{angle=90}
\ra{1.3}
\begin{tabular}{@{}L{0.05\linewidth}L{0.06\linewidth}L{0.06\linewidth}ccccccccC{0.15\linewidth}@{}}
%\begin{tabular}{@{}lllccccccccc@{}}
\toprule
\footnotesize
& \multicolumn{2}{c}{Cache Action} & \phantom{a} & \multicolumn{8}{c}{Coherence Message}\\
\cmidrule{2-3} \cmidrule{5-12}
State & Load & Store && Inv & DATA & STW & WB & TR & ST-WB & ST-TR & ST-TR-WB\\
\midrule
I & ReqRd & ReqWr && & CohAck/X & & & & & &\\\\
S & Hit & ReqWr && InvAck/I & & CohAck/M & & & & &\\\\
E & Hit & Hit/M && & & & NullWB/E & & NullWB/X & DATA/X & DATA, NullWB/X\\
M & Hit & Hit && & & & DirtyWB/M & & DirtyWB/X & DATA/X & DATA, DirtyWB/X\\
O & Hit & ReqWr && & & CohAck/M & DirtyWB/O & DATA/O & DirtyWB/X & DATA/X &\\\\
F & Hit & ReqWr && & & CohAck/M & & DATA/F & & DATA/X &\\
\bottomrule
\end{tabular}
\end{adjustbox}
\caption{\bedrock{} Cache Controller Protocol Table - MOESIF}
\label{table:bedrock-protocol-lce-moesif}
\end{table}

% clear page for protocol table
\clearpage

\begin{table}[t]\centering
\footnotesize
\begin{adjustbox}{angle=90}
\ra{1.3}
\begin{tabular}{@{}L{0.10\linewidth}L{0.13\linewidth}L{0.13\linewidth}L{0.15\linewidth}L{0.15\linewidth}L{0.15\linewidth}lL{0.2\linewidth}@{}}
\toprule
& \multicolumn{5}{c}{Coherence Request} & \phantom{a} & Directory Action\\
\cmidrule{2-6} \cmidrule{8-8}
Directory State & ReqRd & ReqRd (Non-Excl) & ReqWr from Invalid & ReqWr from Sharer & ReqWr from Owner && Replacement\\
\midrule
I & DATA to Req/E & DATA to Req/S & DATA to Req/M &&&&\\\\
S & DATA to Req/S & DATA to Req/S & Inv all S, DATA to Req/M & Inv other S, STW\tss{M} to Req/M &&&\\
E & ST\tss{F}-TR\tss{S}-WB to Owner/F & ST\tss{F}-TR\tss{S}-WB to Owner/F & ST\tss{I}-TR\tss{M} to Owner/M & &&& ST\tss{I}-WB to Req/I\\\\
M & ST\tss{O}-TR\tss{S} to Owner/O & ST\tss{O}-TR\tss{S} to Owner/O & ST\tss{I}-TR\tss{M} to Owner/M &&&& ST\tss{I}-WB to Req/I\\\\
O & TR\tss{S} to Owner/O & TR\tss{S} to Owner/O & Inv all S, ST\tss{I}-TR\tss{M} to Owner/M & Inv other S and Owner, STW\tss{M} to Req/M & Inv all S, STW\tss{M} to Req/M && ST\tss{I}-WB to Req/I\\
F & TR\tss{S} to Owner/F & TR\tss{S} to Owner/F & Inv all S, ST\tss{I}-TR\tss{M} to Owner/M & Inv other S and Owner, STW\tss{M} to Req/M & Inv all S, STW\tss{M} to Req/M &&\\
\bottomrule
\end{tabular}
\end{adjustbox}
\caption{\bedrock{} Coherence Directory Protocol Table - MOESIF}
\label{table:bedrock-protocol-cce-moesif}
\end{table}

% clear page for following text
\clearpage

\subsection{Coherence State Transitions}
\label{sec:protocol-state-transitions}

This section describes the possible coherence state transitions at the cache and directory controllers for the \bedrock{} MOESIF protocol. Each table describes the coherence state transitions for a single cache block given a current starting state and a cache or coherence event that causes a state transition. Events that do not cause a change in the coherence state at the controller are not listed.

\subsubsection{Cache Controller State Transitions}

\begin{table}[t]\centering
\ra{1.3}
\begin{tabular}{@{}L{0.25\linewidth}C{0.18\linewidth}C{0.18\linewidth}@{}}
\toprule
Event & Current State & Next State\\
\midrule
Load & I & S, E\\
Store & I, S, O, F & M\\
Store (Silent Upgrade) & E & M\\
Other Load & E & F\\
& M & O\\
Other Store & S, E, M, O, F & I\\
\bottomrule
\end{tabular}
\caption{\bedrock{} Cache Controller Next State Table - MOESIF}
\label{table:bedrock-states-lce-moesif}
\end{table}

\autoref{table:bedrock-states-lce-moesif} enumerates the possible coherence state transitions as observed by the cache controller for a single cache block as loads and stores occur in the coherence system targeting that block. The Current State column lists the current cache coherence state of the target block and the Next State column provides the next state of the block at the cache controller. The Event column lists the type of load or store event. Events that do not cause a change in the coherence state are not listed (e.g., Load to block in S remains in S).

The first three rows correspond to actions taken by the controller itself, while the last two rows of Other Load and Other Store correspond to load and store actions initiated by some other cache controller. A Store (Silent Upgrade) occurs when the cache performs a store operation on a block cached in the E state. This state has read and write permissions but is considered clean, therefore the store must transition the block to the M state to indicate a write has occurred. This is called a Silent Upgrade since the cache controller does not need to notify the coherence directory of the write because it already has write permissions for the block.

\subsubsection{Coherence Directory State Transitions}

\begin{table}[t]\centering
\ra{1.3}
\begin{tabular}{@{}L{0.2\linewidth}@{}L{0.2\linewidth}@{}C{0.16\linewidth}@{}C{0.16\linewidth}@{}C{0.16\linewidth}@{}}
\toprule
Event & Request Message & Current State (Dir) & Next State (Dir) & Next State (Requestor)\\
\midrule
Load & ReqRd & I & E & E\\
&& S & S & S\\
&& E, F & F & S\\
&& M, O & O & S\\
Load (Non-Excl) & ReqRd-NE & I, S & S & S\\
&& E, F & F & S\\
&& M, O & O & S\\
Store & ReqWr & I, S, O, E, M, F & M & M\\
\bottomrule
\end{tabular}
\caption{\bedrock{} Coherence Directory Next State Table - MOESIF}
\label{table:bedrock-states-cce-moesif}
\end{table}

\autoref{table:bedrock-states-cce-moesif} describes the possible coherence state transitions at the coherence directory for a single cache block as load and store misses are processed. For each event, corresponding \bedrock{} request network message type, and current state of the coherence directory, the resulting next state of the block at both the directory and cache controller that initiated the event are listed. This table fully enumerates the possible state transitions for the MOESIF protocol, covering the cross-product of events and current directory states. In accordance with the SWMR Invariant, a store operation always results in a single cache owning the block and receiving write permissions.

As seen in the table, even when using a MOESIF protocol, the requesting cache will only ever receive a cache block in the S, E, or M states. The O and F states are used when load requests target cache blocks in either the M or E states, respectively. These states allow a single cache controller to retain ownership permissions for read-only cache blocks and allow a read miss to be completed with a cache to cache transfer rather than a LLC or main memory access.

Note that an event may cause the owner cache to change during the transaction, even if the block's state at the directory does not change For example, one cache performing a write to a block that another cache already has write permissions for (i.e., cached in M) results in the state remaining in M but changes ownership of the block.

\section{Uncacheable and Atomic Operations} \label{sec:bedrock-uc-amo}

This section describes the handling of uncached and atomic read-modify-write accesses in \bedrock{}, using the same assumptions and canonical address space layout presented in \autoref{sec:bedrock-protocol}.

\subsection{Uncacheable Accesses}

Uncacheable accesses may target both uncacheable and cacheable memory regions. Uncacheable accesses to uncacheable memory are not a concern for the \bedrock{} coherence protocol because the protocol enforces coherence only for cacheable memory. Uncacheable accesses to uncacheable memory can either bypass the coherence directory or the directory can be augmented to forward the requests and responses to and from memory, respectively. Thus, the only changes required to support uncacheable accesses to uncacheable memory are to add appropriate request and command message types for the \bedrock{} networks and modify the system to handle these requests, assuming that the accesses will travel on the existing \bedrock{} networks. Endpoints that do not participate in coherence must only implement the Request and Command \bedrock{} networks.

Uncacheable accesses to cacheable memory must participate in the \bedrock{} coherence protocol to guarantee coherence within the system. An uncacheable access targeting a cacheable block of memory first must invalidate, and write back, the block from all cache controllers that have it cached. Then, the uncacheable access may me issued to memory with the response from memory being forwarded back to the requesting cache. This access must also be serialized with all coherence requests to the target cache block. Serializing the request is easily enforced using the existing way group and pending bit mechanisms of the coherence directory that serialize cacheable accesses. \bedrock{} must be modified to support the uncacheable request message types and corresponding command message types that deliver uncacheable load data or store complete confirmation to the requesting cache controller. The coherence directory must also be modified to detect an uncacheable request targeting cacheable memory and invoke the invalidation and writeback routines for the target cache block.

\subsection{Atomic Read-Modify-Write Operations}

Atomic read-modify-write style operations, atomics for short, are important operations for multicore processors. Consequently, a multicore processor's memory system, including the coherence system, must support these operations. In the context of the canonical \bedrock{} coherence system, there are two possible locations that atomics can be executed --- at the LLC/memory or within the cache controller managed inclusive cache hierarchy. An atomic to uncacheable memory is assumed to be executed at the LLC/memory while an atomic to cacheable memory is executed by either the LLC/memory or the cache controller's cache hierarchy. This is the target model used by \bedrock{} in \blackparrot{} \cite{petrisko2020}.

\bedrock{} easily supports atomic operations targeting cacheable memory and executed by the cache controller's inclusive cache hierarchy with very minimal modification to the existing cache controller. A cache executing an atomic simply needs write permissions for the target cache block. Once it has write permissions it must complete the read-modify-write sequence as a single, uninterruptible action. The simplest way to accomplish this is for the cache to perform a write request for the target block then briefly "lock" the block while the cache completes the read-modify-write operation. After the atomic executes, the block is unlocked and any coherence commands that arrived targeting the block are processed. Although this seems to violate the spirit of \bedrock{} in that the cache controller momentarily ignores a command from the directory, it is an effective way to implement atomics in practice. As long as the block is locked for only a very short time, there is no risk of deadlocking the coherence protocol. From the point of view of the coherence directory, a controller locking a block for a few cycles is equivalent to the coherence network taking a few extra cycles to deliver a coherence command. Since \bedrock{} does not depend on the coherence networks delivering messages within specific latencies, the few cycles of delay from locking has no impact on correctness, so long as this delay is short and the cache controller resumes processing commands in a timely manner.

Executing atomics at LLC/memory is similar to performing an uncached memory access. The cache controller issues an atomic request with data that must be forwarded to memory for use in the read-modify-write operation, and memory responds with data if the atomic has a return value or an atomic complete message if there is not return value. The memory response is then forwarded to the requesting cache controller on the \bedrock{} command network. This type of operation requires adding a few new message types to the \bedrock{} request and command networks to support atomic requests and atomic data or complete commands. Atomics targeting cacheable memory and executed by the LLC/memory must follow a procedure similar to a regular uncached access to cacheable memory that invalidates and writes back the target cache block from all cache controllers possessing a copy of it. This forces any cache currently using the block to refetch it from memory, and these requests will be serialized by the coherence directory to guarantee the Data Value Invariant holds. Thus, the coherence directory must be modified in a similar manner as it was for uncacheable requests to detect atomic accesses to cacheable memory and enforce serialization of these requests using the existing way group and pending bit mechanisms.

\section{Protocol Verification} \label{sec:bedrock-verification}

The \bedrock{} cache coherence protocol is similar to, yet subtly different from, commonly understood directory-based protocols (e.g., Section 8.3 of ~\cite{nagarajan_primer_2020}). Therefore, it is important that the protocol itself is shown to be correct, especially since it is implemented by the \blackparrot{} multicore. \bedrock{}'s MESI protocol has been verified correct using CMurphi~\cite{cmurphi}, an improved version of the Murphi~\cite{murphi} model checking framework. The \bedrock{} verification model is available in the \blackparrot{} GitHub repository.

\bedrock{}'s CMurphi description assumes that only a single cache block and a single coherence directory are modeled. These assumptions are valid because, by definition, cache coherence is constrained to a single memory location, \bedrock{}'s coherence directories operate independently from one another in a multi-directory system, and every cache block is managed by a single directory. The model uses unordered networks.

\begin{table}[t]\centering
\ra{1.3}
\begin{tabular}{@{}C{0.15\linewidth}C{0.1\linewidth}@{}C{0.1\linewidth}@{}C{0.1\linewidth}@{}C{0.1\linewidth}@{}C{0.1\linewidth}@{}C{0.1\linewidth}@{}}
\toprule
\multirow{2}{*}{Protocol}
& \multicolumn{6}{c}{Cache Count}\\
\cline{2-7}
& 2 & 3 & 4 & 5 & 6 & 8\\
\midrule
\bedrock{} & 0.1s & 0.21s & 3.1s & 47s & 8.9m & 15.2h\\
Traditional & 0.1s & 0.35s & 19.9s & 10.4m & 9.9h & 175d\\
\midrule
Speedup & 1.0x & 1.6x & 6.4x & 13.3x & 66.6x & 1230x\\
\bottomrule
\end{tabular}
\caption{\bedrock{} CMurphi Verification Time and Speedup - MESI}
\label{table:cmurphi-results}
\end{table}

\autoref{table:cmurphi-results} shows the verification time required by CMurphi for \bedrock{} and a traditional MESI coherence protocol, averaged over three runs for each configuration. Verification time for the traditional MESI protocol at 8-caches is a best-fit estimate, as it would take 175 days to complete! Both protocols are verified correct without error by CMurphi for all other configurations. CMurphi explores considerably fewer states and completes verification significantly faster for \bedrock{}, up to 66x faster for a 6-cache system. The verification speedup is a direct consequence of \bedrock{}'s design decisions eliminating protocol races and transient states, which greatly reduces the necessary state-space that must be explored and verified.

\section{Protocol Analysis}
\label{sec:bedrock-protocol-analysis}

\bedrock{} is similar to, but subtly different from, a traditional directory-based coherence protocol. In order to better understand the design tradeoffs of directory-based coherence protocols, this section presents a comparison of the \bedrock{} protocol to a canonical directory protocol. First, a canonical MOESIF directory-based coherence protocol is presented through state transition tables in \autoref{sec:bedrock-protocol-analysis-stable-states}. Next, a discussion of transient states and coherence networks and messages is presented in \autoref{sec:bedrock-protocol-analysis-transient-states} and \autoref{sec:bedrock-protocol-analysis-networks}, respectively. Lastly, \autoref{sec:bedrock-protocol-analysis-state-transition-diagrams} presents state transition diagrams for both protocols and \autoref{sec:bedrock-protocol-analysis-models} presents mathematical models for each protocol, enabling direct comparison of the complexity inherent in each protocol.

\subsection{Stable State Transitions}
\label{sec:bedrock-protocol-analysis-stable-states}

\begin{table}[t]\centering
\ra{1.3}
\begin{tabular}{@{}L{0.25\linewidth}C{0.18\linewidth}C{0.18\linewidth}@{}}
\toprule
Event & Current State & Next State\\
\midrule
Load & I & S, E\\
Store & I, S, O, F & M\\
Store (Silent Upgrade) & E & M\\
Other Load & E & F\\
& M & O\\
Other Store & S, E, M, O, F & I\\
\bottomrule
\end{tabular}
\caption{Canonical Cache Controller Next State Table - MOESIF}
\label{table:canonical-cache-states-moesif}
\end{table}

\begin{table}[t]\centering
\ra{1.3}
\begin{tabular}{@{}L{0.16\linewidth}@{}L{0.2\linewidth}@{}C{0.16\linewidth}@{}C{0.16\linewidth}@{}C{0.16\linewidth}@{}}
\toprule
Event & Request Message & Current State (Dir) & Next State (Dir) & Next State (Requestor)\\
\midrule
Load & GetS & I & E & E\\
&& S & S & S\\
&& E, F & F & S\\
&& M, O & O & S\\
Store & GetM & I, S, O, E, M, F & M & M\\
\bottomrule
\end{tabular}
\caption{Canonical Coherence Directory Next State Table - MOESIF}
\label{table:canonical-directory-states-moesif}
\end{table}

\autoref{table:canonical-cache-states-moesif} and \autoref{table:canonical-directory-states-moesif} describe the cache and directory coherence state transitions, respectively, for a canonical directory-based coherence protocol. The cache controller state transitions are identical to the \bedrock{} protocol. The directory state transitions are very similar to \bedrock{}, with \bedrock{}'s ReqRd and ReqWr messages corresponding to the canonical protocol's GetS and GetM messages, respectively. The canonical protocol does not have a direct equivalence to \bedrock{}'s ReqRd (Non-Excl) message, although it could easily be added, for example by a special GetS message variant that would have similar or identical semantics to ReqRd (Non-Excl).

The similarity of these tables is expected and intuitive since both protocols rely on the same set of stable protocol states and these states have the same semantics in each protocol. The state transitions between the stable states of a coherence protocol depends only on the set of stable states and the possible events. In canonical stored-program~\cite{neumann_1993}, or \emph{von Neumann}, multicore processor architectures utilizing hardware-based cache coherence systems, the set of possible events is effectively only load and store operations. Complex memory operations include load-reserved/store-conditional (LRSC) and atomic read-modify-write (AMO, Atomic RMW) are constructed as special load and store operations or pairs, but the primitive events remain loads and stores. Thus, the canonical protocol described and \bedrock{} are effectively equivalent when viewed at this level because they employ the same set of stable states (MOESIF) and have the same set of possible events (load or store).

The similarity of these protocols at this level also illustrates the importance of understanding the details of a given coherence protocol when determining its advantages, disadvantages, objectives, and non-objectives. The degree to which transient states are employed or the number and ordering of coherence networks required are properties of the protocol's implementation. These details are important in practice because they directly correlate with the complexity, cost, and performance of a coherence protocol's implementation.

\subsection{Transient States}
\label{sec:bedrock-protocol-analysis-transient-states}

Transient states exist in canonical directory-based coherence protocols to handle race conditions or provide additional transaction concurrency within the protocol. Unlike a canonical protocol, \bedrock{} does not expose any transient states in the coherence protocol. \bedrock{} is able to realize a protocol without exposed transient states due to its assumptions that only a single coherence transaction per way group may be active at any time and that each cache controller must not issue duplicate coherence requests. The cache controllers may access cache blocks with their existing permissions, but must request new permissions from the directory as needed. Coherence commands are processed atomically by the cache controllers to guarantee that cache accesses only ever see a cache block in a single stable and consistent state. The coherence directory updates the stable state of a cache block as it issues messages to change the state of the block. Since the directory only processes one request per way group at a time there is no need to expose transient states in the protocol. Implementations may define mechanisms to track the transient behavior of the request processing flow (e.g., invalidations completed, waiting for coherence acknowledgment) in order to enable concurrent processing of independent requests, but at the protocol level all blocks will be in a consistent and stable state at all times.

Canonical protocols also vary in the number and types of transient states defined in the protocol. Simple canonical protocols have fewer transient states and allow less concurrency across transactions targeting the same cache block. This manifests as stall conditions in protocol processing that blocks new requests from targeting a cache block that already has one or more active transactions. A major drawback of transient states is that verification effort typically grows super-linearly or exponentially with the number of protocol states, including both stable and transient states. Thus, while transient states allow more per-block concurrency, the verification complexity cost limits the amount of concurrency that architects are willing to add in practice. As will be seen in \autoref{chap:bp-bedrock} and \autoref{chap:hybrid}, concurrency across cache blocks can be realized through coherence engine implementation decisions that do not alter the coherence protocol specification.

\subsection{Coherence Networks and Protocol Messages}
\label{sec:bedrock-protocol-analysis-networks}

\bedrock{} utilizes four coherence networks instead of the three networks used by the canonical protocol. In \bedrock{}, only the Fill network is bi-directional. The Request, Command, and Response networks are all uni-directional and carry messages between the cache controllers and the coherence directory. In contrast, canonical protocols utilize uni-directional request and forwarded request networks with a bi-directional response network.

\bedrock{} assumes that there may only be one active coherence transaction per way group, that the directory controls cache block replacements, that all response messages return to the directory, and that all networks are unordered. These design constraints necessitate the use of a coherence acknowledgment message from cache to directory to close the transaction and requires a fourth coherence message class and network with higher priority than the canonical directory protocol's response network to carry the coherence acknowledgment message. The fourth network is necessary to accommodate transactions involving cache to cache transfers, which require four phases or hops: Request to directory, Command to owner, Fill from owner to requester, Response to directory.

\begin{table}[t]\centering
\ra{1.3}
\begin{tabular}{@{}L{0.24\linewidth}@{}L{0.24\linewidth}@{}L{0.35\linewidth}@{}}\toprule
\bedrock{} Network & \bedrock{} Message & Canonical Protocol Message\\
\midrule
Request & ReqRd & GetE\\
& ReqRd-NE & GetS\\
& ReqWr & GetM or Upgrade\\
\midrule
Command & Inv & Inv\\
& DATA & Data from Dir or Owner\\
& STW & Data from Dir (ack=0)\\
& WB & No direct equivalence\\
& TR & Fwd-GetX\\
& ST-WB & No direct equivalence\\
& ST-TR & Fwd-GetX\\
& ST-TR-WB & Fwd-GetX\\
\midrule
Fill & DATA & Data from Owner\\
\midrule
Response & InvAck & Inv-Ack\\
& CohAck & No direct equivalence\\
& DirtyWB & PUTM, PUTO\\
& NullWB & PUTS, PUTE, PUTF\\
\bottomrule
\end{tabular}
\caption{\bedrock{} and Canonical Directory Protocol Message Equivalency}
\label{table:protocol-message-comparison}
\end{table}

\autoref{table:protocol-message-comparison} presents a comparison of the message types in \bedrock{} and those used in a canonical protocol. The corresponding message types of the traditional protocol are found in Tables 6.4, 8.3, 8.4, 8.5, and 8.6 of \cite{nagarajan_primer_2020}. A Fwd-GetX message corresponds to a Fwd-Get message with state specified by X, where X is one of the MOESIF states as applicable for a specific instance of the message (e.g., Fwd-GetS or Fwd-GetM). The PUT messages have also been extended to cover all of the MOESIF states. As seen in the table, many \bedrock{} messages have direct equivalences in the traditional protocol. However, a few important differences are worth discussing.

First, \bedrock{} carries coherence messages on four networks instead of three, so there is not a one-to-one correspondence to the three networks of the traditional protocol, which are called Request, Forwarded-Request, and Response. The Request network is similar in each protocol as they carry coherence requests from cache controller to coherence directory. \bedrock{}'s Command network is similar to the Forwarded-Request network as both networks carry messages from the coherence directory to the cache controller. \bedrock{}'s Response and Fill networks carry messages similar to the traditional protocol's Response network. \bedrock{} requires four networks to maintain priority between messages in the protocol due to the use of coherence acknowledgment messages and directory-controlled cache block replacements.

Second, there is no cache controller Replacement message that can be issued on \bedrock{}'s Request network. Unlike a traditional protocol, the coherence directory, not the cache controller, is responsible for performing cache block replacements to make room for new cache blocks in the cache controller. The cache controller provides a replacement way "hint" to the coherence directory with each read or write request. The coherence directory then finalizes the selection of a replacement way, writing back any dirty cache block data as required by issuing a WB message over the Command Network. \bedrock{}'s directory-controlled replacements eliminate a common race between PUT and Fwd-Get messages in the traditional protocol. Thus, one view of the WB message is that it is a directory-initiated PUT and that PUT messages are from directory to cache rather than cache to directory as in the traditional protocol.

Third, the ST-WB message has no direct equivalence in the traditional protocol. ST-WB can be viewed as a directory-initiated PUT message for a dirty cache block. The intent of these messages is to downgrade the permissions of a particular cache block, writing back the dirty block, and then modifying the coherence state of the block. ST-WB is commonly used to combine an invalidate and writeback sequence into a single command/response pair since the DirtyWB or NullWB response serves as an acknowledgment of both the writeback and invalidation actions.

Lastly, the traditional protocol does not require a CohAck message to close a coherence transaction. This message is required to enforce correct serialization of coherence requests targeting the same cache block at the directory controller when using unordered networks. The use of an explicit coherence transaction acknowledgment disallows concurrent transactions to the same block and enables the removal of many transient states from the coherence protocol required to handle races at the cache controller.

% clear page for transition diagrams
\clearpage

\begin{figure}[t]
	\centering
    \includegraphics[width=0.8\linewidth]{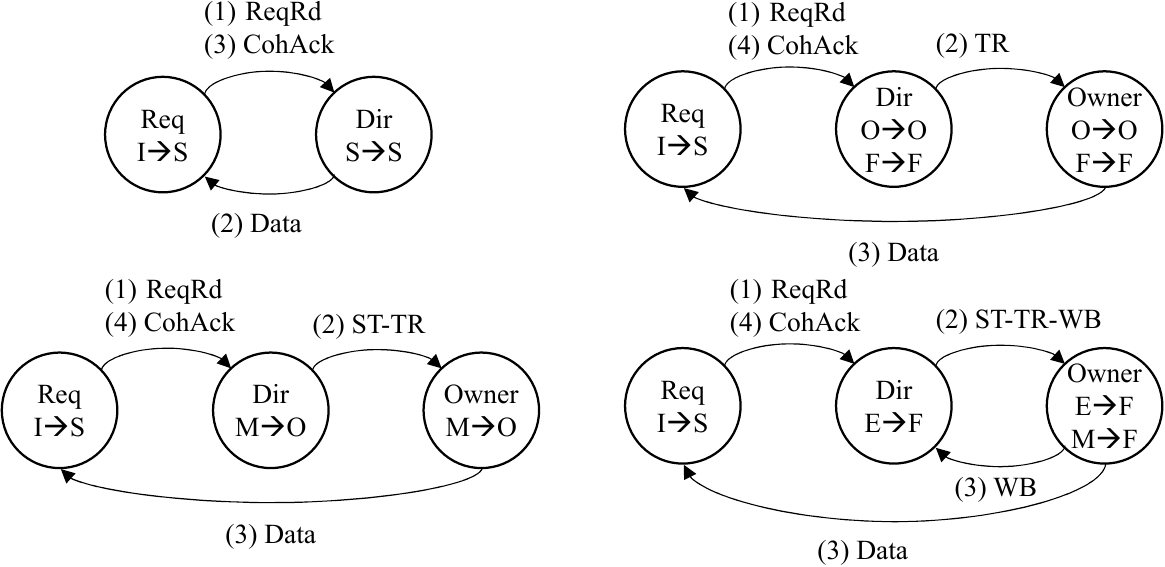}
	\caption{\bedrock{} I$\rightarrow$S Transitions}
    \label{fig:bedrock-i-s}
\end{figure}

\begin{figure}[t]
	\centering
    \includegraphics[width=0.25\linewidth]{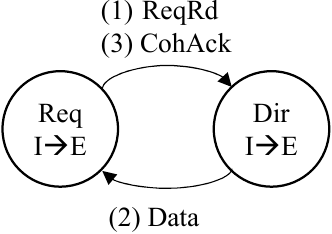}
	\caption{\bedrock{} I$\rightarrow$E Transitions}
    \label{fig:bedrock-i-e}
\end{figure}

\begin{figure}[t]
	\centering
    \includegraphics[width=0.8\linewidth]{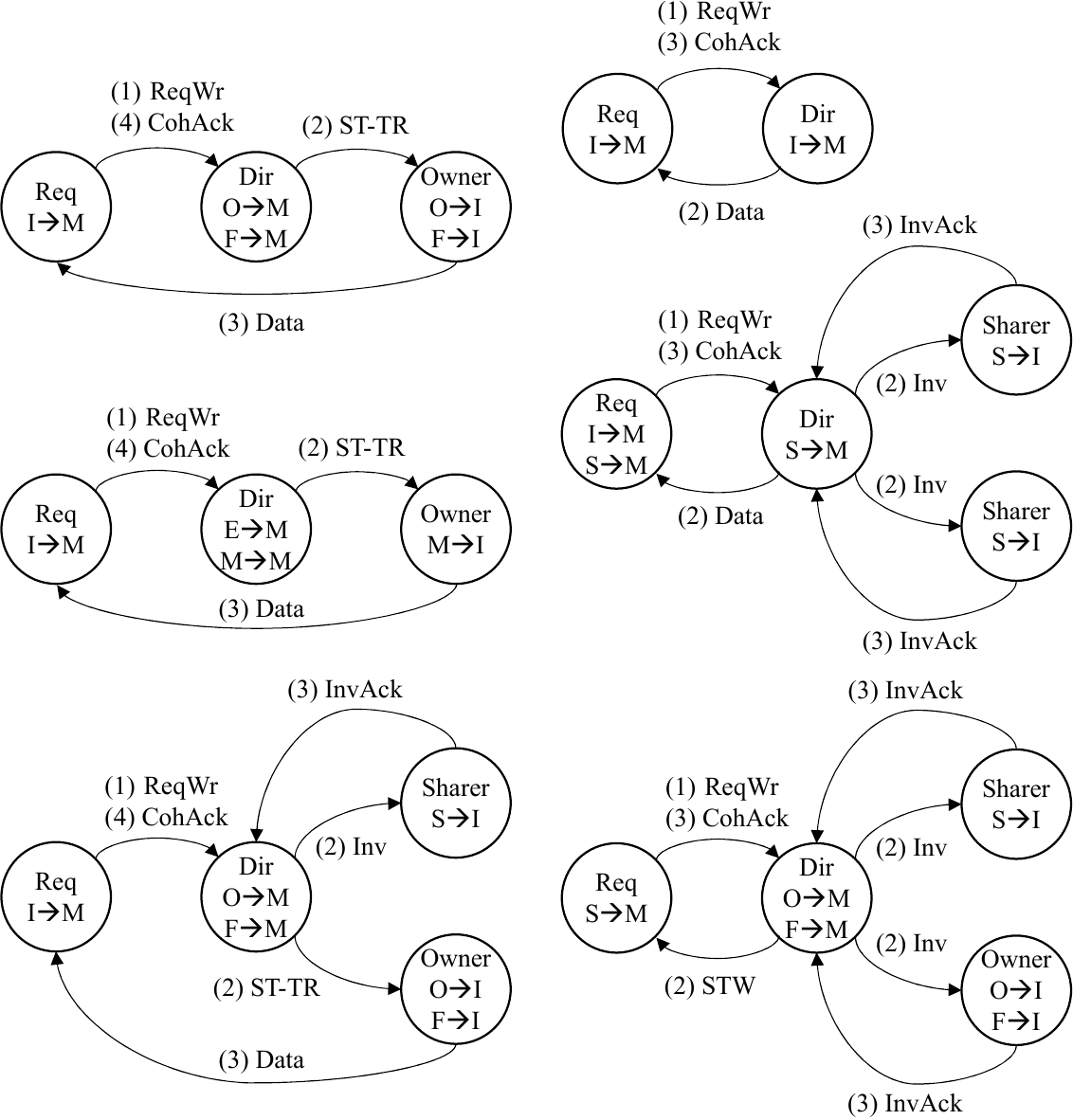}
	\caption{\bedrock{} I/S$\rightarrow$M Transitions}
    \label{fig:bedrock-is-m}
\end{figure}

\begin{figure}[t]
	\centering
    \includegraphics[width=0.7\linewidth]{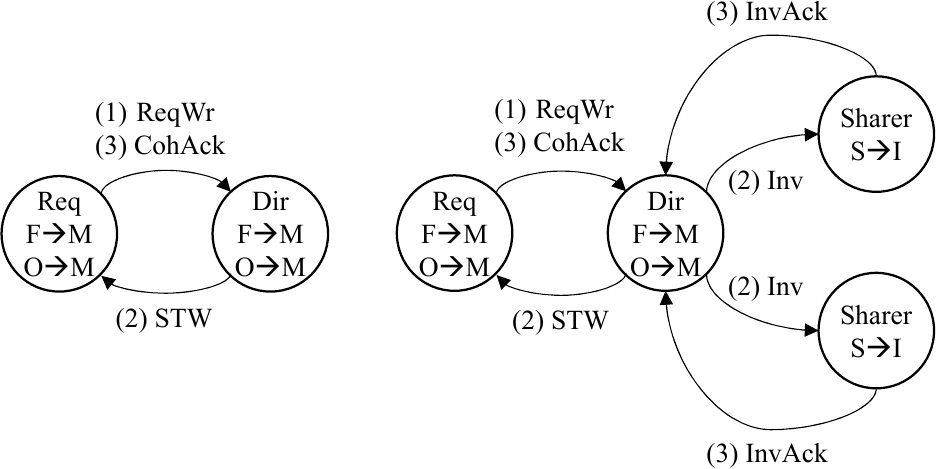}
	\caption{\bedrock{} F/O$\rightarrow$M Transitions}
    \label{fig:bedrock-fo-m}
\end{figure}

\begin{figure}[t]
	\centering
    \includegraphics[width=0.8\linewidth]{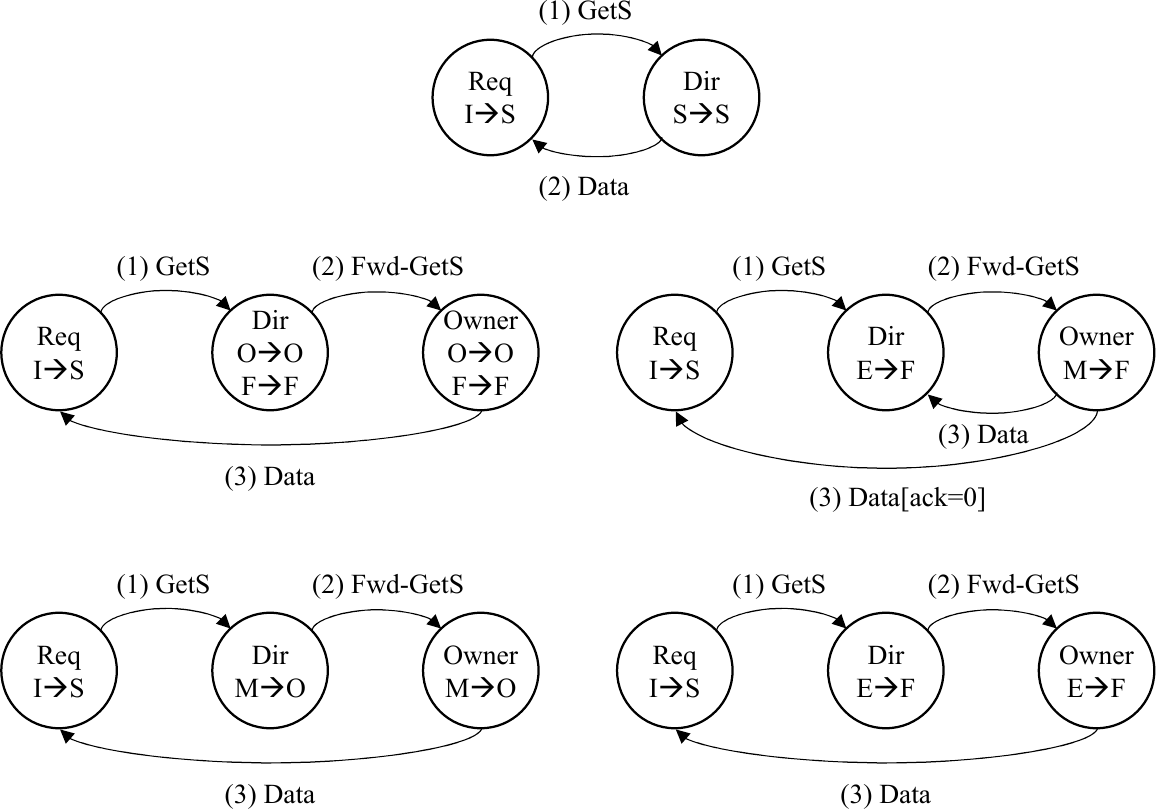}
	\caption{Canonical I$\rightarrow$S Transitions}
    \label{fig:canonical-i-s}
\end{figure}

\begin{figure}[t]
	\centering
    \includegraphics[width=0.25\linewidth]{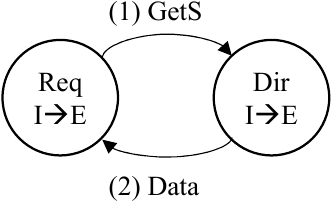}
	\caption{Canonical I$\rightarrow$E Transitions}
    \label{fig:canonical-i-e}
\end{figure}

\begin{figure}[t]
	\centering
    \includegraphics[width=0.8\linewidth]{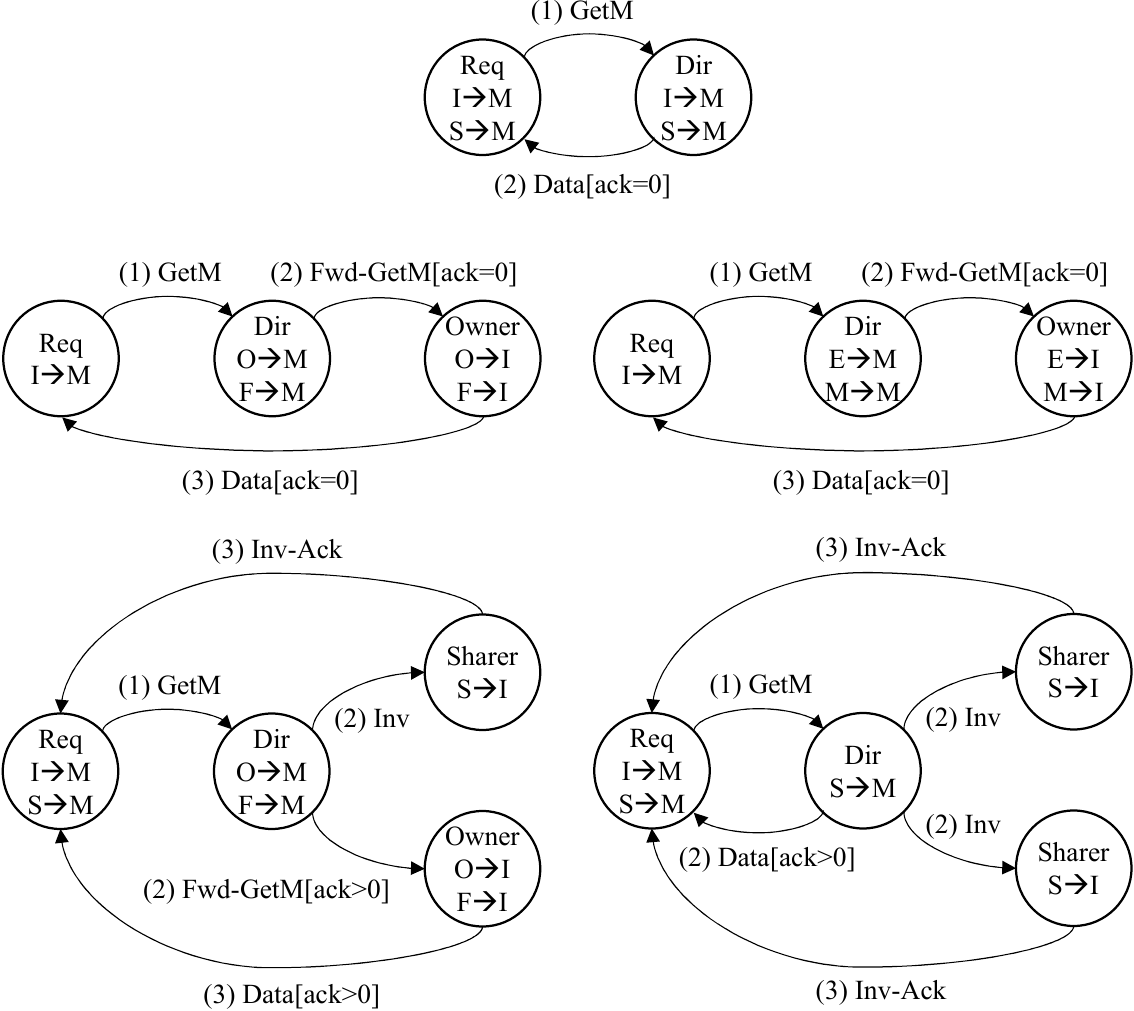}
	\caption{Canonical I/S$\rightarrow$M Transitions}
    \label{fig:canonical-is-m}
\end{figure}

\begin{figure}[t]
	\centering
    \includegraphics[width=0.7\linewidth]{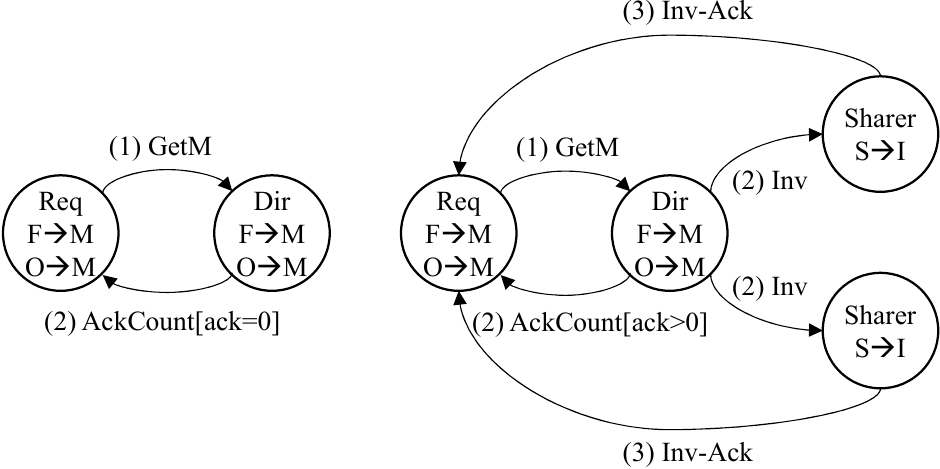}
	\caption{Canonical F/O$\rightarrow$M Transitions}
    \label{fig:canonical-fo-m}
\end{figure}

\clearpage

\subsection{Coherence State Transition Diagrams}
\label{sec:bedrock-protocol-analysis-state-transition-diagrams}

The preceding sections outline how \bedrock{} differs from a canonical directory protocol in terms of stable state transitions, transient states, coherence networks, and protocol messages. In this section, state transition diagrams are presented for both \bedrock{} and the canonical protocol to further illustrate the subtle differences in how specific events are handled given initial cache block and coherence directory state.

In the state transition diagrams, arrows represent messages and circles represent caches or the directory. Each arrow is labeled with a number and the message name. The number indicates the sequence of messages in the depicted transaction. Each circle may be the directory (Dir) or a cache. A cache may be either the requester (Req), the current owner of the cache block (Owner), or a sharer with read-only permissions (Sharer).

\subsubsection{\bedrock{} Protocol}

\autoref{fig:bedrock-i-s}, \autoref{fig:bedrock-i-e}, \autoref{fig:bedrock-is-m}, and \autoref{fig:bedrock-fo-m} present the possible state transitions for the \bedrock{} protocol. These diagrams clearly illustrate the role that the coherence directory plays in orchestrating and completing all coherence requests. All commands are issued by the coherence directory and all responses are sent from the caches to the directory. Additionally, every transaction is closed by the cache sending a coherence acknowledgment (CohAck) response to the directory, which informs the directory that the transaction is complete. The coherence ack is required since \bedrock{} assumes unordered networks are used to implement the protocol.

Read requests that transition the requester from the Invalid (I) state to the Shared (S) state require either three or four phases. The first phase occurs when the requesting cache issues a request to the directory. The second phase is always the directory issuing one or more commands to caches. In the simplest case, the directory reads the requested block from memory and sends a data command to the requester. In cases where the block has an owner, the directory sends some combination of set state, transfer, and writeback commands as a single compound command message to the cache block owner, directing the owner to send the cache block data to the requester. If the directory records the block in the Exclusive (E) state, a writeback must be commanded to ensure that memory and the caches have consistent state. A writeback and the cache to cache transfer can be performed concurrently as the third phase of the transaction. The final phase of all transactions is the coherence acknowledgment phase.

A read request for a block currently in the Invalid (I) state in the directory results in a three-phase transaction comprising the request, data command to the requester from the directory, and the coherence acknowledgment to close the transaction.

Write requests transition the requester from Invalid (I), Shared (S), Owned (O), or Forward (F) states to Modified (M). These requests also require either three or four phases to complete. At the protocol level, commands issued to the requester, sharers, and owners, can, in most cases, occur in the same phase and concurrent to one another. The simplest transactions require three phases. The first phase occurs when the requesting cache issues a request to the directory. The second phase comprises the directory sending data or a coherence state update to the requester, possibly overlapped with the the directory commanding other caches to invalidate the target block. The third phase is the coherence ack from the requester to the directory. The four phase write transactions involve cache to cache transfers, similar to the four phase read request transactions. The second and third phases of these transactions involve the directory commanding the current owner to perform a cache to cache transfer followed by the data being transferred from the current owner to the requester.

\subsubsection{Canonical Protocol}

\autoref{fig:canonical-i-s}, \autoref{fig:canonical-i-e}, \autoref{fig:canonical-is-m}, and \autoref{fig:canonical-fo-m} present the possible state transitions for a canonical directory protocol. As implied by the use of only three coherence networks, the canonical directory protocol requires a maximum of three phases for all coherence transactions. Every transaction begins with the requester sending a coherence request to the coherence directory. Next, the coherence directory either replies directly to the requester with the requested cache block or upgraded permissions or forwards the request to the current cache block owner. The third phase of a transaction comprises data and invalidation messages traveling from other caches to the requesting cache on the response network.

For read transactions, two phases are required when the cache block data is sourced from memory while three phases are required to source cache block data from another cache. In these transactions, the third phase comprises a single data message. Similarly, only two phases are required for write transactions that source cache block data from memory via the directory or only require upgraded permissions from the directory while three phases are required to perform cache to cache transfers or to accumulate invalidations from the other sharer caches. Unlike read transactions, three-phase write transactions may involve multiple response messages being sent to the requesting cache. These messages include both the cache to cache data transfer as well as invalidation acknowledgment messages to inform the requester that all other caches in the system have relinquished permissions for the target block.

\subsubsection{Comparing Protocols}

\bedrock{} and the canonical directory protocol differ in a few important ways, which are illustrated by the protocol transition diagrams. First, \bedrock{}'s assumption of unordered networks necessitates the use of a fourth coherence network and an additional transaction phase for all transactions relative to the canonical protocol. This manifests as a second message from cache controller to coherence directory for every transaction in the form of the coherence acknowledgment response. A canonical protocol implemented using unordered networks may not necessitate the introduction of an additional coherence network or transaction phase, but it would require additional transient states in the protocol to manage race conditions that become possible when point-to-point ordering is not guaranteed by the coherence networks. The fourth phase does not necessarily add any latency to the transaction from the point of view of the requesting cache, since the requested block can be used as soon as the cache data (DATA) or set state and wakeup (STW) message arrives. The coherence acknowledgment message from requester to directory can be sent immediately after the cache processes the third-phase message and its transit over the response network occurs concurrently with the requester resuming execution. Subsequent transactions targeting the same block and originating from other caches may experience small processing delays if they arrive at the coherence directory prior to the coherence acknowledgment returning from the first requester. However, implementations may be able to recover concurrency or hide the coherence acknowledgment latency by overlapping initial processing of the subsequent transaction with waiting for the coherence acknowledgment to return.

Second, \bedrock{}'s decision to centralize state management at the directory results in the requesting cache controller receiving exactly one message to resolve every coherence transaction. This greatly simplifies the protocol specification and implementation at the cache controllers. There is no need for the cache controller to include complex logic to determine the next action to take or how many messages to wait for before closing a transaction, rather the single command or fill provides permissions, and if needed, cache block data that satisfies the request. The controller must only then respond with a coherence acknowledgment concurrent to completing the cache's request.

Third, the coherence directory in \bedrock{} is the recipient of all response messages. In \bedrock{}, the coherence directory exclusively manages the state transitions of the target cache block in all caches, simplifying the implementation of the cache controllers. Additionally, the directory implementation can control the degree of concurrency among invalidating sharers and fulfilling the request by either directly providing data and permissions or commanding a cache to cache transfer. However, the canonical protocol is unable to realize this type of intra-transaction message concurrency. In contrast, the canonical protocol requires the cache controller to manage the accumulation of response messages such as invalidations in order to determine when the requested cache block can safely enter a stable state. The complexity of managing protocol races at the cache controllers can be seen in the canonical protocol's specification tables for the cache controller \cite{nagarajan_primer_2020}. Requiring the cache controller to accumulate all responses, including invalidation acknowledgments, may introduce latency to the transaction as the requester waits for the possibly large number of messages. 

\begin{table}[t]\centering
\ra{1.3}
\begin{tabular}{@{}L{0.08\linewidth}L{0.15\linewidth}L{0.12\linewidth}L{0.57\linewidth}@{}}\toprule
Event & Cache State Transition & Directory State & Mathematical Model\\
\midrule
Load & I $\rightarrow$ S & S & $Req + Dir + Mem + Data + Ack$\\
& I $\rightarrow$ S & F, O & $Req + Dir + Cmd + Fill + Ack$\\
& I $\rightarrow$ S & E, M & $Req + Dir + Cmd + Fill + Ack$\\
& I $\rightarrow$ E & I & $Req + Dir + Mem + Data + Ack$\\
Store & I $\rightarrow$ M & I & $Req + Dir + Mem + Data + Ack$\\
& I, S $\rightarrow$ M & S & $Req + Dir + Max(Inv + InvAck, Mem + Data + Ack)$\\
& I $\rightarrow$ M & E, M & $Req + Dir + Cmd + Fill + Ack$\\
& I $\rightarrow$ M & F, O & $Req + Dir + Cmd + Fill + Ack$\\
& I $\rightarrow$ M & F, O & $Req + Dir + Max(Inv + InvAck, Cmd + Fill + Ack)$\\
& S $\rightarrow$ M & F, O & $Req + Dir + Max(Inv + InvAck, Cmd + Ack)$\\
& F, O $\rightarrow$ M & F, O & $Req + Dir + Cmd + Ack$\\
& F, O $\rightarrow$ M & F, O & $Req + Dir + Max(Inv + InvAck, Cmd + Ack)$\\
\bottomrule
\end{tabular}
\caption{\bedrock{} Protocol Mathematical Model - MOESIF}
\label{table:bedrock-model-moesif}
\end{table}

\begin{table}[t]\centering
\ra{1.3}
\begin{tabular}{@{}L{0.08\linewidth}L{0.15\linewidth}L{0.12\linewidth}L{0.57\linewidth}@{}}\toprule
Event & Cache State Transition & Directory State & Mathematical Model\\
\midrule
Load & I $\rightarrow$ S & S & $Req + Dir + Mem + Data$\\
& I $\rightarrow$ S & F, O & $Req + Dir + FwdGet + Data$\\
& I $\rightarrow$ S & E, M & $Req + Dir + FwdGet + Data$\\
& I $\rightarrow$ E & I & $Req + Dir + Mem + Data$\\
Store & I $\rightarrow$ M & I & $Req + Dir + Mem + Data$\\
& S $\rightarrow$ M & S & $Req + Dir + Mem + Data$\\
& I, S $\rightarrow$ M & S & $Req + Dir + Max(Mem + Data, Inv + InvAck)$\\
& I $\rightarrow$ M & E, M & $Req + Dir + FwdGet + Data$\\
& I $\rightarrow$ M & F, O & $Req + Dir + FwdGet + Data$\\
& I, S $\rightarrow$ M & F, O & $Req + Dir + Max(Inv + InvAck, FwdGet + Data)$\\
& F, O $\rightarrow$ M & F, O & $Req + Dir + AckCount(0)$\\
& F, O $\rightarrow$ M & F, O & $Req + Dir + Max(AckCount(N), Inv + InvAck)$\\
\bottomrule
\end{tabular}
\caption{Canonical Directory Protocol Mathematical Model - MOESIF}
\label{table:canonical-model-moesif}
\end{table}

\subsection{Mathematical Models}
\label{sec:bedrock-protocol-analysis-models}

Using the protocol tables and processing diagrams, it is possible to derive mathematical models of the \bedrock{} and canonical coherence protocols. \autoref{table:bedrock-model-moesif} provides mathematical formulae describing coherence transactions for the \bedrock{} protocol. \autoref{table:canonical-model-moesif} provides mathematical formulae for the same coherence transactions but in the canonical directory protocol. These models follow directly from the transaction processing diagrams shown above.

Examining the formulae, it is clear that \bedrock{} incurs additional latency per transaction to transmit the coherence acknowledgment (CohAck) message from the requester to the directory due to the extra transaction phase required by the protocol's assumption of unordered networks. However, the mathematical models are otherwise effectively equivalent for the two protocols. All commands issued from the \bedrock{} directory can be issued and processed concurrently in the system, as can response messages such as writebacks or invalidations from the owner or sharers, respectively, to the directory. Likewise, commands issued from the directory in the canonical protocol can be executed concurrently in the system. In the most complex transactions, the overall latency is likely determined by the relative costs of these concurrent operations. For \bedrock{} this means that the latency incurred at the directory is typically the worst-case latency of receiving responses to the various commands it has issued while for the canonical protocol the requester experiences the worst-case latency of receiving responses. Thus, \bedrock{} demonstrates a tradeoff in protocol design that can reduce the latency experienced at the cache controllers at the expense of latency incurred by the coherence directory.

These models demonstrate that at the highest level descriptions of the two protocols, it is not clear which may perform better in a given implementation. The impact of particular command and response latencies may manifest differently at the cache controllers and coherence directories in each of the protocols. In \bedrock{}, the requester does not incur the CohAck latency as a cost prior to using the block since this message can be sent concurrently to the cache using the block. In the canonical protocol, the directory does not incur the cost of processing certain response messages as they are sent directly to the requesting cache controller. Further, the latencies experienced for certain phases depends on the amount of contention and sharing for the requested cache block. The models reveal the importance of implementation decisions in determining actual protocol performance. While different protocols exhibit unique latency savings or costs, whether those costs manifest in a real system depends on the realized protocol implementation.
\section{Conclusion}
\label{sec:bedrock-conclusion}

The \bedrock{} cache coherence protocol presented in this chapter is an easy to implement directory-based cache coherence protocol that is well-suited for small- to medium-scale shared-memory multicore processors. The preceding description of \bedrock{} provides a complete specification of the protocol in tabular form along with a description of the necessary system components, coherence states, coherence networks, and coherence messages required by the protocol. \bedrock{}'s emphasis on reducing protocol complexity results in a race-free protocol that requires significantly less verification effort than a canonical directory-based protocol. Additionally, removing races from the protocol results in a simple and straightforward cache controller protocol specification and the elimination of transient states from the entire protocol. A comparison of \bedrock{} to a canonical directory-based protocol shows that \bedrock{} requires an additional transaction phase and may provide less per-block transaction concurrency compared to the canonical protocol. However, since the cache coherence directory explicitly manages all coherence state transitions in the system and is the destination of all response messages, the cache controllers are not required to wait for messages from other caches except during cache to cache transfers. The specification and analysis of \bedrock{} reveal that the design tradeoffs of cache coherence protocols are not always straightforward and different choices may be appropriate for different systems.

\chapter{BlackParrot-BedRock}
\label{chap:bp-bedrock}

Cache coherence protocols define the semantics of how access permissions for a particular block (or word) of memory are acquired and relinquished throughout execution. However, the protocol itself provides little insight into the actual performance and implementation implications of the realized protocol in a shared-memory multicore processor design.

A primary contribution of this dissertation is the open-source implementation of the \bedrock{} cache coherence protocol within the \blackparrot{} shared-memory multicore processor, called \blackparrot{}-\bedrock{} (\bpbedrock{}). In this chapter, the architecture and microarchitecture of \bpbedrock{} is described in detail. First, \autoref{sec:bp-bedrock-stream} describes the on-chip network protocol underpinning the \bedrock{} coherence networks. Next, \autoref{sec:bp-bedrock-lce} describes the design of the cache coherence controller or Local Cache Engine (LCE). \autoref{sec:bp-bedrock-cce-dir}, \autoref{sec:bp-bedrock-cce-fsm}, and \autoref{sec:bp-bedrock-cce-ucode} detail the architecture and microarchitecture design of \bedrock{}'s cache coherence directory and the fixed-function and programmable directory controllers. Lastly, \autoref{sec:bp-bedrock-cce-perf} and \autoref{sec:bp-bedrock-cce-area} compare the two coherence directory controller designs by examining protocol processing performance and area and resource implementation costs.

The implementation of \bpbedrock{} is written in SystemVerilog\cite{systemverilog-1800-2017}, fully open-source, and available at \url{https://github.com/black-parrot/black-parrot}.

\section{BP-BedRock Stream Protocol}
\label{sec:bp-bedrock-stream}

\begin{figure}[t]
	\centering
	\includegraphics[width=0.30\linewidth]{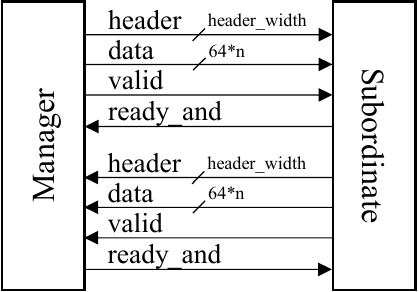}
	\caption{BP-Bedrock Stream Protocol}
	\label{fig:bp-bedrock-stream}
\end{figure}

In an effort to reduce implementation complexity, all \bedrock{} coherence network and \blackparrot{} memory network protocol messages are carried across on-chip networks utilizing a common message format and handshaking protocol called the \bpbedrock{} Stream protocol. \autoref{fig:bp-bedrock-stream} depicts the signals found in the \bpbedrock{} Stream message protocol. Every protocol message comprises one or more message \emph{beats} that transmits both header and data information in parallel. The message handshaking uses a \emph{ready\&valid}\cite{taylor2018} protocol that exchanges the header and data information from manager (sender) to subordinate (receiver) during the cycle in which both the sender asserts the single-bit \emph{valid} signal and the receiver asserts the single-bit \emph{ready\_and} signal.

\begin{lstlisting}[float,floatplacement=H,caption=BedRock Message Header Macro, label=lst:bedrock_header]
`define declare_bp_bedrock_header_s(addr_width_mp, payload_mp, name_mp) \
  typedef struct packed                                    \
  {                                                        \
    payload_mp                                   payload;  \
    bp_bedrock_msg_size_e                        size;     \
    logic [addr_width_mp-1:0]                    addr;     \
    bp_bedrock_wr_subop_e                        subop;    \
    bp_bedrock_msg_u                             msg_type; \
  } bp_bedrock_``name_mp``_header_s;
\end{lstlisting}

\autoref{lst:bedrock_header} shows the SystemVerilog macro code used to define the header component of a \bpbedrock{} coherence network message. Each header comprises a message type, a write sub-operation if necessary, the address associated with the message (or transaction), a message size, and a network-opaque payload that is customized for each particular protocol network.

The message type field is a union that holds a protocol network-specific message type, as described in \autoref{sec:bedrock-protocol} for the \bedrock{} coherence networks or a read, write, or atomic type for a memory network message. The write sub-operation field specifies whether a write request message on the request network is a standard write operation or the type of atomic operation that generated the request. The address field is typically either a cache block- or data word-aligned (32b- or 64b-aligned) address for the block containing the address that caused a cache miss and initiated a coherence transaction. Valid message sizes are between 1 and 128 bytes, by powers of two. The payload field is network-specific. For some networks, this field carries metadata about the transaction while on other networks it contains information that augments the other fields and informs the message receiver how to process a particular message. For example, command network messages that initiate cache to cache data transfers use the payload field to define the which cache is the destination of the transfer and the coherence state that the block for the transfer destination.

\subsection{Stream Pumps}

The \bpbedrock{} implementation of the Stream protocol relies heavily on modules called Stream Pumps to interface protocol processing logic with the protocol network interfaces. Stream pumps are implemented for both message send and receive, or output and input, respectively. Both types of stream pumps provide minimal message buffering to manage backpressure and to decouple the network interface and protocol processing logic. Stream pumps also act as a \emph{gearbox} that can convert message data widths between the network interface and protocol logic, allowing the protocol logic to operate with whichever per-beat data width is most optimal while allowing the on-chip network implementation to be independently sized for system-level power, performance, and area considerations.

The key logical benefit of stream pumps is in the interface they expose to the protocol processing logic. The stream pump generates \emph{new}, \emph{last}, \emph{critical}, and \emph{address} signals in addition to message header and data signals that can be examined by the protocol processing logic to easily take action at important points in messaging processing. The new signal is asserted on the first beat of a message, the last signal is asserted on the final beat of a message, and the critical signal is asserted on the beat of the message that contains the data specified by the message's address. In multi-beat messages that use a word-aligned address, the critical signal may be asserted on any beat. The critical signal enables easy implementation of \emph{critical-word-first} data return for caches, which is a commonly desired performance optimization. The address signal provides the effective per-beat, data channel width-aligned address for every beat in the message, which is incredibly useful when interacting with cache or memory data storage elements. Addressing utilizes a message-size aligned wrapping computation, similar to the WRAP burst type defined in the AMBA AXI4 protocol\cite{axi4}.
\section{Local Cache Engine (LCE)}
\label{sec:bp-bedrock-lce}

\begin{figure}[t]
	\centering
	\includegraphics[width=0.95\linewidth]{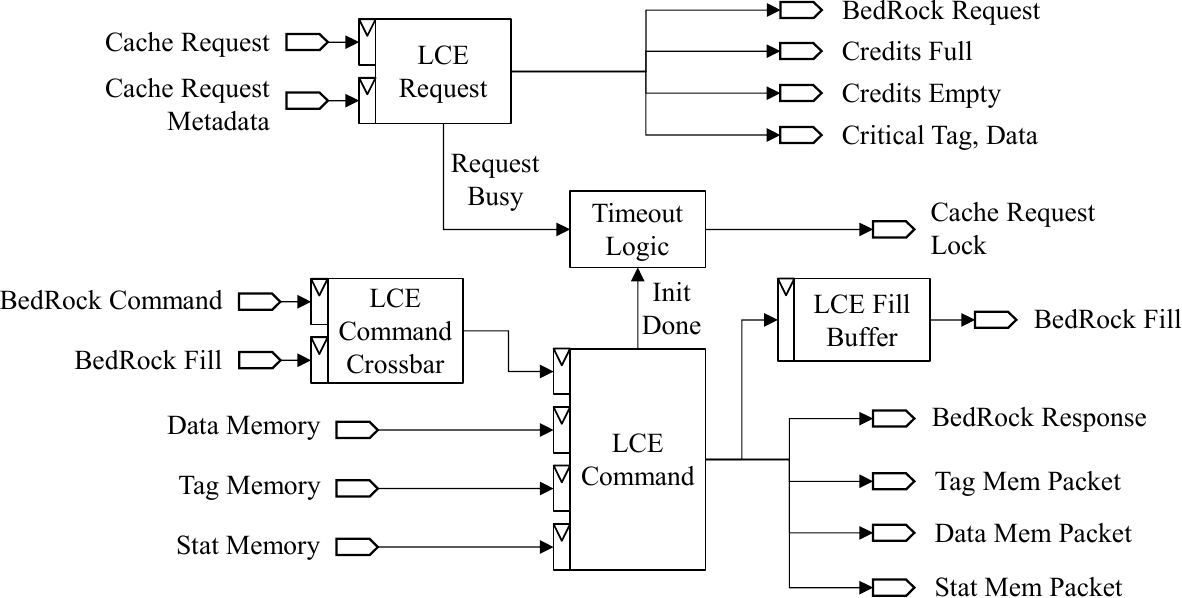}
	\caption{\bpbedrock{} LCE Block Diagram}
	\label{fig:lce-bd}
\end{figure}

The cache coherence controller in \bedrock{} is called a Local Cache Engine (LCE). In \bpbedrock{} there is one LCE attached to each private L1 cache in the multicore processor. The LCE processes cache requests and manages cache coherence for the cache by participating in the \bpbedrock{} coherence system. As discussed in \autoref{sec:background-blackparrot}, the cache and LCE are connected through the cache engine interface, and the LCE connects to the \bedrock{} coherence network using the \bedrock{} Request, Command, Fill, and Response networks.

\autoref{fig:lce-bd} shows a block diagram for the \bpbedrock{} LCE design. The LCE has two concurrently executing finite state machines (FSM) that are responsible for processing cache requests from the attached cache and coherence network messages arriving at the LCE. The two state machines are called the Request FSM and Command FSM, and each FSM is implemented in a separate SystemVerilog module. The top-level LCE module instantiates the Request and Command modules. The LCE also includes timeout logic to guarantee forward progress of the \bpbedrock{} coherence system as a whole and arbitration logic for the cache fill interface ports. The Fill and Command networks are multiplexed into a single interleave stream of commands through the LCE Command Crossbar before being processed by the Command FSM.

The \bpbedrock{} LCE supports cacheable, uncacheable, and atomic read-modify-write operations to lower levels of the memory hierarchy. As discussed in \autoref{sec:background-blackparrot}, the \bpbedrock{} L1 caches are blocking for cacheable requests, uncacheable loads, and atomics, and non-blocking for uncacheable stores. The LCE implementation supports identical behavior. \autoref{fig:lce-req-fsm} and \autoref{fig:lce-cmd-fsm} provide state machine diagrams for the two LCE state machines and are explained in detail below. The initial states for each FSM are shaded blue. States with pink outlines may require more than a single cycle to execute, depending on the type of request or message being processed by the state machine.

Depending on execution behavior, the type of message or event being processed, or resource contention, it is generally possible for any state to take one or more cycles to execute. States outlined in black are expected to take a single cycle to execute under ideal no-contention execution, while states outlined in pink/magenta are commonly expected to take more than one cycle to execute.

\subsection{Request Processing} \label{sec:lce_req}

\begin{figure}[t]
	\centering
	\includegraphics[width=0.95\linewidth]{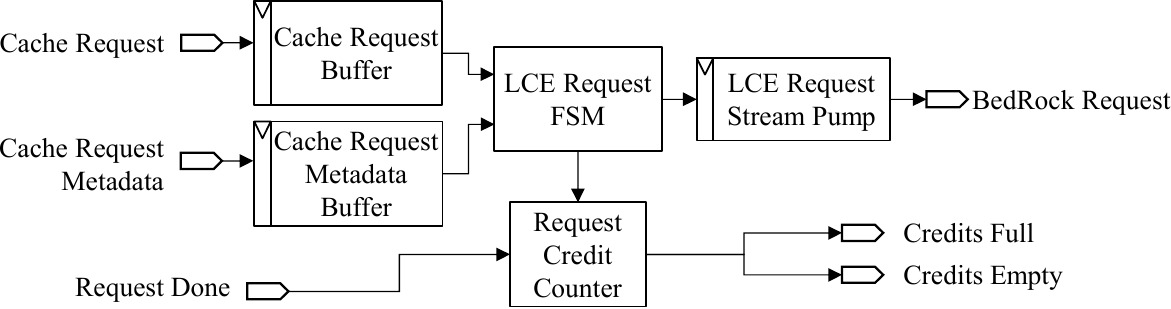}
	\caption{\bpbedrock{} LCE Request Block Diagram}
	\label{fig:lce-req-bd}
\end{figure}

The \bpbedrock{} LCE Request logic processes cache requests arriving on the cache engine request interface. Each cache request results in a single \bedrock{} Request message sent from the LCE to a CCE. \autoref{fig:lce-req-bd} shows the organization of the LCE Request processing module in the LCE. Cache requests and their metadata arriving from the \blackparrot{} L1 cache are buffered and then processed by the LCE Request FSM. The request processing state machine consumes one credit per request issued and limits the total number of outstanding requests to a design-parameterized credit limit. Backpressure is applied to the L1 cache via the cache request buffers and the credit-based flow control, and if the LCE is unable to accept a new cache miss request the \blackparrot{} cache pipeline will stall and attempt to replay the access at a later cycle. In the current \bpbedrock{} implementation, this counter only limits the number of uncacheable stores, as all other operations are blocking and limited to one per LCE.

\begin{figure}[t]
	\centering
	\includegraphics[width=0.40\linewidth]{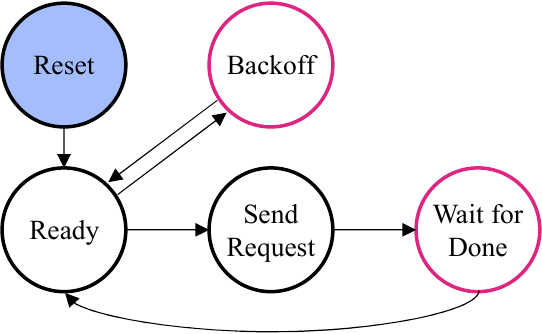}
	\caption{\bpbedrock{} LCE Request FSM}
	\label{fig:lce-req-fsm}
\end{figure}

The request state machine is a multi-cycle FSM, and is depicted in \autoref{fig:lce-req-fsm}. The state machine waits for new requests from the cache in the Ready state. Non-blocking uncached store operations are issued from the Ready state as soon as the request and its metadata are available. All blocking requests proceed to the Send Request state, which issues the appropriate LCE Request message to the CCE before transitioning to the Wait for Done state, which waits for a signal from the cache indicating the cache miss has been resolved and a new cache miss may be processed. The backoff state is used to apply backpressure to the L1 cache when the LCE is already processing a cache request.

\subsubsection{Request Occupancy}

\begin{table}[t]\centering
\ra{1.3}
\begin{tabular}{@{}L{0.21\linewidth}C{0.15\linewidth}C{0.2\linewidth}L{0.35\linewidth}@{}}\toprule
Cache Request & Occupancy (cycles) & Initiation Interval (cycles) & Description \\
\midrule
Cacheable Load, Cacheable Store & $2 + N$ & \textbf{\textendash} & Block-based cache miss, send cacheable miss request\\
Uncacheable Load & $2 + N$ & \textbf{\textendash} & 1, 2, 4, or 8-byte load, send uncached load request\\
Uncacheable Store & $1$ & $1$ & 1, 2, 4, or 8-byte store, send uncached store request\\
Uncacheable Atomic & $2 + N$ & \textbf{\textendash} & 4 or 8-byte atomic read-modify-write, send uncached amo request\\
\bottomrule
\end{tabular}
\caption{\bpbedrock{} LCE Request State Machine Occupancy}
\label{table:lce-req-occupancy}
\end{table}

\autoref{table:lce-req-occupancy} describes the request state machine's no-contention occupancy and throughput for each possible cache request, both given in cycles. The occupancy is the number of cycles required to process a cache request, beginning with the cycle that the cache request is valid and available for processing by the state machine. The initiation interval describes the achievable throughput for consecutive non-blocking operations, as the number of cycles required to issue an additional request of the same type.

In the request state machine, uncacheable stores have an occupancy of one cycle, and all other operations have an occupancy of two plus N cycles. Uncacheable stores require a single cycle to issue the request header and data on the outbound network via the request stream pump, and uncacheable store requests can be issued in consecutive cycles with an initiation interval of one cycle, or one request every cycle. Cacheable requests, uncacheable loads, and uncacheable atomics all require two cycles to receive the request from the cache and issue the coherence request message via the request stream pump. These requests require an additional N cycles of waiting for the transaction to complete before a new transaction can be processed.

The \blackparrot{} L1 cache provides the request metadata in the cycle following the request packet, however the \bpbedrock{} LCE actually supports receiving the metadata as soon as the same cycle as the request packet. This one cycle of latency is accounted for by registering the valid cache request in the Ready state and then both receiving the valid metadata and sending the outbound coherence requst in the Send Request state in the same cycle. Thus, the extra cycle of occupancy is a consequence of the \bpbedrock{} L1 cache design.

\subsection{Command Processing} \label{sec:lce_cmd}

\begin{figure}[t]
	\centering
	\includegraphics[width=0.95\linewidth]{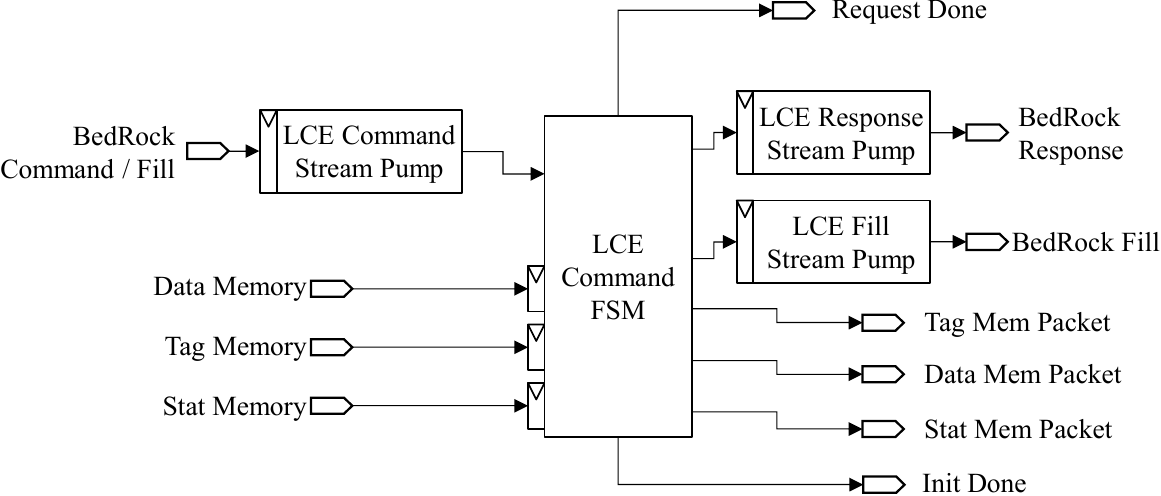}
	\caption{\bpbedrock{} LCE Command Block Diagram}
	\label{fig:lce-cmd-bd}
\end{figure}

The \bpbedrock{} LCE Command logic is responsible for processing all \bedrock{} Command and Fill network messages arriving at the LCE. \autoref{fig:lce-cmd-bd} depicts the organization of the LCE Command module implementation. This module includes the LCE Command state machine, stream pumps for sending \bpbedrock{} Response and Fill messages, a stream pump for receiving Command and Fill messages, and signals that implement the cache fill interfaces with the associated \blackparrot{} L1 cache.

\begin{figure}[t]
	\centering
	\includegraphics[width=0.5\linewidth]{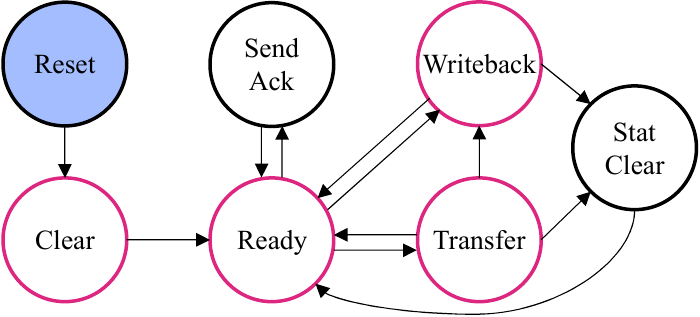}
	\caption{\bpbedrock{} LCE Command State Machine}
	\label{fig:lce-cmd-fsm}
\end{figure}

The \bpbedrock{} LCE Command state machine is a multi-cycle state machine responsible for processing all \bedrock{} Command network messages arriving at the LCE. \autoref{fig:lce-cmd-fsm} shows the states and possible state transitions in the command FSM. At system startup, the command FSM first clears the cache's tag and stat memories, ensuring that all cache blocks are invalidated, LRU and replacement information is reset, and all dirty bits are cleared. The state machine then transitions to its ready state and waits for \bedrock{} Command and Fill messages from the CCE or other LCEs. The command state machine is more complex than the request state machine, due to the variety of command messages carried on the \bedrock{} Command network. There are three broad command message classes that are processed by the state machine.

The first class of messages resolves coherence miss requests previously issued by the request FSM. These messages consume cache block or word data and respond with coherence acknowledgment messages to the CCE.

The second class of message is responsible for coherence protocol management, such as updating the state of a cache block currently present in the cache, initiating writebacks of dirty cache blocks, or performing cache to cache block transfers. Whether any or all of these actions occurs is determined by the command arriving at the LCE, however the state update always happens first, followed by the transfer, and lastly, the writeback. Processing these compound commands is handled atomically by the LCE, and performing the state update first ensures that any future accesses to the target cache block performed by the cache are serialized with the command.

The third class of messages resolves previously issued uncached load and store requests. These commands are processed exclusively in the Ready state, but may take multiple cycles to process. The arrival of either type of command results in the command module raising a request completion signal to the request module, which maintains the credit-based flow control logic.

\begin{table}[ht!]\centering
	\ra{1.3}
	\begin{tabular}{@{}L{0.24\linewidth}C{0.15\linewidth}L{0.55\linewidth}@{}}\toprule
		Message & Occupancy (cycles) & Description\\
		\midrule
		Sync & $1$ & Increment sync received counter, send sync ack response \\
		Set Clear & $1$ & Invalidate all blocks in cache set specified by address \\
		
        Set State & $1$ & Write state to tag memory \\
		Set State \& Wakeup & $2$ & Write state to tag memory and send coherence ack response \\

        Writeback (clean) & $2$ & Read stat memory, send null writeback response \\
        Writeback (dirty) & $2 + N$ & Read stat memory, read data memory, and send writeback response\\
        
        Set State \& Writeback (clean) & $2$ & Write tag to tag memory, read stat memory, and send null writeback response\\
		Set State \& Writeback (dirty) & $2 + N$ & Read stat memory and write state to tag memory, read data memory, and send writeback response\\

        Invalidate & $1$ & Invalidate block by writing state I to tag memory \\
        
        Data & $1 + N$ & Write tag and coherence state to tag memory, data to data memory, signal request complete, and send coherence ack response\\
		
		Transfer & $1 + N$ & Read data memory and send fill data message to another LCE\\
		Set State \& Transfer & $1 + N$ & Write state to tag memory, read data memory, and send fill data message to another LCE\\
		Set State \& Transfer \& Writeback (clean) & $2 + N$ & Write state to tag memory, read data memory, send fill data message to another LCE, and send null writeback response\\
		Set State \& Transfer \& Writeback (dirty) & $2 + (2*N)$ & Write state to tag memory, read data memory, send fill data message to another LCE, read stat and data memory, and send writeback response\\
		
		Uncached Data & $1$ & send data to cache and request complete to request FSM\\
		Uncached Store Done & $1$ & sink command and send request complete to request FSM\\
		\bottomrule
	\end{tabular}
	\caption{\bpbedrock{} LCE Command State Machine Occupancy}
	\label{table:lce-cmd-occupancy}
\end{table}

\subsubsection{Command Occupancy}

\autoref{table:lce-cmd-occupancy} details the no-contention state machine occupancy, in cycles, to process \bedrock{} command and fill messages. All arriving commands and data are buffered by the command stream pump. The command state machine begins processing commands as soon as the stream pump output is valid. The state machine is ready to process the next command immediately following completion of the previous command.

In general, all commands have a base cost of one cycle coming from initial processing in the Ready state. Sending a coherence acknowledgment or a null writeback response requires an additional cycle. Sending a cache to cache transfer on the fill network requires N cycles, as does consuming incoming data from a data command or fill network message. The value of N is equal to the cache block width divided by the cache fill width and is the number of message beats required to transmit a complete cache block. A few commands require a final cycle to update the cache's stat memory, which tracks whether a cache block is clean or dirty.

The Sync command is used at system startup by the CCE to confirm that all LCEs are ready to execute the coherence protocol. The LCE increments a sync counter when the command is received and then responds with a Sync Ack response message. Processing a Sync command takes a single cycle. The Set Clear command invalidates all cache blocks in a single cache set, which is specified by the command address, and has an occupancy of one cycle. The state machine issues set clear operations to both the tag and stat memories, but sends no response to the CCE. The Set Clear command is currently unused by the \bpbedrock{} coherence protocols.

Set State and Set State \& Wakeup commands modify the coherence state of a cache block and have a base cost of one cycle to write the tag memory. Set State \& Wakeup requires a second cycle to send the coherence acknowledgment response message.

Data, Transfer, and Set State \& Transfer commands all require one plus N cycles to process and send responses. Data commands require N cycles to consume the arriving cache block data and write it to the cache before sending a coherence acknowledgment message to the CCE. Transfer and Set State \& Transfer commands require one cycle to read the cache's data memory and write the cache's tag memory, if needed, plus N cycles to transfer the cache block data on the outbound fill network via a stream pump.

The uncached commands finalize uncached load and store requests. Both commands have an occupancy of one cycle in the best case. Uncached Store Done commands inform the LCE that the a previously issued uncached store request has been committed to memory, and require a single cycle to process and send a request completion signal to the request state machine. Uncached data commands provide uncached load data to the cache alongside sending a request completion signal to the request state machine.

Writeback commands take either two or two plus N cycles, depending on whether the cache block is clean or dirty, respectively. One cycle is required to read the cache's stat and data memories. If the block is clean, a single cycle is then required to send the null writeback response. A dirty block requires N cycles to send the writeback of the block plus an additional cycle to update the cache's stat memory.

The remaining commands are effectively combinations of their components. Their occupancies can be computed using the general rules above and are shown in the table. These commands perform a combination of modifying the cache block state, sending a cache to cache transfer, and writing back a cache block to the CCE or sending a null writeback response. The order of operations matches the names of the commands, for example Set State \& Transfer \& Writeback commands first update the state of the target cache block before performing a cache to cache transfer and lastly performing a writeback, whether a null or dirty writeback.

\subsection{Address Alignment} \label{sec:lce_alignment}

The \bpbedrock{} LCE assumes that addresses and data are aligned, but the alignment requirements depend on the interface and message type. All data sent or received by the LCE is sent with little-endian ordering where the least significant byte of the load or store data is placed into the least significant byte of the data message. The data channel width of all four \bpbedrock{} coherence networks is the same at all LCEs. On the cache request interface, all addresses and data are aligned to the size of the request for both cacheable and uncacheable requests. On the \bpbedrock{} networks, uncached request and command messages require addresses and data to be naturally aligned to the size of the operation. All other messages are block-based and require that addresses and data are aligned to the coherence network data channel width.

The LCE implements critical-word first behavior and issues the coherence miss request by aligning the cache request address to the data channel width. Arriving Fill and Data commands are expected to return data beginning with the data channel width-aligned sub-block of cache block data that contains the coherence miss request address. In other words, cache block data may be left-rotated to make the data channel width-aligned sub-block containing the requested data the first data beat of the message. The width of the data channel is always at least 64-bits, which guarantees that the cache request data will arrive in the first data beat. If the data channel width is less than the cache block width, more than one data beat is required to send or receive data. Data is transmitted in data channel width sub-blocks, wrapping at the cache block boundary.

\section{Coherence Directory}
\label{sec:bp-bedrock-cce-dir}

\begin{figure}[t]
	\centering
	\includegraphics[width=0.9\linewidth]{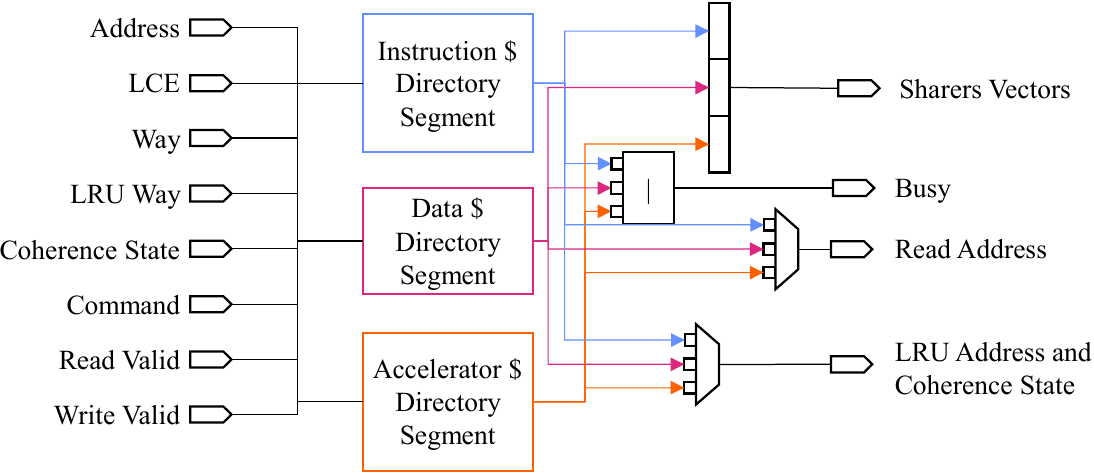}
	\caption{\bpbedrock{} Coherence Directory Architecture}
	\label{fig:cce-dir}
\end{figure}

The \bpbedrock{} implementation of the \bedrock{} cache coherence protocol \cite{bedrock} relies on a full-duplicate tag directory to track coherence state of every block cached by every cache participating in the coherence system. \bpbedrock{} includes multiple coherence engine implementations that utilize a single coherence directory implementation. The directory is constructed from multiple directory segments, and there is one directory segment per cache type in the system. \bpbedrock{}'s cache types are instruction, data, and optional coherent accelerator caches. \autoref{fig:cce-dir} shows a block diagram of the coherence directory. All inputs are routed to each directory segment with minimal modification, and each segment's outputs are either combined or multiplexed to the output ports of the coherence directory. If no coherent accelerators are present in the design, the accelerator cache directory segment is not instantiated.

\subsection{Tag Sets and Way Groups} \label{sec:cce_dir_wg}

\bpbedrock{} relies on the concepts of tag sets and way groups to track coherence state and enforce ordering among related coherence transactions. A tag set is the collection of address tag and coherence state for each cache block (way) within a single cache set of a single cache. All of the tag sets in the system are grouped into way groups that provide coherence transaction ordering for related addresses.

\subsubsection{Tracking State - Tag Sets}

\begin{figure}[t]
	\centering
	\includegraphics[width=0.50\linewidth]{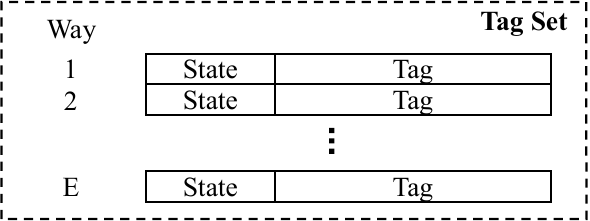}
	\caption{\bpbedrock{} Tag Set}
	\label{fig:cce-tag-set}
\end{figure}

The coherence state of every block cached in the system is tracked using the concept of a tag set, which is depicted in \autoref{fig:cce-tag-set}. A tag set is simply the collection of address tag and coherence state for each cache block (way) within a single cache set. The pair of address tag and coherence state is called a Tag Set Entry. The cache controller (LCE) has one tag set per cache set in the cache it manages, and typical implementations integrate the tracking of coherence state into the existing address tag metadata associated with each cache block. The coherence directory tracks all of the tag sets for every cache set at each cache controller in the coherence system. The directory's collection of tag sets comprises the full-duplicate tag directory.

At all times, the tag sets tracked at the coherence directory are considered to be the \emph{golden} copies of the tag sets, which hold the current state of the coherence system across all cached blocks. The directory updates its tag sets during request processing, primarily when sending commands that change the coherence state of blocks at the cache controllers\footnote{Since the \bpbedrock{} networks guarantee message delivery, the state updates at the directory occur when the directory sends coherence commands}. In contrast to the directory, the cache controllers maintain \emph{shadow} copies of the tag sets that are read-only by the cache controller\footnote{The sole exception to this read-only property is a silent upgrade from E to M to record that a write occurred and a cache block has become dirty.}. The cache controller tag sets are updated by coherence commands received by the controller.

\subsubsection{Ordering Transactions - Way Groups}

\begin{figure}[t]
	\centering
	\includegraphics[width=0.80\linewidth]{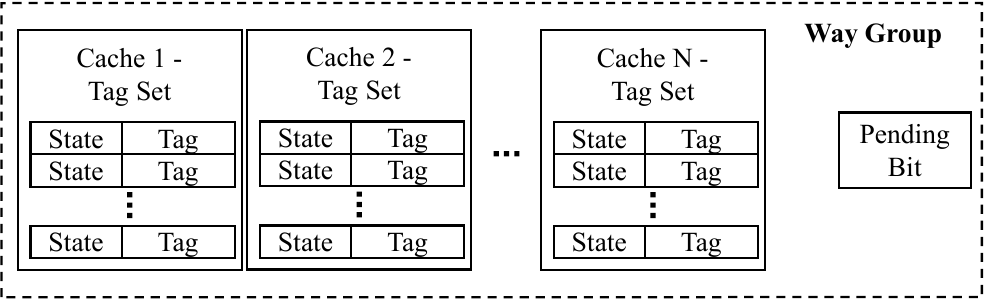}
	\caption{\bpbedrock{} Way Group}
	\label{fig:cce-way-group}
\end{figure}

\begin{figure}[t]
	\centering
	\includegraphics[width=0.90\linewidth]{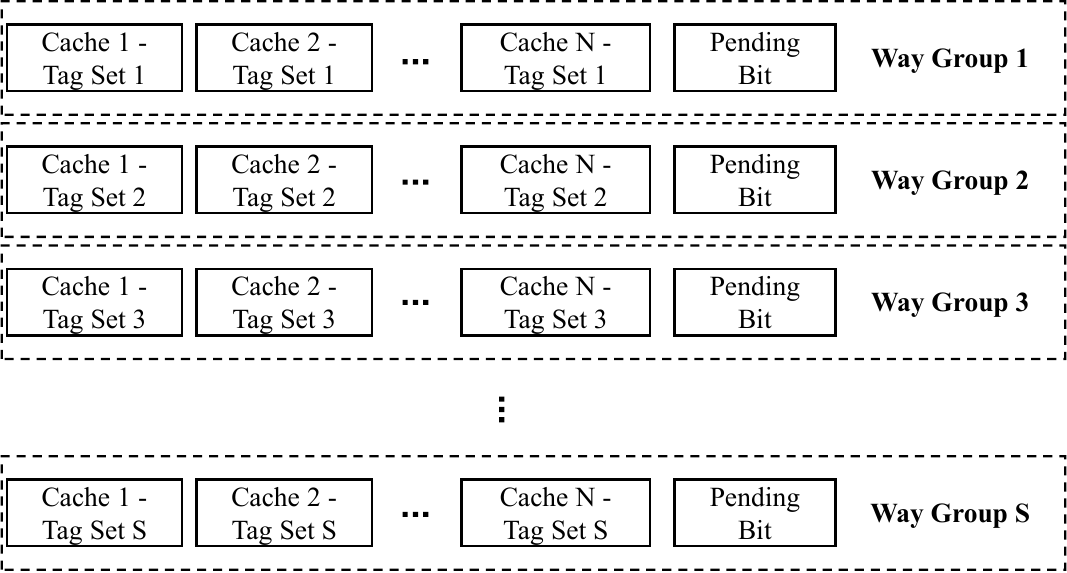}
	\caption{\bpbedrock{} Way Groups}
	\label{fig:cce-way-group-full}
\end{figure}

\autoref{fig:cce-way-group} depicts the contents of a way group, which contains one tag set from every cache controller in the coherence system and a Pending Bit. In a \bpbedrock{} system where every cache controller has the same organization and there are S sets per cache, there are S way groups at the coherence directory with way group X including tag set X from every cache controller. In other words, a cache block that maps to cache set X (equivalently, tag set X), is a member of way group X. \autoref{fig:cce-way-group-full} shows the way groups of a canonical system with N caches, S sets per cache, and S way groups.

The pending bit in each way group is used to enforce transaction ordering for requests that target the same way group. This bit is set when the coherence directory begins processing a coherence request targeting a cache block belonging to the associated way group and is cleared when the coherence acknowledgment (CohAck) message for the transaction is received by the directory. Each way group allows for a single active coherence transaction at a time. Any newly arriving request at the coherence directory must check the pending bit of the target way group and stall if the bit is set. Coherence transactions targeting separate way groups are, by definition, independent, and may be processed concurrently because all cache blocks that map to the same cache set are a member of the same way group. Thus requests to a single way group have no possibility of causing any coherence state change to a block in any other way group.

\subsubsection{Mapping Addresses to Way Groups}

\begin{figure}[t]
	\centering
	\includegraphics[width=0.85\linewidth]{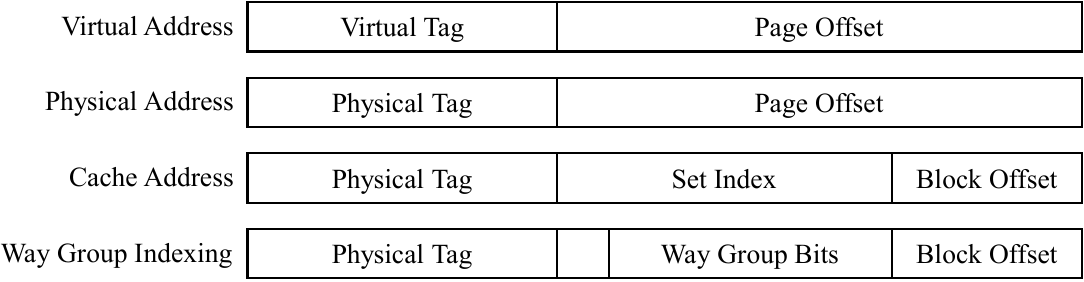}
	\caption{\bpbedrock{} Address Breakdown}
	\label{fig:cce-wg-address}
\end{figure}

\autoref{fig:cce-wg-address} shows how addresses are deconstructed at various stages of a memory access in \bpbedrock{}. A program issues a virtual address, which contains a tag and page offset. This address is translated to a Physical Address containing a physical tag and a page offset that is identical to the virtual address' page offset. The cache is accessed by further dividing the page offset field into set index and block offset fields. In the event of a cache miss, the LCE issues a miss request containing the physical address to the CCE that is responsible for the address.

A subset of the set index bits, called the way group bits with width $\log_{2}(way groups)$, is used as the input to a hash module that maps an address to a single way group. The hash module implements a semi-generic hash bank function that takes an address and constant number of banks as inputs and outputs a bank and index. Hash bank functions are commonly used to distribute addresses across cache banks, but in \bpbedrock{} one is used to spread addresses across CCEs. In \bpbedrock{}, the number of banks is the number of CCEs and the input address is the way group bits. The hash function then provides the ID of the CCE responsible for managing coherence for the address as the bank output. The index output provides a CCE-local index for the way group, which is used by the CCE but unused by the LCE. The hash function ensures that way groups are spread evenly among the CCEs, and the number of way groups managed by any given CCE will differ by at most one from the number of way groups managed by all other CCEs.

\subsubsection{Number of Way Groups}

\begin{figure}[ht]
	\centering
	\includegraphics[width=0.75\linewidth]{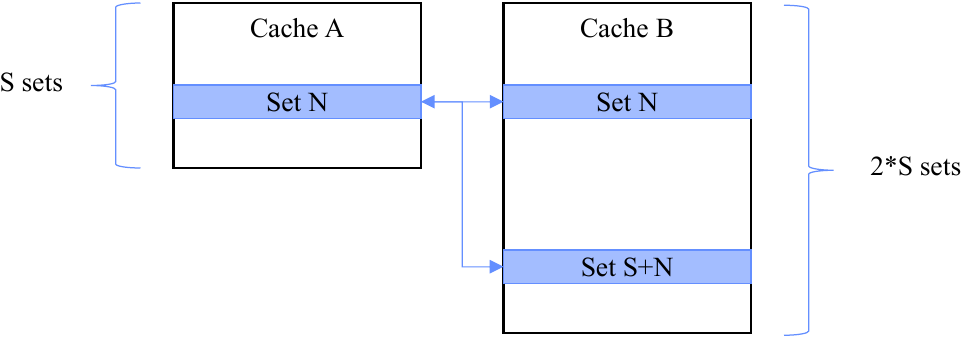}
	\caption{Mapping Cache Blocks to Cache Sets}
	\label{fig:cce-set-map}
\end{figure}

Every physical address maps to exactly one way group that is managed by exactly one CCE. The total number of way groups in a \bpbedrock{} system is equal to the \emph{minimum} number of cache sets across all cache types participating in coherence. There are three types of caches that participate in coherence: L1 instruction and data caches attached to each BlackParrot core and coherent accelerator caches. Addresses are related if they may map to the same cache block in \emph{any} cache participating in the coherence system.

\autoref{fig:cce-set-map} illustrates how cache blocks in two caches with different organizations may be related. Cache A has S sets and Cache B has 2*S sets. A cache block that maps to set N in Cache A may map to either set N or S+N in Cache B. Conversely, a block that maps to either set N or S+N in Cache B will map to set N in Cache A. Therefore, the collection of related cache blocks (equivalently, addresses) in Cache A and B is the collection of blocks that map to set N in Cache A. Therefore, the number of way groups in \bpbedrock{} is computed as the \emph{minimum} number of cache sets across all cache organizations in the coherence system\footnote{Note that the number of sets in any cache must be a power-of-two, and there is a power-of-two relationship between the number of cache sets in any two caches.}.

\subsection{Coherence Directory Segment Architecture} \label{sec:cce_dir_segment_arch}

\begin{figure}[t]
	\centering
	\includegraphics[width=0.75\linewidth]{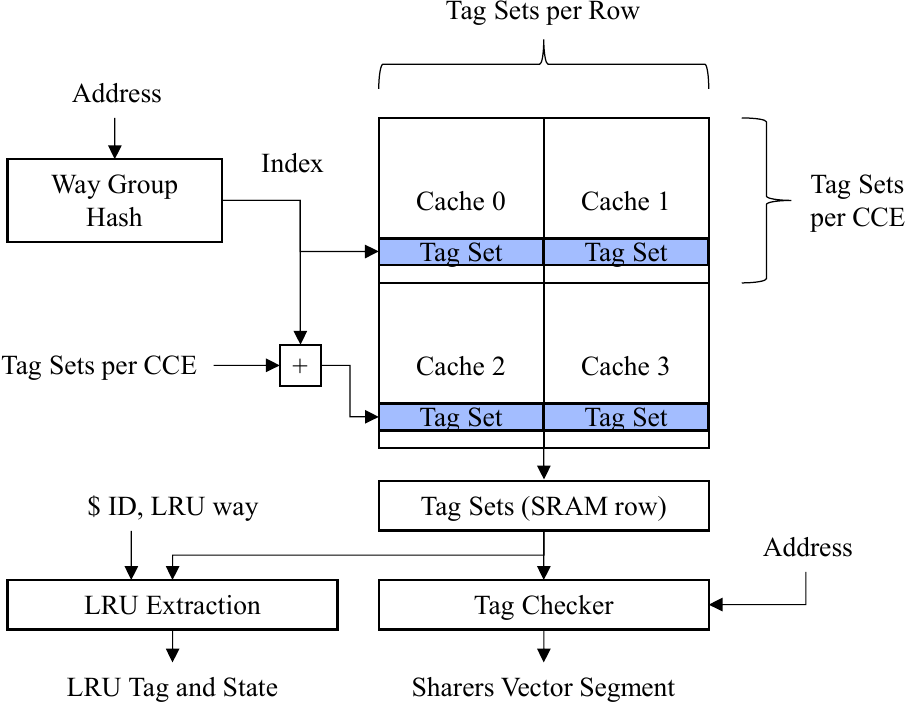}
	\caption{\bpbedrock{} Coherence Directory Segment}
	\label{fig:cce-dir-segment}
\end{figure}

\begin{table}[t]\centering
	\ra{1.3}
	\begin{tabular}{@{}L{0.3\linewidth}L{0.5\linewidth}@{}}\toprule
		Property & Value \\
		\midrule
		Tag Set Entry Width & $Coherence State Width + Address Tag Width$ \\
		Tag Set Width & $Tag Set Entry Width * Cache Associativity$ \\
		Tag Sets Per Row & $2$ \\
		Tag Sets Per CCE & $\lceil{Tag Sets Per Cache / Num CCE}\rceil$ \\
		Rows Per Set & $\lceil{Num Caches / Tag Sets Per Row}\rceil$ \\
		SRAM Rows & $Rows Per Set * Tag Sets Per CCE$ \\
		SRAM Size & $Tag Set Width * Tag Sets Per Row * SRAM Rows$ \\
		Entry Read Latency & $2$\\
		Way Group Read Latency & $Rows Per Set + 1$\\
		\bottomrule
	\end{tabular}
	\caption{\bpbedrock{} Directory Segment Properties}
	\label{table:cce-dir-segment}
\end{table}

\subsubsection{Directory Segment Storage}

\autoref{fig:cce-dir-segment} shows the organization of a \bpbedrock{} directory segment and \autoref{table:cce-dir-segment} describes the properties of a segment. Each segment stores a subset of all tag sets for all caches of a single type. The Tag Sets of all caches are spread evenly across the CCEs, and each directory segment allocates enough storage for $Tag Sets Per CCE$ sets for each cache it tracks. The tag sets of all caches tracked are stored as shown in \autoref{fig:cce-dir-segment}. The tag sets of a single cache stored in sequential rows, and each row stores tag sets of a single cache set from one or more caches. If the number of tag sets per row is less than the number of caches being tracked by the directory segment, additional blocks of $Tag Sets Per CCE$ rows are added to the directory to track all caches. The directory segment SRAM is a single-ported synchronous read-write memory.

The number of tag sets per row is a parameter of the directory segment, but it must be a power-of-two, which greatly simplifies the directory lookup logic. Prior physical design analysis has shown that a value of two tag sets per row is PPA-efficient for \bpbedrock{}. If the tag sets per row is not an even divisor of the total number of caches tracked by the segment, the last group of directory rows will not be fully utilized. For example, if the total number of caches tracked by the segment shown in \autoref{fig:cce-dir-segment} was only three, the section of directory rows outlined by the Cache 3 label would be unused. In this case, the selection of two tag sets per row minimizes unused storage space in the directory SRAM, and in the worst case the unused space is exactly the amount required to track a single cache. The location of a cache's tag set is easily computed using the cache's (LCE's) ID bits. The least significant bits determine the cache's position within a row, and the remaining bits index a lookup table that provides the row within the directory memory. The lookup table is computed at compilation and provides fast-access with minimal hardware overhead.

\subsubsection{Directory Operations and Access}

A small FSM controls each directory segment. Each segment supports reading a single tag set entry and reading all tag sets from all caches for a single cache set. Reading the tag sets across all caches is also called a way group read. Supported write operations are clearing an entire physical SRAM row, writing the coherence state to a single tag set entry, and writing the tag and coherence state to a single tag set entry.

The directory segment organization provides single cycle writes and multiple cycle reads. Write operations are immediately processed and initiated to the directory memory by the FSM, and a new write can be processed every cycle. Reading a single tag set entry requires two cycles. The read address is computed and presented to the directory memory in the first cycle, and the memory produces the read data in the second cycle, from which the requested tag set entry is extracted and output. Way group reads require two or more cycles, depending on the number of caches being tracked by the segment, and generate the LRU information and Sharers Vectors as outputs. A way group read is initialized by the FSM in the cycle that it receives the read command, and valid read data emerges from the memory beginning in the second cycle. As read data emerges, it is sent to both the LRU Extraction and Tag Checker modules for processing. The directory output becomes valid one cycle after the last read data emerges from the directory memory, and is valid for at least one cycle and until the next read or write operation occurs. The total latency of a way group read is equal to the number of $Rows Per Set + 1$.

\subsubsection{Tag Checker and Sharers Vectors}

\begin{figure}[t]
	\centering
	\includegraphics[width=0.65\linewidth]{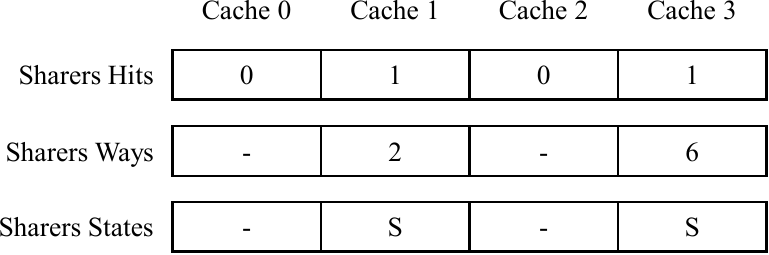}
	\caption{\bpbedrock{} Sharers Vectors}
	\label{fig:cce-dir-sharers}
\end{figure}

The Tag Checker module processes directory rows during way group reads, and produces the Sharers Vectors. The Sharers Vectors are a collection of three vector outputs containing a cache hit bit, coherence state, and cache way for each cache tracked by the segment. As discussed above, each directory row contains a full tag set from one or more caches. The tag checker examines each tag set and determines if the cache block containing the directory read address is present in any tag set entry within the tag set. If a matching block is found, the tag checker sets the cache hit bit for that cache and outputs the cache way within the tag set that the hit occurred in and the currently recorded coherence state of the matching block. \autoref{fig:cce-dir-sharers} depicts an example Sharers Vectors output for the directory segment of \autoref{fig:cce-dir-segment}, where a read to address A found the block containing A cached in a valid coherence state in caches 0 and 2. Cache 0 has the block in way 6 and state S, and cache 2 has the block in way 2 and state S.

\subsubsection{LRU Extraction}

The LRU extraction module processes directory rows during way group reads, and extracts the tag set entry at the LRU way for the specified LCE. If the directory segment requires multiple rows to store all cache's tag sets for a single cache set, the LRU Extraction module outputs the LRU information only for the row containing the requesting LCE's tag set. The module determines when to output valid information using the cache (LCE) ID and a row count provided by the directory.

\subsection{Coherence Directory Storage Overhead}

\begin{figure}[t]
	\centering
	\includegraphics[width=1\linewidth]{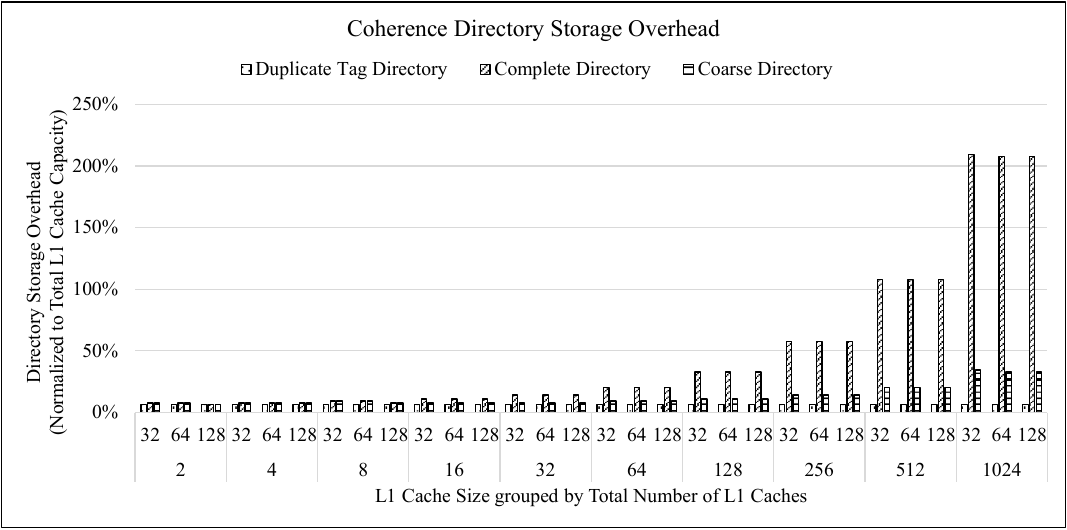}
	\caption{Coherence Directory Storage Overhead Comparison}
	\label{fig:cce-dir-overheads}
\end{figure}

\bpbedrock{} utilizes a standalone duplicate tag coherence directory, but this is not the only choice of directory organization that could have been made. Here, \bpbedrock{} is compared to standalone complete \cite{nagarajan_primer_2020} and coarse \cite{weber1992} directories to understand how the coherence storage directory overheads scale as the size of the system scales.

The analysis assumes 8-way associative L1 caches with 64-byte cache blocks, which are the defaults for \bpbedrock{}, with private instruction and data caches per core. The system uses 28-bit physical address tags and 3-bit coherence states. The number of caches is swept from 2 to 1024 by powers of two and L1 cache size is varied from 32 KiB, to 64 KiB, to 128 KiB. \bpbedrock{}'s duplicate tag directory stores the cache block tag and coherence state bits for every cache block in every L1 cache in the system. A standalone complete directory must store the cache block tag, coherence state, and owner ID bits in addition to a complete sharers bit vector for every cache block. The size of the sharers vector scales linearly with the number of caches in the system as each bit represents the presence of the cache block in a single cache. The coarse directory requires similar information as the complete directory per entry, however the sharers vector is encoded as one bit per N caches, where the bit is set if any of the corresponding N caches contains the block. In this analysis, a coarse vector bit can represent eight caches at most, i.e., there is an 8:1 encoding of sharers to bits.

A directory's storage overhead, assuming coherence is maintained across the L1 caches in the design using a standalone directory design, can be computed as
\begin{flalign}
Overhead = \frac{\lceil(\frac{Directory\ Bits)}{8}\rceil * Total\ L1\ Cache\ Blocks}{Total\ L1\ Cache\ Capacity}
\end{flalign}

where \emph{Directory Bits} is the number of bits required per directory entry as determined by the type of directory employed. \autoref{eqn:duplicate-tag-dir}, \autoref{eqn:complete-dir}, and \autoref{eqn:coarse-dir} list the formulae for computing the number of directory bits for a duplicate tag, complete, and coarse directory, respectively.

\begin{flalign}
Directory\ Bits =&\ Tag\ Bits + State\ Bits\label{eqn:duplicate-tag-dir}\\
Directory\ Bits =&\ Tag\ Bits + State\ Bits + Owner\ Bits + Sharers\ Vector\ Bits\label{eqn:complete-dir}\\
Directory\ Bits =&\ Tag\ Bits + State\ Bits + Owner\ Bits + Coarse\ Vector\ Bits\label{eqn:coarse-dir}
\end{flalign}

\autoref{fig:cce-dir-overheads} and \autoref{table:directory-overhead} show that \bpbedrock{}'s duplicate tag directories have a constant storage overhead of 6.25\%, which is less than both the complete and coarse directories. The coarse directory is able to achieve fairly low storage overheads up to about 64 caches. However, with larger numbers of caches, the size of the coarse sharers vector continues to grow as cache counts increase since each bit represents a maximum of eight caches. This overhead can be lessened by allowing each bit to represent a greater number of caches, however this coarsens the directory's sharers knowledge and results in more coherence messages to manage the coherence state of a block, many of which may be unnecessary when a block is cached by one or a small number of caches covered by each bit. The complete directory's overhead is modest through 16 caches, but becomes excessive at larger cache counts as the number of bits per sharers vector grows linearly with the cache count.

The constant storage overhead of \bpbedrock{}'s duplicate tag directory is greatly beneficial for the physical design of \bpbedrock{}, where a slice of the coherence directory is instantiated on each multicore tile and the multicore is constructed by instantiating tiles in a 2D mesh. The constant overhead results in a fixed directory size per tile, regardless of core count, which enables the use of a hierarchical, tile-based backend design flow for ASIC implementations.

\begin{table}[H]\centering
\begin{tabular}{@{}C{0.1\linewidth}@{}C{0.15\linewidth}@{}C{0.15\linewidth}@{}C{0.15\linewidth}@{}C{0.2\linewidth}@{}}
\toprule
Caches & Cache Size & \bpbedrock{} & Complete & Coarse (8:1) \\
\midrule
2 & 32 KiB & 6.25\% & 7.81\% & 7.81\%\\
& 64 KiB & 6.25\% & 7.81\% & 7.81\%\\
& 128 KiB & 6.25\% & 6.25\% & 6.25\%\\
\midrule
4 & 32 KiB & 6.25\% & 7.81\% & 7.81\%\\
& 64 KiB & 6.25\% & 7.81\% & 7.81\%\\
& 128 KiB & 6.25\% & 7.81\% & 7.81\%\\
\midrule
8 & 32 KiB & 6.25\% & 9.38\% & 9.38\%\\
& 64 KiB & 6.25\% & 9.38\% & 9.38\%\\
& 128 KiB & 6.25\% & 7.81\% & 7.81\%\\
\midrule
16 & 32 KiB & 6.25\% & 10.94\% & 7.81\%\\
& 64 KiB & 6.25\% & 10.94\% & 7.81\%\\
& 128 KiB & 6.25\% & 10.94\% & 7.81\%\\
\midrule
32 & 32 KiB & 6.25\% & 14.06\% & 7.81\%\\
& 64 KiB & 6.25\% & 14.06\% & 7.81\%\\
& 128 KiB & 6.25\% & 14.06\% & 7.81\%\\
\midrule
64 & 32 KiB & 6.25\% & 20.31\% & 9.38\%\\
& 64 KiB & 6.25\% & 20.31\% & 9.38\%\\
& 128 KiB & 6.25\% & 20.31\% & 9.38\%\\
\midrule
128 & 32 KiB & 6.25\% & 32.81\% & 10.94\%\\
& 64 KiB & 6.25\% & 32.81\% & 10.94\%\\
& 128 KiB & 6.25\% & 32.81\% & 10.94\%\\
\midrule
256 & 32 KiB & 6.25\% & 57.81\% & 14.06\%\\
& 64 KiB & 6.25\% & 57.81\% & 14.06\%\\
& 128 KiB & 6.25\% & 57.81\% & 14.06\%\\
\midrule
512 & 32 KiB & 6.25\% & 107.81\% & 20.31\%\\
& 64 KiB & 6.25\% & 107.81\% & 20.31\%\\
& 128 KiB & 6.25\% & 107.81\% & 20.31\%\\
\midrule
1024 & 32 KiB & 6.25\% & 209.38\% & 34.38\%\\
& 64 KiB & 6.25\% & 207.81\% & 32.81\%\\
& 128 KiB & 6.25\% & 207.81\% & 32.81\%\\
\bottomrule
\end{tabular}
\caption{Coherence Directory Storage Overhead Comparison}
\label{table:directory-overhead}
\end{table}

\section{Fixed-Function CCE (FSM CCE)}
\label{sec:bp-bedrock-cce-fsm}

\begin{figure}[t]
	\centering
	\includegraphics[width=0.95\linewidth]{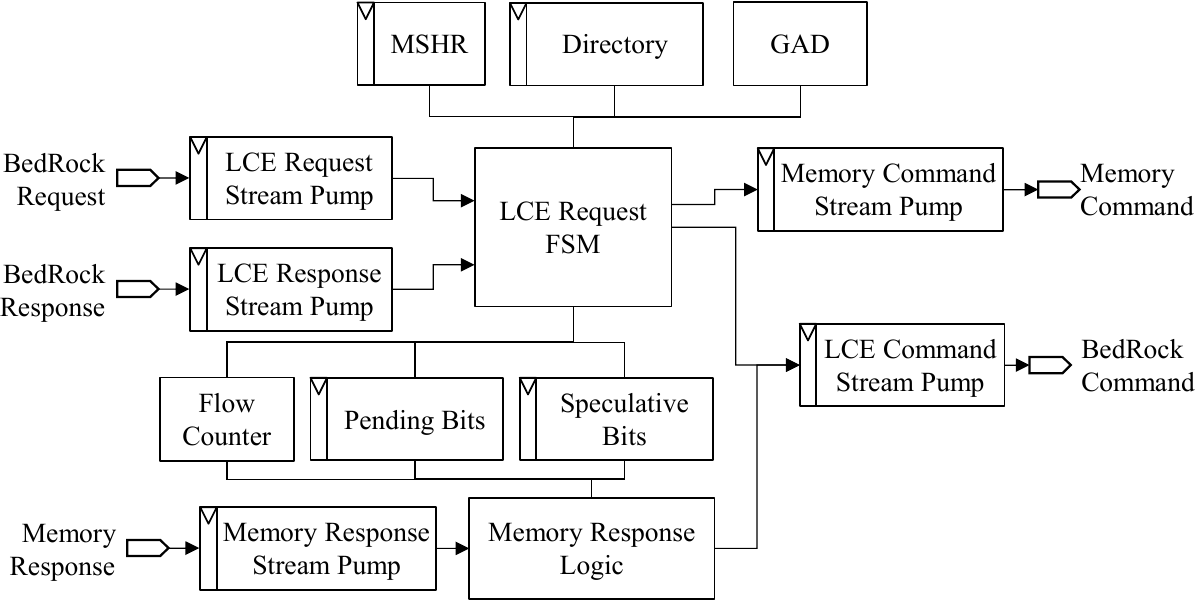}
	\caption{BP-Bedrock FSM CCE Block Diagram}
	\label{fig:fsm-cce-bd}
\end{figure}

The \bpbedrock{} Fixed-Function CCE (FSM CCE) is a hardware-based finite state machine (FSM) implementation of the \bedrock{} Cache Coherence Engine (CCE). The FSM CCE implements the \bedrock{} MOESIF cache coherence protocol with full support for coherent uncacheable loads and stores to cacheable memory and uncacheable access to uncacheable memory. The FSM CCE is the default coherence engine employed by \bpbedrock{}. \autoref{fig:fsm-cce-bd} shows a block diagram of the FSM CCE. The FSM CCE has an LCE Request state machine and Memory Response state machine that work together to execute the \bedrock{} coherence protocol. The CCE includes Speculative Bits to track speculative memory reads issued during request processing, Pending Bits to enforce coherence transaction ordering for each way group, and a Flow Counter to provide network flow control on the memory network. The LCE Request state machine instantiates the \bpbedrock{} coherence directory and a GAD module that processes the output of the coherence directory for use by the state machine logic. It also includes a Miss Status Handling Register (MSHR) that accumulates state during request processing. The remainder of this section describes these modules in detail before describing the functionality of the two state machines.

In this section, the terms cache and LCE are used interchangeably to reference a pair of LCE and attached cache. Strictly speaking, the cache processes requests from the \bpbedrock{} cores (or coherent accelerators) and issues miss requests to the LCE, while the LCE participates in the coherence protocol and maintains coherence for the cache by manipulating its state. Practically speaking, from the point of view of the CCE the LCE and its attached cache are a single entity\footnote{A non-\bpbedrock{} implementation of the \bedrock{} coherence protocol may implement the cache and LCE as a single module.}.

\subsection{GAD (Generate Auxiliary Directory Information)} \label{sec:cce_fsm_gad}

The \bpbedrock{} coherence protocol implementation enforces cache coherence for all requests that target cacheable memory. As the LCE Request state machine processes a request to cacheable memory it first reads the coherence directory to determine the state of the target cache block. In order to facilitate efficient control flow in the request FSM, the directory first processes the tags and coherence state stored in its memory into the Sharers Vectors and LRU Information, as explained in \autoref{sec:bp-bedrock-cce-dir}. This information is then further processed by the Generate Auxiliary Directory Information, or GAD, module for use by the request FSM.

The GAD module consumes the Sharers Vectors and LRU Information and outputs a a set of flag bits that can be used for efficient control flow decisions; the owner, location, and coherence state of the target cache block, if an owner exists; and the cache way of the target block within the requesting LCE's cache if it is already cached in a valid coherence state. The GAD module takes a single cycle to execute, but greatly reduces the complexity of control flow decisions in the request FSM.

\begin{table}[t]\centering
	\ra{1.3}
	\begin{tabular}{@{}L{0.25\linewidth}@{}L{0.75\linewidth}@{}}\toprule
		Output & Description \\
		\midrule
		Replacement Flag & Cache block in replacement way of requesting LCE needs to be evicted to make room for requested block\\
		Upgrade Flag & Cache block exists in a read-only coherence state at requesting LCE\\
		Cached Shared Flag & Block is cached in S in at least one other LCE\\
		Cached Exclusive Flag & Block is cached in E in at least one other LCE\\
		Cached Modified Flag & Block is cached in M in at least one other LCE\\
		Cached Owned Flag & Block is cached in O in at least one other LCE\\
		Cached Forward Flag & Block is cached in F in at least one other LCE\\
		Owner LCE & LCE ID of cache block owner\\
		Owner Way & Cache way of block at owner LCE\\
		Owner Coh State & Coherence state of block at owner LCE\\
		Req Addr Way & Cache way of block in requesting LCE\\
		\bottomrule
	\end{tabular}
	\caption{\bpbedrock{} GAD Outputs}
	\label{table:cce-fsm-gad}
\end{table}

\autoref{table:cce-fsm-gad} describes the outputs of the GAD module. There are seven single bit output flags that are used by the request FSM to make efficient control flow decisions when determining the specific actions required to maintain coherence. Five of these flags (Cache * Flag) are set if the requested cache block is cached in the specified MOESIF coherence state in \emph{any} cache other than the requesting cache. The Upgrade Flag is set if the requesting cache already has a valid copy of the requested cache block in a read-only state and the LCE request is a write-miss, indicating the cache needs read-write permissions for the block. The Replacement Flag is set if the replacement (LRU) way provided by the requesting cache is in a valid and possibly dirty state. This flag is also overloaded for uncacheable accesses to mean that the cache block containing the address in the uncacheable request exists in any valid state at the requesting cache block.

The Owner outputs from the GAD module provide the LCE ID, cache way, and coherence state of the requested cache block at the LCE that currently owns the block, if such an LCE exists. In the MOESIF protocol, a block cached in any of the E, M, O, or F states has a specified owner. If a block has an owner, the CCE is able to perform a cache-to-cache transfer of the block rather than performing a memory read of the requested block. Cache-to-cache transfers are often significantly lower latency than memory reads, which improves the overall performance of the \bpbedrock{} multicore system.

The Req Addr Way specifies the cache way of the target cache block within the requesting LCE's cache. This is used in combination with the upgrade\_flag when a cache is requesting write permissions for a block that it has cached in a read-only state. The Req Addr Way is also used for uncached requests when the block must be evicted from the requesting cache.

\subsection{Pending Bits} \label{sec:cce_fsm_pending}

The Pending Bits track outstanding coherence transactions for each way group managed by the CCE and serialize coherence requests targeting the same way group. There is one pending bit per way group. Each pending bit is implemented as a small counter, and a way group's pending bit is considered set if the counter is non-zero and unset if the counter is zero. When a new coherence request arrives at the CCE, the pending bits are checked and request processing begins only when the pending bit is not set. The pending bits serialize all coherence requests targeting the same way group.

The Pending Bits module implements independent read and write ports. Both ports require an input address that is sent through a hash function to determine the local way group index at the CCE. A write operations increments or decrements the pending bit counter by one, and a clear operation resets the counter to zero. Read operations output a single bit that is set if the counter is non-zero and unset if the counter is zero. Write to read forwarding is supported for concurrent read and writes.

The CCE increments the pending bit when it begins processing a new request targeting a cacheable block of memory and whenever a memory command is issued during request processing. Pending bits are decremented when memory responses are consumed, when coherence acknowledgment responses are received, and when the CCE finishes processing uncached requests to coherent memory. Uncached requests do not generate coherence acknowledgment messages and are considered complete when the CCE sends the load data or store complete command to the LCE.

\subsection{Speculative Bits} \label{sec:cce_fsm_spec}

The Speculative Bits record information about speculative memory reads issued to memory. When the CCE processes a new coherence request it may issue a speculative memory read of the target cache block in an effort to reduce the total request latency observed by the LCE. There is one Speculative Bits entry per way group. Each entry includes a coherence state, a speculative bit, a squash bit, and a forward-modified bit. At startup, all bits are cleared and the coherence state is set to Invalid.

The CCE sets an entry's speculative bit when it issues a speculative memory read for the way group, and clears the bit at the end of the request processing flow once it has determined the next coherence state and source of the cache block. The speculative memory read message payload contains the ID of the LCE whose request caused the memory read, the cache way the block will occupy, and the CCE's best guess of the final coherence state for the cache block.

When the CCE resolves the state and source of the block it may also set the other three fields of the entry. The squash bit is set if the speculative memory read is not needed to fulfill the request. This happens when the block will be provided by a cache to cache transfer or the requesting LCE already has the block and only needs upgrading coherence permissions. The forward-modified bit and coherence state field are set if the request will be fulfilled with the block read from memory, but the coherence state required differs from the state issued in the payload of the memory message. When the memory response returns and is processed, the CCE will observe the set forward-modified bit and replace the coherence state in the message payload with the coherence state stored in the speculative bits entry. If neither the squash or forward-modified bits are set, the CCE will forward the speculative memory read response to the LCE recorded in the payload as a Data command to fulfill the request.

\subsection{Miss Status Handling Register (MSHR)}
\label{sec:cce_fsm_mshr}

\begin{table}[t]\centering
	\ra{1.3}
	\begin{tabular}{@{}L{0.2\linewidth}@{}L{0.2\linewidth}@{}L{0.6\linewidth}@{}}
        \toprule
		Field & Source & Description\\
		\midrule
		Msg Type & LCE Request & Request message type\\
		Msg Sub Op & LCE Request & Request message sub-operation\\
		Msg Size & LCE Request & Request message size\\
		LCE ID & LCE Request & Requesting LCE ID\\
		Address & LCE Request & Physical address of request\\
		
        LRU Way ID & LCE Request & Replacement way from requesting LCE\\
        LRU Address & Directory & Physical address of block in LRU way at requesting LCE\\
        LRU Coh State & Directory & Coherence state of block in LRU way at requesting LCE\\
        
        Way ID & GAD & Cache way of block in requesting LCE\\
        
		Owner LCE & GAD & LCE ID of cache block owner\\
		Owner Way & GAD & Cache way of block at owner LCE\\
		Owner Coh State & GAD & Coherence state of block at owner LCE\\
		
		Flags & Multiple & Control flow flags\\
		
		Next Coh State & FSM & Next coherence state of requested block\\
		\bottomrule
	\end{tabular}
	\caption{\bpbedrock{} MSHR State}
	\label{table:cce-fsm-mshr}
\end{table}

The FSM CCE contains a Miss Status Handling Register (MSHR) that accumulates information related to the current LCE request. \autoref{table:cce-fsm-mshr} shows the fields of the MSHR, their source within the FSM CCE, and a brief description of each field. The information stored in the MSHR come from the LCE Request being processed, the coherence directory, the GAD module outputs, and the FSM logic. The MSHR is referenced throughout the FSM CCE's LCE request processing flow. Most fields are self-explanatory, and the rows of the table are ordered from top to bottom in approximately the order that the fields are populated. The first set of fields come from the LCE Request message header and include the requesting LCE, the request address, and the specific request message type and size. The cache way to use for a replacement, if required, is provided by the LRU Way ID field. Most of the remaining fields are generated by reading and processing the directory, indicated by the Directory and GAD sources. These include information about the owner LCE, if one exists, the address and coherence state of the block in the LRU way of the requesting cache, the cache way of the requested block in the requesting cache, and a subset of the control flow flags. The final field of the MSHR stores the next coherence state for the requested block at the requesting LCE. 

\begin{table}[t]\centering
	\ra{1.3}
	\begin{tabular}{@{}L{0.22\linewidth}@{}L{0.2\linewidth}@{}L{0.58\linewidth}@{}}
        \toprule
		Flag & Source & Description \\
		\midrule
		Write Not Read & LCE Request & Write Miss or Uncached Store request message type\\
		Uncached & LCE Request & Uncached request message type\\
		Non Exclusive & LCE Request & Non-Exclusive bit set in LCE Request\\
		Atomic & LCE Request & Atomic operation request message type\\
		Atomic No Return & LCE Request & Atomic no return bit set in request\\
		Cacheable Address & FSM & Request is to a cacheable address\\
		
        Null Writeback & LCE Response & LCE Response message is a Null Writeback\\
		
		Pending & FSM & Pending bit was set in last pending bit read\\
		Speculative & FSM & Speculative bit was set in last speculative bits read\\
		
		Cached Shared & GAD & Block is cached in S in at least one other LCE\\
		Cached Exclusive & GAD & Block is cached in E in at least one other LCE\\
		Cached Modified & GAD & Block is cached in M in at least one other LCE\\
		Cached Owned & GAD & Block is cached in O in at least one other LCE\\
		Cached Forward & GAD & Block is cached in F in at least one other LCE\\
		Replacement & GAD & Cache block in replacement way of requesting LCE needs to be evicted to make room for requested block\\
		Upgrade & GAD & Cache block exists in a read-only coherence state at requesting LCE\\
		\bottomrule
	\end{tabular}
	\caption{\bpbedrock{} MSHR Flags}
	\label{table:cce-fsm-mshr-flags}
\end{table}

\autoref{table:cce-fsm-mshr-flags} describes the control flow flags present in the MSHR. These flags are used to make efficient control flow decisions in the FSM CCE. The first set of flags come directly from the LCE Request message that is being processed. These indicate if the request is a write or read request, cached or uncached, atomic with or without a return value, and whether or not the Non-Exclusive hint bit is set in the request message. The cacheable address flag is derived from the LCE Request message address and is set if the requested address is in the cacheable memory address space. The null writeback flag records whether the last LCE writeback response message was a null writeback or a data-carrying writeback. The pending bit is set whenever a pending bit module read occurs, and indicates if the specified way group has an open coherence transaction. The speculative bit is set whenever the speculative bits module is read, and indicates if there is an unresolved speculative memory read outstanding for the specified way group. The remaining flags are generated by the GAD module as explained above in \autoref{sec:cce_fsm_gad}.

\subsection{Memory Response Logic}
\label{sec:cce_fsm_mem_resp}

\begin{figure}[t]
	\centering
	\includegraphics[width=0.45\linewidth]{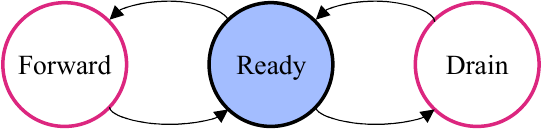}
	\caption{BP-Bedrock FSM CCE Memory Response Abstract State Machine}
	\label{fig:fsm-cce-mem-resp-fsm}
\end{figure}

The memory response logic processes \bpbedrock{} Memory Response messages returning to the CCE from the L2 cache or the I/O devices. Every memory command issued by the CCE results in a single memory response back to the CCE. \autoref{fig:fsm-cce-mem-resp-fsm} depicts a logical representation of the memory response logic as a three state FSM. In \bpbedrock{}, this state machine is implemented without any explicit encoding of the three discrete states shown in the figure. Logically, as each memory response arrives at the FSM CCE, it either forwards the message to the appropriate LCE or sinks the response from the memory network. Every memory response with an address in the cacheable memory address space also decrements the pending bit counter of the associated way group when it is consumed by the CCE. Memory responses are divided into speculative and non-speculative responses, as indicated by the speculative bit in the message header payload. This bit is set for all memory response messages generated by speculative memory read commands.

If the response is speculative, the response logic reads the speculative bits to determine if the LCE Request FSM has completed request processing and resolved the speculation. The response logic stalls until speculation has been resolved, and then processes the response according to the speculative bits entry. As explained in \autoref{sec:cce_fsm_spec}, there are three outcomes to speculation. The memory response may be squashed because it was not needed, it may be forwarded unmodified to the specified LCE using the coherence state supplied in the response message payload, or it may be forwarded to the specified LCE using the coherence state stored in the speculative bits entry. Squashing a memory response is achieved by draining the entire message from the memory response stream pump and sinking it in the CCE without sending any message to an LCE. Forwarding a memory response sends a \bedrock{} Command message on the outbound command network to the LCE specified in the memory response message's header payload data.

Non-speculative memory responses are either forwarded directly to the LCE specified in the response message header payload or are sunk at the CCE, depending on the message type. Uncached load (\lstinline{uc_rd}) and cached block read responses (\lstinline{rd}) are forwarded directly to the \bedrock{} command network using multi-beat data command messages. Uncached store responses (\lstinline{uc_wr}) are consumed by the response logic and transformed into a single-beat uncached store done command, which is sent to the LCE that initiated the store operation. Cacheable write (\lstinline{wr}) responses inform the CCE that a cache block writeback to memory has completed. The memory response logic sinks write responses as they arrive without sending any command messages to an LCE.

\subsubsection{Memory Response FSM Occupancy}

\begin{table}[t]\centering
	\ra{1.3}
	\begin{tabular}{@{}L{0.18\linewidth}@{}C{0.24\linewidth}@{}L{0.58\linewidth}@{}}
        \toprule
		Message & Occupancy (cycles) & Description \\
		\midrule
		Read & $N$ & Cache block read data; forward to LCE\\
		Write & $1$ & Cache block writeback complete; sink message\\
		Uncached Read & $N$ & Uncached load data; forward to LCE\\
		Uncached Write & $1$ & Uncached store commited to memory; send Uncached Store Done to LCE\\
		\bottomrule
	\end{tabular}
	\caption{\bpbedrock{} FSM CCE Memory Response State Machine Occupancy}
	\label{table:fsm-cce-mem-resp-occupancy}
\end{table}

\autoref{table:fsm-cce-mem-resp-occupancy} provides the no-contention memory response logic processing occupancies for the supported memory response message types. Write and Uncached Write responses each require a single cycle to process, which includes writing the pending bit and sending a command message, if required. Uncached Read and Read responses each require $N$ cycles to process. The number of cycles required to send the \bedrock{} command message is determined by the the data width of the \bedrock{} network channels and the cache block size in the coherence system.

\subsection{LCE Request FSM} \label{sec:cce_fsm_req}

\begin{figure}[t]
	\centering
	\includegraphics[width=0.9\linewidth]{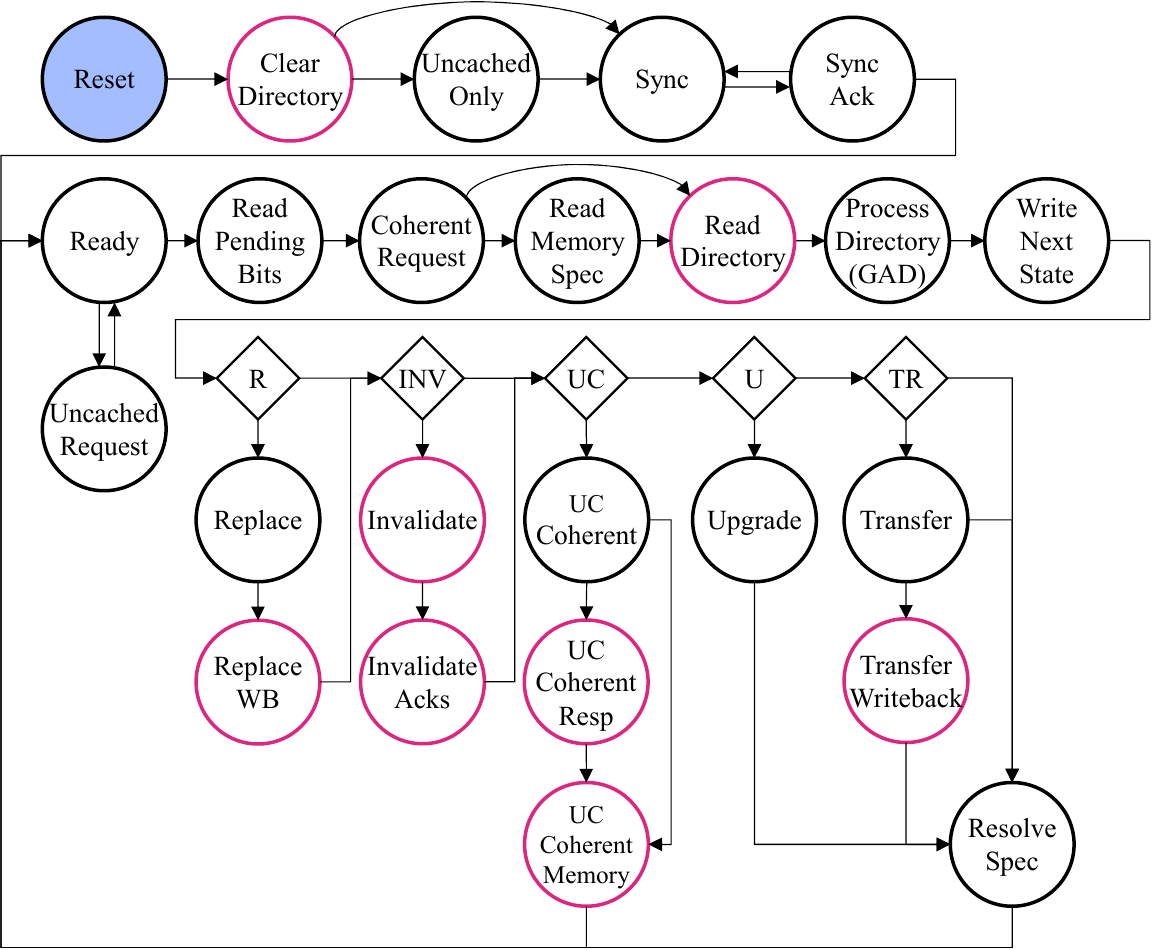}
	\caption{BP-Bedrock FSM LCE Request State Machine}
	\label{fig:fsm-cce-req-fsm}
\end{figure}

The LCE Request FSM processes \bpbedrock{} LCE Request messages, and is depicted in \autoref{fig:fsm-cce-req-fsm}. The Ready state, highlighted in blue, is the state machine's initial state. The CCE processes a single request at a time, and the request state machine runs without interruption. Every request is classified as either coherent or uncacheable based on the request address. All requests that target cacheable memory are coherent requests and invoke the cache coherence protocol. Requests to cacheable memory may be either cacheable or uncacheable, as issued by the LCE. Requests targeting uncacheable memory are limited to uncached load, store, and atomic operations, and are processed outside the coherence protocol.

An uncached request to uncacheable memory is received in the Ready state, which sets the request address, message type, and a few flags in the MSHR. The FSM then moves to the Uncached Request state which issues the uncached access to memory. Uncached stores may require additional cycles to issue message data to memory, which is handled by the Send Uncached Data state. The state machine returns to the Ready state after the memory command has been sent.

Cached and Uncached Requests to cacheable memory participate in the coherence protocol and require the CCE to read the directory and possibly adjust the coherence state of one or more LCEs. All requests to cacheable memory move through an initial set of states that read the pending bits, optionally issue a speculative memory read, read and process the directory, and update the coherence state of the requested block at the requesting LCE. The CCE first reads the pending bits to check if there is an active coherence transaction for the target way group. Once the previous transaction completes, the state machine consumes the request and increments the target way group's pending bit to open a new coherence transaction. Cacheable requests then issue a speculative memory read, which reduces total processing latency for requests that will be fulfilled from memory by overlapping the memory read with request processing. All requests next read the coherence directory, process the output using the GAD module, and determine the next state of the targeted block at the requesting LCE. The initial request processing finishes by updating the coherence directory with the block's next coherence state.

After the initial request processing completes, the request state machine branches to execute only those steps necessary to complete the request and maintain coherence. These decisions are represented by the diamonds in the state machine diagram. Each diamond has a latency of zero cycles and the next state selection occurs as each step completes. All coherent requests may require a cache block replacement and invalidations. Cacheable requests may require replacing the block specified in the LRU Way ID field of the MSHR to make room for the requested block, while uncacheable requests may require evicting the cache block containing the requested address and found in the cache way given by the Way ID field of the MSHR. Both types of requests may require invalidating the target cache block from all non-owner LCEs.

Cacheable requests are resolved by upgrading the coherence permissions of the block at the requesting LCE, initiating a cache to cache transfer of the block from the current owner to the requester, or fulfilling the cache block request from memory. Regardless of the action required, all cacheable requests are finalized in the resolve speculation state, which updates the speculative bits entry for the target way group. The speculative bits entry update communicates the action to take in the memory response state machine when processing the speculative memory read response. After the requesting LCE receives the necessary cache block data and coherence state, it sends a Coherence Acknowledgment response to the CCE. The CCE immediately sinks the coherence acknowledgment message and decrements the target way group's pending bit. Processing coherence acknowledgments happens outside of the request state machine because

Uncacheable requests to cacheable memory are resolved by first invalidating and possibly writing back the target cache block from the LCE that owns the block, if one exists and if the block may be dirty. The state machine then forwards the dirty writeback to memory, if needed, before issuing the uncached load or store operation to memory.

\subsubsection{LCE Request FSM Occupancy}

\begin{table}[t]\centering
	\ra{1.3}
	\begin{tabular}{@{}L{0.22\linewidth}@{}C{0.18\linewidth}@{}L{0.6\linewidth}@{}}
        \toprule
		State & Occupancy (cycles) & Description \\
		\midrule
		Read Directory & $(C/2) + 1$ & One cycle setup, plus one cycle per two cores\\
		Replacement Response & $1\ \textrm{or}\ N$ & One cycle for null writeback, N for dirty writeback\\
		Invalidation Commands & $S$ & One cycle per Sharer\\
		Invalidation Response & $S$ & One cycle per Sharer\\
		Transfer Writeback Response & $1\ \textrm{or}\ N$ & One cycle for null writeback, N for dirty writeback\\
		Uncached INV/WB Response & $1\ \textrm{or}\ N$ & One cycle for null writeback, N for dirty writeback\\
		Uncached Coherent Memory Command & $1\ \textrm{to}\ N$ & One cycle per data beat for store, or one cycle for load\\
		Send Uncached Data & $N - 1,\ max$ & One cycle per data beat\\
		All other states & $1$ & \\
		\bottomrule
	\end{tabular}
	\caption{\bpbedrock{} FSM CCE Request FSM State Machine Occupancy}
	\label{table:fsm-cce-req-state-occupancy}
\end{table}

\autoref{table:fsm-cce-req-state-occupancy} provides the no-contention, best-case processing occupancy for each state in the request state machine. As discussed in \autoref{sec:bp-bedrock-cce-dir}, the coherence directory is organized such that the read latency of a given directory segment is equal to the total number of directory rows required to store the tag sets for all caches plus one cycle to initiate the first read. In \bpbedrock{}, where two cache's tag sets are stored per directory row, the directory read latency is equal to the number of cores divided by two plus one cycles. Each directory segment stores the tag sets for a single cache type, and each core has one instruction and data cache. Therefore, there are two directory segments, and each segment tracks exactly C caches, where C is the number of cores.

The remaining states that require multiple cycles to execute involve message send and receive operations. An LCE request may require the CCE to invalidate all LCEs with target block in the Shared (S) state, which requires S cycles to send the invalidations and S cycles to consume the invalidation acknowledgment responses. If S is large, it is possible for the first invalidation responses to arrive while the final invalidation commands are being issued, in which case the request CCE is able to perform both a command send and response sink in the same cycle, thereby overlapping invalidation send and receive.

All states that wait for writeback responses from an LCE require either one or N cycles, where N is the number of data beats required to transmit the cache block data. Uncached store commands, whether targeting cacheable or uncacheable memory require at most N cycles to consume the uncached store request and convert it to an uncached memory write message.

States not listed in \autoref{table:fsm-cce-req-state-occupancy} require a single cycle to execute under no-contention, best-case conditions. These states collectively perform reads or writes to the pending or speculative bits, issue header-only command or memory messages, or write the coherence directory, which are all single-cycle operations.

% Base cost = 7 + (C/2) for Ready through Write Next State
\begin{table}[t]\centering
	\ra{1.3}
	\begin{tabular}{@{}L{0.1\linewidth}@{}C{0.15\linewidth}@{}C{0.2\linewidth}@{}C{0.3\linewidth}@{}L{0.25\linewidth}@{}}
        \toprule
		Request & LCE State & Directory State & Occupancy (cycles) & Notes\\
		\midrule
		\multirow{4}{*}{Read} & \multirow{4}{*}{I} & I & $8 + (C/2)$ & Block from Memory\\ % read mem
		& & S & $8 + (C/2)$ & Block from Memory\\ % read mem
		& & E (clean) & $10 + (C/2)$ & Transfer and Writeback \\ % ST_TR_WB
        & & E (dirty) & $9 + (C/2) + N$ & Transfer and Writeback \\ % ST_TR_WB
		& & M, O, F & $9 + (C/2)$ & Transfer\\ % ST_TR
		\midrule
		\multirow{4}{*}{Write} & \multirow{4}{*}{I} & I & $8 + (C/2)$ & Block from Memory\\ % read mem
		& & S & $8 + (C/2) + (2*S)$ & Block from Memory\\ % Inv all S, read mem
        & & E, M & $9 + (C/2)$ & Transfer\\ % ST_TR
		& & O, F & $9 + (C/2) + (2*S)$ & Invalidate and Transfer\\ % Inv all S, ST_TR
		\midrule
		\multirow{2}{*}{Write} & \multirow{2}{*}{S} & S & $9 + (C/2) + (2*(S-1))$ & Invalidate and Upgrade\\ % Inv all S, STW to Req
		& & O, F & $9 + (C/2) + (2*(S-1))$ & Invalidate and Upgrade\\ % Inv all S and owner, STW to Req
		\midrule
		Write & O, F & O, F & $9 + (C/2) + (2*S)$ & Invalidate and Upgrade\\ % Inv all S, STW to owner
		\bottomrule
	\end{tabular}
	\caption{\bpbedrock{} FSM CCE Request Occupancy}
	\label{table:fsm-cce-req-occupancy}
\end{table}

\autoref{table:fsm-cce-req-occupancy} provides the no-contention, best-case processing occupancy for various cacheable LCE requests given an initial coherence state for the target cache block. These occupancies are computed by progressing through the state machine in \autoref{fig:fsm-cce-req-fsm} and summing the latency of each state visited. All requests assume that a cache block replacement is not required. The addition of a replacement adds either two or $1 + N$ cycles to the processing latency for null and dirty writebacks, respectively.

From the request state machine diagram, all requests have a base processing cost of $8 + (C/2)$ cycles to move from the Ready state through the Write Next State state, which includes reading and processing the coherence directory, and then to resolve the speculative memory read. Read and write requests targeting a cache block in the Invalid state are fulfilled using the block from memory and require no additional cycles. Invalidating all caches in the S state requires a total of $(2*S)$ cycles. Issuing a Set State and Wakeup command (STW) to upgrade the cache block from read-only to read-write permissions at the requesting LCE requires a single cycle. The only remaining special case is a write request issued when the block is in the Shared (S) state at the directory and in the Shared state at the requesting LCE. In this case, the requesting LCE is not invalidated, slightly reducing the cost of invalidations.

Applying the formulae from \autoref{table:fsm-cce-req-occupancy}, the no-contention request processing occupancy at the CCE in an eight-core \bpbedrock{} multicore design is between 12 and 27 cycles. Directory reads require four cycles ($(C/2) = 4$), and there are at most seven sharer caches ($S <= 7$) that must be invalidated to complete any given transaction. Direct substitution into the derived formulae provide best-case estimates for request processing occupancy.

\subsection{Memory Consistency Model}

\bpbedrock{}'s coherence protocol belongs to a class of protocols that are called \emph{consistency-agnostic}. These protocols support the implementation of many different memory consistency models on top of the provided coherence protocol that maintains coherence for each shared-memory location. Importantly, cache coherence is maintained for each memory location and defines the semantics and ordering for accesses to a single location or address. Additionally, most consistency-agnostic cache coherence protocols, including \bedrock{}, operate invisibly to the programmer. Memory consistency defines the allowable orderings for memory accesses across all memory locations and is visible to the programmer. Due to the presence of microarchitectural optimizations such as cache write buffers or support for inter-access concurrency, it may be possible for certain memory accesses to be observed in an order that either differs from the given program order or that differs among observes (i.e., caches and cores) in the system.

The current \bpbedrock{} implementation realizes a memory consistency model called Sequential Consistency (SC). In this model, the execution and memory accesses of each processor core are observed in the order given by the program. Further, the execution of all operations across all processors is simply an interleaving of each processors sequentially consistent execution. All \bpbedrock{} multicore processors, wether using the FSM CCE described in this section, the microcode-programmable CCE described in the following section, or the hybrid coherence engine described in \autoref{chap:hybrid}, implement the SC memory consistency model. Readers are referred to the literature for additional details and formalizations of memory consistency. Nagarajan et al. provide an excellent overview for those with backgrounds in computer architecture \cite{nagarajan_primer_2020}.
\section{Microcode-Programmable CCE (ucode CCE)}
\label{sec:bp-bedrock-cce-ucode}

\begin{figure}[t]
	\centering
	\includegraphics[width=0.95\linewidth]{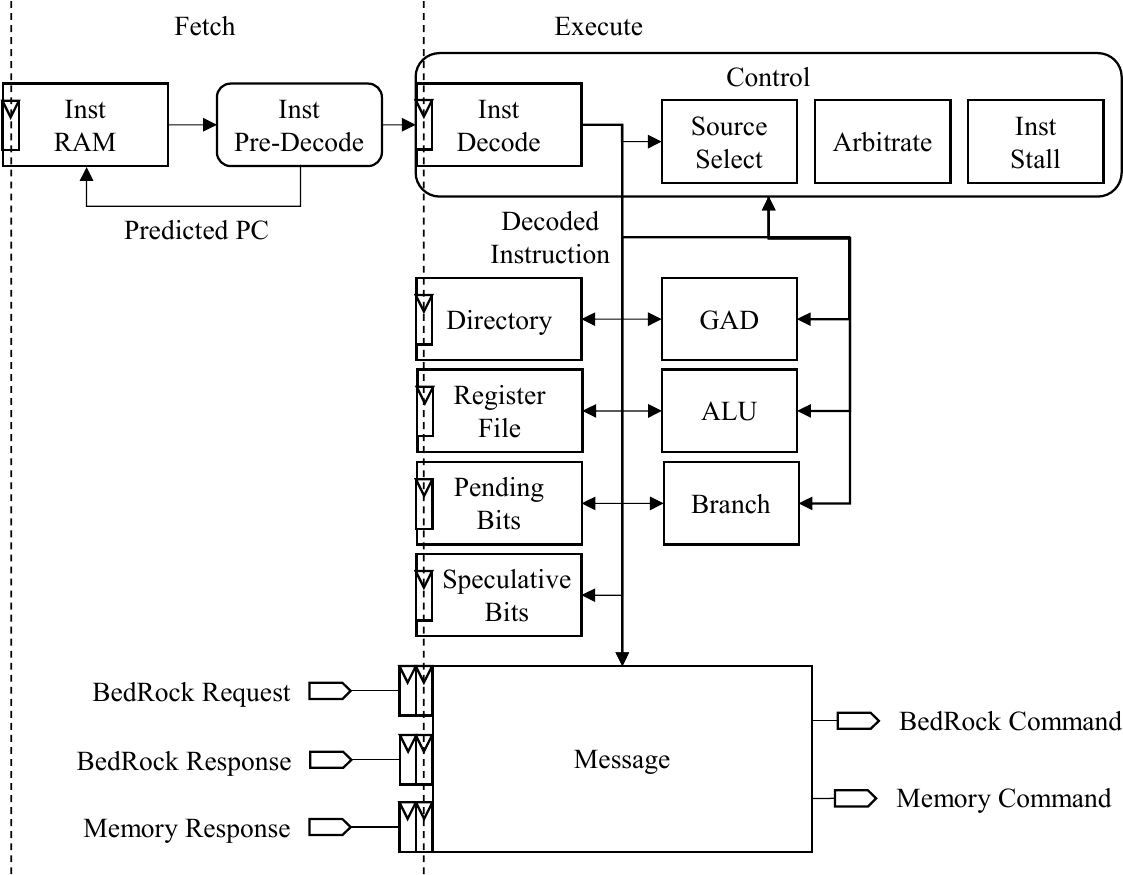}
	\caption{\bpbedrock{} Microcode-Programmable CCE Block Diagram}
	\label{fig:ucode-cce-bd}
\end{figure}

The \bpbedrock{} microcode-programmable CCE (ucode CCE) is an experimental \bedrock{} cache coherence engine implementation featuring a user-programmable coherence engine. The ucode CCE implements a two-stage fetch-execute pipeline with 64-bit general purpose registers and datapath, specialized coherence protocol processing logic, and a custom instruction set architecture. \autoref{fig:ucode-cce-bd} shows a block diagram of the \bpbedrock{} ucode CCE. The programmable CCE re-uses a number of modules from the FSM CCE, including the Coherence Directory, GAD, Pending Bits, and Speculative Bits that are explained in \autoref{sec:bp-bedrock-cce-fsm}. The functionality of the ucode CCE attempts to match that of the FSM CCE. The programmable nature of the ucode CCE allows it to execute any variant of the \bedrock{} coherence protocol simply by changing the microcode that it executes. \bpbedrock{} includes microcode for the MOESIF, MESI, MSI, and EI protocols, along with variations of the MESI and MOESIF protocols that implement the speculative memory fetch behavior of the FSM CCE. This section details the implementation and instruction set of the ucode CCE and provides details on the modules that are unique to its design.

\begin{table}[H]\centering
    \ra{1}
	\begin{tabular}{@{}L{0.2\textwidth}@{}L{0.35\textwidth}@{}L{0.3\textwidth}@{}C{0.15\textwidth}@{}}
        \toprule
		Op & Format & Function & Pseudo-Op\\
		\midrule
        nop & \texttt{nop} & $r0 = r0 + 0$ & \ding{51}\\
		add & \texttt{add ra rb rd} & $rd = ra+rb$ &\\
        addi & \texttt{addi ra imm rd} & $rd = ra + imm$ &\\
        inc & \texttt{inc rd} & $rd = rd + 1$ & \ding{51}\\
        sub & \texttt{sub ra rb rd} & $rd = ra-rb$ &\\
        subi & \texttt{subi ra imm rd} & $rd = ra - imm$ &\\
        dec & \texttt{dec rd} & $rd = rd - 1$ & \ding{51}\\
        not & \texttt{not rd} & $rd = !rd$ &\\
        lsh & \texttt{lsh ra rb rd} & $rd = ra<<rb$ &\\
        lshi & \texttt{lshi ra imm rd} & $rd = ra << imm$ &\\
        rsh & \texttt{rsh ra rb rd} & $rd = ra>>rb$ &\\
        rshi & \texttt{rshi ra imm rd} & $rd = ra >> imm$ &\\
        and & \texttt{and ra rb rd} & $rd = ra\ {\&}\ rb$ &\\
        or & \texttt{or ra rb rd} & $rd = ra\ |\ rb$ &\\
        xor & \texttt{xor ra rb rd} & $rd = ra \oplus rb$ &\\
        neg & \texttt{neg rd} & $rd = {\sim}rd$ & \\
		\bottomrule
	\end{tabular}
	\caption{\bpbedrock{} ucode CCE Base ISA - ALU}
	\label{table:bedrock-isa-base-alu}
\end{table}

\begin{table}[H]\centering
    \ra{1}
	\begin{tabular}{@{}L{0.2\textwidth}@{}L{0.35\textwidth}@{}L{0.3\textwidth}@{}C{0.15\textwidth}@{}}
        \toprule
		Op & Format & Function & Pseudo-Op\\
		\midrule
		bi & \texttt{bi tgt} & $pc = tgt$ & \ding{51}\\
        beq & \texttt{beq ra rb tgt [pt]} & $pc = tgt\ \textrm{if}\ ra == rb$ &\\
        bne & \texttt{bne ra rb tgt [pt]} & $pc = tgt\ \textrm{if}\ ra != rb$ &\\
        blt & \texttt{blt ra rb tgt [pt]} & $pc = tgt\ \textrm{if}\ ra < rb$ &\\
        bgt & \texttt{bgt ra rb tgt [pt]} & $pc = tgt\ \textrm{if}\ ra > rb$ & \ding{51}\\
        ble & \texttt{ble ra rb tgt [pt]} & $pc = tgt\ \textrm{if}\ ra \leq rb$ &\\
        bge & \texttt{bge ra rb tgt [pt]} & $pc = tgt\ \textrm{if}\ ra \geq rb$ & \ding{51}\\
        beqi & \texttt{beqi ra imm tgt [pt]} & $pc = tgt\ \textrm{if}\ ra == imm$ &\\
        bneqi & \texttt{bneqi ra imm tgt [pt]} & $pc = tgt\ \textrm{if}\ ra != imm$ &\\
        bz & \texttt{bz ra tgt [pt]} & $pc = tgt\ \textrm{if}\ ra == 0$ & \ding{51}\\
        bnz & \texttt{bnz ra tgt [pt]} & $pc = tgt\ \textrm{if}\ ra != 0$ & \ding{51}\\
        bs & \texttt{bs rspc rb tgt [pt]} & $pc = tgt\ \textrm{if}\ rspc == rb$ &\\
        bss & \texttt{bss rspca rspcb tgt [pt]} & $pc = tgt\ \textrm{if}\ rspc_a == rspc_b$ &\\
        bsi & \texttt{bsi rspc imm tgt [pt]} & $pc = tgt\ \textrm{if}\ rspc == imm$ &\\
		\bottomrule
	\end{tabular}
	\caption{\bpbedrock{} ucode CCE Base ISA - Branch}
	\label{table:bedrock-isa-base-branch}
\end{table}

\begin{table}[H]\centering
    \ra{1}
	\begin{tabular}{@{}L{0.15\textwidth}@{}L{0.25\textwidth}@{}L{0.45\textwidth}@{}C{0.15\textwidth}@{}}
        \toprule
		Op & Format & Function & Pseudo-Op\\
		\midrule
        mov & \texttt{mov ra rd} & $rd = ra$ &\\
        movsg & \texttt{movsg rspc rd} & $rd = rspc$ &\\
        movgs & \texttt{movgs ra rspc} & $rspc = ra$ &\\
        movfg & \texttt{movfg flag rd} & $rd[0] = flag$ &\\
        movgf & \texttt{movgf ra flag} & $flag = ra[0]$ &\\
        movpg & \texttt{movsg param gpr} & $gpr = param$ &\\
        movgp & \texttt{movgs gpr param} & $param = gpr$ &\\
        movi & \texttt{movi imm rd} & $rd = imm$ &\\
        movis & \texttt{movis imm rspc} & $rspc = imm$ &\\
        movip & \texttt{movis imm param} & $param = imm$ &\\
        clm & \texttt{clm} & $clear\ MSHR$ &\\
        clf & \texttt{clf} & $clear\ MSHR.flags$ & \ding{51}\\
        ldflags & \texttt{ldflags ra} & $MSHR.flags = ra[0+:num\_flags]$ & \ding{51}\\
        ldflagsi & \texttt{ldflagsi imm} & $MSHR.flags = imm[0+:num\_flags]$ & \ding{51}\\
		\bottomrule
	\end{tabular}
	\caption{\bpbedrock{} ucode CCE Base ISA - Data Movement}
	\label{table:bedrock-isa-base-data}
\end{table}

\begin{table}[H]\centering
    \ra{1}
	\begin{tabular}{@{}L{0.2\textwidth}@{}L{0.4\textwidth}@{}L{0.4\textwidth}@{}}
        \toprule
		Op & Format & Function\\
		\midrule
        sf & \texttt{sf flag} & $flag = 1$\\
        sfz & \texttt{sfz flag} & $flag = 0$\\
		andf, orf & \texttt{op flag1 flag2 gpr} & $rd = flag1\ op\ flag2$\\
		nandf, notf & \texttt{op flag1 flag2 gpr} & $rd = flag1\ op\ flag2$\\
		notf & \texttt{notf flag gpr} & $rd = {\sim}flag$\\
		bf & \texttt{bf tgt flag [flag...] [pt]} & $pc = tgt\ \textrm{if}\ all\ specified\ flags == 1$\\
		bfnot & \texttt{bfnot tgt flag [flag...] [pt]} & $pc = tgt \ \textrm{if}\ all\ specified\ flags == 0$\\
		bfz & \texttt{bfz tgt flag [flag...] [pt]} & $pc = tgt\ \textrm{if}\ any\ specified\ flag == 1$\\
		bfnz & \texttt{bfnz tgt flag [flag...] [pt]} & $pc = tgt\ \textrm{if}\ any\ specified\ flag == 0$\\
		\bottomrule
	\end{tabular}
	\caption{\bpbedrock{} ucode CCE Coherence ISA - Flag}
	\label{table:bedrock-isa-coherence-flag}
\end{table}

\begin{table}[H]\centering
    \ra{1}
	\begin{tabular}{@{}L{0.1\textwidth}@{}L{0.59\textwidth}@{}L{0.31\textwidth}@{}}
        \toprule
		Op & Format & Function\\
		\midrule
		rdp & \texttt{rdp addr=a} & $pf = pending\_bits[addr]$ \\
		rdw & \texttt{rdw addr=a lce=l lru\_way=w [src=ra]} & $\textrm{produce sharers, lru info, etc.}$\\
		rde & \texttt{rde addr=a lce=l way=w [src=ra] dst=rd} & $rd = addr, sh\_st[lce] = state$\\
		wdp & \texttt{wdp addr=a p=0,1} & $pending\_bits[addr] +/- 1$\\
		clp & \texttt{clp addr=a} & $pending\_bits[addr] = 0$\\
		clr & \texttt{clr addr=a lce=l} & $\textrm{clear directory row}$\\
		wde & \texttt{wde addr=a lce=l way=w [src=ra] state=s [state]} & $dir[addr, lce] = [tag, state]$\\
		wds & \texttt{wds addr=a lce=l way=w [src=ra] state=s [state]} & $dir[addr, lce].state = state$\\
		gad & \texttt{gad} & $\textrm{execute GAD unit}$\\
		\bottomrule
	\end{tabular}
	\caption{\bpbedrock{} ucode CCE Coherence ISA - Directory}
	\label{table:bedrock-isa-coherence-dir}
\end{table}

\begin{table}[H]\centering
    \ra{1}
	\begin{tabular}{@{}L{0.1\textwidth}@{}L{0.5\textwidth}@{}L{0.4\textwidth}@{}}
        \toprule
		Op & Format & Function\\
		\midrule
		wfq & \texttt{wfq queue [queue...]} & wait for message on queue(s)\\
		pushq & \texttt{pushq queue cmd addr=a lce=l way=w [src=ra] wp=0,1 spec=0,1} & push message to queue\\
		popq & \texttt{popq queue [wp]} & dequeue message, write pending bit\\
		poph & \texttt{poph queue rd} & capture message header\\
		specq & \texttt{specq spec\_cmd addr\_sel [state]} & speculation bits operation\\
		inv & inv & send invalidations\\
		\bottomrule
	\end{tabular}
	\caption{\bpbedrock{} ucode CCE Coherence ISA - Queue}
	\label{table:bedrock-isa-coherence-queue}
\end{table}

\subsection{Instruction Set Architecture (ISA)}

The ucode CCE executes a custom instruction set architecture (ISA) designed to efficiently execute the \bedrock{} coherence protocol. The ISA is divided into a \emph{Base ISA} and a \emph{Coherence ISA}. The Base ISA includes standard, general-purpose RISC ISA instructions such as arithmetic, branching, and data movement operations. The Coherence ISA includes the \bedrock{}-specialized instructions that enable efficient execution and processing of the \bedrock{} coherence protocol. The Coherence ISA instructions are divided into Flag, Directory, and Queue operations, and include operations to perform coherence directory reads and writes, message send and receive, and complex coherence protocol control flow execution.

Across the ISA, all branch instructions are tagged with a static taken/not-taken prediction bit and the branch mispredict penalty is one cycle. Directory read and queue operations may take more than one cycle to execute depending on functional unit conflicts and latencies, while all other instructions execute in a single cycle.

\subsubsection{Base ISA}

The Base ISA comprises ALU (\autoref{table:bedrock-isa-base-alu}), Branch (\autoref{table:bedrock-isa-base-branch}), and Data Movement (\autoref{table:bedrock-isa-base-data}) instructions. Many of the instructions in these groups are commonly found in general-purpose RISC instruction sets. The ALU instructions include basic arithmetic and bitwise operations. The Base ISA tables describe each op, its microcode format, the operation's function, and whether it is implemented in hardware or is a software pseudo-operation. Pseudo-operations are indicated by a \ding{51} symbol and are available for the programmer to use in the microcode. These operations are transformed into hardware operations by the microcode assembler, which allows for a richer programmer interface while reducing hardware implementation complexity.

\subsubsection{Flag Instructions}

Flag instructions can set or clear flags, perform logic operations on pairs of flags, and make control flow decisions based on the state of a programmer-selected set of flags. \autoref{table:bedrock-isa-coherence-flag} lists the available flag instructions. The most important of these are the flag-based branch instructions (bf, bfz, bfnz, bfnot). Each flag-based branch examines a set of programmer-selected MSHR flags, encoded in a bitmask within the instruction, and branches the microcode PC to the supplied target PC if the branch condition is met. A single flag-based branch instruction is able to replace a sequence of regular branch instructions, thereby accelerating common protocol processing control flow decisions.

\subsubsection{Directory Instructions}

Directory instructions, listed in \autoref{table:bedrock-isa-coherence-dir}, accelerate directory read, write, and processing operations by invoking the coherence directory and GAD modules. Directory way group reads require only $1 + (C/2)$ cycles to execute, compared to tens or hundreds of cycles that would be required by a general-purpose implementation of the same routine using \lstinline{for} loops. Pending bit and directory entry reads require one and two cycles, respectively. Directory writes execute in a single cycle. The GAD module executes in a single cycle, compared to a cost of tens of instructions to implement equivalent logic in general-purpose RISC code. Additionally, the flag outputs of the GAD module never need to be recomputed by the microcode program, saving many additional cycles for every flag-based branch instruction.

\subsubsection{Queue Instructions}

Queue instructions enable efficient sending and receiving of coherence protocol and memory messages and are listed in \autoref{table:bedrock-isa-coherence-queue}. The ucode CCE is able to send and receive messages with a cost of one cycle per message header or data beat. The invalidate (inv) instruction further accelerates coherence protocol processing by invoking a small hardware-implemented state machine within the ucode CCE's message unit to efficiently send invalidation commands to all caches with a Shared (S) copy of the specified cache block at a rate of one message per cycle. A general-purpose RISC routine for invalidations would require at least a few instructions per invalidation sent if executed in a tight \lstinline{for} loop.

\subsubsection{Programming the CCE}

The CCE is programmed at the microcode level. A custom assembler applies a limited set of instruction transformations to map available software pseudo-ops into hardware-implemented microcode instructions. \bpbedrock{}'s MOESIF protocol microcode is only 125 instructions, which includes support for uncacheable access to both cacheable and uncacheable memory and system initialization.

\subsection{Fetch - Instruction RAM and Predecode}

The Fetch stage of the ucode CCE includes the Instruction RAM and Predecode modules. The Instruction RAM module contains the microcode instruction memory, the fetch program counter (fetch PC), and next PC logic. After system reset, an external configuration bus loads the CCE microcode into the instruction memory and then transitions the CCE into its microcode execution mode. Once regular execution begins, a new instruction is fetched every cycle unless the Execute stage raises the stall signal. If a stall occurs, the previously fetched instruction is held valid on the output of the instruction memory.

The instruction RAM module outputs the fetched instruction and the fetch PC, which are fed to the Instruction Predecode module. The predecoder determines if the instruction is a branch instruction, whether the instructions predict taken bit is set, and the branch target encoded in the instruction. It then outputs a predicted fetch PC that is either the current PC plus one or the branch target. The instruction RAM uses the predicted fetch PC to fetch the next instruction, unless the execute stage reports a branch misprediction. Branch mispredictions unconditionally redirect the fetch PC to the resolved PC provided by the Branch module in the Execute stage.

\subsection{Execution Control}

The Instruction Decode, Source Select, Arbitrate, and Instruction Stall modules make up the Execute stage's control logic. Collectively, these modules create the necessary control signals for the ucode CCE's functional units, abritrate access to functional units shared by the microcode and the Message unit, and detect execution hazards.

\subsubsection{Instruction Decode}

The Instruction Decode unit expands the narrow microcode instruction into a much wider decoded instruction that contains functional unit control signals. The decode module also contains the current instruction and PC registers. Instruction Stalls cause the current instruction to be replayed in the next cycle, and branch mispredictions cause a single cycle bubble in execution while the Execute stage waits for a new instruction to be fetched. The output of the decoder is the decoded instruction and the current Execute stage PC.

\subsubsection{Source Select}

The source select module routes operands to the ucode CCE's functional units based on the current instruction. A source operand may come from the general purpose registers, MSHR fields, inbound network message fields, or directory outputs, depending on the specific instruction being executed. This module primarily exists to centralize the source selection logic and de-clutter the \bpbedrock{} implementation code.

\subsubsection{Arbitration}

The arbitration unit controls access to the coherence directory, pending bits write port, and speculative bits read port. In a given cycle, each of these three resources may be used by either the microcode instruction or the message unit. Conflicting use of a resource results in the current instruction stalling and the message unit winning arbitration.

\subsubsection{Instruction Stall}

The Instruction Stall unit controls whether the current instruction executes and commits or must be replayed in the following cycle. Stalls occur due to functional unit hazards and when attempting to send or receive \bedrock{} messages when the target network is either busy or does not have a valid message available for processing, respectively. The unit takes the decoded instruction as input, examines its control signals, and stalls execution if any of the possible stall conditions are met. The stall signal is routed to the Fetch stage to retain the previously fetched instruction, and to the instruction decoder to replay the current instruction in the next cycle. Functional unit hazards arise when the Message unit and the current microcode instruction attempt to use the same functional unit. To ensure forward progress and provide higher message processing throughput in the coherence protocol, the message unit has priority over the microcode instruction.

\subsection{Register File}

The Register File stores the internal state of the ucode CCE. There are four registers stored in the register file that hold the CCE's Miss Status Handling Register (MSHR), eight 64-bit general purpose registers (GPRs), a coherence state register, and an auto-forward control register. The MSHR register is identical to the one in the FSM CCE and is described in \autoref{sec:cce_fsm_mshr}. Registers are primarily written directly by microcode instructions, but the LRU Address and LRU Coherence State fields of the MSHR are also written by the directory during way group reads. The eight 64-bit general purpose registers are used by the microcode program to store temporary variables and values during execution. The coherence state register is a special register that holds a default coherence state that can be applied to coherence or memory commands and used as a source operand by microcode instructions. The auto-forward control register is a single bit register that controls whether the ucode CCE's Message unit will automatically process memory response messages. This bit is set by default, but can be disabled for debugging purposes or to allow the microcode full control over memory response processing.

\subsection{Functional Units}

The ucode CCE includes a handful of Functional Units that execute operations requested by the current microcode instruction. The Coherence Directory, Pending Bits, Speculative Bits, and GAD modules are all re-used without modification from the FSM CCE design, and are described in \autoref{sec:bp-bedrock-cce-fsm}.

\subsubsection{Branch}

The Branch unit resolves branches and validates the Fetch stage's speculative fetch. The branch unit takes the current instruction's two operands, branch operation, valid bit, predict taken bit, PC, and branch target as inputs. It then computes the result of the branch operation and determines if a misprediction occurred by comparing the branch outcome to the predict taken bit. Mispredictions result in a single cycle bubble in the execute stage and redirect the Fetch stage to the proper fetch PC. The next fetch PC will either be the current Execute stage PC plus one or the branch target, depending on the outcome of the branch comparison.

\subsubsection{ALU}

The Arithmetic Logic Unit (ALU) is a simple, 64-bit wide ALU supporting addition, subtraction, logical shifts, and bitwise operations. The supported bitwise operations are AND, OR, XOR, NAND, NOR, and negation. The ALU also supports logical negation of a single operand. The hardware ALU is purposefully simplistic to reduce complexity. Additional common operations are supported at the software level by the assembler. Software supported operations include increment, decrement, add immediate, subtract immediate, and shift immediate.

\subsubsection{Message}

The Message unit is responsible for sending and receiving all \bedrock{} network messages. The message unit can write the pending bits, read the speculative bits, and write the coherence directory. It also contains the memory credit flow counter that limits the number of outstanding memory commands issued by the CCE. The message unit has two state machines to process memory response messages and send or receive messages according to the execution of microcode instructions.

The memory response state machine is effectively identical to the memory response FSM of the FSM CCE. It processes arriving memory response messages, reads the speculative bits if the response is speculative, and then squashes the message, forwards the message to the appropriate , or sinks the response at the CCE. The memory response state machine can be disabled by clearing the auto-forward control register stored in the register file.

The other state machine sends and receives messages based on the currently executing instruction. This state machine is activated by a push or pop instruction, and the instruction specifies the network, message type, and message information required.

\subsection{Uncached-Only Mode}

The ucode CCE contains logic sufficient to support uncached requests immediately following system reset. This logic is implemented primarily in the Message unit, which consumes LCE request messages and forwards them to memory as uncached loads or stores. Memory responses are auto-forwarded from the memory network to the LCE command network. Very minimal processing occurs in the uncached-only mode, and it is meant to support system debugging. The CCE exits uncached-only mode when commanded to by a mode change via the configuration bus. The external configuration device must guarantee that it is safe to transition from uncached-only to normal mode, and must ensure that the microcode program required by the ucode CCE has been loaded into the instruction memory.

\subsection{LCE Request Processing} \label{sec:ucode_cce_req}

\begin{figure}[t]
	\centering
	\includegraphics[width=0.90\linewidth]{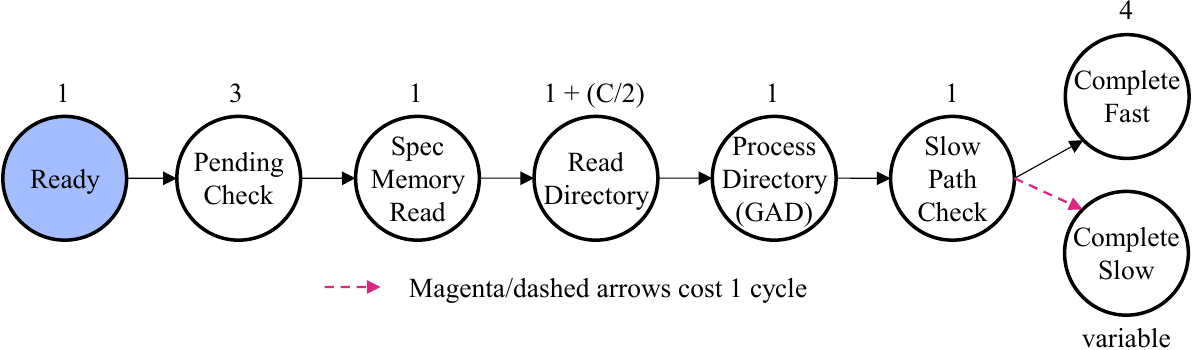}
	\caption{\bpbedrock{} MOESIF Microcode Processing Flow - Initial and Fast Path}
	\label{fig:ucode-cce-moesif-front}
\end{figure}

\begin{figure}[t]
	\centering
	\includegraphics[width=0.90\linewidth]{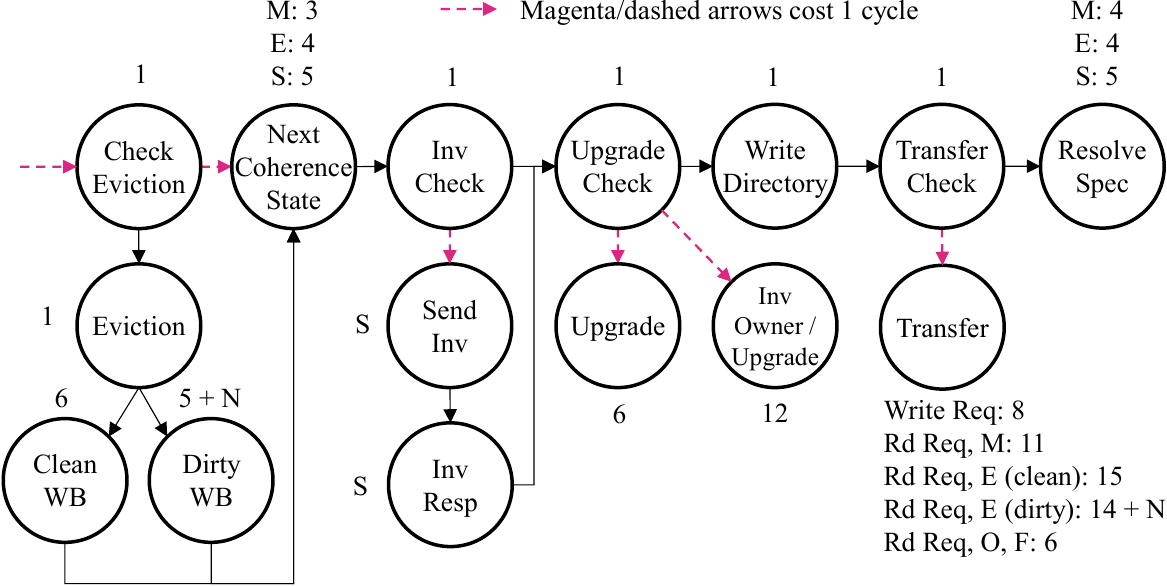}
	\caption{\bpbedrock{} MOESIF Microcode Processing Flow - Slow Path}
	\label{fig:ucode-cce-moesif-slow}
\end{figure}

The ucode CCE's microcode programs implement a similar execution flow as the FSM CCE. Arriving requests trigger a coherence directory read before the microcode examines the outputs to determine whether any replacement, invalidations, upgrade, transfer, or memory read is required to complete the request. This section describes the processing flow of the MOESIF cache coherence protocol microcode implementation.

\subsubsection{LCE Request Processing Diagrams}

\autoref{fig:ucode-cce-moesif-front} and \autoref{fig:ucode-cce-moesif-slow} depict the MOESIF protocol microcode processing flow. Each circle represents one or more microcode instructions and is called a subroutine. Transitions between subroutines are colored either black or magenta/dashed, with black transitions having no cost and magenta/dashed transitions having a cost of one cycle due to a branch misprediction penalty. The best-case, no contention occupancy in cycles is shown as a number of expression adjacent to each subroutine. The occupancy is typically one cycle per instruction, however certain instructions require multiple cycles to execute, as noted above. Most subroutines have a fixed-cost occupancy, but the value may depend on the specific type of request.

\autoref{fig:ucode-cce-moesif-front} shows the initial subroutines executed by the MOESIF protocol for all coherence requests. The microcode is optimized to handle load requests to blocks in the Invalid state (not cached anywhere in the system), which is called the \emph{Fast Path}. All other requests are handled by the \emph{Slow Path}. A fast path request is fulfilled by a memory access, which is issued speculatively and as soon as possible, before the directory read occurs. The directory read latency depends on the number of Cores (C), as explained in \autoref{table:fsm-cce-req-state-occupancy} and \autoref{sec:cce_fsm_req}. Following the directory read, a single cycle is required to confirm the request does not require the slow path for processing before finalizing the past path processing.

Slow path requests follow the same initial processing as fast path requests, but then branch to the processing flow shown in \autoref{fig:ucode-cce-moesif-slow}. Note that the magenta arrow between "Slow Path Check" and "Slow Path Completion" subroutines in \autoref{fig:ucode-cce-moesif-front} is the same arrow that enters the  "Replacement Check" subroutine in \autoref{fig:ucode-cce-moesif-slow}. A request processed by the slow path either targets a block that is already cached somewhere in the system, requires a cache block replacement at the requesting LCE, or is a write request. The occupancy of the \emph{Compute Next Coherence State} and \emph{Resolve Speculation} subroutines depend on the coherence state of the requested block that will be assigned to the requesting LCE. Invalidations require S cycles to send the invalidation commands and another S cycles to receive the invalidation responses. Writebacks during cache block replacement or from a transfer require N cycles to forward the N cache block data beats from the LCE response to the memory network. The \emph{Transfer} subroutine occupancy depends both on the type of request (read or write) and the coherence state of the current block owner. Request processing completes after performing an Upgrade, Transfer, or Resolving Speculation.

\subsubsection{LCE Request Processing Occupancy}

\begin{table}[t]\centering
	\ra{1.3}
	\begin{tabular}{@{}L{0.35\linewidth}C{0.15\linewidth}L{0.43\linewidth}@{}}\toprule
		Routine & Occupancy (cycles) & Notes \\
		\midrule
		Ready through Directory Read & $7 + (C/2)$ & Ready through Process Directory (GAD)\\
		Fast path Completion & $4$ & Load request to block in I\\
		Branch to slow path & $2$ & 1 cycle branch mispredict penalty \\
		Skip replacement & $2$ & 1 cycle branch mispredict penalty \\
		Replacement (Dirty) & $7 + N$ & Dirty writeback\\
		Replacement (Clean) & $8$ & Null writeback\\
		Compute Next State & $3\ \textrm{to}\ 5$ & Next state of M (3), E (4), S (5)\\
		Skip invalidations & $1$ & Branch predicted taken\\
		Invalidation & $2 + (2*S)$ & Invalidation commands and responses\\
		Skip upgrade & $1$ & Branch predicted taken\\
		Upgrade & $7$ & No owner\\
		Upgrade (Owner in O/F) & $13$ & Invalidate Owner\\
		Write Directory & $1$ & Write tag and state to directory\\
		Skip Transfer & $1$ & Branch predicted taken\\
		Transfer (Read, O/F) & $7$ & Owner in O or F\\
		Transfer (Read, M) & $12$ & Owner in M\\
		Transfer (Read, E and clean) & $16$ & Owner in E and clean\\
		Transfer (Read, E and dirty) & $15 + N$ & Owner in E and dirty\\
		Transfer (Write) & $9$ & Write request\\
		Resolve Speculation & $4\ \textrm{or}\ 5$ & Next state of E, M (4), S (5)\\
		\bottomrule
	\end{tabular}
	\caption{\bpbedrock{} ucode CCE Subroutine Occupancy - MOESIF}
	\label{table:ucode-cce-subroutine-occupancy-moesif}
\end{table}

\autoref{table:ucode-cce-subroutine-occupancy-moesif} lists the processing occupancy in cycles for subroutines found in the MOESIF coherence protocol microcode program. The cycle counts assume best-case, no-contention execution of the microcode program. Most rows in the table correspond directly to the subroutines shown in \autoref{fig:ucode-cce-moesif-front} and \autoref{fig:ucode-cce-moesif-slow}. Routines named starting with \emph{Skip} correspond to performing one of the \emph{Check} subroutines followed by a horizontal transition in \autoref{fig:ucode-cce-moesif-slow}.

\begin{table}[t]\centering
	\ra{1.3}
	\begin{tabular}{@{}L{0.12\linewidth}C{0.07\linewidth}C{0.12\linewidth}C{0.25\linewidth}L{0.32\linewidth}@{}}\toprule
		Request & LCE State & Directory State & Occupancy (cycles) & Notes\\
		\midrule
		Read Excl & \multirow{2}{*}{I} & \multirow{2}{*}{I} & $12 + (C/2)$ & Block from Memory\\ % read mem, fast path
		Read NE & & & $26 + (C/2)$ & Block from Memory\\ % read mem, slow path
		\midrule
		\multirow{5}{*}{Read} & \multirow{5}{*}{I} & S & $26 + (C/2)$ & Block from Memory\\ % read mem
		& & E (clean) & $36 + (C/2)$ & Transfer and Null Writeback \\ % ST_TR_WB
		& & E (dirty) & $35 + (C/2) + N$ & Transfer and Dirty Writeback \\ % ST_TR_WB
		& & M & $32 + (C/2)$ & Transfer\\ % ST_TR
		& & O, F & $27 + (C/2)$ & Transfer\\ % ST_TR
		\midrule
		\multirow{4}{*}{Write} & \multirow{4}{*}{I} & I & $23 + (C/2)$ & Block from Memory\\ % read mem
		& & S & $24 + (C/2) + (2*S)$ & Block from Memory\\ % Inv all S, read mem
		& & E, M & $27 + (C/2)$ & Transfer\\ % ST_TR
		& & O, F & $28 + (C/2) + (2*S)$ & Invalidate and Transfer\\ % Inv all S, ST_TR
		\midrule
		\multirow{2}{*}{Write} & \multirow{2}{*}{S} & S & $24 + (C/2) + (2*(S-1))$ & Invalidate and Upgrade\\ % Inv all S, STW to Req
		& & O, F & $30 + (C/2) + (2*(S-1))$ & Invalidate and Upgrade\\ % Inv all S and owner, STW to Req
		\midrule
		Write & O, F & O, F & $24 + (C/2) + (2*S)$ & Invalidate and Upgrade\\
		\bottomrule
	\end{tabular}
	\caption{\bpbedrock{} ucode CCE Request Occupancy - MOESIF}
	\label{table:ucode-cce-req-occupancy-moesif}
\end{table}

\autoref{table:ucode-cce-req-occupancy-moesif} details the request processing occupancy for the MOESIF coherence protocol implementation. The occupancy cycles shown in the tables are derived directly from the MOESIF protocol processing flow diagrams. Given a request type, the current state of the block at the LCE, and the state of the block at the coherence directory, the processing occupancy of the request can be computed by starting at the \emph{Ready} subroutine in \autoref{fig:ucode-cce-moesif-front} and progressing through the two diagrams. All requests assume that a cache block replacement is not required. The addition of a replacement adds either six or $5 + N$ cycles to the processing latency for null and dirty writebacks, respectively.

All requests have a bast processing cost of $8 + (C/2)$ cycles to move from Ready through directory read and processing and then execute the slow path check. The slow patch check branch is predicted taken to the the fast path, which requires an additional four cycles to finish processing exclusive read requests for invalid blocks. All other requests incur a branch mispredict penalty of one cycle between the slow path check and the \emph{Replacement Check} subroutine in \autoref{fig:ucode-cce-moesif-slow}. Slow path requests then perform a replacement, if required, compute the next coherence state of the requested block for the requesting LCE, and invalidate the requested block from any LCE with the block in the Shared (S) state as needed. Request processing is then finalized by performing a cache block upgrade if the requestor already has a valid copy of the block, a cache to cache transfer if another cache owns the block, or simply resolving the speculative memory access if the block will be filled from memory.

\section{FSM and ucode CCE Performance Comparison}
\label{sec:bp-bedrock-cce-perf}

The fixed-function and microcode-programmable CCEs implement the same coherence protocol using two very different approaches. Therefore, it is important to compare the performance of the two designs to fully understand the implications of using a programmable protocol processing engine. In this section, the performance of the two \bpbedrock{} coherence engines is compared by first examining the best-case no-contention request processing occupancy in the two coherence engine designs before evaluating the impact of request processing occupancy at the application level.

\subsection{Comparison of ucode and FSM CCE}

% FSM Base cost = 7 + (C/2) for Ready through Write Next State
\begin{table*}[t]\centering
    \footnotesize
	\begin{tabular}{@{}L{0.12\linewidth}@{}C{0.12\linewidth}@{}C{0.14\linewidth}@{}C{0.31\linewidth}@{}C{0.31\linewidth}@{}}
        \toprule
		Request & LCE State & Directory State & FSM CCE Occupancy (cycles) & ucode CCE Occupancy (cycles) \\
		\midrule
		Read Excl & \multirow{2}{*}{I} & \multirow{2}{*}{I} & \multirow{2}{*}{$8 + (C/2)$} & $12 + (C/2)$ \\ % read mem, fast path
		Read NE & & & & $26 + (C/2)$ \\ % read mem, slow path
		\midrule
		\multirow{5}{*}{Read} & \multirow{5}{*}{I} & S & $8 + (C/2)$ & $26 + (C/2)$ \\ % read mem
		& & E (clean) & $10 + (C/2)$ & $36 + (C/2)$ \\ % ST_TR_WB
		& & E (dirty) & $9 + (C/2) + N$ & $35 + (C/2) + N$ \\ % ST_TR_WB
		& & M & $9 + (C/2)$ & $32 + (C/2)$\\ % ST_TR
		& & O, F & $9 + (C/2)$ & $27 + (C/2)$\\ % ST_TR
		\midrule
		\multirow{4}{*}{Write} & \multirow{4}{*}{I} & I & $8 + (C/2)$ & $23 + (C/2)$ \\ % read mem
		& & S & $8 + (C/2) + (2*S)$ & $24 + (C/2) + (2*S)$ \\ % Inv all S, read mem
		& & E, M & $9 + (C/2)$ & $27 + (C/2)$ \\ % ST_TR
		& & O, F & $9 + (C/2) + (2*S)$ & $28 + (C/2) + (2*S)$ \\ % Inv all S, ST_TR
		\midrule
		\multirow{2}{*}{Write} & \multirow{2}{*}{S} & S & $9 + (C/2) + (2*(S-1))$ & $24 + (C/2) + (2*(S-1))$ \\ % Inv all S, STW to Req
		& & O, F & $9 + (C/2) + (2*(S-1))$ & $30 + (C/2) + (2*(S-1))$ \\ % Inv all S and owner, STW to Req
		\midrule
		Write & O, F & O, F & $9 + (C/2) + (2*S)$ & $24 + (C/2) + (2*S)$ \\
		\bottomrule
	\end{tabular}
	\caption{\bpbedrock{} CCE Request Occupancy Comparison - MOESIF}
	\label{table:cce-occupancy-moesif}
\end{table*}

\autoref{table:cce-occupancy-moesif} presents the request processing occupancy for both coherence engines for \bedrock{}'s MOESIF protocol. Processing occupancy, given in cycles, is the number of cycles required in a best-case, no-contention execution to process a coherence request. Three constants are used in the processing occupancy computations: C is the number of cores in the multicore processor, N is the number of data beats required to send a full cache block across the coherence network data channels, and S is the number of caches holding a block in the Shared (S) coherence state, called the sharers. The data presented are the number of cycles that the coherence engine is busy processing a single request. The numbers presented assume that a cache block eviction (replacement) is not required. Occupancy provides insight into the maximum achievable throughput of the coherence engine designs. The request occupancy does not include the time required to process memory responses, which are handled by a separate state machine in both designs that operates concurrent to request processing. Network time is also excluded as the time for messages to transit networks is the same for both designs.

The FSM CCE has a base request processing occupancy of $7 + (C/2)$ cycles, incurred by all requests. During this initial processing, the request is consumed, the directory is read and processed, and the directory entry for the requesting cache is updated with the final next state for the block. Then, depending on the specific request and state of the target block in the system, the FSM executes only those steps required to complete the transaction. The key performance advantage of the FSM-based design is that control flow decisions are effectively free; in any given state, the next state is computed concurrently with the protocol processing occurring in the state. Thus, after executing the initial processing, the added cost to complete a request is simply the cost of the remaining states visited. The worst-case request, in terms of occupancy, is a write request to a block in the O or F state, which is owned by a single cache but shared by many caches and may be present in every single cache in the system. Additionally, a cache block replacement adds either two or $1+N$ cycles of processing time for clean and dirty blocks, respectively.

The ucode CCE incurs execution overheads relative to the FSM-based CCE primarily due to its inability to execute protocol processing and control flow in the same instruction and the fact that each control flow decision requires a separate instruction. As described in \autoref{sec:ucode_cce_req}, the MOESIF microcode program includes a fast path to process regular reads for blocks in the Invalid state. This path has an execution overhead of only four cycles compared to the FSM-based coherence engine. The fast path is effectively a single basic-block of microcode, and therefore can be executed at a rate matching that of the FSM-based engine. However, all other requests must branch to the full path, which is capable of performing replacements, invalidations, and cache to cache transfers. The base occupancy for both paths is only one cycle greater than the FSM-based engine at $8 + (C/2)$ cycles. Requests processed by the full path have occupancy overheads between 15 to 25 cycles. Significant overheads are incurred for subroutines that require multiple control flow decisions. In particular, determining the proper next coherence state for the block, resolving the outcome of the speculative memory access, and initiating cache to cache transfers all add significant latency to request processing. Additionally, a cache block replacement adds either seven or $6+N$ cycles of processing time for clean and dirty blocks, respectively.

As noted in \autoref{sec:bp-bedrock-cce-fsm}, the no-contention request processing occupancy at the CCE in an eight-core \bpbedrock{} multicore design is between 12 and 27 cycles. The microcode-programmable engine's occupancy overheads of 15 to 25 cycles effectively results in occupancy overheads of approximately 100\% relative to the fixed-function coherence engine for such a design. However, the manifested impact of this overhead on application or system performance may still be minimal, depending on the frequency of cache coherence operations and transactions during execution.

\subsection{Splash-3 Application-Level Performance}

\begin{figure}[t]
	\centering
	\includegraphics[width=0.5\linewidth]{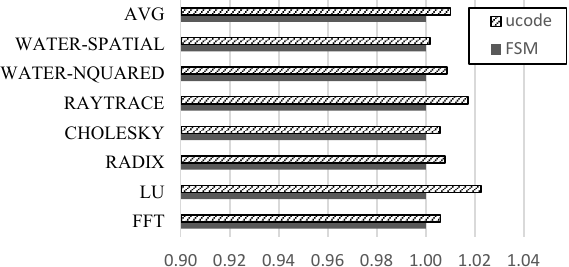}
	\caption{\bpbedrock{} Splash-3 Normalized Execution Time - 8 core}
	\label{fig:splash3}
\end{figure}

To compare the impact of coherence engine design on system performance, a collection of benchmarks from the Splash-3 \cite{sakalis2016} suite were run on an FPGA-based 8-core \bpbedrock{} systems with the FSM-based and microcode programmable coherence engines. The \bpbedrock{} FPGA designs instantiate an 8-core \blackparrot{} multicore with 32 KiB L1 instruction and data caches, a 512 MiB shared L2 cache, 2 GiB of HBM-based main memory, and a core clock frequency of 50 MHz. The benchmarks were compiled for the RISC-V ISA using \lstinline{gcc} targeting a Linux environment, and invoked to execute on all 8 available processor cores. The programs were run in a Linux-based OS environment constructed using BuildRoot \cite{buildroot} and Busybox \cite{busybox} with Linux kernel v5.15 \cite{linux} and OpenSBI v1.0 \cite{opensbi}. The FFT, LU, RADIX, and CHOLESKY programs are smaller kernel programs, while the remaining three programs are larger application programs. Readers are referred to \cite{sakalis2016} and \cite{woo1995} for more details on the synchronization and memory characteristics of these programs. Wall-clock execution time for all benchmarks ranged between tens of seconds and tens of minutes.

\autoref{fig:splash3} shows total execution time measured using the \lstinline{time} utility for each benchmark, averaged over three runs, and normalized to the FSM-based multicore design. Despite having a 15 to 25 cycle best-case processing occupancy overhead, the ucode CCE-based design experiences a very small performance degradation, being within 1\% of the hardware-based coherence engine's performance on average, and only 2.3\% slower at worst. Intuitively, this result makes sense as any program with reasonably good cache utilization and low miss rates will only invoke the coherence system on a cache miss. If misses are infrequent, the overall impact of the programmable coherence engine's increased processing occupancy latencies will be small, which follows directly from the standard average memory access time computation. This result indicates a promising path forward for further exploration of programmable coherence engines. Careful design of the protocol processing paths can keep a programmable coherence engine competitive with a fixed-function engine, while the flexibility of a programmable system can unlock exciting new system features.

\section{FSM and ucode CCE Area Comparison}
\label{sec:bp-bedrock-cce-area}

\begin{table}[t]\centering
\begin{tabular}{@{}L{0.1\linewidth}@{}C{0.2\linewidth}@{}C{0.2\linewidth}@{}C{0.2\linewidth}@{}}
\toprule
Design & Component & Resource & Overhead \\
\midrule
\multirow{3}{*}{ASIC} & Multicore & \multirow{3}{*}{Die Area} & $4.08\%$ \\
& Tile & & 4.28\% \\
& CCE & & 31.08\%\\
\midrule
\multirow{5}{*}{FPGA} & \multirow{2}{*}{Multicore} & Logic LUTs & 6.32\% \\
& & BRAM & 1.54\%\\
\cmidrule{2-4}
& \multirow{2}{*}{Tile} &  Logic LUTs & 7.08\% \\
& & BRAM & 1.54\%\\
\cmidrule{2-4}
& \multirow{2}{*}{CCE} &  Logic LUTs & 66.19\%\\
& & BRAM & 1 per CCE\\
\bottomrule
\end{tabular}
\caption{\bpbedrock{} ucode CCE Resource Overheads}
\label{table:area-comparison}
\end{table}

\blackparrot{}, including \bpbedrock{}, has been silicon validated using GlobalFoundries 12nm FinFET process and FPGA validated in an 8-core configuration for each coherence engine using a Xilinx Ultrascale+ VCU128 development platform\cite{xilinx-ultrascale}. \autoref{table:area-comparison} provides area and resource utilization overheads for ASIC and FPGA-based designs using the ucode CCE, normalized to designs using the FSM CCE. The more efficient ASIC implementations show the introduction of programmability into the coherence system comes at a small area cost of only 4.08\% extra die area for the entire multicore and a 4.28\% increase per \blackparrot{} Tile. Each \blackparrot{} Tile comprises a \blackparrot{} core, its 32 KiB L1 D\$ and I\$, a 64 KiB slice of the distributed L2 cache, the on-chip networks and routers to connect tiles, and an instance of the \bpbedrock{} coherence engine and directory. All SRAM macros are hardened in the ASIC flow. The area overheads of the ucode CCE designs are largely due to the addition of the microcode instruction SRAM. In the FPGA implementations, the logic utilization increases by only 6.32\% and 7.08\% for the entire multicore and per tile, respectively, when using the ucode CCE. Each programmable CCE additionally requires a single 18 Kib block RAM resource, which amounts to a 1.54\% increase in 18 Kib block RAM resources\footnote{36 Kib block RAMs are counted as two 18 Kib block RAMs for this analysis.}. Additionally, both coherence engine implementations meet the same design target frequency in both the ASIC and FPGA implementations.
\section{Conclusion}
\label{sec:bp-bedrock-conclusion}

In this chapter, a complete, fully open-source implementation of the \bedrock{} cache coherence protocol within the \blackparrot{} 64-bit RISC-V shared-memory multicore processor, called \bpbedrock{}, is described. \bpbedrock{} provides a fully functional implementation of the \bedrock{} protocol including its Local Cache Engines (LCE), Cache Coherence Engines (CCE), and coherence networks. The coherence directories in \bpbedrock{} are complete duplicate-tag directories and rely on an innovative directory segment architecture to provide constant-sized coherence engines and directory storage, regardless of the number of cores in the tiled multicore design. The coherence directory storage overhead in \bpbedrock{} is a constant 6.25\% relative to the capacity of the coherent L1 caches. Two coherence directory implementations are provided, one that is hardware-based and fixed-function and a second that is microcode programmable. An analysis of the two coherence engine designs show that it is possible to introduce programmability into the cache coherence system of modern shared-memory multicore processors with minimal area and performance overheads. The key to realizing programmability at low overheads is the use of highly-specialized coherence processing modules and instruction set extensions that offload the core of the coherence protocol processing from general purpose code. Consequently, the microcode programmable coherence engine implementation has only single-digit percentage area and resource costs at the multicore design level while incurring only a 1\% average (2.3\% worst-case) performance overhead for the Splash-3 benchmarks. Using programmability to implement the cache coherence protocol has an area or resource overhead of 4-7\% at the multicore tile level.

\chapter{Hybrid CCE}
\label{chap:hybrid}

The two coherence directory implementations of \bedrock{} presented in \autoref{chap:bp-bedrock} demonstrate the tradeoffs involved in introducing programmability into the cache coherence system at the coherence directory. While the fixed-function directory offers superior performance, the programmable engine can be leveraged to introduce system-specific functionality. Despite architectural and microarchitectural optimization, the microcode-programmable coherence engine is unable to match the coherence protocol processing performance of the fixed-function coherence engine.

In this chapter, a hybrid fixed-function and programmable coherence engine architecture (Hybrid CCE) is described that attempts to preserve the protocol processing performance of the fixed-function coherence engine and the system-dependent flexibility of the programmable engine with minimal cost and overhead. First, \autoref{sec:hybrid-fsm} describes the baseline coherence protocol processing architecture of the hybrid coherence engine, including key learnings that are incorporated from the fixed-function coherence engine described in \autoref{chap:bp-bedrock}. Next, \autoref{sec:hybrid-prog-pipe} introduces a programmable pipeline to the hybrid CCE architecture, drawing from the learnings of implementing the microcode-programmable coherence engine described in \autoref{chap:bp-bedrock}. \autoref{sec:hybrid-perf-analysis} presents a performance analysis and comparison of the hybrid coherence architecture by comparing its processing occupancy latencies and microbenchmark performance to that of the fixed-function and programmable architectures. \autoref{sec:hybrid-area-analysis} presents a comparison of resource utilization for the hybrid coherence engine architecture to that of the fixed-function and programmable coherence engines.

\section{Fixed-Function Protocol Processing Architecture}
\label{sec:hybrid-fsm}

\begin{figure}[t]
	\centering
	\includegraphics[width=0.90\linewidth]{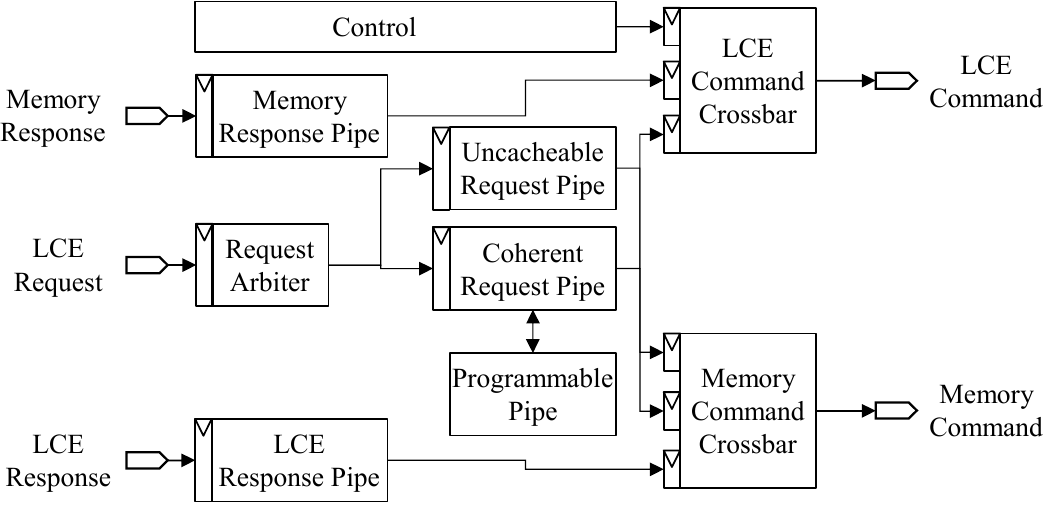}
	\caption{Hybrid CCE Block Diagram}
	\label{fig:hybrid-full-bd}
\end{figure}

In \autoref{chap:bp-bedrock}, a fixed-function coherence engine architecture is presented that correctly implements the \bedrock{} cache coherence protocol. The FSM CCE architecture employs specialized hardware to accelerate coherence protocol operations, but its initial design relies on one large and complex state machine to handle LCE request and response message processing and memory command issue. Therefore, the first step in defining the hybrid coherence engine is to revisit the fixed-function coherence protocol processing logic architecture. \autoref{fig:hybrid-full-bd} presents the outcome of this design iteration. The hybrid CCE's coherence protocol functionality is decomposed into a set of independent pipes. All pipes execute concurrently and manage a single type of incoming protocol or memory message. Memory response messages are processed by the Memory Response Pipe, LCE Response messages are processed by the LCE Response Pipe, and LCE Request messages are processed by either the Uncacheable Request Pipe or the Coherent Request Pipe after first being classified by the Request Arbiter block. The rest of this section describes the functionality of each processing pipe and the other major modules in the coherence engine.

\subsection{Coherence State Management}

\begin{figure}[t]
	\centering
	\includegraphics[width=0.75\linewidth]{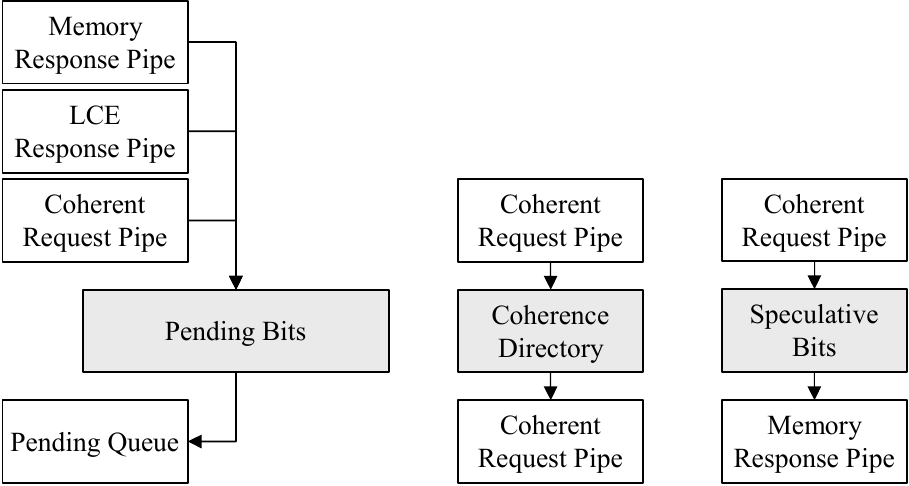}
	\caption{Hybrid CCE Coherence State and Pipe Interaction}
	\label{fig:hybrid-coherence-state}
\end{figure}

\autoref{fig:hybrid-coherence-state} shows the three coherence state blocks in the hybrid CCE and their interaction with the various functional pipes in the design. As with the FSM and ucode CCE designs, the three pieces of coherence state used to implement the coherence protocol are the coherence directory, pending bits, and speculative bits. The pending bits are used to enforce intra-way-group ordering of requests. They are written by the memory response, LCE response, and coherent request pipes and read by the pending queue. New coherence requests and outgoing memory commands increment the associated pending bit while LCE and memory responses decrement the associated pending bit. The coherence directory is used exclusively by the coherent request pipe to track and manage the coherence state of all cache blocks in the coherence system. The speculative bits are written by the coherent request as speculative memory reads are issued. The memory response pipe reads the speculative bits when memory response messages return to the coherence engine to determine how to process the message.

\subsection{Command Crossbars}
\label{sec:hybrid-cmd-xbars}

The LCE Command and Memory Command Crossbars arbitrate access to the outbound LCE and Memory Command network interfaces, respectively, for the various hybrid CCE modules that issue commands into the coherence and memory systems. Each crossbar provides minimal input buffering and round-robin arbitration among input sources for fair and efficient interconnect utilization.

The LCE Command crossbar arbitrates messages sent from the coherent request pipeline, the memory response pipeline, and the control unit. The control unit only issues commands when the CCE performs initialization and mode switches from an uncached request only mode to its normal fully-coherent mode. The coherent request and memory response pipelines frequently issue commands throughout the course of normal operation.

The Memory Command crossbar arbitrates access to the outbound memory command interface from the uncacheable request, coherent request, and LCE response pipelines. During normal operation, the majority of memory commands are issued by the coherent request pipe as it issues speculative or non-speculative memory block reads to the LLC or main memory, with commands also coming from the LCE response pipe to perform writebacks of dirty cache blocks. The uncacheable request pipe is primarily used during startup or to interact with I/O devices through uncached accesses.

\subsection{Control}
\label{sec:hybrid-control-pipe}

\begin{figure}[t]
	\centering
	\includegraphics[width=0.45\linewidth]{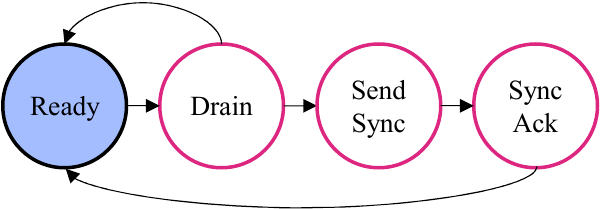}
	\caption{Hybrid CCE Control State Machine}
	\label{fig:hybrid-control-fsm}
\end{figure}

The Control unit is primarily responsible for managing the mode switch from the initial uncached-only request processing mode to the CCE's fully coherent normal mode. At system boot, the hybrid CCE begins execution in an uncached-only mode that processes all requests as if they were targeting uncacheable and uncoherent memory space. All requests are forward to the uncacheable request pipeline by the request arbiter. The uncacheable request pipe then forwards the request to the memory system (LLC or main memory) without performing any coherence checks or operations. This mode is intended to facilitate system boot operations, such as preloading memory, bootroms, and other configuration registers within the multicore and system.

At some time after boot, the system or the multicore itself is expected to perform a software-managed mode switch to enable the cache coherence system. This occurs by writing a register in \blackparrot{}'s configuration bus device, which is then observed by the control unit. Upon seeing the mode switch register write, the control state machine, depicted in \autoref{fig:hybrid-control-fsm}, halts the processing of all new requests and drains all outstanding requests from the CCE's various pipes by asserting a \texttt{drain\_then\_stall} signal. Once all requests have been drained, it issues a sync command to each LCE in the system before waiting for all sync acknowledgments to return, indicating that all LCEs in the system have entered coherent mode. Once the mode switch is complete, the control unit deasserts the \texttt{drain\_then\_stall} signal, allowing all of the hybrid CCE's pipes to resume operation in coherent mode.

\subsection{Request Arbitration}
\label{sec:hybrid-req-arb}

The Request Arbiter module splits the inbound LCE request messages into two logical streams for processing during normal coherent operation. Uncacheable, uncoherent requests are sent to the uncacheable request pipe while all requests targeting cacheable and coherent memory are sent to the coherent request pipe. The CCE itself is not responsible for enforcing ordering between the coherent and uncoherent streams. Ordering must be enforced by the system or software running on the individual cores using mechanisms such as fencing.

Arbitration adds one cycle of latency, but can sustain one request per cycle assuming downstream resources are available. The added cycle of latency comes from a request FIFO buffer used to decouple the inbound request consumption from the arbitration logic. The size of the request buffer is parameterizable and has a default size of two elements.

It is possible for one stream to stall the other stream if the downstream module, either the uncacheable or coherent request pipe, is unable to accept a new request message. While the request arbiter does not include message buffers for each downstream module, the downstream modules both provide request message buffers to reduce the likelihood of cross-stream stall events.

\subsection{Coherent Request Pipe}
\label{sec:hybrid-coh-req-pipe}

\begin{figure}[t]
	\centering
	\includegraphics[width=0.95\linewidth]{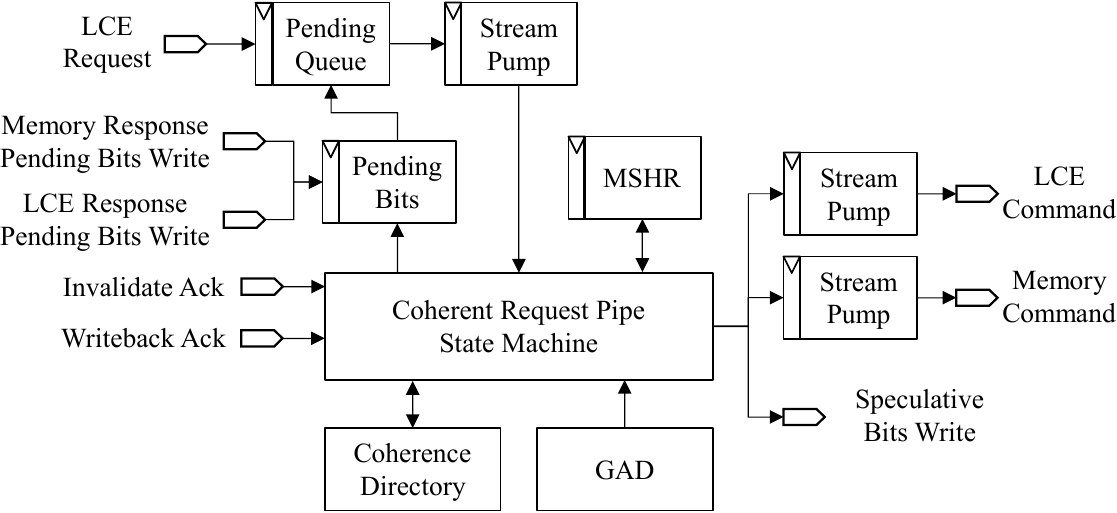}
	\caption{Hybrid CCE Coherent Request Pipe Block Diagram}
	\label{fig:hybrid-coh-req-pipe-bd}
\end{figure}

The Coherent Request Pipe implements the directory request processing portion of the \bedrock{} MOESIF cache coherence protocol using logic that is very similar to that of the FSM CCE described in \autoref{chap:bp-bedrock}. As shown in \autoref{fig:hybrid-coh-req-pipe-bd}, the coherent request pipe comprises a central state machine that implements the protocol's request processing logic and interacts with the coherence state and message send and receive interfaces. As in the FSM and ucode CCE designs, coherence state is tracked using the Pending Bits and Coherence Directory. The GAD (Generate Auxiliary Directory Information) module accelerates processing of coherence directory reads, and the MSHR (Miss Status Handling Register) tracks state data for the current request being processed by the pipe.

A key difference between the logic of the coherent request pipe and the FSM CCE's state machine is that the coherent request pipe does not directly include the logic to process LCE and Memory Response messages. Instead, single-bit signals from the respective message processing pipes are routed as inputs to the coherent request pipe to indicate when important protocol messages, such as cache block writebacks and invalidation acknowledgments, have been received and processed.

A second important difference between the coherent request pipe and the FSM CCE's microarchitecture is the presence of the Pending Queue. This module buffers coherence requests that are stalled due to the associated pending bit being set, indicating that a coherence request targeting the same way group is still active. Stalled requests are pushed to the pending queue, which is a simple first-in first-out (FIFO) ordered buffer, allowing the next request in the CCE's request stream to be examined for readiness. This allows newer, younger, and independent coherence transactions to bypass older, stalled transactions, thereby increasing the realized inter-transaction concurrency and the hybrid coherence engine's effective transaction processing throughput. Throughout execution, if both the pending queue and the incoming LCE request stream have valid requests, the pending queue is prioritized, allowing older requests to make forward progress as soon as the prior request in the same way group has resolved. This preserves the ordering of related requests that is determined by the interconnection network and minimizes the complexity of managing request starvation.

\begin{figure}[t]
	\centering
	\includegraphics[width=0.95\linewidth]{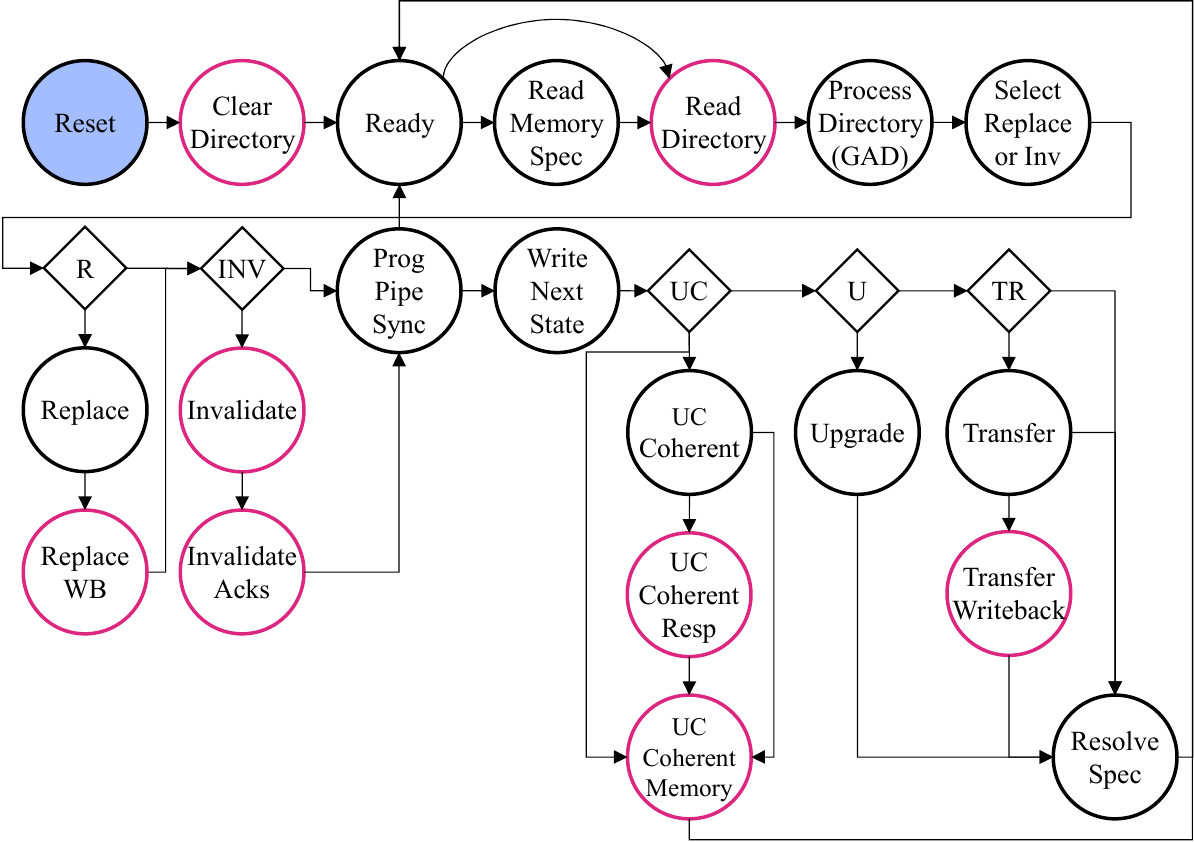}
	\caption{Hybrid CCE Coherent Request Pipe State Machine}
	\label{fig:hybrid-coh-req-pipe-fsm}
\end{figure}

\autoref{fig:hybrid-coh-req-pipe-fsm} shows the state machine implemented by the coherent request pipe. In the hybrid CCE, the request state machine is logically divided into two halves separated by the coherence directory update. All processing that occurs prior to the coherence directory update is effectively idempotent. These actions can only revoke access permissions for the target cache block at caches other than the requester or invalidate a block in the requesting cache to make room for the target cache block. Since the cache coherence system operates invisibly to software, the effect of revoking access permissions can only affect performance. If a cache block is invalidated from a cache but program execution requires it to be re-accessed, the cache coherence system will issue a new coherence transaction to acquire permissions for the block. Software remains oblivious to whether the block is currently cached or must be re-fetched, however re-acquiring the block will incur a latency cost that may increase execution latency of the program overall.

After the coherence directory is written to update the coherence state of the requesting LCE the request is considered to be \emph{committed} within the scope of the coherence protocol. The directory write commits the block's coherence state change and alters permissions for the requesting LCE. This change may allow irreversible changes to the memory data within the target cache block or may allow an irrevocable access to the memory data within the cache block. Thus, the coherence protocol processing logic must guarantee that committing the transaction is allowed by the protocol and any other relevant system constraints to maintain program execution correctness. The organization of the request processing state machine into pre- and post-commit halves is a key difference between the hybrid and fixed-function CCE designs.

\subsubsection{Coherent Request Pipe State Machine Occupancy}

\begin{table}[t]\centering
	\ra{1.3}
	\begin{tabular}{@{}L{0.22\linewidth}@{}C{0.18\linewidth}@{}L{0.6\linewidth}@{}}
        \toprule
		State & Occupancy (cycles) & Description \\
		\midrule
		Read Directory & $(C/2) + 1$ & One cycle setup, plus one cycle per two cores\\
		Replacement WB Response & $1$ & One cycle per ack from LCE Response Pipe\\
		Invalidation Commands & $S$ & One cycle per Sharer\\
		Invalidation Response & $S$ & One cycle per ack from LCE Response Pipe\\
		Transfer WB Response & $1$ & One cycle per ack from LCE Response Pipe\\
		Uncached INV/WB Response & $1$ & One cycle per ack from LCE Response Pipe\\
		Uncached Coherent Memory Command & $1\ \textrm{to}\ N$ & One cycle per data beat for store, or one cycle for load\\
		All other states & $1$ & \\
		\bottomrule
	\end{tabular}
	\caption{\bpbedrock{} Hybrid CCE Request FSM State Machine Occupancy}
	\label{table:hybrid-cce-req-state-occupancy}
\end{table}

\autoref{table:hybrid-cce-req-state-occupancy} describes the coherent request pipe's no-contention occupancy, in cycles, for states in the coherent request pipe's state machine. As with the fixed-function and programmable engine occupancy tables, the numbers provided assume a best case processing latency with no contention for resources and no waiting for inbound messages or outbound network availability. As in the other designs, reading the coherence directory requires one cycle of setup plus $C/2$ cycles to read the tag sets from the directory SRAM's rows, where C is the number of cores in the multicore.

In general, most states require only one cycle to execute, assuming the required resource or message is available. This includes states that process responses to commands such as writebacks, since the actual response message is processed by the LCE Response pipe's state machine that uses single-bit signals to the request pipe to indicate that specific responses have returned and been processed. Invalidation command and response processing requires one cycle per command and response in the general case, to issue the S commands and process the S responses, where S is the number of caches holding the block in the Shared (S) state. Other states, such as Write Next State, Resolve Speculation, and states that issue single LCE or memory commands all require one cycle to execute, as is the case in the fixed-function coherence engine design.

% 5 + (C/2) cycles base cost for Ready through Select RIS
% 6 + (C/2) including resolving speculation
\begin{table}[t]\centering
	\ra{1.3}
	\begin{tabular}{@{}L{0.1\linewidth}@{}C{0.15\linewidth}@{}C{0.2\linewidth}@{}C{0.3\linewidth}@{}L{0.25\linewidth}@{}}
        \toprule
		Request & LCE State & Directory State & Occupancy (cycles) & Notes\\
		\midrule
		\multirow{4}{*}{Read} & \multirow{4}{*}{I} & I & $8 + (C/2)$ & Block from Memory\\ % read mem
		& & S & $8 + (C/2)$ & Block from Memory\\ % read mem
		& & E & $10 + (C/2)$ & Transfer and Writeback \\ % ST_TR_WB
		& & M, O, F & $9 + (C/2)$ & Transfer\\ % ST_TR
		\midrule
		\multirow{4}{*}{Write} & \multirow{4}{*}{I} & I & $8 + (C/2)$ & Block from Memory\\ % read mem
		& & S & $8 + (C/2) + (2*S)$ & Block from Memory\\ % Inv all S, read mem
		& & E, M & $9 + (C/2)$ & Transfer\\ % ST_TR
		& & O, F & $9 + (C/2) + (2*S)$ & Invalidate and Transfer\\ % Inv all S, ST_TR
		\midrule
		\multirow{2}{*}{Write} & \multirow{2}{*}{S} & S & $9 + (C/2) + (2*(S-1))$ & Invalidate and Upgrade\\ % Inv all S, STW to Req
		& & O, F & $9 + (C/2) + (2*(S-1))$ & Invalidate and Upgrade\\ % Inv all S and owner, STW to Req
		\midrule
		Write & O, F & O, F & $9 + (C/2) + (2*S)$ & Invalidate and Upgrade\\ % Inv all S, STW to owner
		\bottomrule
	\end{tabular}
	\caption{\bpbedrock{} Hybrid CCE Request Occupancy}
	\label{table:hybrid-cce-req-occupancy}
\end{table}

\autoref{table:hybrid-cce-req-occupancy} provides the no-contention, best-case processing occupancy for various cacheable and coherent LCE requests given an initial coherence state for the target cache block. These occupancies are computed by stepping through the state machine in \autoref{fig:hybrid-coh-req-pipe-fsm} and summing the latency of each state visited. All requests assume that a cache block replacement is not required. The addition of a replacement adds two cycles to issue the replacement and then process the response signal from the LCE Response pipe.

Examining the coherent request pipe's state machine diagram, every request has an initial cost of $5 + (C/2)$ cycles to move from the Ready state through the Select Replacement or Invalidation state. An additional cycle is required for all requests to resolve the speculative memory read that is issued before the directory is read, which is incurred for all transactions. Requests that require invalidating the target block in other caches require $(2*S)$ cycles to issue the invalidation commands and then process the invalidation acknowledgment messages. Processing the invalidation acks may overlap with issuing invalidation commands, depending on the network and LCE processing latencies in the system. Additionally, every request requires one cycle to update the coherence directory and one cycle to sync with the programmable pipe. Therefore, the total base latency cost for processing every request is $8 + (C/2)$ cycles, which is equivalent to the base cost of the fixed-function state machine.

\subsection{Uncacheable Request Pipe}
\label{sec:hybrid-uc-req-pipe}

The Uncacheable Request Pipe implements basic request processing logic to handle requests targeting uncacheable, uncoherent memory and for forwarding all requests during uncached-only mode at system startup. Due to the use of the \bpbedrock{} Stream message protocol and message formats on both the coherence and memory networks, the translation of messages from the request network to the memory command network is straightforward. All of the complex message handling logic is implemented by the stream pumps attached to the LCE request and memory command network interfaces, while the uncacheable request pipe is responsible for coordinating the network handshaking, setting the memory command message type, and populating message payload fields so the memory response pipe can process the returning memory response message correctly.

\subsection{Memory Response Pipe}
\label{sec:hybrid-mem-resp-pipe}

\begin{figure}[t]
	\centering
	\includegraphics[width=0.80\linewidth]{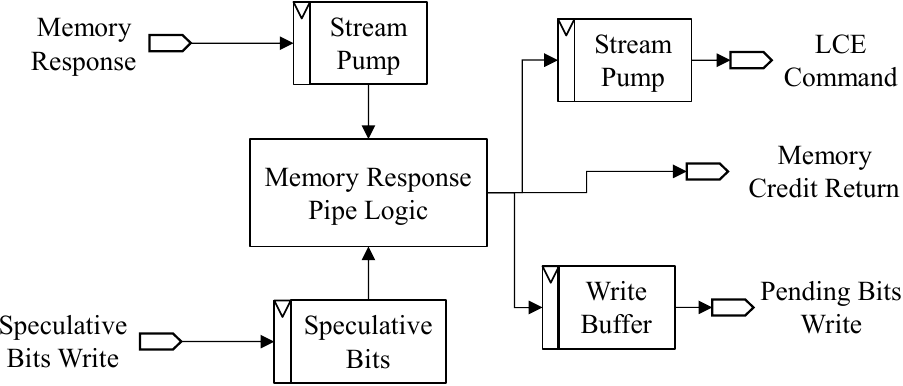}
	\caption{BP-Bedrock Hybrid CCE Memory Response Pipe Block Diagram}
	\label{fig:hybrid-cce-mem-resp-bd}
\end{figure}

\begin{figure}[t]
	\centering
	\includegraphics[width=0.45\linewidth]{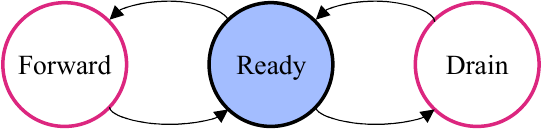}
	\caption{BP-Bedrock Hybrid CCE Memory Response Pipe Abstract State Machine}
	\label{fig:hybrid-cce-mem-resp-fsm}
\end{figure}

The memory response pipe implements the same response processing logic as the fixed-function coherence engine. The organization of the memory response pipe is depicted in \autoref{fig:hybrid-cce-mem-resp-bd} while \autoref{fig:hybrid-cce-mem-resp-fsm} depicts a logical abstraction of the memory response logic as a three state FSM. As with the fixed-function coherence engine, this state machine is implemented without any explicit encoding of the three discrete states shown in the figure. The following text briefly summarizes the functionality of the memory response pipe, which is outlined in full in \autoref{sec:cce_fsm_mem_resp}.

Logically, as every returning memory response message is processed by the memory response pipe it is either forwarded to the appropriate LCE or squashed and sunk in the pipe. Responses with addresses in the cacheable memory address space also decrement the pending bit counter of the associated way group when consumed. Speculative responses read the speculative bits to determine if the request processing logic has finished and resolved the speculation, with processing stalling until speculation has been resolved in the event that the memory read response returns prior to the coherence request processing logic resolving speculation. Speculative responses can either be forwarded without modification, forwarded after modifying the coherence state supplied in the data command to the LCE, or squashed if the read is not required by the protocol. Non-speculative memory responses are either forwarded directly to the LCE specified in the message header or sunk by the memory response pipe, depending on the type of response message.

\subsubsection{Memory Response FSM Occupancy}

\begin{table}[t]\centering
	\ra{1.3}
	\begin{tabular}{@{}L{0.18\linewidth}@{}C{0.24\linewidth}@{}L{0.58\linewidth}@{}}
        \toprule
		Message & Occupancy (cycles) & Description \\
		\midrule
		Read & $N$ & Cache block read data; forward to LCE\\
		Write & $1$ & Cache block writeback complete; sink message\\
		Uncached Read & $N$ & Uncached load data; forward to LCE\\
		Uncached Write & $1$ & Uncached store commited to memory; send Uncached Store Done to LCE\\
		\bottomrule
	\end{tabular}
	\caption{\bpbedrock{} Hybrid CCE Memory Response State Machine Occupancy}
	\label{table:hybrid-cce-mem-resp-occupancy}
\end{table}

\autoref{table:hybrid-cce-mem-resp-occupancy} provides the no-contention memory response logic processing occupancies for the supported memory response message types. Write and Uncached Write responses each require a single cycle to process, which includes writing the pending bit and sending a command message, if required. Uncached Read and Read responses each require $N$ cycles to process. The number of cycles required to send the \bedrock{} command message is determined by the the data width of the \bedrock{} network channels and the cache block size in the coherence system.

\subsection{LCE Response Pipe}
\label{sec:hybrid-lce-resp-pipe}

\begin{figure}[t]
	\centering
	\includegraphics[width=0.8\linewidth]{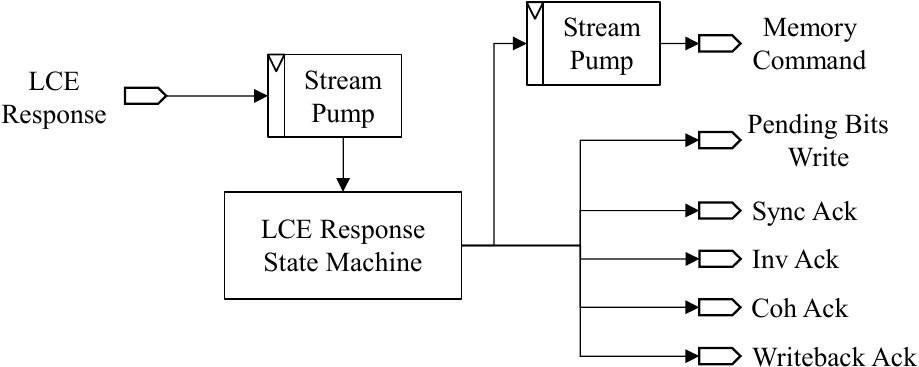}
	\caption{Hybrid CCE LCE Response Pipe Block Diagram}
	\label{fig:hybrid-lce-resp-pipe-bd}
\end{figure}

The LCE response pipe processes response messages from the LCEs, including command acknowledgments and cache block writebacks. \autoref{fig:hybrid-lce-resp-pipe-bd} depicts the organization of the LCE response pipe. LCE response messages arrive from the coherence network interface into a message stream pump, which presents the messages for processing to the state machine. The response state machine then issues memory commands, updates the pending bits, or signals message arrival to the coherent request pipe, depending on the specific type of response message being processed. The LCE response pipe operates completely independently from the coherent request or uncached request pipes. However, output signals from the LCE response pipe are consumed by other pipes in the hybrid coherence engine, which requires the other pipes to be able to process these signals at any cycle, independent of any other processing occurring in those pipes.

At system startup, synchronization acknowledgment messages are sunk by the response pipe, with each message raising the sync ack signal to the control pipe. Invalidation acknowledgments are also sunk by the response pipe, with each ack causing the invalidation ack signal to be raised for one cycle to inform the coherent request pipe that an invalidation ack has returned. Writeback responses are forwarded to the last-level cache or memory controller by issuing memory command messages, while null writebacks are sunk by the response pipe. In both cases, the writeback acknowledgment signal is raised to notify the coherent request pipe that the writeback has returned and been processed. Writebacks that are forwarded as memory commands also increment the pending bit of the associated way group to maintain ordering across related coherence transactions. As coherence transactions complete, coherence acknowledgment messages return from the LCEs. Each coherence ack decrements the pending bit of the way group associated with the transaction and raises the coherence ack signal in case it is needed by one of the other pipes in the design.

\begin{figure}[t]
	\centering
	\includegraphics[width=0.25\linewidth]{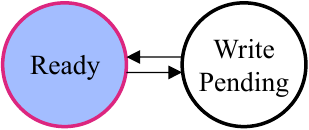}
	\caption{Hybrid CCE LCE Response Pipe State Machine}
	\label{fig:hybrid-lce-resp-pipe-fsm}
\end{figure}

\autoref{fig:hybrid-lce-resp-pipe-fsm} shows the state machine implemented in the LCE response pipe. Due to the use of stream pumps, the state machine is very simple. Every arriving message is processed over one or more cycles in the Ready state. Response messages that require writing the pending bits transition to the Write Pending state after the message has been consumed and the appropriate single-bit signal has been raised. The pending bit write takes a single cycle, assuming there is no contention for the port from the memory response pipe.

\subsubsection{LCE Response Pipe FSM Occupancy}

\begin{table}[t]\centering
	\ra{1.3}
	\begin{tabular}{@{}L{0.18\linewidth}@{}C{0.24\linewidth}@{}L{0.58\linewidth}@{}}
        \toprule
		Message & Occupancy (cycles) & Description \\
		\midrule
		Sync Ack & $1$ & Sink message, raise synchronization ack signal\\
		Invalidate Ack & $1$ & Sink message, raise invalidation ack signal\\
        Coherence Ack & $2$ & Sink message, raise coherence ack signal, write pending bits\\
        Writeback & $1 + N$ & Forward as memory command, raise writeback ack signal, write pending bits\\
        Null Writeback & $1$ & Sink message, raise writeback ack signal\\
		\bottomrule
	\end{tabular}
	\caption{\bpbedrock{} Hybrid CCE LCE Response State Machine Occupancy}
	\label{table:hybrid-cce-lce-resp-occupancy}
\end{table}

\autoref{table:hybrid-cce-lce-resp-occupancy} lists the message processing occupancy, in cycles, for the LCE response state machine. All messages require at least one cycle to process. Synchronization ack, invalidation ack, and null (clean) writeback messages require exactly one cycle to sink the message and raise the corresponding signal to notify other pipes that the message has returned. Coherence acknowledgments require two cycles to first sink the message and raise the coherence ack signal and then write the pending bit associated with the coherence transaction. Full cache block writeback messages require $N$ cycles to forward the writeback as a memory command, where N is the number of beats per message as determined by the implementation's network channel width, plus one cycle to write the pending bit associated with the cache block. For writeback messages, the writeback ack signal is raised when the last beat of the memory command is sent.
\section{Programmable Pipe Architecture}
\label{sec:hybrid-prog-pipe}

\begin{figure}[t]
	\centering
	\includegraphics[width=0.80\linewidth]{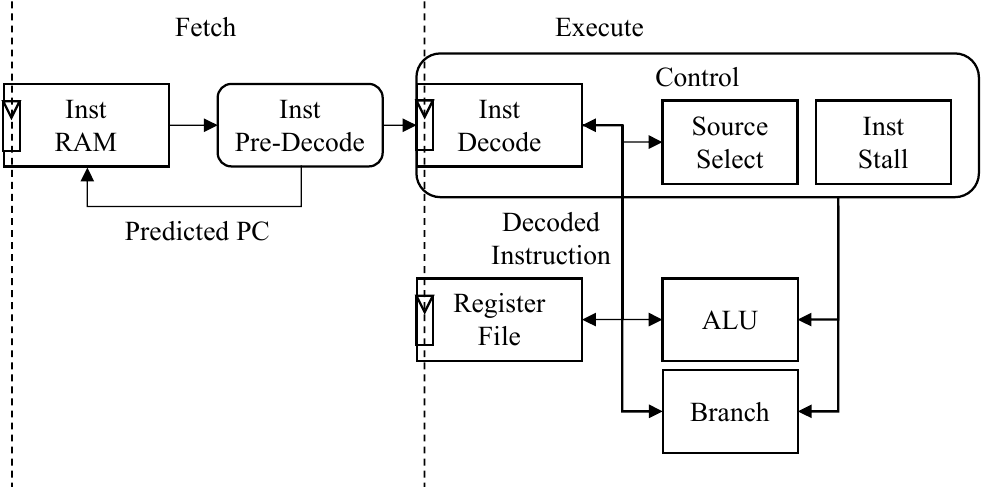}
	\caption{Hybrid CCE Programmable Pipe Block Diagram}
	\label{fig:hybrid-prog-pipe-bd}
\end{figure}

\autoref{fig:hybrid-prog-pipe-bd} shows the organization of the hybrid CCE's programmable pipe. This pipe is a trimmed-down version of the full microcode-programmable coherence engine architecture that retains the general-purpose microcode-programmable execution logic but removes most of the coherence protocol-specific logic, which is handled by the fixed-function hardware in the hybrid coherence engine. As with the microcode-programmable coherence engine, the programmable pipe is organized as a two-stage fetch-execute pipeline with 64-bit general purpose registers and datapath.

\subsection{Fetch - Instruction RAM and Predecode}

The fetch stage comprises the instruction storage RAM and fetch logic plus an instruction predecoder. The Instruction RAM module contains the microcode instruction memory, the fetch program counter (PC), and next PC logic. After system reset, an external configuration bus loads the programmable pipe's microcode into the instruction memory and starts the pipe's execution. Once execution begins, a new instruction is fetched every cycle unless the execute stage raises the stall signal. If a stall occurs, the previously fetched instruction is held valid on the output of the instruction memory.

The instruction RAM module outputs the fetched instruction and the fetch PC, which are fed to the Instruction Predecode module. The predecoder determines if the instruction is a branch instruction, whether the instructions predict taken bit is set, and the branch target encoded in the instruction. It then outputs a predicted fetch PC that is either the current PC plus one or the branch target. The instruction RAM uses the predicted fetch PC to fetch the next instruction, unless the execute stage reports a branch misprediction. Branch mispredictions unconditionally redirect the fetch PC to the resolved PC provided by the Branch module in the execute stage.

\subsection{Execution Control}

The Instruction Decode, Source Select, and Instruction Stall modules make up the execute stage's control logic. Collectively, these modules create the necessary control signals for the programmable pipe's functional units and detect execution hazards that require the pipe to stall and replay instructions.

\subsubsection{Instruction Decode}

The Instruction Decode unit expands the narrow microcode instruction into a wider decoded instruction that contains functional unit control signals. The decode module also contains the current instruction and PC registers. Instruction Stalls cause the current instruction to be replayed in the next cycle, and branch mispredictions cause a single cycle bubble in execution while the execute stage waits for a new instruction to be fetched. The output of the decoder is the decoded instruction and the current execute stage PC.

\subsubsection{Source Select}

The source select module routes operands to the programmable pipe's functional units based on the current instruction. A source operand may come from the general purpose registers or special registers associated with the coherent request pipe to programmable pipe interface, depending on the specific instruction being executed.

\subsubsection{Instruction Stall}

The Instruction Stall unit controls whether the current instruction executes and commits or must be replayed in the following cycle. Stalls occur due to functional unit hazards. The unit takes the decoded instruction as input, examines its control signals, and stalls execution if any of the possible stall conditions are met. The stall signal is routed to the fetch stage to retain the previously fetched instruction, and to the instruction decoder to replay the current instruction in the next cycle. Stalls occur when the programmable pipe must wait for a new coherence transaction from the coherent request pipe.

\subsection{Register File}

The Register File stores the architectural state of the programmable pipe in eight 64-bit general purpose registers (GPRs). These registers can be read and written by the microcode instructions and can be used to store temporary variables and values during execution of the microcode program.

\subsection{Functional Units}

The programmable pipe includes two functional units that implement instruction execution. The Branch unit manages control flow while the ALU unit implements standard RISC-style general purpose arithmetic instructions.

\subsubsection{Branch}

The Branch unit resolves branches and validates the fetch stage's speculative fetch. The branch unit takes the current instruction's two operands, branch operation, valid bit, predict taken bit, PC, and branch target as inputs. It then computes the result of the branch operation and determines if a misprediction occurred by comparing the branch outcome to the predict taken bit. Mispredictions result in a single cycle bubble in the execute stage and redirect the fetch stage to the proper fetch PC. The next fetch PC will either be the current execute stage PC plus one or the branch target, depending on the outcome of the branch comparison.

\subsubsection{ALU}

The Arithmetic Logic Unit (ALU) is a simple, 64-bit wide ALU supporting addition, subtraction, logical shifts, and bitwise operations. The supported bitwise operations are AND, OR, XOR, NAND, NOR, and negation. The ALU also supports logical negation of a single operand. The hardware ALU is purposefully simplistic to reduce complexity. Additional common operations are supported at the software level by the assembler. Software supported operations include increment, decrement, add immediate, subtract immediate, and shift immediate.

\subsection{Instruction Set Architecture (ISA)}

The programmable pipe effectively executes a restricted subset of the microcode-programmable coherence engine's Base ISA, listed in \autoref{table:bedrock-isa-base-alu}, \autoref{table:bedrock-isa-base-branch}, and \autoref{table:bedrock-isa-base-data}. All of the basic ALU, Branch, and Data Movement instructions operating on the general purpose registers are supported, as are limited operations on some special registers associated with the coherent request pipe to programmable pipe interface, which carries an LCE request message header. Some of the instructions from the Coherence ISA's Queue subset, listed in \autoref{table:bedrock-isa-coherence-queue}, are also implemented to handle processing of the LCE request message header provided by the coherent request pipe.

\subsection{Programmable Pipe Interface}

The programmable pipe interfaces with the coherent request pipe using a simple status message interface. In the initial implementation of the hybrid coherence engine, the status message is a one-bit wire that is driven by bit zero of GPR zero (r0). A value of zero on this signal tells the coherent request pipe to squash the current coherence request being processed, while a value of one tells the coherent request pipe to proceed processing the request and commit the transaction in the coherence protocol. The microcode program executing in the programmable pipe must set bit zero of GPR zero appropriately for every new coherence request being processed. As shown in \autoref{fig:hybrid-coh-req-pipe-fsm}, the coherent request pipe state machine stalls in the Programmable Pipe Sync state until the programmable pipe sends the status message for the transaction.

The information that is communicated from the programmable pipe to the coherent request pipe and the capabilities of the programmable pipe is an important design decision with many possibilities and tradeoffs. The interface can be synchronous or asynchronous and stalling (blocking) or non-stalling (non-blocking). As implemented, the interface is synchronous and stalling, requiring the programmable pipe to provide a squash or proceed status message to the coherent request pipe for every coherence request while the coherent request pipe stalls waiting for the status message.

The decision to implement a synchronous status message interface results in two major practical considerations for the coherence engine's functionality. First, the programmable pipe's microcode must make a binary decision about every transaction, determining whether the transaction is allowable or not. Thus, while the mechanics of processing the coherence transaction are handled by the fixed-function protocol processing logic, the programmable pipe retains full control over the system's functional behavior and can squash any transaction. Second, the programmable logic has a strict processing latency that must be met to avoid incurring additional transaction processing overheads. As shown in \autoref{fig:hybrid-coh-req-pipe-fsm}, the fixed-function protocol processing logic spends only $5 + (C/2)$ cycles to perform the initial pre-commit processing of each request, assuming no replacement or invalidations. Thus, the programmable pipe has only a handful of cycles to perform its processing to avoid stalling the state machine. In practice, this may limit the complexity of programmable processing that can be performed by the programmable pipe within the initial hybrid coherence engine design.

If the programmable pipe is intended to support functionality that does not control whether each transaction should proceed or be squashed, an asynchronous non-stalling interface can be used. In this situation, it may not even be necessary for the programmable pipe to return any type of status or other message back to the coherent request pipe. Rather, the programmable pipe would simply perform its processing for each coherence request independent of the coherent request pipe, including possibly accessing memory or sending messages to other system components. If the programmable pipe's program is unable to execute with a throughput that matches the coherence protocol processing, the coherent request pipe can be made to stall until the programmable pipe, or a buffer that feeds it requests, has space available. Alternatively, processing of some coherence requests can be skipped by the programmable pipe while the coherent request pipe continues to process all requests for protocol correctness. However, having the programmable pipe process only a subset of coherence requests means that it cannot be used to implement logic that requires visibility of every coherence request.

\section{Performance Comparison}
\label{sec:hybrid-perf-analysis}

The hybrid coherence engine implements identical protocol processing logic as the fixed-function coherence engine (FSM CCE), however the organization of the two designs differs significantly. Therefore, it is important to compare the request processing occupancies and protocol processing performance of the two designs to understand the implications of these design decisions.

\subsection{Request Processing Occupancy}

\begin{table*}[t]\centering
    \footnotesize
	\begin{tabular}{@{}L{0.12\linewidth}@{}C{0.12\linewidth}@{}C{0.14\linewidth}@{}C{0.31\linewidth}@{}C{0.31\linewidth}@{}}
        \toprule
		Request & LCE State & Directory State & FSM CCE Occupancy (cycles) & Hybrid CCE Occupancy (cycles) \\
		\midrule
		\multirow{4}{*}{Read} & \multirow{4}{*}{I} & I, S & $8 + (C/2)$ & $8 + (C/2)$ \\
		& & E (clean) & $10 + (C/2)$ & $10 + (C/2)$ \\ % ST_TR_WB
		& & E (dirty) & $9 + (C/2) + N$ & $10 + (C/2)$ \\ % ST_TR_WB
		& & M, O, F & $9 + (C/2)$ & $9 + (C/2)$\\ % ST_TR
		\midrule
		\multirow{4}{*}{Write} & \multirow{4}{*}{I} & I & $8 + (C/2)$ & $8 + (C/2)$ \\ % read mem
		& & S & $8 + (C/2) + (2*S)$ & $8 + (C/2) + (2*S)$ \\ % Inv all S, read mem
		& & E, M & $9 + (C/2)$ & $9 + (C/2)$ \\ % ST_TR
		& & O, F & $9 + (C/2) + (2*S)$ & $9 + (C/2) + (2*S)$ \\ % Inv all S, ST_TR
		\midrule
		Write & S & S, O, F & $9 + (C/2) + (2*(S-1))$ & $9 + (C/2) + (2*(S-1))$ \\ % Inv all S, STW to Req
		\midrule
		Write & O, F & O, F & $9 + (C/2) + (2*S)$ & $9 + (C/2) + (2*S)$ \\
		\bottomrule
	\end{tabular}
	\caption{\bpbedrock{} CCE Request Occupancy Comparison - MOESIF}
	\label{table:all-cce-occupancy-moesif}
\end{table*}

\autoref{table:all-cce-occupancy-moesif} presents a comparison of the coherence request processing occupancies for the hybrid and fixed-function coherence engines. As seen in the table, the two coherence engines have nearly identical theoretical request processing occupancies. Both designs have baseline request occupancy costs of $8 + (C/2)$ cycles for every request and require $(2*S)$ cycles to perform invalidations of cache block sharers. The two designs differ only in the case where a writeback of the target coherence block is required, which occurs when a read request is made to a block cached in the Exclusive (E) state in another cache and that cache has performed a write that silently upgraded the block to the Modified (M) state. This results in the owning cache sending a dirty writeback response to the coherence engine as the block transitions to the Owned (O) state, which is dirty and shared with a single owner. The fixed-function coherence engine incurs a one cycle cost to issue the command message and then N cycles to process a dirty writeback response while the hybrid coherence engine incurs a cost of two cycles to issue the command message and then observe the writeback has been processed by the LCE response pipe.

This analysis, however, is potentially misleading and illuminates the challenges of inferring protocol processing throughput and application performance from processing occupancies. While the coherent request pipe of the hybrid coherence engine appears to have a lower processing occupancy in this particular situation, if the state machine stalls waiting for the writeback ack signal from the LCE response pipe then it will in practice experience an overhead of one plus N cycles to issue the writeback and see the writeback response, just as the fixed-function engine experiences. This happens because the LCE response pipe will take at least N cycles to process the N beats of the writeback and forward them to the memory command network.

\subsection{Micro-benchmark Performance}

\begin{table*}[t]\centering
    %\footnotesize
	\begin{tabular}{@{}L{0.25\linewidth}@{}L{0.75\linewidth}@{}}
        \toprule
		Program & Description \\
		\midrule
        Sanity & Coherence protocol sanity check program with deliberate false sharing.\\
        Atomic Add & Atomic (amoadd.d) increment of shared global variable by all cores.\\
        LR/SC Add & LR/SC-based increment of shared global variable by all cores.\\
        Random Walk & Epoch-based synchronization with no data sharing.\\
        Work Sharing Sort & Cooperative sorting of a large collection of arrays with synchronization for array selection.\\
		\bottomrule
	\end{tabular}
	\caption{\bpbedrock{} Microbenchmark Programs}
	\label{table:hybrid-ci-programs}
\end{table*}

The coherence engine request processing occupancy comparison presented above illuminates the challenges of comparing coherence system implementations at the component level. To better understand the holistic performance of the \bpbedrock{} hybrid coherence engine, a set of experiments are run using microbenchmarks that stress different aspects of the coherence system. Five microbenchmark programs are tested in RTL simulation of the \bpbedrock{} multicore processor design with core counts ranging from one to sixteen, in powers of two. At each core count, designs with all three \bpbedrock{} coherence engines are tested. The five microbenchmarks are listed in \autoref{table:hybrid-ci-programs}.

The Sanity program was designed as a simple smoke test of the \bpbedrock{} coherence system. It stripes accesses to shared global memory by all cores at a data word granularity, for example core 0 access word 0, core 1 accesses word 1, core 2 accesses word 2, and so on. This creates deliberate false sharing of cache blocks, which stresses cache block sharing in the coherence system. Every core accumulates a sum of the data words it accesses into a local variable, meaning that there are no writes occurring to the shared global memory array. This program primarily stresses the throughput of the coherence directory as it continually processes and services read requests from the caches.

The Atomic Add and LR/SC Add programs both use all cores to increment a single shared global variable a large number of times. This program stresses cache to cache transfers since every increment requires write permissions, thereby causing the cache block containing the shared global variable to be continually passed around among the cores' data caches.

The Random Walk program performs epoch-based synchronization among cores. Program execution is segmented into a large number of identical epochs during which each core executes a trivial local computation. At the end of each epoch's local computation, the cores perform a global synchronization with each other, waiting for for all cores to finish the current epoch before proceeding to the next epoch. This program stresses infrequent bursts of demand for a shared block.

The Work Sharing Sort microbenchmark uses all of the cores to collaboratively sort a large collection of small to moderate sized arrays. Each core executes an acquire-then-sort loop that first performs synchronization on a shared global variable to acquire a pointer to one of the unsorted arrays. The core then sorts the array, reading and writing the array from the shared global memory. After the array is sorted, the core repeats the loop, attempting to acquire another unsorted array for sorting. All cores execute until there are no more arrays to sort. This program stresses random synchronization overlapped with constant read and write accesses to the shared global memory. While there is no sharing of array data among the cores, each core is constantly issuing requests to the coherence directory, which puts pressure on the processing throughput of the coherence system.

\begin{figure}[t]
	\centering
	\includegraphics[width=1\linewidth]{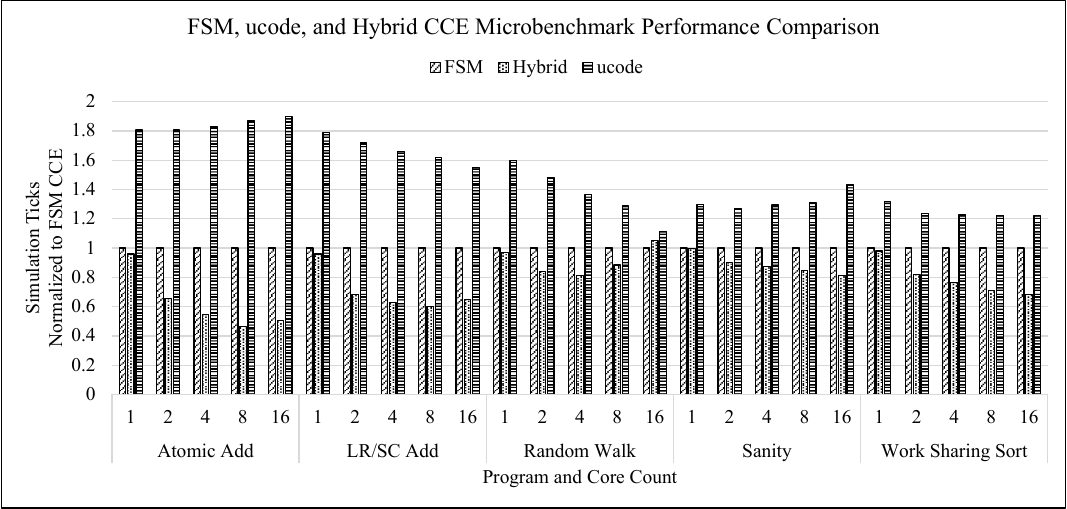}
	\caption{Hybrid CCE Performance Comparison}
	\label{fig:hybrid-ci-microbench}
\end{figure}

\autoref{fig:hybrid-ci-microbench} shows the results of these microbenchmark simulation experiments with results normalized to the fixed-function coherence engine design. For each microbenchmark, total simulation time is collected and then normalized to the FSM CCE design. These results show that the microcode-programmable coherence engine-based multicore designs exhibit consistently lower program-level performance than the FSM-based design. The hybrid coherence engine-based multicore designs exhibit consistently greater program-level performance than the FSM-based design. However, there are not clear trends across the set of microbenchmarks regarding the scalability of the coherence engines. Atomic Add and Sanity show generally degrading performance for the microcode-programmable coherence engine-based designs while LR/SC Add, Random Walk, and Work Sharing Sort demonstrate favorable performance scaling as the multicore core count increases. For the hybrid coherence engine, the Random Walk microbenchmark initially shows improving performance as core count increases, but then regresses when a core count of 16 is reached. The other four programs generally show favorable scaling relative to the fixed-function coherence engine design.

The results generally show that using the microcode-programmable coherence engine in the multicore results in lower application-level performance while using the hybrid engine generally results in improved performance. For the microcode-programmable coherence engine-based multicores, these results generally follow from the request processing occupancy analysis in \autoref{sec:bp-bedrock-cce-perf}. The request processing occupancy analysis shows that the microcode-programmable engine has nearly 100\% request processing latency overheads, which can have a significant impact on total application performance for applications that stress the cache coherence system with frequent cache block sharing and cache to cache transfers. In contrast, the hybrid coherence engine implements nearly identical logic as the fixed-function coherence engine and has effectively identical theoretical request processing occupancy latencies. However, the microbenchmark experiments show that the hybrid coherence engine-based multicore designs have a significant performance advantage over the FSM-based designs. The hybrid coherence engine's implementation differs in subtle but important ways from the FSM coherence engine, which likely accounts for the differences in application-level performance. For example, the hybrid engine includes more buffering for inbound requests due to the use of independent coherent and uncacheable pipes and the additional request buffering provided by the request arbiter. This allows more requests to be queued up at a given coherence engine. Additionally, the hybrid engine implements a pending request queue that holds requests blocked by the pending bits while allowing newer requests targeting other way groups to proceed ahead of the blocked request. This mechanism preserves the ordering of requests targeting the same cache block but enables greater request processing throughput relative to the FSM-based coherence engine that will stall ready requests behind requests blocked by the pending bits. Collectively, these implementation details reinforce the importance of the coherence engine architecture and organization in determining realized protocol processing and application-level performance.

\section{Resource Comparison}
\label{sec:hybrid-area-analysis}

\begin{figure}[t]
	\centering
	\includegraphics[width=1\linewidth]{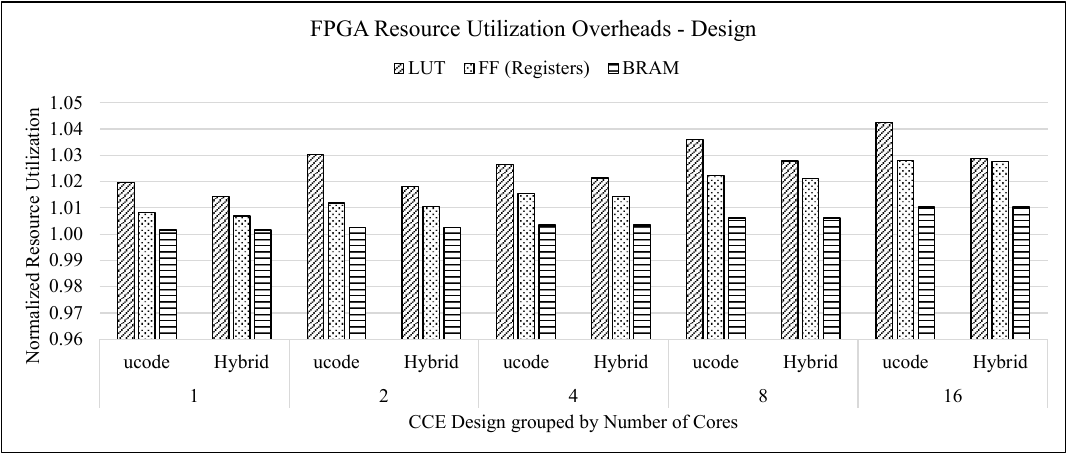}
	\caption{FPGA Resource Utilization Overheads - Full Design}
	\label{fig:fpga-util-design}
\end{figure}

\begin{figure}[t]
	\centering
	\includegraphics[width=1\linewidth]{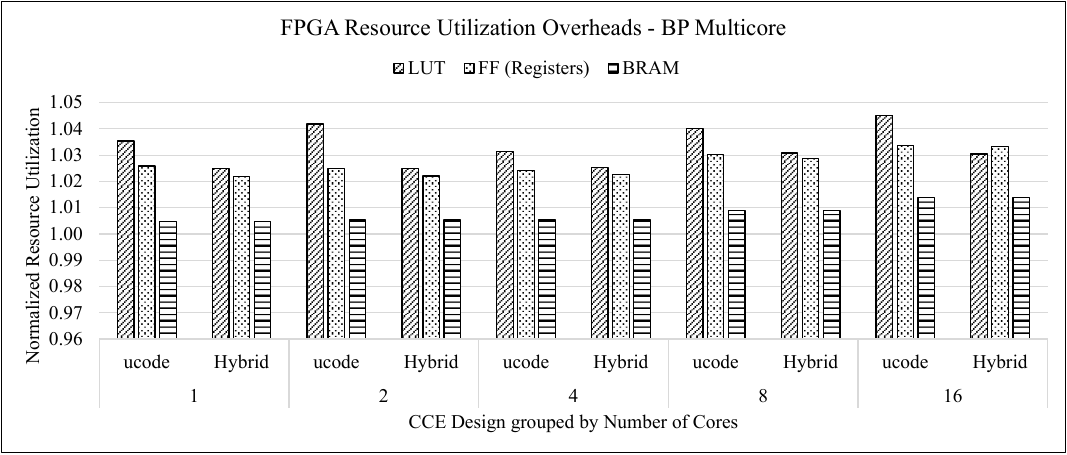}
	\caption{FPGA Resource Utilization Overheads - BlackParrot Multicore}
	\label{fig:fpga-util-bp}
\end{figure}

Alongside protocol processing and application-level performance, area and resource utilization are important metrics for any processor design. As discussed in \autoref{sec:bp-bedrock-cce-area}, the fixed-function and microcode-programmable coherence engine designs were compared in both ASIC and FPGA implementations, revealing that programmability has a small, but non-trivial resource cost. After implementing the hybrid coherence engine design, FPGA-based implementations of multicore designs leveraging all three coherence engine designs were compared again. Overall, this updated comparison reveals similar trends and overheads as the initial area and resource utilization comparisons for the FSM and ucode designs.

\autoref{fig:fpga-util-design} shows the resource utilization overheads of the ucode CCE and hybrid CCE designs at core counts ranging from 1 to 16, in powers of two. Resource utilization is recorded at the FPGA design level, which includes all logic required in the FPGA design. The complete FPGA design comprises the \bpbedrock{} multicore processor with one of the three coherence engine designs, PCI Express (PCIe) interface logic, AXI network interconnect blocks, FPGA host logic to control the multicore, and HBM memory controller logic. The three FPGA resources compared are the number of Lookup Tables (Total LUT), the number of flip-flop elements or registers (FF), and the number of hardened memories or block RAM elements (BRAM). The hybrid coherence engine resource utilization is normalized to a design including the hybrid coherence engine with the programmable pipe logic removed while the microcode-programmable design resource utilization is normalized to a design using the fixed-function coherence engine. The figure shows that the cost of programmability in both the microcode-programmable and hybrid coherence engines is relatively small at the overall design level. Both designs have single-digit percentage overheads for all three resource classes, which is consistent with earlier results evaluating the microcode-programmable coherence engine.

\autoref{fig:fpga-util-bp} shows the resource utilization overheads at the \bpbedrock{} multicore design level when using either the microcode-programmable or hybrid coherence engine designs. As above, the ucode CCE resource utilization is normalized to the FSM CCE resource utilization while the hybrid CCE resource utilization is normalized to a hybrid CCE design without any programmable logic (excludes the programmable pipe logic). At the multicore level of the design hierarchy, which excludes all of the peripheral system logic such as memory controllers, on-chip interconnects, and the FPGA-based processor host logic, the resource utilization overheads remain small. Overheads are still in the single-digit percentage range and are marginally lower than the overheads at the FPGA design level. This makes sense intuitively, as the peripheral logic, while providing important functionality, is still relatively minimal compared to the \bpbedrock{} multicore.

One important note about the FPGA resource utilization results is that FPGA implementation tools use non-deterministic algorithms. These algorithms result in different optimizations and resource utilization decisions being made depending on the total design complexity. For example, in some designs, the same logical storage element may be implemented using either LUT, flip-flop, or hardened block RAM (BRAM) resources. Therefore, the results presented do not necessarily show definitive trends of growing or shrinking overheads as the multicore processor's core count grows. In particular, \autoref{fig:fpga-util-bp} shows how LUT and FF resource utilization may grow or shrink across various core counts at the \bpbedrock{} multicore level.

Overall, these resource utilization results confirm the earlier findings of \autoref{sec:bp-bedrock-cce-area}. The area and resource overheads required to introduce programmability into the cache coherence system are on the order of single-digit percentages when viewed at the multicore design level. Despite the significant importance of the coherence directory within the multicore in ensuring the correctness of the shared-memory system, the logic required to implement the coherence protocol is small compared to the logic and storage resources required to implement the processor pipeline and its data and instruction caches.

Data tables containing the full resource utilization numbers are provided in \autoref{app:fpga-util} for reference. Additionally, \autoref{app:fpga-layout} provides screen captures of the FPGA-based implementation layouts.

\section{Conclusion}
\label{sec:hybrid-conclusion}

The hybrid coherence engine presented in this chapter expands upon the investigations of \autoref{chap:bp-bedrock} to realize a coherence engine that maintains the performance of a fixed-function coherence engine design while introducing programmability to accommodate domain-specific functionality. The hybrid coherence engine design evolves the fixed-function protocol processing logic of the FSM-based coherence engine to provide additional inter-transaction concurrency. It leverages the innovations of the initial \bpbedrock{} coherence engines in directory organization and transaction management to preserve the low directory storage overheads and scalability of a tiled multicore implementation, while decomposing the protocol processing logic into a set of independently operating pipelines. The hybrid coherence engine architecture presents one possible method of integrating programmable logic with the coherence processing pipelines using a synchronous status message interface and discusses the tradeoffs involved when architecting the interface and interaction among the protocol processing logic and the domain-specific programmable logic. An analysis of the hybrid coherence engine in comparison to the fixed-function and microcode-programmable engines presented earlier shows that the hybrid engine's architectural enhancements result in improved protocol processing and application-level performance while retaining the low, single-digit percentage resource overheads for adding programmability to the coherence engine. These results further motivate investigating the integration of useful programmability into the cache coherence engines of shared-memory multicore processors.
\chapter{Related Work}
\label{chap:related}

A significant amount of research has been published on the topic of cache coherence protocols, coherence systems, and multicore RISC-V processors. Cache coherence has existed since the first multiprocessor computers included caches, and hardware-based cache coherence remains the solution of choice for nearly all modern shared-memory multicore processors. Software-based and hybrid coherence systems have been explored to a lesser extent but have received more attention recently due to the emerging popularity of GPU, manycore, and heterogeneous architectures and systems that place additional demands on the coherence and memory systems, while programmable coherence engines have been explored by a handful of prior research projects. The growing open-source hardware movement and the emergence of the RISC-V architecture has further driven contemporary research in efficient multicore processor design.

In the rest of this chapter, \autoref{sec:related_coherence} first discusses prior work on cache coherence protocols. \autoref{sec:related_hw_sw_hybrid} describes related research spanning hardware, software, and hybrid coherence systems. \autoref{sec:related_programmable} describes past efforts involving the use of programmability within the cache coherence system and its coherence engines, the memory hierarchy, or the network. \autoref{sec:related_open_source} concludes the chapter discussing contemporary open-source RISC-V multicore processor designs and systems. Some prior works appear in more than one section as they are related to \bedrock{} and \bpbedrock{} in multiple ways.

\section{Cache Coherence Protocol Design}
\label{sec:related_coherence}

A large body of work has studied the problem of cache coherence protocol design. The topic has been of critical importance since the first multiprocessor computers with private caches attached to the processor cores were introduced. This dissertation describes the \bedrock{} directory-based cache coherence protocol, however the focus of this dissertation is not an in-depth investigation into optimal cache coherence protocol design. \bedrock{} draws heavily from prior research on cache coherence protocols, which makes this area highly relevant.

\subsection{The Cache Coherence Problem}

An excellent contemporary survey of cache coherence is provided by Nagarajan et al. \cite{nagarajan_primer_2020}. The terminology used in this dissertation is largely drawn from their overview and the reader is referred to this reference for a thorough description of the memory consistency problem, the cache coherence problem, and their canonical solutions. Early works that described the cache coherence problem include those of Tang \cite{tang_cache_1976} and Censier and Feautrier \cite{censier_new_1978}. Both of these papers describe a cache coherence protocol with three states, which are canonically referred to today as the MSI states. Censier and Feautrier provide an informal yet effective definition of the coherence problem, and perhaps more importantly, describe the coherence system as a set of two processes running asynchronously and in collaboration to implement the protocol. Viewing the set of coherence controllers as a collection of asynchronous processes is a common and powerful abstraction when dealing with coherence systems and remains in wide use in the literature today. This abstraction is applied to describe \bedrock{} through its state transition and coherence protocol tables, which specify how both the cache controllers and coherence directories respond to various messages and requests.

\subsection{Protocol States}

The canonical set of stable coherence states used in most protocols are called the \textit{MOESIF} states, standing for Modified (M), Owned (O), Exclusive (E), Shared (S), Invalid (I), and Forward (F). As mentioned above, both Tang \cite{tang_cache_1976} and Censier and Feautrier \cite{censier_new_1978} effectively described three-state protocols with the MSI states. Papamarcos introduced the four-state MESI protocol, which is at times called the "Illinois Protocol" due to its author's association with the University of Illinois \cite{papamarcos_low-overhead_1984}. A year later, Katz described the Berkeley Ownership Protocol, which is equivalent to a four-state MOSI protocol \cite{katz_implementing_1985}. The five-state MOESI protocol was proposed by Sweazey and Smith \cite{sweazey_class_1986}. Lastly, Intel introduced the Forward state and the five-state MESIF protocol used in Intel QPI \cite{intel_qpi_2009}. The MOESIF states can also be defined a set of four basic properties: validity, dirtiness, exclusivity, and ownership \cite{papamarcos_low-overhead_1984, sweazey_class_1986}. In relation to these works, \bedrock{} defines a family of coherence protocols using subsets of the MOESIF coherence states, where each state is described by the three properties of validity, dirtiness, and non-exclusivity.

\subsection{Snooping versus Directory Protocols}

Most protocols in use today can be classified as either \textit{snooping} or \textit{directory} protocols. Snooping protocols rely on all caches participating in coherence being able to observe all coherence transactions \cite{goodman_using_1983, papamarcos_low-overhead_1984, katz_implementing_1985, thacker_firefly_1987}. In other words, transactions are broadcast across the communication network or shared bus to all caches in the system. Each cache takes an action in response to every transaction to maintain the overall correctness of the system. Snooping protocols have limited scalability as they typically require a shared communication bus, which quickly becomes expensive to implement as the system scales up in size. Directory protocols were introduced to overcome the scalability and bandwidth limitations of snooping protocols \cite{agarwal_evaluation_1988, zhang_fractal_2010, ferdman_cuckoo_2011, gupta1990, tao_li_adirsub_2001, censier_new_1978, conway_amd_2007, conway_cache_2010, martin_token_2003, kuskin_stanford_1994, menezo_rainbow_2020, sanchez_scd_2012, franques_widir_2021, james_distributed-directory_1990, lenoski_stanford_1992}. They utilize unicast, point-to-point messaging and leverage indirection through a coherence directory to maintain protocol correctness. These protocols tend to be more complex than their snooping counterparts, but their scalability and bandwidth improvements make them the preferred approach for most modern medium- to large-scale multicore processors. Directory protocols also naturally extend to inter-chip, inter-socket, and inter-node coherence systems. There are also schemes that mix snooping and directory designs \cite{pugsley_swel_2010, menezo_flask_2015}.

\bedrock{} defines a directory protocol that is well suited for implementation on the tile-based \blackparrot{} multicore architecture with small to moderate core counts. \bedrock{}'s directory design and protocol are influenced by and share similarities with the OpenSparc T1 \cite{noauthor_opensparc_2008} and Piranha \cite{barroso_piranha_2000}. \bedrock{} utilizes a complete duplicate-tag directory, assumes that store misses allocate into a writeback L1 cache, and assumes the interconnection network is unordered. Additionally, the L2 cache does not participate in coherence and acts as a memory-side buffer.

\section{Hardware, Software, and Hybrid Coherence}
\label{sec:related_hw_sw_hybrid}

Cache coherence systems can often be defined as either hardware, software, or hybrid schemes. The emergence of heterogeneous systems and bespoke accelerators, brought about by the stalling of Moore's Law \cite{moore_1965, moore_1998} and the end of Dennard Scaling \cite{dennard_1974}, has resulted in the introduction of coherence systems tailored to heterogeneous systems. The driving factors in choosing a coherence system approach are the desired performance and memory system architecture of the system components. Adve et al. \cite{adve_comparison_1991}, Grahn et al. \cite{grahn_comparative_2004}, and Komuravelli et al. \cite{komuravelli_revisiting_2015} provide useful comparisons of hardware and software approaches to cache coherence. The appearance of this type of paper in each of the past three decades reveals the prevailing importance of cache coherence as researchers constantly reevaluate the best solution to this complex problem as the computing landscape changes.

\subsection{Hardware-Managed Coherence}

Hardware coherence schemes are by far the most common approach and are used in virtually all modern shared-memory multicore processor systems. These systems are highly specialized for the implemented coherence protocol and provide excellent performance with acceptable power and area overheads. Martin et al. effectively argue why hardware-based cache coherence has been and will continue to be the dominant coherence paradigm \cite{martin_why_2012}. However, as computing systems become more heterogeneous and a broader range of applications are developed, the dominance of hardware-based coherence may not be as certain as they argue. Information about the cache coherence systems of most modern commercial multicore processors remains a tightly held secret by industry, however, from the limited published information, they all use hardware-managed coherence. Conway et al. detailed the coherence system of the AMD Opteron processor, providing rare insight into the workings of a commercial coherence system \cite{conway_amd_2007, conway_cache_2010}. An exhaustive listing of hardware schemes, protocols, and optimizations is beyond the scope of this dissertation.

\subsection{Software-Managed Coherence and Scratchpads}

At the other extreme, some systems manage memory and implement cache coherence, if required, completely in software. An early approach implementing fine-grain access control for shared memory was Blizzard-S \cite{schoinas_fine-grain_1994}. This was preceded by the VMP Multiprocessor \cite{cheriton_1986}, which implemented coherence entirely in software, invoking software cache miss handlers when the network bus detected a coherence action was required. Similarly, Grahn et al. \cite{grahn_1995} investigate implementing the coherence directory logic entirely in software executed on a single processor without multi-threading, overlapped with hardware-based data transfers. SMTp \cite{chaudhuri_smtp_2004, chaudhuri_integrated_2007} extends this idea by using a hardware thread of an SMT-enabled processor to execute coherence protocol processing logic on cache misses.

The Intel Single-chip Cloud Computer (SCC) project is perhaps the most well known contemporary example of software-managed coherence \cite{gries_scc_2011, IntelSCC, gschwandtner_performance_2011, lee_opencl_2011, howard_48-core_2010}. Like SCC, most systems without hardware coherence rely on some form of explicit message passing for data sharing among processor cores. Another interesting example is the COMIC runtime system for the Cell BE processor, which was used in the Sony PlayStation 3 gaming console \cite{lee_comic_2008}. COMIC provides coherent shared memory across the two processing element complexes of the Cell BE. Software-managed coherence has also been explored for GPU devices, which have very different memory access and usage patterns than traditional CPUs \cite{pai_fast_2012}. Implementing coherent shared memory in software incurs large overheads or requires significant intervention from either the compiler or programmer to guarantee correctness. Reasoning about memory correctness, both for coherence and the consistency model built atop it, remains extremely difficult and out of reach for the average programmer. Combined with the long history of hardware-based solutions in commercial products providing seamless backwards compatibility, software-managed coherence remains uncommon. \bedrock{} proposes introducing programmability into the coherence system to enable system- or application-specific functionality as opposed to using programmability to enable arbitrary coherence protocols.

A second architecture trend that has regained popularity recently are manycore designs with scratchpad memories \cite{davidson_celerity_2018, zaruba_manticore_2021}. These memories are managed purely in software and require explicit management of data movement. However, when data movement can be successfully coordinated, these architectures are highly performant and efficient. Recent work has explored techniques to introduce hardware, software, and hybrid coherence schemes for manycore architectures \cite{wang_efficiently_2020, kommrusch_optimizing_2021, cilardo_lightweight_2019, alvarez_coherence_2015, fu_coherence_2015, vianes_case_2022}. In contrast to these efforts, \bedrock{}'s coherence protocol targets small- to medium-scale shared-memory multicore processors, which rely on hardware-managed caches rather than software-managed scratchpad memories.

\subsection{Hybrid Coherence Schemes}

Hybrid coherence schemes employ dedicated coherence hardware and coherence-specific software mechanisms that either inform or control aspects of the coherence system. Compiler analysis is commonly employed and includes techniques that dictate when invalidations are required \cite{min1989, zhiyuan_li_software_1994}, when coherence can be omitted for data that remains private to a single cache \cite{derebasoglu_coherency_2022}, or provide specialized hardware to accelerate a software protocol \cite{ashby_software-based_2011}. The MIT Alewife system can track between zero and five sharer caches in its limited pointer directory, falling back to software when the pointers are exhausted \cite{agarwal_mit_1995, agarwal_mit_1999, chong_application_1996, chaiken_software-extended_1994, chaiken_limitless_1991}. Another common approach relies on programmer annotations or restricted memory models \cite{hill_cooperative_1992, wood_mechanisms_1993, hill_cooperative_1993, choi_denovo_2011}. In comparison to existing hybrid hardware-software schemes, \bedrock{} utilizes software to control the coherence directory. Application and runtime code can remain completely unaware of the microcode program controlling the directory. The programmability of the directory can be exposed to software, whether firmware, operating system, or runtime/application, to implement system- or application-specific features, however it can also be completely hidden from any level.

\section{Programmability in the Coherence System}
\label{sec:related_programmable}

% Focus: programmability in/near the coherence system
% programmable coherence engines
% processing invoked on cache accesses / misses (Tako, ...)

The core topic of this dissertation is the feasibility and design of a programmable cache coherence engine for modern shared-memory multicore processors. Programmability within the cache coherence system has been explored in the past, with the majority of prior work occurring during the emergence of distributed shared-memory multiprocessor computers. A primary motivation for revisiting this topic is that the computing landscape has changed significantly, in terms of both applications and technology, since the topic was last examined in depth. Compared to the programmable protocol engine research of thirty years ago that was primarily focused on supporting multiple communication models, the design described in this dissertation focuses on supporting system- or application-specific customization through the programmable coherence engine. Some of the benefits and drawbacks of programmable controllers are evaluated in \cite{michael_coherence_1997, michael_coherence_1999}. Their key findings are reconfirmed by the work in this dissertation, namely that reducing protocol processing occupancy in the coherence controllers and specializing the controllers for coherence are key to achieving performance competitiveness with hardware-only implementations.

\subsection{Software-Based Protocol Handlers}

As noted in \autoref{sec:related_hw_sw_hybrid}, cache coherence protocols have been implemented using both hardware and software mechanisms. The Blizzard-S machine \cite{schoinas_fine-grain_1994} relies on instructions inserted into programs before shared-memory access to provide fine-grain access control. The VMP Multicore processor \cite{cheriton_1986} relied on simple hardware state machines to invoke software protocol handlers when coherence actions were required as detected by communication on the shared bus. Similarly, Grahn et al. \cite{grahn_1995} and SMTp \cite{chaudhuri_smtp_2004, chaudhuri_integrated_2007} implement coherence protocol handlers in software that execute on the main processor core, whether single- or multi-threaded. Later systems, like MIT Alewife \cite{agarwal_mit_1995, agarwal_mit_1999, chong_application_1996} with LimitLESS directories \cite{chaiken_limitless_1991, chaiken_software-extended_1994}, relied on software protocol handlers to manage exceptional conditions in the protocol, for example when the available encoding for limited sharers becomes oversubscribed. Despite the use of software protocol handlers, these projects did not focus on how the programmability and flexibility of software handlers could be used to support a wide variety of coherence protocols or to enable system-specific functionality. In contrast, the research described in this dissertation investigates the feasibility of including programmability in the coherence system for the explicit purpose of enabling non-protocol functionality.

\subsection{Shared-Memory Multiprocessors}

During the emergence of shared-memory multiprocessors, a number of projects investigated the use of programmability in the coherence, memory, and interconnect systems of multiprocessor designs. These included the Sun S3.mp \cite{nowatzyk_s3mp_1993, nowatzyk_s3mp_1994, fong_pong_design_1998}, Wisconsin Typhoon \cite{reinhardt_tempest_1994}, Sequent STiNG \cite{lovett_1996_sting}, Stanford FLASH \cite{heinrich_performance_1994, kuskin_flash_1997}, and Piranha \cite{barroso_piranha_2000} machines. These machines primarily focused on solving one of two problems: providing inter-node coherence or enabling multiple inter-node communication paradigms within a single system.

\subsubsection{Sun S3.mp}

The Sun Microsystems Sun Scalable Shared-memory MultiProcessor (S3.mp) \cite{nowatzyk_s3mp_1994, nowatzyk_s3mp_1993, nowatzyk_1995, fong_pong_design_1998} is a cache-coherent non-uniform memory access (CC-NUMA) multiprocessor with distributed shared memory. It is constructed by interconnecting commodity workstations using distributed directories and point-to-point communication between workstation nodes. Each S3.mp node includes multiple processor cores that are kept coherent using snooping-based coherence mechanisms, a portion of the multiprocessor's main memory, a memory controller, and an interconnect controller. Unique characteristics of the S3.mp machine are that it has no preferred or fixed network topology, an internode cache of programmable size that is carved out of the main memory storage, and it relies on microprogrammed and multithreaded coherence protocol engines.

The memory controller on each node includes a remote memory handler (RMH) and remote access server (RAS) that handle requests to and from other nodes in the system, respectively, and collectively implement the cache coherency protocol among nodes in the multiprocessor. Both the RMH and RAS are implemented using identical microcode-programmable protocol engines. Each engine includes input and output state machines and buffers that accelerate message send and receive operations, as well as managing the creation of request threads based on the arriving messages. The microcode engine, called a microcode sequencer, executes a program that operates on one request thread at a time, and threads can be context switched with no latency penalty since all state for all threads is stored in separate hardware registers. Like \bpbedrock{}, the microcode engines execute an instruction set customized for cache coherence protocol operations. S3.mp's protocol engines include message send instructions that offload the arbitration and data transfer of outbound protocol messages from the microcode engine to the message send hardware. Unlike \bpbedrock{}, the coherence directory is stored in main memory rather than dedicated hardware directory storage.

\subsubsection{Wisconsin Typhoon}

Wisconsin Typhoon is a hardware platform that implements the Tempest interface for low-level communication and memory system mechanisms \cite{reinhardt_tempest_1994}. Tempest allows programmers and compilers to directly control hardware-provided communication mechanisms such that both message passing and shared-memory can exist on the same system and be used in the same program. Typhoon is a proposed hardware system architecture that implements message passing and shared-memory communication using a fully-programmable, user-level processor in the network interface within each processor of the multiprocessor system.

Typhoon, like S3.mp, comprises workstation-like nodes connected with a point-to-point interconnection network. Each node includes processors, a memory controller, main memory, and a custom network interface processor (NP). The network processor is an off-the-shelf commodity SPARC integer processor with its own instruction and data caches and translation lookaside buffer (TLB). The NP connects the processor node and its portion of main memory to the interconnect network and the other nodes in the system. This processor is tightly-coupled to the network interface, enabling efficient handling of inbound network messages. The code running on the NP is managed using a dispatch loop that invokes a protocol or message handler based on the arriving message and handlers run to completion once invoked, similar to how \bpbedrock{}'s microcode examines arriving request messages and runs to completion based on the message type and current directory state. It achieves low-overhead message send and receive using memory-mapped register accesses from code executing on the NP, however, unlike \bpbedrock{} it lacks specialized or integrated message send and receive functional units. Typhoon also does not include any coherence-specific hardware, for example a coherence directory.

\subsubsection{Stanford FLASH and the MAGIC Node Controller}

\begin{table}[t]\centering
	\begin{tabular}{@{}L{0.32\linewidth}C{0.29\linewidth}C{0.29\linewidth}@{}} 
        \toprule
		Property & MAGIC & \bpbedrock{} \\
        \midrule
		Base ISA & MIPS & Custom RISC \\
		\multirow{3}{*}{ISA Extensions}  & Bitfield Op & Directory Rd/Wr \\
		& Set/Test Bit  & Flag Op \\
		& Tx/Rx Message & Tx/Rx Message \\
		GPRs & 32 x 64-bit & 8 x 64-bit \\
		Data Cache & 32 KiB off-chip & none \\
		Instruction Cache & 16 KiB on-chip & 1.5 KiB \\
		Data Buffers & 2 KiB 6-port SRAM & none \\
		Directory Memory & none & 3.625 KiB \\
		Protocol Agnostic & yes & no \\
		Message Passing & yes & no \\
		Coherence Type & Distributed Directory & Distributed Directory \\
		Coherence Domain & Inter-node & Multicore \\
		Coherence Model & All memory blocks & Only cached blocks \\
		HW Address Translation & no & no \\
		Interrupts & no & no \\
		Open Source & no & yes \\
        \bottomrule
	\end{tabular}
	\caption{Architectural Comparison of \bpbedrock{} and MAGIC}
	\label{table:magic-properties}
\end{table}

\begin{table}[t]\centering
\footnotesize
\begin{subtable}[t]{0.45\textwidth}
    \begin{tabular}{@{}L{0.32\linewidth}C{0.29\linewidth}C{0.29\linewidth}@{}} 
    	\toprule
    	Operation & \bpbedrock{} & MIPS (MAGIC) \\
    	\midrule
    	Directory Read & $2 + C/2$ & $20*C$ \\
    	Invalidation & $2+(2*S)$ & $15*S$ \\
    	Branch & 1 & 1 \\
    	Flag Branch & 1 & -- \\
    	\bottomrule
    \end{tabular}
    \caption{Selected operation latency in cycles. C is the number of cores and S is the number of sharer caches.}
    \label{table:magic-dir-latency}
\end{subtable}
\hfill
\begin{subtable}[t]{0.45\textwidth}
	\begin{tabular}{@{}L{0.28\linewidth}C{0.30\linewidth}C{0.30\linewidth}@{}} 
		\toprule
		Request (Directory State) & \bpbedrock{} & MIPS (MAGIC) \\
		\midrule
		Read (I) & 16 & 184 \\
		Read (S) & 30 & 184 \\
		Read (M) & 36 & 219 \\
		Write (I) & 27 & 184 \\
		Write (S) & $28 + (2*S)$ & $184 + (15*S)$ \\
		Write (M) & 31 & 198 \\
		\bottomrule
	\end{tabular}
	\caption{Request Occupancy in cycles, assuming 8-cores and an invalid block at the requester. The coherence state in parentheses indicates the state of the block at the directory. S is the number of sharer caches.}
	\label{table:magic-dir-occupancy}
\end{subtable}
\caption{Processing Latency and Occupancy Comparison of \bpbedrock{} and MAGIC}
\label{table:magic-overheads}
\end{table}

\bpbedrock{} is most similar to the MAGIC node controller of the Stanford FLASH multiprocessor \cite{kuskin_stanford_1994,heinlein_integration_1994,kuskin_flash_1997,heinrich_performance_1994,heinrich_hardwaresoftware_1997,heinrich_performance_1998,heinrich_quantitative_1999}. MAGIC is a protocol-processing specialized MIPS processor and includes ISA extensions similar to those found in \bpbedrock{}. Both \bpbedrock{} and MAGIC are effectively small, specialized integer-only RISC ISA engines. Unlike MAGIC, which is designed as a generic protocol processor, \bpbedrock{}'s programmable engine is designed to efficiently implement the \bedrock{} coherence protocol while enabling unique system- and application-specific functionality via programmable routines executing alongside protocol processing. \bpbedrock{} is not designed to support arbitrary coherence protocols or shared-memory solutions. \bpbedrock{} includes dedicated directory storage and a microcode instruction memory instead of general purpose instruction and data caches. Both designs use specialized RISC instruction sets with similar extensions for bit manipulations and message send and receive operations, however, \bpbedrock{} also includes specialized instructions for reading and processing the coherence directory and to perform efficient flag-based control flow. Neither \bpbedrock{} or MAGIC supports virtual memory or interrupts. \autoref{table:magic-properties} provides a qualitative comparison of \bpbedrock{} and MAGIC.

\autoref{table:magic-dir-latency} and \autoref{table:magic-dir-occupancy} provide quantitative comparisons between \bpbedrock{} and a MIPS-based protocol processor like MAGIC. \autoref{table:magic-dir-latency} compares the latency of selected directory operations such as reading and processing a duplicate-tag directory, issuing invalidations, and control flow operations. \bpbedrock{}'s specialized functional blocks enable highly efficient coherence directory reads, while a MIPS-based protocol processor such as MAGIC requires a significant number of instructions to execute the same operation. Likewise, \bpbedrock{}'s specialization allows it to issue one invalidation command per cycle and consume one response per cycle, whereas a MIPS-based processor would require executing these routines as tight loops with approximately 10 instructions per send or receive operation. \autoref{table:magic-dir-occupancy} shows the processing occupancy in cycles at the coherence directory for common requests. The table assumes the requesting cache does not have a valid copy of the block, which is currently in the coherence state listed in parentheses at the directory. \bpbedrock{}'s specialized logic for reading and processing the directory, issuing invalidations, and executing control flow decisions based on the coherence-specific MSHR control flags give \bpbedrock{} a significant advantage over the MIPS-based execution of MAGIC.

The Stanford FLASH machine was also used to investigate the inclusion of non-coherence logic within coherence protocols. FlashPoint \cite{martonosi_integrating_1996} incorporates performance monitoring functionality into a cache coherence protocol by leveraging the existing programmable protocol handlers provided in the MAGIC node controller. Similar to \bpbedrock{}, FlashPoint argues that programmability in the cache coherence or memory system can be used to implement system-specific functionality. The authors find that memory performance monitoring can be introduced with less than 10\% slowdown over an unmonitored execution. \bpbedrock{}'s hybrid coherence engine design illustrates how the interface between the coherence processing logic and the programmable engine impacts whether system-specific functionality, such as memory performance monitoring, may impact protocol processing or application performance.

\subsubsection{Sequent STiNG}

STiNG \cite{lovett_1996_sting} is a cache-coherent non-uniform memory access (CC-NUMA) multiprocessor from Sequent Computer Systems, Inc. A complete STiNG system comprises multiple processor nodes, each of which contains four processor cores. The nodes are interconnected with a scalable coherent interconnect based on the Scalable Coherence Interface (SCI). Coherence within a single node is provided by a snooping-based MESI coherence protocol, while coherence between nodes is implemented using a directory-based protocol. The inter-node coherence controller is implemented using a programmable protocol engine within the SCI Cache Link Interface Controller (SCLIC) ASIC. STiNG's use of a programmable protocol engine was motivated in part by the MAGIC node controller in the Stanford FLASH multiprocessor, as well as the desire to implement protocol bug-fixes or more efficient protocols post-implementation. 

The programmable protocol engine is a three-stage pipelined processor with 64-bit wide instructions. The instruction set includes custom bit field operations. The engine supports twelve independent tasks that are time-multiplexed for execution on the pipeline. At high-level, the microarchitecture of the SCLIC programmable protocol processor is quite similar to both MAGIC and \bpbedrock{}'s microcode-programmable engine. Like \bpbedrock{}, each node includes dedicated storage for coherence directory data. To reduce tag access overheads, the SCLIC includes a small cache that holds directory tag information for transactions in progress. In contrast, \bpbedrock{} reads the directory once per transaction and then processes the output into an encoded format that can be easily operated on for control flow operations during protocol processing. The researchers also come to similar conclusions as prior work and \bpbedrock{}, namely that increased protocol processing occupancy results in reduced application performance. Additionally, they identify techniques to reduce occupancy, such as optimized microcode, co-designing coherence protocols with the microcode to reduce program size, and specialized hardware logic to accelerate common instruction sequences. \bpbedrock{} applies similar techniques through its specialized hardware blocks for coherence protocol processing that are accessed via the coherence ISA.

\subsubsection{Piranha}

Piranha \cite{barroso_piranha_2000} is primarily a single-chip multiprocessor architecture, however it includes on-chip features that enable building scalable multi-chip multiprocessor systems. Each single-chip multiprocessor includes eight processor cores, memory controllers, and specialized Home and Remote Engines to accelerate inter-chip shared-memory and cache coherence operations. A multiprocessor Piranha system also includes I/O nodes, which participate in the global coherence protocol and contain home and remote engines, too. The home engine manages requests for memory that resides on the node, and the remote engine handles requests for memory on other nodes in the system. The organization of Piranha's home and remote protocol engines is derived directly from S3.mp. They are microcode-programmable engines that comprise input and output buffers and state machines as well as a microcode-programmed execution unit. The protocol engines execute a custom instruction set including message send and receive, data movement, and test and set operations. Threads of execution are multiplexed on the protocol engine using an interleaved execution paradigm that switches between active threads every cycle.

Piranha's inter-node coherence protocol is an invalidation-based directory protocol, similar to \bedrock{}. Like \bedrock{}, Piranha's protocol avoids the use of negative acknowledgment messages and supports unordered networks. However, like canonical protocols, invalidation messages are sent to the requesting node rather than to the managing directory. The microcode-programmable engine design is similar to \bpbedrock{} in that it executes a custom instruction set tailored for protocol operations. The instruction set also includes instructions that can behave as multi-way conditional branches that accelerate control flow decisions, similar to how \bpbedrock{}'s flag-based branches enable efficient control flow during request processing. As with S3.mp, Piranha does not include dedicated coherence directory storage and instead relies on storing the directory information in unused ECC bits of the main memory. Overall, Piranha's protocol engines, like S3.mp's are more specialized and less general-purpose than \bpbedrock{}'s or MAGIC's node controller.

\subsection{Programmability in the Core, Memory, or Network}

Beyond research into programmable engines within shared-memory multiprocessors as described above, other recent projects have investigated programmability throughout the processor core, memory hierarchy, and interconnect.

\subsubsection{Programmability in the Core}

Programmability has been explored within traditional processor cores, for example to support application-specific functionality or execution profiling. Zilles et al. \cite{zilles_programmable_2001} proposed the inclusion of a microcode-programmable co-processor for execution profiling alongside an out-of-order CPU core pipeline microarchitecture. Instructions of interest are tagged in the core pipeline and then pushed to a sample buffer at retirement before being processed by routines executing on the profiling co-processor. DySER (dynamically specialized execution resources) \cite{govindaraju_2011} integrates a specialized circuit-switched heterogeneous array of compute elements as a functional unit in the CPU's core pipeline, enabling application-specific functionality to be configured and invoked using special instructions. The programmable compute array can be reconfigured dynamically during runtime to accelerate specific execution patterns that are repeated within an application. Similarly, PSM (post-silicon microarchitecture) \cite{kumar_post-silicon_2020} and PFM (post-fabrication microarchitecture) \cite{kumar_post-fabrication_2021} propose the integration of reconfigurable fabrics, like FPGA or CGRA, with a CPU's core pipeline. The reconfigurable fabric can be programmed to realize new instruction or accelerator functionality after design or fabrication, enabling the device to be specialized for applications or domains as needed. The custom logic is interfaced directly with the core pipeline and appears much like existing hardened functional units from the programmer's point-of-view. Despite focusing on different subsystems than \bpbedrock{}, these works share common insights and motivation with \bpbedrock{}. The inclusion of user-defined, application-specific functionality is an important feature for future computing systems, whether in the core itself or in the memory or interconnect systems.

\subsubsection{Programmability in the Memory Hierarchy}

A common theme in contemporary research focuses on accelerating data movement and bringing compute closer to memory. A significant body of work focuses on prefetching data from memory, and a subset of work investigates the use of programmability within the prefetch engines. A few approaches include event-triggered programmable prefetching \cite{ainsworth_event-triggered_2018}, DROPLET (data-aware decoupled prefetcher) \cite{basak_2019}, RnR (Record-and-Replay) \cite{zhang_rnr_2020}, and Prodigy \cite{talati_2021}. A common theme across this research is using programmability to improve prefetching for irregular and graph applications for which existing hardware-only prefetchers are ill-suited. These works also propose various hardware-software interfaces that expose the programmable prefetch engines to software, including user-level applications, allowing programmers to direct the prefetch algorithms or behavior. 

Another interesting direction of research is exemplified by t{\"{a}}k\={o}\cite{schwedock_tako_2022}, which leverages programmability to accelerate data movement. t{\"{a}}k\={o} is a polymorphic cache hierarchy enabling programmers to define software callbacks that are triggered on cache misses, evictions, or writebacks to manage data movement between the cache and other levels of the memory hierarchy. The callbacks are executed on a programmable spatial dataflow engine located near a cache, for example at the private L2 cache of each core. These callbacks can be used to manipulate data as it is moved throughout the memory hierarchy, such as compressing or decompressing data items or reshaping data structures to optimize for memory access locality.

Like \bpbedrock{}, these works share the goal of exposing programmability to the system- or application-programmer to enable custom functionality. An interesting avenue of research for \bpbedrock{} is investigating whether programmability in the coherence engine could be used to perform prefetching or data movement and manipulation, either directly or by issuing commands to other specialized engines like those proposed in the related works.

\subsubsection{Programmability in the Network}

Research into programmable packet processing engines for networks and interconnects has continued beyond the emergence of shared-memory multiprocessors. Two examples of this research are PSM (programmable state machine) \cite{wangyang_lai_programmable_2003} and FPE (FSM-based Processing Engine) \cite{septinus_fully_2010}. These works make similar observations as \bpbedrock{} related to the tradeoffs between fixed-function or programmable hardware. PSM employs a four-stage pipelined RISC processor executing an instruction set customized for packet processing and including special registers that expose packet information directly to the programmable engine. This processor can interact with other state machines or specialized packet processing hardware using register-based control interfaces. These design decisions are also found in \bpbedrock{}, as ISA specialization and hardware acceleration of critical path operations is necessary for both generic network packet processing and coherence protocol processing. FPE is a programmable packet processor designed for integration in a network co-processor. As in \bpbedrock{}, it implements a 2-stage fetch-execute pipeline and specialization of the packet processor results in significant performance improvement compared to a general-purpose RISC processor implementing the same functionality. FPE also enables multi-way branch evaluation, recognizing the importance of minimizing control flow overheads similar to \bpbedrock{}'s use of flag-based branching.

Programmable packet processors share many similarities with \bpbedrock{} in terms of both purpose and design. In one view, a directory-based coherence protocol is simply a specialized communication protocol. Coherence protocols are typically carried as payloads of the underlying interconnect packets and generally do not consider routing or control flow processing. However, every coherence message requires packet processing to detect the type of protocol message and take appropriate actions as dictated by the protocol rules. Therefore, while the type of processing may differ between a generic network packet processor and \bpbedrock{}'s coherence message processing, the underlying architectural and microarchitectural designs are often quite similar.
\section{Open-Source Multicore RISC-V Processors}
\label{sec:related_open_source}

The emergence and rapid growth of the open-source hardware movement and the RISC-V instruction set architecture are driving a new age of computer architecture. \blackparrot{} and \bpbedrock{} are an important part of this movement, and are closely related to many ongoing efforts within it. This section outlines how \bpbedrock{} relates to many contemporary open-source processor and system efforts, with a focus on those implementing the RISC-V architecture. All of these projects make important contributions within the open-source hardware and RISC-V processor communities. However, \bpbedrock{} stands apart for its investigation into both cache coherence for open-source multicore processors and programmability within the cache coherence system.

\subsection{RISC-V, Rocket Chip, BOOM, and Chipyard}

The RISC-V Instruction Set Architecture (ISA) \cite{asanovic_riscv, waterman_2016_thesis, waterman_2016_riscv} emerged from the University of California, Berkeley in 2010 and has since become a driving innovation within the open-source processor and hardware movement. The RISC-V ISA defines a processor architecture, including allowable memory consistency models, however, it does not prescribe how to implement the architecture or consistency model.

The creators of RISC-V also developed multiple open-source processor implementations and surrounding infrastructure to support researchers adopting the new architecture. Rocket Chip \cite{asanovic2016rocket} is an open-source System-on-Chip (SoC) generator that provides implementations of both in-order and out-of-order processor cores called called Rocket and BOOM \cite{celio_2017_thesis, celio2015berkeley, celio_2017_boomv2, zhao_2020_sonicboom}, respectively, as well as generators for the uncore and on-chip networks required to construct an SoC. Chipyard \cite{amid_chipyard_2020} expands on the work of Rocket Chip and provides a framework enabling designers to construct and evaluate complete hardware systems and SoCs. Chipyard relies on RTL generators, including Rocket Chip, to provide implementations for a system's hardware components. Rocket Chip and Chipyard implement on-chip networks using the TileLink network architecture \cite{sifive_2021_tilelink_181}. TileLink comprises a set of links to communicate between two agents, where each link is a collection of one-way channels carrying messages of the same priority. The TileLink Cached (TL-C) protocol supports coherent caches. Chipyard also includes the Constellation \cite{zhao_2022_constellation} network-on-chip (NoC) generator, which provides virtual-channel wormhole-routed NoC implementations. The generated network is a protocol-independent transport layer capable of carrying arbitrary system- or application-level protocols, including cache coherence protocols. No ordering is maintained between packets carried on a single flow in a Constellation network, which may break ordering assumptions in coherence protocols, requiring endpoint buffering to recover ordering.

\bpbedrock{} builds on the contributions of RISC-V while taking a different approach than Rocket Chip and Chipyard, differing from these works in a few important ways. First, \bpbedrock{} is implemented entirely in SystemVerilog, making it easy to understand, integrate, and debug in system designs using standard open-source or commercial EDA tools. In contrast, Rocket Chip and Chipyard rely on Chisel \cite{bachrach_2012_chisel}, an open-source hardware construction language that designers can specify entire SoC configurations that are compiled into synthesizable Verilog RTL. While Chisel raises the level of abstraction for hardware designers, it is not yet widely adopted by hardware designers and the generated Verilog RTL is not as easily understood as human-written SystemVerilog RTL.

Second, \bpbedrock{} focuses on building a single-chip shared-memory cache coherent multicore while Rocket Chip and Chipyard focus on generating cores and complete SoCs. \bpbedrock{} provides a parameterized multicore architecture and explores the cache coherence system in depth, including the introduction of programmability. In comparison, Rocket Chip and Chipyard generate cores and SoCs, using TileLink to provide shared memory with cache coherence across tiles. However, they do not investigate cache coherence or programmability in the coherence system in depth. The default network generator connects tiles using a crossbar network, which becomes very costly as system size increases, however the integration of Constellation provides an interesting avenue for exploring SoCs with unique network topologies. \bpbedrock{} relies on \blackparrot{}'s tile-based architecture and implements a 2-D mesh network across tiles with wormhole-routed networks.

Third, \bedrock{} provides a complete coherence protocol while TileLink Cached (TL-C) provides a specification of only the network channels and messages to support an implementation-defined coherence protocol. TL-C specifies an interconnect protocol defining the available memory access operations and channel messages, but it requires an implementation-defined coherence policy defining how cache blocks and permissions are transferred among agents in the system. TL-C employs five-channel links that can be used to implement five-phase messaging and, therefore, five-phase coherence protocols. It also does not prescribe the use of a coherence directory to maintain coherence among agents. An Inclusive Cache \cite{rocketchip_inclusive_cache} generator for Rocket Chip provides an inclusive last-level cache that provides coherence using invalidation-based coherence with a complete coherence directory integrated with the cache tag storage. In contrast, \bedrock{} is a fully-defined four-phase, four-network invalidation-based directory protocol, and \bpbedrock{} provides a complete implementation of the \bedrock{} policy within a tiled shared-memory architecture. \bpbedrock{} provides a four-phase protocol and the network implementation to carry protocol messages between the cache controllers and coherence directory.

\subsection{Ariane/CVA6 and PULP}

CVA6 (formerly known as Ariane) \cite{zaruba_cost_2019} is a 64-bit RISC-V application-class processor implementing the RV64GC ISA variant using a single-issue in-order commit pipeline. It was originally developed as part of the PULP (Parallel Ultra Low Power) Platform \cite{pulp_platform} at ETH Zürich. CVA6 itself does not define any cache coherence mechanisms, rather it focuses on the design and implementation of the processor core and its private caches. \blackparrot{} and CVA6 are very similar in implementation and philosophy. Both cores are open-source, implemented in SystemVerilog, have private virtually-indexed physically-tagged (VIPT) L1 instruction and data caches, and have AXI interfaces to memory. However, CVA6 does not natively support cache coherence or multicore implementations, whereas \bpbedrock{} is explicitly focused on the design and implementation of a cache-coherent \blackparrot{} shared-memory multicore processor.

The PULP (Parallel Ultra Low Power) Platform \cite{pulp_platform} is an open hardware platform started by ETH Zürich, originally focused on low-power, energy-efficient architecture research. Over time, the scope of the project has grown to include high-performance platform design and research. Whereas \bpbedrock{} focuses on the implementation of a single-chip shared-memory cache-coherent multicore, most of the PULP Platform projects focus on heterogeneous platform and domain-specific accelerator research. Both the HERO \cite{kurth_2017, kurth_2022} and Occamy \cite{paulin_2024_occamy} projects, for example, include a single CVA6 core that acts as a host processor for the chip's accelerator fabric. The Manticore \cite{zaruba_manticore_2021} manycore conceptual architecture, which preceded the Occamy project, proposed using a quad-core CVA6 multicore on each chiplet as a host core, however this was never realized.

CVA6 has also been included in a number of other projects, ESP \cite{carloni_2016_esp, mantovani_2020_esp}, OpenPiton \cite{balkind_openpiton_2016, balkind_2022_thesis}, BYOC \cite{balkind_byoc_2020, balkind_2022_thesis}, and Chipyard \cite{amid_chipyard_2020}, demonstrating its success as a high-quality, user-friendly RISC-V implementation. Many of these projects are described in this section.

\subsubsection{A Consistency-Directed Cache-Coherent CVA6 Multicore}

Recently, CVA6 was used to construct a multicore processor that relies on a consistency-directed coherence algorithm \cite{miceli_2023_msthesis}. In contrast to the consistency-agnostic coherence protocol employed by \bpbedrock{}, consistency-directed protocols rely on explicit self-invalidation and writeback of cache blocks and memory data at programmer-defined synchronization points. The promoted benefit of these protocols is reduced coherence protocol complexity, since the burden of orchestrating data synchronization and coherence is shifted from the hardware-implemented protocol to the software executing in the system. However, there is often a performance tradeoff associated with this approach because it generally requires all cache lines to be invalidated and any dirty lines to be written back to the shared LLC or memory whenever a synchronization point is reached. In many applications, synchronization is frequently required. 

In comparison to \bpbedrock{}'s sequentially consistent model, the consistency-directed CVA6 multicore aims to implement the RISC-V Weak Memory Ordering (RVWMO) memory consistency model. This model allows more load and store reorderings than are permitted in \bpbedrock{}, and notably, the SWMR invariant need not be maintained. Like \bpbedrock{}, the CVA6 multicore maintains coherence among the private L1 caches, however the implemented design was only explored up to four cores whereas \bpbedrock{} can scale to at least 16 or 32 cores. Each CVA6 core is connected to a shared AXI crossbar, and all cores access a unified shared memory that is also connected to the crossbar. A dual-core platform was synthesized on an FPGA to test booting a Linux operating system and running programs from the Splash-3 \cite{sakalis2016} benchmark suite. This design was compared to a similarly configured OpenPiton-based design using CVA6 cores, which maintains coherence using a directory-based protocol that shares similarities with \bpbedrock{}. Their experiments show an application-level performance overhead between 0\% and 66\% for the Splash-3 programs relative to the OpenPiton-based design, demonstrating the high performance overheads of consistency-directed coherence when synchronization events are common. Relative to a CVA6-based design without coherence state tracking, consistency-directed coherence has an area cost between 1.6\% and 3.2\% of total core area, although it is likely possible to reduce this further with better memory macro optimization. These results validate \bpbedrock{}'s decision to employ a consistency-agnostic coherence protocol.

\subsubsection{Culsans}

Culsans \cite{tedeschi_2024_culsans} is another recent project investigating the design of a CVA6-based multicore processor. Culsans defines a tightly-coupled shared-memory multicore with a snooping-based cache coherence protocol. The cache coherence system implements a MOESI protocol using Arm's AMBA ACE \cite{axi4} specification. A Cache Coherency Unit (CCU) designed for small core counts, between two and four cores in the multicore, snoops the AXI crossbar connecting the processor cores with the last-level cache and main memory. Coherence is maintained among the private caches of the two to four cores in the system, and the LLC does not participate in the coherence protocol. The design is implemented in industry-standard SystemVerilog RTL and is entirely open-source.

Culsans provides an interesting point of comparison for \bpbedrock{} since it investigates the overheads of directory-based coherence in low core count multicores. Culsans claims significant speedups relative to a dual-core CVA6/OpenPiton-based multicore, with up to 33\% performance uplift for programs from the Splash-3 benchmark suite, and a 16\% performance uplift on average. The performance uplift comes from the tight coupling of snooping-based coherence, which reduces indirection and latency in the coherence protocol compared to the directory-based OpenPiton implementation. The CCU has an area overhead of less than 2\% of total design area. Similar to \bpbedrock{}, Culsans is open-source, implemented in SystemVerilog RTL, and focuses on the design and implementation of a tightly-coupled single-chip shared-memory multicore architecture. However, Culsans' snooping-based coherence is limited to a maximum of four cores, while \bpbedrock{} easily scales to 16 to 32 cores. Culsans' performance uplift over an OpenPiton-based design is not surprising, as integrating CVA6 into OpenPiton requires an additional layer of cache hierarchy and the OpenPiton tiles are not tightly-coupled. \bpbedrock{} occupies a middle point between these two designs, using directory-based coherence to scale to larger core counts than Culsans and with cores that are more tightly coupled than the manycore tiles in OpenPiton.

\subsection{ESP}

ESP \cite{carloni_2016_esp} is an open-source heterogeneous SoC research platform, enabling full-stack heterogeneous SoC research using agile design methodologies. ESP relies on a tile-based architecture to compose SoCs with various processor core, accelerator, memory, and peripheral or auxiliary device tiles. Tiles are connected by a multiplane NoC \cite{yoon_2013}, and the NoC can be auto-generated to construct a system with processors, accelerators, and a distributed memory hierarchy. Tiles are encapsulated into shells and sockets that manage the tile's NoC interfaces and implement platform services. ESP aims to realize system-level design using SystemC rather than RTL, thereby raising the level of design abstraction for low-level hardware design to higher-level system design. It relies on high-level synthesis (HLS) techniques to explore implementations of the IP blocks and system components. The use of system-level design techniques and SystemC provides fast full-system simulation of virtual platforms, enabling software design and bring-up.

ESP has been extended to support cache-coherent multicore SoC designs using CVA6-based processor tiles \cite{mantovani_2020_esp}. The baseline ESP design is extended to enable coherence between the private write-back L2 caches on processor and accelerator tiles and the distributed shared LLC on memory tiles. The AXI Coherenecy Extensions (ACE) were used to implement invalidations between the private L2 cache and the CVA6 processor's L1 caches. The coherence protocol is a directory-based MESI protocol extended to support accelerators sending requests directly to the LLC either with or without coherence enabled. It assumes and requires point-to-point ordering of messages on the NoC and uses a three network protocol with request, forward, and response message classes. The platform has been further extended to explore cache coherence and memory hierarchies for heterogeneous systems and accelerators \cite{giri_2018, giri_2019}. Three common types of coherence for accelerators are defined, including non-coherent, LLC-coherent, and fully-coherent, and the baseline coherence protocol is extended with accelerator-specific actions.

\bpbedrock{} and ESP differ largely in their project scope and focus. Both projects employ agile design methodologies, are open-source, and create cache coherent multicore systems. The cache coherence systems of both ESP and \bpbedrock{} employ directory-based protocols. While \bpbedrock{} maintains coherence among the private L1 caches of each processor core, ESP provides coherence among a distributed LLC in the memory tiles and private L2 caches in the core and accelerator tiles. ESP explores implementing cache coherence among heterogeneous components, but does not explore any aspects of programmability within the coherence system. The ESP approach is more similar to OpenPiton than \bpbedrock{}, and a key focus of the project is providing infrastructure to integrate cores or accelerators into large, tiled, heterogeneous manycore SoCs. ESP also relies on SystemC implementations and HLS methodologies to raise the level of design abstraction, whereas \bpbedrock{} is implemented entirely in SystemVerilog RTL. Although HLS and system-level design methodologies are attractive, they remain less supported than RTL-based design flows in open-source and commercial EDA tools.

\subsection{OpenPiton and BYOC}

OpenPiton \cite{balkind_openpiton_2016, balkind_2022_thesis, openpiton_mas} is an open-source manycore research platform from Princeton University. The project's original motivation was providing a framework for building large, scalable manycore prototypes in academia. The OpenPiton architecture is a tiled manycore, relying on a 2D-mesh NoC and supporting a distributed directory-based coherence protocol. At a high-level, the manycore comprises chips and chipsets. A chipset includes I/O, DRAM, and NoC routers, and each chip may be a grid of tiles. Each tile includes a processor core, private cache, slice of the distributed L2 cache, and connections to the manycore NoC. The original design employed modified OpenSparc T1 cores \cite{noauthor_opensparc_2008}, but other cores including CVA6 and \blackparrot{} have since been integrated with OpenPiton.

Each tile in OpenPiton includes a private L1.5 cache and a slice of the distributed L2 cache, with the L1.5 cache originally serving to convert from the OpenSparc T1 L1 cache's writethroug interface to a writeback interface. Cache coherence is maintained among the L1.5 and L2 caches in the system using an invalidation-based directory protocol with the MESI coherence states. The memory subsystem implements a TSO memory consistency model, as is found in the OpenSparc T1. The coherence directory is integrated into the L2 cache, and the L2 cache is inclusive of the private L1.5 and L1 caches found on each processor tile. The coherence protocol is implemented on top of three physical NoCs that provide point-to-point ordering guarantees. The NoC and coherence protocol are co-designed, with the NoC using 64-bit physical channels and protocol messages defined as sequences of 64-bit NoC flits. The protocol utilizes 4-step message communication, supports silent eviction of clean (Exclusive or Shared) cache blocks, and does not provide acknowledgments to dirty writebacks.

BYOC (Bring Your Own Core) \cite{balkind_byoc_2020, balkind_2022_thesis} extends the OpenPiton architecture to enable the construction of heterogeneous-ISA manycores. Supported ISAs include SPARCv9, RISC-V, and x86. Like OpenPiton, BYOC is fully open source and implemented in SystemVerilog RTL. The core building blocks of BYOC are the same as OpenPiton, however BYOC precisely defines an interface between arbitrary-ISA processor cores and the memory system through a Transaction-Response Interface (TRI). Any core or accelerator that conforms with the TRI receives a private cache and inclusion in the system-wide cache coherence protocol. The BYOC memory system provides the NoC, routers, last-level cache, and a BYOC Private Cache (BPC) implementing the TRI. The coherence protocol and system implemented by BYOC is largely identical to those found in OpenPiton with coherence maintained among the BPC and LLC caches. The LLC and BPC in BYOC correspond to the L2 and L1.5 caches found in OpenPiton, respectively.

OpenPiton and BYOC share many similarities with ESP, and differ from \bpbedrock{} in many of the same ways. While the focus of \bpbedrock{} is on the design and implementation of a single-chip multicore processor, OpenPiton and BYOC focus on the design of highly-scalable manycore systems. However, \bpbedrock{} and OpenPiton/BYOC share similarities in regard to their coherence protocols. Both designs implement directory-based coherence and have distributed directories. In OpenPiton/BYOC, the directory is integrated into the L2/LLC, while in \bpbedrock{} the directory is a standalone block that maintains cohernence among the L1 caches. The \bpbedrock{} directory is a duplicate-tag directory whereas the OpenPiton directory integrates 64-bits of directory storage per L2 cache entry. \bpbedrock{} and OpenPiton/BYOC rely on similar coherence network priority schemes to avoid protocol deadlock, but \bpbedrock{} supports unordered networks while OpenPiton/BYOC use point-to-point ordered networks. Another difference with the protocols and implementations is that OpenPiton/BYOC does not allow direct cache to cache transfers, while \bpbedrock{} supports this operation using the dedicated Fill network. Additionally, OpenPiton/BYOC allow for cache-initiated eviction and writeback of cache blocks, whereas \bpbedrock{}'s coherence directory explicitly manages all coherence state transitions. OpenPiton/BYOC do not explore the use of programmability within the coherence system.

\subsection{RISC-V Manycore Processors}

Manticore \cite{zaruba_manticore_2021}, CIFER \cite{chang_2023_cifer, li_2023_cifer}, DECADES \cite{gao_2023_decades}, HammerBlade \cite{jung_2024_hammerblade}, and Occamy \cite{paulin_2024_occamy} are a few examples of RISC-V-based manycore processors and systems developed over the past few years. Like \bpbedrock{}, these projects contribute to the growing open-source hardware movement by providing high-quality processor, accelerator, and systems designs. In contrast to the cache-coherent \bpbedrock{} multicore, manycore processors typically comprise a large collection of loosely-coupled processing elements. Although there may be a large shared memory or shared caches, coherence among cores is generally maintained using software mechanisms rather than hardware-based cache coherence. In the open-source hardware movement, both cache-coherent multicores and loosely-coupled manycores are needed, with the former often employed as a host processor for the latter, which is one potential use for a \bpbedrock{} multicore.

\chapter{Conclusion}
\label{chap:conclusion}

This dissertation revisits the topic of programmability within the cache coherence system of shared-memory multicore processors against the backdrop of the modern computing landscape. In the two decades since single-chip multicore processors first emerged and the nearly three decades since programmability in the cache coherence system of multiprocessors was last investigated, the computing landscape has changed dramatically. Computing systems have become more diverse and applied across an ever-expanding set of domains. The breakdown of long-relied upon transistor and technology scaling laws and the rise of novel computing applications, especially in the areas of machine learning and artificial intelligence, have driven computer architects toward both domain-specific and highly adaptable computer architectures. At the same time, the emergence and rapid growth of the open-source hardware movement and open-source RISC-V instruction set architecture have further democratized computer processor and system design. Collectively, the confluence of trends in applications, technology scaling, and open-source hardware and software have fundamentally changed the computing landscape.

Taking a bottom-up, architecture-first approach, the feasibility of introducing programmability into the cache coherence system at the coherence directory controller is explored. This investigation first presents the \bedrock{} cache coherence protocol, an easy to implement race-free protocol suitable for small- to medium-scale modern shared-memory multicore processors. \bedrock{} is a family of directory-based invalidate coherence protocols using subsets of the common MOESIF coherence states. A complete specification of the \bedrock{} protocol is presented in tabular form long with a description of the necessary system components, coherence states, coherence networks, and coherence messages required by the protocol. Comparing \bedrock{} to a canonical directory-based coherence protocol reveals that the design tradeoffs of cache coherence protocols are not always straightforward and that implementations often dictate the realizable concurrency and performance of a given protocol.

Following the description and analysis of \bedrock{}, a complete open-source implementation of the protocol within the \blackparrot{} 64-bit RISC-V shared-memory multicore processor, called \bpbedrock{}, is described. \bpbedrock{} includes both fixed-function and microcode-programmable coherence directory implementations and demonstrates the feasibility of introducing programmability at the coherence directory. Utilizing a complete duplicate tag directory organization, \bpbedrock{} maintains a constant 6.25\% coherence directory storage overhead relative to the capacity of the coherent L1 caches, regardless of the number of processor tiles in the tiled \blackparrot{} multicore architecture. The microcode-programmable coherence engine realizes programmability at low overheads due to its use of highly-specialized coherence processing modules and instruction set extensions that offload the core of the coherence protocol processing from general purpose code. Consequently, the microcode-programmable coherence engine implementation has only single-digit percentage area and resource costs at the multicore design level while incurring only a 1\% average (2.3\% worst-case) performance overhead for the Splash-3 benchmarks. Applying programmability to implement the cache coherence protocol has an area or resource overhead of 4-7\% at the multicore tile level.

Drawing on learnings from \bpbedrock{}'s initial implementation, a hybrid coherence engine design is presented that blends the protocol processing performance of a hardware-based fixed-function coherence engine with the programmability and adaptability of a microcode-programmable coherence engine. The hybrid coherence engine design evolves the fixed-function protocol processing logic of the FSM-based coherence engine to provide additional inter-transaction concurrency, while integrating a programmable processing pipeline to handle application- or system-specific processing of coherence requests. The hybrid coherence engine architecture presents one possible method of integrating programmable logic with the coherence processing pipeline and illustrates the breadth of possibility in the design space. This investigation further motivates research into the integration of useful programmability into the cache coherence engines of shared-memory multicore processors.

\section{Future Work}
\label{sec:future-work}

The research presented in this dissertation reveals and motivates numerous promising avenues of future research. Possible research spans from continued development of both \bedrock{} and the \bpbedrock{} implementation, to cache coherence protocols and their adaptability to new system architectures or application domains, to open-source hardware and processor development.

\subsection{Cache Coherence Protocols and Systems}

Cache coherence protocols remain a niche area of research, despite being the backbone of nearly all contemporary shared-memory multicore processor systems. As evidenced by the lack of publications from industry, the inner workings of cache coherence systems, which are often tightly coupled with the architecture and implementation of on-chip networks, remain closely guarded secrets. However, the performance and efficiency of cache coherence systems directly impacts the achievable performance of all modern shared-memory multicore processors. Therefore, it is imperative that research continues into a wide variety of cache coherence protocols and implementations.

At the coherence protocol level, most contemporary research focuses on developing protocols for large, high-performance multicore processor architectures. However, as computing systems are applied in more and more domains, the use of small- to medium-scale processors is increasing. Thus, investigations into protocols optimized for small- to medium-scale processors is a promising direction for future research.

The investigations described in this dissertation also show that protocol design decisions are not always straightforward or obvious. There are complex relationships between the coherence protocol, its implementation, and the system's processing elements. Thus, building real cache coherent multicore systems is critical to understanding the realizable benefits and drawbacks of any given coherence protocol and system. These questions become increasingly relevant as systems become more heterogeneous and incorporate an ever growing variety of specialized processing elements and accelerators that can access the same memory as a host processor. Contemporary accelerator offload models largely rely on bulk-synchronous offload or direct memory access (DMA) models where there is little active sharing of data between the host and accelerator or among accelerators. Enabling fine-grained cache coherent access to shared memory in an efficient manner is an open research challenge.

\subsection{Open-Source Cache-Coherent RISC-V Multicore Processors}

The rapid growth of the open-source hardware ecosystem and the RISC-V instruction set architecture has democratized computer processor and system design. However, there exist few high-quality open-source cache-coherent multicore processor implementations. \bpbedrock{} contributes one design to this space, allowing researchers to explore new and novel coherence system designs. Due to \bpbedrock{}'s use of latency-insensitive interfaces between the caches and cache controllers and between the coherence directories and last-level caches and main memory, it is possible to completely swap out the coherence system in a \bpbedrock{} implementation. Future research can leverage the infrastructure and design methodologies from \bpbedrock{} to explore alternative coherence protocols and their implementations.

The availability of high-quality open-source RISC-V multicore processors can also drive research into heterogeneous system architecture. Researchers are constantly proposing new accelerator architectures for domain-specific applications, but evaluating these accelerators remains challenging. Simulation is fast but often overlooks important implementation details that may have non-trivial impacts on the accelerator's architecture, performance, power, or area. Hardware implementation provides real world evaluation, but requires significant infrastructure development. \bpbedrock{} and \blackparrot{} enable researchers to build real systems with tightly- or loosely-coupled accelerators. One promising avenue of future research is leveraging the \bpbedrock{} infrastructure to explore cache coherence within heterogeneous systems including one or more domain-specific accelerators.

\subsection{Use Cases for Programmability in the Cache Coherence System}

This dissertation takes a bottom-up, architecture-first approach to investigating the feasibility of introducing programmability into the cache coherence system of shared-memory multicore processors. Equally important is investigating the use cases for this programmability. An important avenue for future research is taking a top-down, application-first approach to discover the types of systems and applications that may benefit most from programmability in the coherence system. This research is likely to span both system and application software.

At the system level, programmability could be exploited to provide security, virtualization, or debugging capabilities. One obvious application of programmability at the coherence directory is to provide address space isolation verification. In some systems, whether virtualized or not, it may be desirable to isolate certain processing elements from the rest of the system by restricting their access to specific regions of physical memory. Programmability in the coherence system can be used to implement memory access checks beyond the capabilities of existing mechanisms like the virtual memory and address translation systems. In other cases, it may be desirable to simply monitor and report on physical memory accesses made by specific processor cores or accelerators. Tracing memory accesses or implementing watchpoints are common techniques for application debugging, which may be implementable within a programmable coherence system. In all cases, the exact needs and demands placed on a coherence system with programmability for any of these applications are open research questions.

User software and applications may have other use cases for programmability within the coherence system. For example, a user application may wish to monitor accesses to memory it has allocated from its heap or memory pages that have been shared with other processes or among threads of a multi-threaded application. A critical unanswered question in this area is how to support and execute untrusted application functionality or code on the programmable portion of the coherence system. By definition, cache coherence systems are among the most trusted components of a multicore processor design as they have access to the entire physical address space. Since hardware-based caching operates invisibly to the user and system software, they have been developed assuming the entire coherence system is privileged and trusted. Developing a safe and secure mechanism for executing untrusted routines within the coherence domain is an important open challenge to delivering programmability that can be leveraged by user-space software.

%% Back Matter
% Appendix
\appendix
\label{chap:appendix}

%%% Cache Controller Protocol Tables

%\newgeometry{hmargin=1in,bmargin=1in,tmargin=0.5in}
%\begin{landscape}

\chapter{BedRock Cache Controller (LCE) Coherence Protocol Tables}
\label{app:protocol-lce}
%\section{Cache Controller Protocol Tables} \label{app:protocol-lce}

\clearpage

\begin{table}[H]\centering
\footnotesize
\begin{adjustbox}{angle=90}
\ra{1.3}
\begin{tabular}{@{}lllccccccccC{0.1\linewidth}@{}}\toprule
& \multicolumn{2}{c}{Cache Action} & \phantom{a} & \multicolumn{8}{c}{Coherence Message}\\
\cmidrule{2-3} \cmidrule{5-12}
State & Load & Store && Inv & DATA & STW & WB & TR & ST-WB & ST-TR & ST-TR-WB\\
\midrule
I & ReqRd & ReqWr && & CohAck/X & & & & & &\\\\
M & Hit & Hit && & & & DirtyWB/M & & DirtyWB/X & DATA/X & DATA, DirtyWB/X\\
\bottomrule
\end{tabular}
\end{adjustbox}
\caption{BedRock Cache Controller Protocol Table - MI}
\label{table:bedrock-protocol-lce-mi}
\end{table}

\begin{table}[H]\centering
\footnotesize
\begin{adjustbox}{angle=90}
\ra{1.3}
\begin{tabular}{@{}lllccccccccC{0.1\linewidth}@{}}\toprule
& \multicolumn{2}{c}{Cache Action} & \phantom{a} & \multicolumn{8}{c}{Coherence Message}\\
\cmidrule{2-3} \cmidrule{5-12}
State & Load & Store && Inv & DATA & STW & WB & TR & ST-WB & ST-TR & ST-TR-WB\\
\midrule
I & ReqRd & ReqWr && & CohAck/X & & & & & &\\\\
S & Hit & ReqWr && InvAck/I & & CohAck/M & & & & &\\\\
M & Hit & Hit && & & & DirtyWB/M & & DirtyWB/X & DATA/X & DATA, DirtyWB/X\\
\bottomrule
\end{tabular}
\end{adjustbox}
\caption{BedRock Cache Controller Protocol Table - MSI}
\label{table:bedrock-protocol-lce-msi}
\end{table}

\begin{table}[H]\centering
\footnotesize
\begin{adjustbox}{angle=90}
\ra{1.3}
\begin{tabular}{@{}lllccccccccC{0.1\linewidth}@{}}\toprule
& \multicolumn{2}{c}{Cache Action} & \phantom{a} & \multicolumn{8}{c}{Coherence Message}\\
\cmidrule{2-3} \cmidrule{5-12}
State & Load & Store && Inv & DATA & STW & WB & TR & ST-WB & ST-TR & ST-TR-WB\\
\midrule
I & ReqRd & ReqWr && & CohAck/X & & & & & &\\\\
S & Hit & ReqWr && InvAck/I & & CohAck/M & & & & &\\\\
E & Hit & Hit/M && & & & NullWB/E & & NullWB/X & DATA/X & DATA, NullWB/X\\
M & Hit & Hit && & & & DirtyWB/M & & DirtyWB/X & DATA/X & DATA, DirtyWB/X\\
\bottomrule
\end{tabular}
\end{adjustbox}
\caption{BedRock Cache Controller Protocol Table - MESI}
\label{table:bedrock-protocol-lce-mesi}
\end{table}

\begin{table}[H]\centering
\footnotesize
\begin{adjustbox}{angle=90}
\ra{1.3}
\begin{tabular}{@{}lllccccccccC{0.1\linewidth}@{}}\toprule
& \multicolumn{2}{c}{Cache Action} & \phantom{a} & \multicolumn{8}{c}{Coherence Message}\\
\cmidrule{2-3} \cmidrule{5-12}
State & Load & Store && Inv & DATA & STW & WB & TR & ST-WB & ST-TR & ST-TR-WB\\
\midrule
I & ReqRd & ReqWr && & CohAck/X & & & & & &\\\\
S & Hit & ReqWr && InvAck/I & & CohAck/M & & & & &\\\\
E & Hit & Hit/M && & & & NullWB/E & & NullWB/X & DATA/X & DATA, NullWB/X\\
M & Hit & Hit && & & & DirtyWB/M & & DirtyWB/X & DATA/X & DATA, DirtyWB/X\\
F & Hit & ReqWr && & & CohAck/M & & DATA/F & & DATA/X &\\
\bottomrule
\end{tabular}
\end{adjustbox}
\caption{BedRock Cache Controller Protocol Table - MESIF}
\label{table:bedrock-protocol-lce-mesif}
\end{table}

\begin{table}[H]\centering
\footnotesize
\begin{adjustbox}{angle=90}
\ra{1.3}
\begin{tabular}{@{}lllccccccccC{0.1\linewidth}@{}}\toprule
& \multicolumn{2}{c}{Cache Action} & \phantom{a} & \multicolumn{8}{c}{Coherence Message}\\
\cmidrule{2-3} \cmidrule{5-12}
State & Load & Store && Inv & DATA & STW & WB & TR & ST-WB & ST-TR & ST-TR-WB\\
\midrule
I & ReqRd & ReqWr && & CohAck/X & & & & & &\\\\
S & Hit & ReqWr && InvAck/I & & CohAck/M & & & & &\\\\
M & Hit & Hit && & & & DirtyWB/M & & DirtyWB/X & DATA/X & DATA, DirtyWB/X\\
O & Hit & ReqWr && & & CohAck/M & DirtyWB/O & DATA/O & DirtyWB/X & DATA/X &\\
\bottomrule
\end{tabular}
\end{adjustbox}
\caption{BedRock Cache Controller Protocol Table - MOSI}
\label{table:bedrock-protocol-lce-mosi}
\end{table}

\begin{table}[H]\centering
\footnotesize
\begin{adjustbox}{angle=90}
\ra{1.3}
\begin{tabular}{@{}lllccccccccC{0.1\linewidth}@{}}\toprule
& \multicolumn{2}{c}{Cache Action} & \phantom{a} & \multicolumn{8}{c}{Coherence Message}\\
\cmidrule{2-3} \cmidrule{5-12}
State & Load & Store && Inv & DATA & STW & WB & TR & ST-WB & ST-TR & ST-TR-WB\\
\midrule
I & ReqRd & ReqWr && & CohAck/X & & & & & &\\\\
S & Hit & ReqWr && InvAck/I & & CohAck/M & & & & &\\\\
M & Hit & Hit && & & & DirtyWB/M & & DirtyWB/X & DATA/X & DATA, DirtyWB/X\\
O & Hit & ReqWr && & & CohAck/M & DirtyWB/O & DATA/O & DirtyWB/X & DATA/X &\\\\
F & Hit & ReqWr && & & CohAck/M & & DATA/F & & DATA/X &\\
\bottomrule
\end{tabular}
\end{adjustbox}
\caption{BedRock Cache Controller Protocol Table - MOSIF}
\label{table:bedrock-protocol-lce-mosif}
\end{table}

\begin{table}[H]\centering
\footnotesize
\begin{adjustbox}{angle=90}
\ra{1.3}
\begin{tabular}{@{}lllccccccccC{0.1\linewidth}@{}}\toprule
& \multicolumn{2}{c}{Cache Action} & \phantom{a} & \multicolumn{8}{c}{Coherence Message}\\
\cmidrule{2-3} \cmidrule{5-12}
State & Load & Store && Inv & DATA & STW & WB & TR & ST-WB & ST-TR & ST-TR-WB\\
\midrule
I & ReqRd & ReqWr && & CohAck/X & & & & & &\\\\
S & Hit & ReqWr && InvAck/I & & CohAck/M & & & & &\\\\
E & Hit & Hit/M && & & & NullWB/E & & NullWB/X & DATA/X & DATA, NullWB/X\\
M & Hit & Hit && & & & DirtyWB/M & & DirtyWB/X & DATA/X & DATA, DirtyWB/X\\
O & Hit & ReqWr && & & CohAck/M & DirtyWB/O & DATA/O & DirtyWB/X & DATA/X &\\
\bottomrule
\end{tabular}
\end{adjustbox}
\caption{BedRock Cache Controller Protocol Table - MOESI}
\label{table:bedrock-protocol-lce-moesi}
\end{table}

%\end{landscape}
%\restoregeometry

%% Coherence Directory Protocol Tables

\chapter{BedRock Coherence Directory (CCE) Coherence Protocol Tables}
\label{app:protocol-cce}

\clearpage

%\newgeometry{hmargin=1in,bmargin=1in,tmargin=0.5in}
%\begin{landscape}

\begin{table}[H]\centering
\footnotesize
\begin{adjustbox}{angle=90}
\ra{1.3}
\begin{tabular}{@{}L{0.10\linewidth}L{0.13\linewidth}L{0.13\linewidth}L{0.15\linewidth}L{0.15\linewidth}L{0.15\linewidth}lL{0.2\linewidth}@{}}
\toprule
& \multicolumn{5}{c}{Coherence Request} & \phantom{a} & Directory Action\\
\cmidrule{2-6} \cmidrule{8-8}
Directory State & ReqRd & ReqRd (Non-Excl) & ReqWr from Invalid & ReqWr from Sharer & ReqWr from Owner && Replacement\\
\midrule
I & DATA to Req/M & DATA to Req/M & DATA to Req/M &&&&\\
M & ST\tss{I}-TR\tss{M} to Owner/M & ST\tss{I}-TR\tss{M} to Owner/M & ST\tss{I}-TR\tss{M} to Owner/M &&&& ST\tss{I}-WB to Req/I\\
\bottomrule
\end{tabular}
\end{adjustbox}
\caption{BedRock Coherence Directory Protocol Table - MI}
\label{table:bedrock-protocol-cce-mi}
\end{table}

\begin{table}[H]\centering
\footnotesize
\begin{adjustbox}{angle=90}
\ra{1.3}
\begin{tabular}{@{}L{0.10\linewidth}L{0.13\linewidth}L{0.13\linewidth}L{0.15\linewidth}L{0.15\linewidth}L{0.15\linewidth}lL{0.2\linewidth}@{}}
\toprule
& \multicolumn{5}{c}{Coherence Request} & \phantom{a} & Directory Action\\
\cmidrule{2-6} \cmidrule{8-8}
Directory State & ReqRd & ReqRd (Non-Excl) & ReqWr from Invalid & ReqWr from Sharer & ReqWr from Owner && Replacement\\
\midrule
I & DATA to Req/S & DATA to Req/S & DATA to Req/M &&&&\\
S & DATA to Req/S & DATA to Req/S & Inv all S, DATA to Req/M & Inv other S, STW\tss{M} to Req/M &&&\\
M & ST\tss{S}-TR\tss{S}-WB to Owner/S & ST\tss{S}-TR\tss{S}-WB to Owner/S & ST\tss{I}-TR\tss{M} to Owner/M &&&& ST\tss{I}-WB to Req/I\\
\bottomrule
\end{tabular}
\end{adjustbox}
\caption{BedRock Coherence Directory Protocol Table - MSI}
\label{table:bedrock-protocol-cce-msi}
\end{table}

\begin{table}[H]\centering
\footnotesize
\begin{adjustbox}{angle=90}
\ra{1.3}
\begin{tabular}{@{}L{0.10\linewidth}L{0.13\linewidth}L{0.13\linewidth}L{0.15\linewidth}L{0.15\linewidth}L{0.15\linewidth}lL{0.2\linewidth}@{}}
\toprule
& \multicolumn{5}{c}{Coherence Request} & \phantom{a} & Directory Action\\
\cmidrule{2-6} \cmidrule{8-8}
Directory State & ReqRd & ReqRd (Non-Excl) & ReqWr from Invalid & ReqWr from Sharer & ReqWr from Owner && Replacement\\
\midrule
I & DATA to Req/E & DATA to Req/S & DATA to Req/M &&&&\\\\
S & DATA to Req/S & DATA to Req/S & Inv all S, DATA to Req/M & Inv other S, STW\tss{M} to Req/M &&&\\
E & ST\tss{S}-TR\tss{S}-WB to Owner/S & ST\tss{S}-TR\tss{S}-WB to Owner/S & ST\tss{I}-TR\tss{M} to Owner/M & &&& ST\tss{I}-WB to Req/I\\\\
M & ST\tss{S}-TR\tss{S}-WB to Owner/S & ST\tss{S}-TR\tss{S}-WB to Owner/S & ST\tss{I}-TR\tss{M} to Owner/M &&&& ST\tss{I}-WB to Req/I\\\\
\bottomrule
\end{tabular}
\end{adjustbox}
\caption{BedRock Coherence Directory Protocol Table - MESI}
\label{table:bedrock-protocol-cce-mesi}
\end{table}

\begin{table}[H]\centering
\footnotesize
\begin{adjustbox}{angle=90}
\ra{1.3}
\begin{tabular}{@{}L{0.10\linewidth}L{0.13\linewidth}L{0.13\linewidth}L{0.15\linewidth}L{0.15\linewidth}L{0.15\linewidth}lL{0.2\linewidth}@{}}
\toprule
& \multicolumn{5}{c}{Coherence Request} & \phantom{a} & Directory Action\\
\cmidrule{2-6} \cmidrule{8-8}
Directory State & ReqRd & ReqRd (Non-Excl) & ReqWr from Invalid & ReqWr from Sharer & ReqWr from Owner && Replacement\\
\midrule
I & DATA to Req/E & DATA to Req/S & DATA to Req/M &&&&\\\\
S & DATA to Req/S & DATA to Req/S & Inv all S, DATA to Req/M & Inv other S, STW\tss{M} to Req/M &&&\\
E & ST\tss{F}-TR\tss{S}-WB to Owner/F & ST\tss{F}-TR\tss{S}-WB to Owner/F & ST\tss{I}-TR\tss{M} to Owner/M & &&& ST\tss{I}-WB to Req/I\\\\
M & ST\tss{F}-TR\tss{S}-WB to Owner/F & ST\tss{F}-TR\tss{S}-WB to Owner/F & ST\tss{I}-TR\tss{M} to Owner/M &&&& ST\tss{I}-WB to Req/I\\\\
F & TR\tss{S} to Owner/F & TR\tss{S} to Owner/F & Inv all S, ST\tss{I}-TR\tss{M} to Owner/M & Inv other S and Owner, STW\tss{M} to Req/M & Inv all S, STW\tss{M} to Req/M &&\\
\bottomrule
\end{tabular}
\end{adjustbox}
\caption{BedRock Coherence Directory Protocol Table - MESIF}
\label{table:bedrock-protocol-cce-mesif}
\end{table}

\begin{table}[H]\centering
\footnotesize
\begin{adjustbox}{angle=90}
\ra{1.3}
\begin{tabular}{@{}L{0.10\linewidth}L{0.13\linewidth}L{0.13\linewidth}L{0.15\linewidth}L{0.15\linewidth}L{0.15\linewidth}lL{0.2\linewidth}@{}}
\toprule
& \multicolumn{5}{c}{Coherence Request} & \phantom{a} & Directory Action\\
\cmidrule{2-6} \cmidrule{8-8}
Directory State & ReqRd & ReqRd (Non-Excl) & ReqWr from Invalid & ReqWr from Sharer & ReqWr from Owner && Replacement\\
\midrule
I & DATA to Req/S & DATA to Req/S & DATA to Req/M &&&&\\\\
S & DATA to Req/S & DATA to Req/S & Inv all S, DATA to Req/M & Inv other S, STW\tss{M} to Req/M &&&\\
M & ST\tss{O}-TR\tss{S} to Owner/O & ST\tss{O}-TR\tss{S} to Owner/O & ST\tss{I}-TR\tss{M} to Owner/M &&&& ST\tss{I}-WB to Req/I\\\\
O & TR\tss{S} to Owner/O & TR\tss{S} to Owner/O & Inv all S, ST\tss{I}-TR\tss{M} to Owner/M & Inv other S and Owner, STW\tss{M} to Req/M & Inv all S, STW\tss{M} to Req/M && ST\tss{I}-WB to Req/I\\
\bottomrule
\end{tabular}
\end{adjustbox}
\caption{BedRock Coherence Directory Protocol Table - MOSI}
\label{table:bedrock-protocol-cce-mosi}
\end{table}

\begin{table}[H]\centering
\footnotesize
\begin{adjustbox}{angle=90}
\ra{1.3}
\begin{tabular}{@{}L{0.10\linewidth}L{0.13\linewidth}L{0.13\linewidth}L{0.15\linewidth}L{0.15\linewidth}L{0.15\linewidth}lL{0.2\linewidth}@{}}
\toprule
& \multicolumn{5}{c}{Coherence Request} & \phantom{a} & Directory Action\\
\cmidrule{2-6} \cmidrule{8-8}
Directory State & ReqRd & ReqRd (Non-Excl) & ReqWr from Invalid & ReqWr from Sharer & ReqWr from Owner && Replacement\\
\midrule
I & DATA to Req/F & DATA to Req/S & DATA to Req/M &&&&\\\\
S & DATA to Req/S & DATA to Req/S & Inv all S, DATA to Req/M & Inv other S, STW\tss{M} to Req/M &&&\\
M & ST\tss{O}-TR\tss{S} to Owner/O & ST\tss{O}-TR\tss{S} to Owner/O & ST\tss{I}-TR\tss{M} to Owner/M &&&& ST\tss{I}-WB to Req/I\\\\
O & TR\tss{S} to Owner/O & TR\tss{S} to Owner/O & Inv all S, ST\tss{I}-TR\tss{M} to Owner/M & Inv other S and Owner, STW\tss{M} to Req/M & Inv all S, STW\tss{M} to Req/M && ST\tss{I}-WB to Req/I\\
F & TR\tss{S} to Owner/F & TR\tss{S} to Owner/F & Inv all S, ST\tss{I}-TR\tss{M} to Owner/M & Inv other S and Owner, STW\tss{M} to Req/M & Inv all S, STW\tss{M} to Req/M &&\\
\bottomrule
\end{tabular}
\end{adjustbox}
\caption{BedRock Coherence Directory Protocol Table - MOSIF}
\label{table:bedrock-protocol-cce-mosif}
\end{table}

Note: it may be beneficial to make F the next state of a ReqRd to a block in S.

\begin{table}[H]\centering
\footnotesize
\begin{adjustbox}{angle=90}
\ra{1.3}
\begin{tabular}{@{}L{0.10\linewidth}L{0.13\linewidth}L{0.13\linewidth}L{0.15\linewidth}L{0.15\linewidth}L{0.15\linewidth}lL{0.2\linewidth}@{}}
\toprule
& \multicolumn{5}{c}{Coherence Request} & \phantom{a} & Directory Action\\
\cmidrule{2-6} \cmidrule{8-8}
Directory State & ReqRd & ReqRd (Non-Excl) & ReqWr from Invalid & ReqWr from Sharer & ReqWr from Owner && Replacement\\
\midrule
I & DATA to Req/E & DATA to Req/S & DATA to Req/M &&&&\\\\
S & DATA to Req/S & DATA to Req/S & Inv all S, DATA to Req/M & Inv other S, STW\tss{M} to Req/M &&&\\
E & ST\tss{O}-TR\tss{S} to Owner/O & ST\tss{O}-TR\tss{S} to Owner/O & ST\tss{I}-TR\tss{M} to Owner/M & &&& ST\tss{I}-WB to Req/I\\\\
M & ST\tss{O}-TR\tss{S} to Owner/O & ST\tss{O}-TR\tss{S} to Owner/O & ST\tss{I}-TR\tss{M} to Owner/M &&&& ST\tss{I}-WB to Req/I\\\\
O & TR\tss{S} to Owner/O & TR\tss{S} to Owner/O & Inv all S, ST\tss{I}-TR\tss{M} to Owner/M & Inv other S and Owner, STW\tss{M} to Req/M & Inv all S, STW\tss{M} to Req/M && ST\tss{I}-WB to Req/I\\
\bottomrule
\end{tabular}
\end{adjustbox}
\caption{BedRock Coherence Directory Protocol Table - MOESI}
\label{table:bedrock-protocol-cce-moesi}
\end{table}

%\end{landscape}
%\restoregeometry

\chapter{BedRock Cache Controller (LCE) Coherence State Transition Tables} \label{app:state-tables-lce}

\clearpage

% Cache
%\begin{tabular}{@{}L{0.25\linewidth}C{0.18\linewidth}C{0.18\linewidth}@{}}

\begin{table}[H]\centering
\ra{1.3}
\begin{tabular}{@{}L{0.25\linewidth}C{0.18\linewidth}C{0.18\linewidth}@{}}
\toprule
Event & Current State & Next State\\
\midrule
Load & I & M\\
Store & I & M\\
Other Load & M & I\\
Other Store & M & I\\
\bottomrule
\end{tabular}
\caption{BedRock Cache Controller Next State Table - MI}
\label{table:bedrock-states-lce-mi}
\end{table}

\begin{table}[H]\centering
\ra{1.3}
\begin{tabular}{@{}L{0.25\linewidth}C{0.18\linewidth}C{0.18\linewidth}@{}}
\toprule
Event & Current State & Next State\\
\midrule
Load & I & S\\
Store & I, S & M\\
Other Load & M & S\\
Other Store & S, M & I\\
\bottomrule
\end{tabular}
\caption{BedRock Cache Controller Next State Table - MSI}
\label{table:bedrock-states-lce-msi}
\end{table}

\begin{table}[H]\centering
\ra{1.3}
\begin{tabular}{@{}L{0.25\linewidth}C{0.18\linewidth}C{0.18\linewidth}@{}}
\toprule
Event & Current State & Next State\\
\midrule
Load & I & S, E\\
Store & I, S & M\\
Store (Silent Upgrade) & E & M\\
Other Load & E, M & S\\
Other Store & S, E, M & I\\
\bottomrule
\end{tabular}
\caption{BedRock Cache Controller Next State Table - MESI}
\label{table:bedrock-states-lce-mesi}
\end{table}

\begin{table}[H]\centering
\ra{1.3}
\begin{tabular}{@{}L{0.25\linewidth}C{0.18\linewidth}C{0.18\linewidth}@{}}
\toprule
Event & Current State & Next State\\
\midrule
Load & I & S, E\\
Store & I, S, F & M\\
Store (Silent Upgrade) & E & M\\
Other Load & E, M & F\\
Other Store & S, E, M, F & I\\
\bottomrule
\end{tabular}
\caption{BedRock Cache Controller Next State Table - MESIF}
\label{table:bedrock-states-lce-mesif}
\end{table}

\begin{table}[H]\centering
\ra{1.3}
\begin{tabular}{@{}L{0.25\linewidth}C{0.18\linewidth}C{0.18\linewidth}@{}}
\toprule
Event & Current State & Next State\\
\midrule
Load & I & S\\
Store & I, S, O & M\\
Other Load & M & O\\
Other Store & S, O, M & I\\
\bottomrule
\end{tabular}
\caption{BedRock Cache Controller Next State Table - MOSI}
\label{table:bedrock-states-lce-mosi}
\end{table}

\begin{table}[H]\centering
\ra{1.3}
\begin{tabular}{@{}L{0.25\linewidth}C{0.18\linewidth}C{0.18\linewidth}@{}}
\toprule
Event & Current State & Next State\\
\midrule
Load & I & S, F\\
Store & I, S, O, F & M\\
Other Load & M & O\\
Other Store & S, O, M, F & I\\
\bottomrule
\end{tabular}
\caption{BedRock Cache Controller Next State Table - MOSIF}
\label{table:bedrock-states-lce-mosif}
\end{table}

\begin{table}[H]\centering
\ra{1.3}
\begin{tabular}{@{}L{0.25\linewidth}C{0.18\linewidth}C{0.18\linewidth}@{}}
\toprule
Event & Current State & Next State\\
\midrule
Load & I & S, E\\
Store & I, S, O & M\\
Store (Silent Upgrade) & E & M\\
Other Load & E & S\\
& M & O\\
Other Store & S, E, M, O & I\\
\bottomrule
\end{tabular}
\caption{BedRock Cache Controller Next State Table - MOESI}
\label{table:bedrock-states-lce-moesi}
\end{table}
\chapter{BedRock Coherence Directory (CCE) Coherence State Transition Tables} \label{app:state-tables-cce}

\clearpage

% Directory
%\begin{tabular}{@{}L{0.2\linewidth}@{}L{0.2\linewidth}@{}C{0.16\linewidth}@{}C{0.16\linewidth}@{}C{0.16\linewidth}@{}}

\begin{comment}
\begin{table}[t]\centering
\ra{1.3}
\begin{tabular}{@{}L{0.2\linewidth}@{}L{0.2\linewidth}@{}C{0.16\linewidth}@{}C{0.16\linewidth}@{}C{0.16\linewidth}@{}}
\toprule
Event & Request Message & Current State (Dir) & Next State (Dir) & Next State (Requestor)\\
\midrule
Load & ReqRd & I & E & E\\
&& S & S & S\\
&& E, F & F & S\\
&& M, O & O & S\\
Load (Non-Excl) & ReqRd-NE & I, S & S & S\\
&& E, F & F & S\\
&& M, O & O & S\\
Store & ReqWr & I, S, O, E, M, F & M & M\\
\bottomrule
\end{tabular}
\caption{BedRock Coherence Directory Next State Table - MOESIF}
\label{table:bedrock-states-cce-moesif-template}
\end{table}
\end{comment}

%\begin{tabular}{@{}L{0.2\linewidth}@{}C{0.16\linewidth}@{}C{0.16\linewidth}@{}C{0.16\linewidth}@{}}

\begin{table}[H]\centering
\ra{1.3}
\begin{tabular}{@{}L{0.2\linewidth}@{}L{0.2\linewidth}@{}C{0.16\linewidth}@{}C{0.16\linewidth}@{}C{0.16\linewidth}@{}}
\toprule
Event & Request Message & Current State (Dir) & Next State (Dir) & Next State (Requestor)\\
\midrule
Load (Any) & ReqRd, ReqRd-NE & I, M & M & M\\
Store & ReqWr & I, M & M & M\\
\bottomrule
\end{tabular}
\caption{BedRock Coherence Directory Next State Table - MI}
\label{table:bedrock-states-cce-mi}
\end{table}

\begin{table}[H]\centering
\ra{1.3}
\begin{tabular}{@{}L{0.2\linewidth}@{}L{0.2\linewidth}@{}C{0.16\linewidth}@{}C{0.16\linewidth}@{}C{0.16\linewidth}@{}}
\toprule
Event & Request Message & Current State (Dir) & Next State (Dir) & Next State (Requestor)\\
\midrule
Load (Any) & ReqRd, ReqRd-NE & I, S, M & S & S\\
Store & ReqWr & I, S, M & M & M\\
\bottomrule
\end{tabular}
\caption{BedRock Coherence Directory Next State Table - MSI}
\label{table:bedrock-states-cce-msi}
\end{table}

\begin{table}[H]\centering
\ra{1.3}
\begin{tabular}{@{}L{0.2\linewidth}@{}L{0.2\linewidth}@{}C{0.16\linewidth}@{}C{0.16\linewidth}@{}C{0.16\linewidth}@{}}
\toprule
Event & Request Message & Current State (Dir) & Next State (Dir) & Next State (Requestor)\\
\midrule
Load & ReqRd & I & E & E\\
&& S, E, M & S & S\\
Load (Non-Excl) & ReqRd-NE & I, S, E, M & S & S\\
Store & ReqWr & I, S, E, M & M & M\\
\bottomrule
\end{tabular}
\caption{BedRock Coherence Directory Next State Table - MESI}
\label{table:bedrock-states-cce-mesi}
\end{table}

\begin{table}[H]\centering
\ra{1.3}
\begin{tabular}{@{}L{0.2\linewidth}@{}L{0.2\linewidth}@{}C{0.16\linewidth}@{}C{0.16\linewidth}@{}C{0.16\linewidth}@{}}
\toprule
Event & Request Message & Current State (Dir) & Next State (Dir) & Next State (Requestor)\\
\midrule
Load & ReqRd & I & E & E\\
&& S, M & S & S\\
&& E, F & F & S\\
Load (Non-Excl) & ReqRd-NE & I, S, M & S & S\\
&& E, F & F & S\\
Store & ReqWr & I, S, E, M, F & M & M\\
\bottomrule
\end{tabular}
\caption{BedRock Coherence Directory Next State Table - MESIF}
\label{table:bedrock-states-cce-mesif}
\end{table}

\begin{table}[H]\centering
\ra{1.3}
\begin{tabular}{@{}L{0.2\linewidth}@{}L{0.2\linewidth}@{}C{0.16\linewidth}@{}C{0.16\linewidth}@{}C{0.16\linewidth}@{}}
\toprule
Event & Request Message & Current State (Dir) & Next State (Dir) & Next State (Requestor)\\
\midrule
Load (Any) & ReqRd, ReqRd-NE & I, S & S & S\\
&& M, O & O & S\\
Store & ReqWr & I, S, O, M & M & M\\
\bottomrule
\end{tabular}
\caption{BedRock Coherence Directory Next State Table - MOSI}
\label{table:bedrock-states-cce-mosi}
\end{table}

\begin{table}[H]\centering
\ra{1.3}
\begin{tabular}{@{}L{0.2\linewidth}@{}L{0.2\linewidth}@{}C{0.16\linewidth}@{}C{0.16\linewidth}@{}C{0.16\linewidth}@{}}
\toprule
Event & Request Message & Current State (Dir) & Next State (Dir) & Next State (Requestor)\\
\midrule
Load (Any) & ReqRd, ReqRd-NE & I & F & F\\
&& S & S & S\\
&& M, O & O & S\\
&& F & F & S\\
Store & ReqWr & I, S, O, M, F & M & M\\
\bottomrule
\end{tabular}
\caption{BedRock Coherence Directory Next State Table - MOSIF}
\label{table:bedrock-states-cce-mosif}
\end{table}

\begin{table}[H]\centering
\ra{1.3}
\begin{tabular}{@{}L{0.2\linewidth}@{}L{0.2\linewidth}@{}C{0.16\linewidth}@{}C{0.16\linewidth}@{}C{0.16\linewidth}@{}}
\toprule
Event & Request Message & Current State (Dir) & Next State (Dir) & Next State (Requestor)\\
\midrule
Load & ReqRd & I & E & E\\
&& S, E & S & S\\
&& M, O & O & S\\
Load (Non-Excl) & ReqRd-NE & I, S, E & S & S\\
&& M, O & O & S\\
Store & ReqWr & I, S, O, E, M & M & M\\
\bottomrule
\end{tabular}
\caption{BedRock Coherence Directory Next State Table - MOESI}
\label{table:bedrock-states-cce-moesi}
\end{table}

\chapter{\bpbedrock{} FPGA Resource Utilization Tables}
\label{app:fpga-util}

\begin{comment}
\begin{table}[t]
    \centering
    \ra{1.3}
    \begin{tabular}{@{}L{0.08\linewidth}@{}L{0.16\linewidth}@{}C{0.12\linewidth}@{}C{0.14\linewidth}@{}C{0.16\linewidth}@{}C{0.14\linewidth}@{}C{0.1\linewidth}@{}C{0.1\linewidth}@{}}
        \toprule
        Cores & CCE & Total LUT & Logic LUTs & Memory LUTs & FF (Regs) & BRAM & DSP\\
        \midrule
        1 & FSM & & & & & &\\
        & ucode & & & & & &\\
        & Hybrid-FSM & & & & & &\\
        & Hybrid & & & & & &\\
        \midrule
        2 & FSM & & & & & &\\
        & ucode & & & & & &\\
        & Hybrid-FSM & & & & & &\\
        & Hybrid & & & & & &\\
        \midrule
        4 & FSM & & & & & &\\
        & ucode & & & & & &\\
        & Hybrid-FSM & & & & & &\\
        & Hybrid & & & & & &\\
        \midrule
        8 & FSM & & & & & &\\
        & ucode & & & & & &\\
        & Hybrid-FSM & & & & & &\\
        & Hybrid & & & & & &\\
        \midrule
        16 & FSM & & & & & &\\
        & ucode & & & & & &\\
        & Hybrid-FSM & & & & & &\\
        & Hybrid & & & & & &\\
        \midrule
        \multicolumn{2}{c}{Total Available} & 1303680 & 1303680 & 600960 & 2607360 & 2016 & 9024\\
        \bottomrule
    \end{tabular}
    \caption{FPGA Design Utilization Template}
    \label{table:fpga-util-template}
\end{table}
\end{comment}

The tables in this appendix list the resource utilization of FPGA-implemented \bpbedrock{} designs. Each table shows resource utilization for designs with 1, 2, 4, 8, and 16 cores with the FSM, ucode, and Hybrid CCE designs. The Hybrid-FSM design rows use the Hybrid CCE implementation with the programmable pipe logic removed. The columns in each table indicate the core count, CCE design, and FPGA resources. The FPGA resources include the number of Lookup Tables (Total LUT) used, the number of LUTs used for logic (Logic LUTs) or memory (Memory LUTs), the number of flip-flop elements or registers (FF), the number of hardened memories (BRAM), and the number of digital signal processing blocks (DSP).

\autoref{table:fpga-util-design} and \autoref{table:fpga-util-design-percentage} list the total number of FPGA resources per type as counts and percentages, respectively, for the complete FPGA-based \blackparrot{} multicore design, including all necessary support logic to connect the multicore to memory and the host computer. This additional logic includes HBM memory controllers, PCI Express (PCIe) interface logic, AXI interconnect blocks, and FPGA host logic that provides a register-based control interface to software running on the host computer.

\autoref{table:fpga-util-bp} and \autoref{table:fpga-util-bp-percentage} list the FPGA resource utilization per type as counts and percentages, respectively, for only the \blackparrot{} multicore component of the complete FPGA design. These tables exclude resources used by the support logic described above.

\begin{table}[t]
    \centering
    \ra{1.3}
    \begin{tabular}{@{}L{0.08\linewidth}@{}L{0.16\linewidth}@{}C{0.12\linewidth}@{}C{0.14\linewidth}@{}C{0.16\linewidth}@{}C{0.14\linewidth}@{}C{0.1\linewidth}@{}C{0.1\linewidth}@{}}
        \toprule
        Cores & CCE & Total LUT & Logic LUTs & Memory LUTs & FF (Regs) & BRAM & DSP\\
        \midrule
        1 & FSM & 110043 & 93336 & 16707 & 87291 & 306 & 11\\
        & ucode & 112218 & 95511 & 16707 & 88001 & 306.5 & 11\\
        & Hybrid-FSM & 111060 & 93369 & 17691 & 87393 & 306 & 11\\
        & Hybrid & 112644 & 94953 & 17691 & 87995 & 306.5 & 11\\
        \midrule
        2 & FSM & 173483 & 148579 & 24904 & 113723 & 389.5 & 22\\
        & ucode & 178723 & 153819 & 24904 & 115071 & 390.5 & 22\\
        & Hybrid-FSM & 175629 & 148709 & 26920 & 113928 & 389.5 & 22\\
        & Hybrid & 178825 & 151905 & 26920 & 115119 & 390.5 & 22\\
        \midrule
        4 & FSM & 296004 & 257710 & 38294 & 165589 & 578.5 & 44\\
        & ucode & 303814 & 265520 & 38294 & 168157 & 580.5 & 44\\
        & Hybrid-FSM & 300376 & 257954 & 42422 & 166007 & 578.5 & 44\\
        & Hybrid & 306786 & 264364 & 42422 & 168398 & 580.5 & 44\\
        \midrule
        8 & FSM & 489766 & 428548 & 61218 & 226461 & 644.5 & 88\\
        & ucode & 507454 & 446172 & 61282 & 231493 & 648.5 & 88\\
        & Hybrid-FSM & 498240 & 428702 & 69538 & 227271 & 644.5 & 88\\
        & Hybrid & 512092 & 442554 & 69538 & 232096 & 648.5 & 88\\
        \midrule
        16 & FSM & 869480 & 764590 & 104890 & 348964 & 776.5 & 176\\
        & ucode & 906422 & 801404 & 105018 & 358740 & 784.5 & 176\\
        & Hybrid-FSM & 886897 & 765279 & 121618 & 350758 & 776.5 & 176\\
        & Hybrid & 912497 & 790871 & 121626 & 360433 & 784.5 & 176\\
        \midrule
        \multicolumn{2}{c}{Total Available} & 1303680 & 1303680 & 600960 & 2607360 & 2016 & 9024\\
        \bottomrule
    \end{tabular}
    \caption{FPGA Design Utilization}
    \label{table:fpga-util-design}
\end{table}

\begin{table}[t]
    \centering
    \ra{1.3}
    \begin{tabular}{@{}L{0.08\linewidth}@{}L{0.16\linewidth}@{}C{0.12\linewidth}@{}C{0.14\linewidth}@{}C{0.16\linewidth}@{}C{0.14\linewidth}@{}C{0.1\linewidth}@{}C{0.1\linewidth}@{}}
        \toprule
        Cores & CCE & Total LUT & Logic LUTs & Memory LUTs & FF (Regs) & BRAM & DSP\\
        \midrule
        1 & FSM & 8.44\% & 7.16\% & 2.78\% & 3.35\% & 15.18\% & 0.12\%\\
        & ucode & 8.61\% & 7.33\% & 2.78\% & 3.38\% & 15.20\% & 0.12\%\\
        & Hybrid-FSM & 8.52\% & 7.16\% & 2.94\% & 3.35\% & 15.18\% & 0.12\%\\
        & Hybrid & 8.64\% & 7.28\% & 2.94\% & 3.37\% & 15.20\% & 0.12\%\\
        \midrule
        2 & FSM & 13.31\% & 11.40\% & 4.14\% & 4.36\% & 19.32\% & 0.24\%\\
        & ucode & 13.71\% & 11.80\% & 4.14\% & 4.41\% & 19.37\% & 0.24\%\\
        & Hybrid-FSM & 13.47\% & 11.41\% & 4.48\% & 4.37\% & 19.32\% & 0.24\%\\
        & Hybrid & 13.72\% & 11.65\% & 4.48\% & 4.42\% & 19.37\% & 0.24\%\\
        \midrule
        4 & FSM & 22.71\% & 19.77\% & 6.37\% & 6.35\% & 28.70\% & 0.49\%\\
        & ucode & 23.30\% & 20.37\% & 6.37\% & 6.45\% & 28.79\% & 0.49\%\\
        & Hybrid-FSM & 23.04\% & 19.79\% & 7.06\% & 6.37\% & 28.70\% & 0.49\%\\
        & Hybrid & 23.53\% & 20.28\% & 7.06\% & 6.46\% & 28.79\% & 0.49\%\\
        \midrule
        8 & FSM & 37.57\% & 32.87\% & 10.19\% & 8.69\% & 31.97\% & 0.98\%\\
        & ucode & 38.92\% & 34.22\% & 10.20\% & 8.88\% & 32.17\% & 0.98\%\\
        & Hybrid-FSM & 38.22\% & 32.88\% & 11.57\% & 8.72\% & 31.97\% & 0.98\%\\
        & Hybrid & 39.28\% & 33.95\% & 11.57\% & 8.90\% & 32.17\% & 0.98\%\\
        \midrule
        16 & FSM & 66.69\% & 58.65\% & 17.45\% & 13.38\% & 38.52\% & 1.95\%\\
        & ucode & 69.53\% & 61.47\% & 17.48\% & 13.76\% & 38.91\% & 1.95\%\\
        & Hybrid-FSM & 68.03\% & 58.70\% & 20.24\% & 13.45\% & 38.52\% & 1.95\%\\
        & Hybrid & 69.99\% & 60.66\% & 20.24\% & 13.82\% & 38.91\% & 1.95\%\\
        \bottomrule
    \end{tabular}
    \caption{FPGA Design Utilization (Percentage)}
    \label{table:fpga-util-design-percentage}
\end{table}

\begin{table}[t]
    \centering
    \ra{1.3}
    \begin{tabular}{@{}L{0.08\linewidth}@{}L{0.16\linewidth}@{}C{0.12\linewidth}@{}C{0.14\linewidth}@{}C{0.16\linewidth}@{}C{0.14\linewidth}@{}C{0.1\linewidth}@{}C{0.1\linewidth}@{}}
        \toprule
        Cores & CCE & Total LUT & Logic LUTs & Memory LUTs & FF (Regs) & BRAM & DSP\\
        \midrule
        1 & FSM & 62152 & 54715 & 7405 & 27419 & 106 & 11\\
        & ucode & 64347 & 56910 & 7405 & 28129 & 106 & 11\\
        & Hybrid-FSM & 63202 & 54781 & 8389 & 27521 & 106 & 11\\
        & Hybrid & 64767 & 56346 & 8389 & 28123 & 106 & 11\\
        \midrule
        2 & FSM & 125609 & 109975 & 15570 & 53851 & 189 & 22\\
        & ucode & 130865 & 115231 & 15570 & 55199 & 190 & 22\\
        & Hybrid-FSM & 127768 & 110118 & 17586 & 54056 & 189 & 22\\
        & Hybrid & 130937 & 113287 & 17586 & 55247 & 190 & 22\\
        \midrule
        4 & FSM & 248139 & 219115 & 28896 & 105717 & 378 & 44\\
        & ucode & 255938 & 226914 & 28896 & 108285 & 380 & 44\\
        & Hybrid-FSM & 252496 & 219344 & 33024 & 106135 & 378 & 44\\
        & Hybrid & 258895 & 225743 & 33024 & 108526 & 380 & 44\\
        \midrule
        8 & FSM & 441878 & 389930 & 51692 & 166589 & 444 & 88\\
        & ucode & 459570 & 407558 & 51756 & 171621 & 448 & 88\\
        & Hybrid-FSM & 450336 & 390068 & 60012 & 167399 & 444 & 88\\
        & Hybrid & 464203 & 403935 & 60012 & 172224 & 448 & 88\\
        \midrule
        16 & FSM & 821603 & 725983 & 95108 & 289092 & 576 & 176\\
        & ucode & 858618 & 762870 & 95236 & 298868 & 584 & 176\\
        & Hybrid-FSM & 839059 & 726703 & 111844 & 290886 & 576 & 176\\
        & Hybrid & 864602 & 752246 & 111844 & 300561 & 584 & 176\\
        \midrule
        \multicolumn{2}{c}{Total Available} & 1303680 & 1303680 & 600960 & 2607360 & 2016 & 9024\\
        \bottomrule
    \end{tabular}
    \caption{FPGA BP Multicore Utilization}
    \label{table:fpga-util-bp}
\end{table}

\begin{table}[t]
    \centering
    \ra{1.3}
    \begin{tabular}{@{}L{0.08\linewidth}@{}L{0.16\linewidth}@{}C{0.12\linewidth}@{}C{0.14\linewidth}@{}C{0.16\linewidth}@{}C{0.14\linewidth}@{}C{0.1\linewidth}@{}C{0.1\linewidth}@{}}
        \toprule
        Cores & CCE & Total LUT & Logic LUTs & Memory LUTs & FF (Regs) & BRAM & DSP\\
        \midrule
        1 & FSM & 4.77\% & 4.20\% & 1.23\% & 1.05\% & 5.23\% & 0.12\%\\
        & ucode & 4.94\% & 4.37\% & 1.23\% & 1.08\% & 5.26\% & 0.12\%\\
        & Hybrid-FSM & 4.85\% & 4.20\% & 1.40\% & 1.06\% & 5.23\% & 0.12\%\\
        & Hybrid & 4.97\% & 4.32\% & 1.40\% & 1.08\% & 5.26\% & 0.12\%\\
        \midrule
        2 & FSM & 9.63\% & 8.44\% & 2.59\% & 2.07\% & 9.38\% & 0.24\%\\
        & ucode & 10.04\% & 8.84\% & 2.59\% & 2.12\% & 9.42\% & 0.24\%\\
        & Hybrid-FSM & 9.80\% & 8.45\% & 2.93\% & 2.07\% & 9.38\% & 0.24\%\\
        & Hybrid & 10.04\% & 8.69\% & 2.93\% & 2.12\% & 9.42\% & 0.24\%\\
        \midrule
        4 & FSM & 19.03\% & 16.81\% & 4.81\% & 4.05\% & 18.75\% & 0.49\%\\
        & ucode & 19.63\% & 17.41\% & 4.81\% & 4.15\% & 18.85\% & 0.49\%\\
        & Hybrid-FSM & 19.37\% & 16.82\% & 5.50\% & 4.07\% & 18.75\% & 0.49\%\\
        & Hybrid & 19.86\% & 17.32\% & 5.50\% & 4.16\% & 18.85\% & 0.49\%\\
        \midrule
        8 & FSM & 33.89\% & 29.91\% & 8.60\% & 6.39\% & 22.02\% & 0.98\%\\
        & ucode & 35.25\% & 31.26\% & 8.61\% & 6.58\% & 22.22\% & 0.98\%\\
        & Hybrid-FSM & 34.54\% & 29.92\% & 9.99\% & 6.42\% & 22.02\% & 0.98\%\\
        & Hybrid & 35.61\% & 30.98\% & 9.99\% & 6.61\% & 22.22\% & 0.98\%\\
        \midrule
        16 & FSM & 63.02\% & 55.69\% & 15.83\% & 11.09\% & 28.57\% & 1.95\%\\
        & ucode & 65.86\% & 58.52\% & 15.85\% & 11.46\% & 28.97\% & 1.95\%\\
        & Hybrid-FSM & 64.36\% & 55.74\% & 18.61\% & 11.16\% & 28.57\% & 1.95\%\\
        & Hybrid & 66.32\% & 57.70\% & 18.61\% & 11.53\% & 28.97\% & 1.95\%\\
        \bottomrule
    \end{tabular}
    \caption{FPGA BP Multicore Utilization (Percentage)}
    \label{table:fpga-util-bp-percentage}
\end{table}

\chapter{\bpbedrock{} FPGA Implementation Layouts}
\label{app:fpga-layout}

The figures in this appendix depict FPGA layouts of the four \bpbedrock{} designs across various core counts. The highlighted colors indicate resources consumed by individual \bpbedrock{} core tiles for each design. Colors do not correlate to specific core IDs within the multicore, i.e., the core tile highlighted in pink for one design may not represent the same core tile that is highlighted pink in another design, even when the core count is the same for both designs. The FSM design is a \bpbedrock{} multicore employing the fixed-function hardware coherence engine. The ucode designs uses the microcode-programmable coherence engine. The Hybrid FSM design uses the hybrid coherence engine design with the programmable pipe removed from the design, while the Hybrid design employs the complete hybrid coherence engine including the programmable pipe.

\clearpage

\begin{figure}
     \centering
     \begin{subfigure}[b]{0.4\textwidth}
         \centering
         \includegraphics[width=\textwidth]{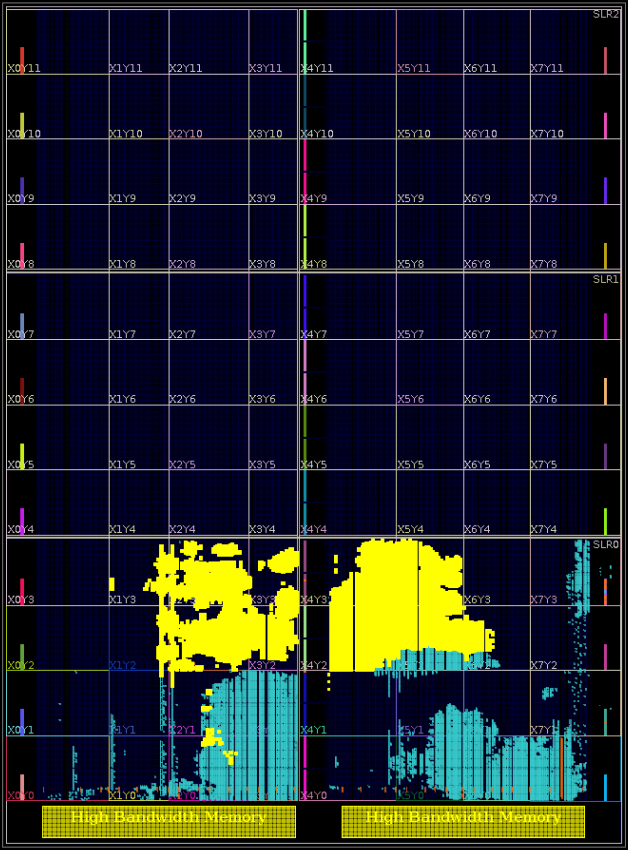}
         \caption{FSM}
         \label{fig:fpga-x1-fsm}
     \end{subfigure}
     \hspace{10mm}
     \begin{subfigure}[b]{0.4\textwidth}
         \centering
         \includegraphics[width=\textwidth]{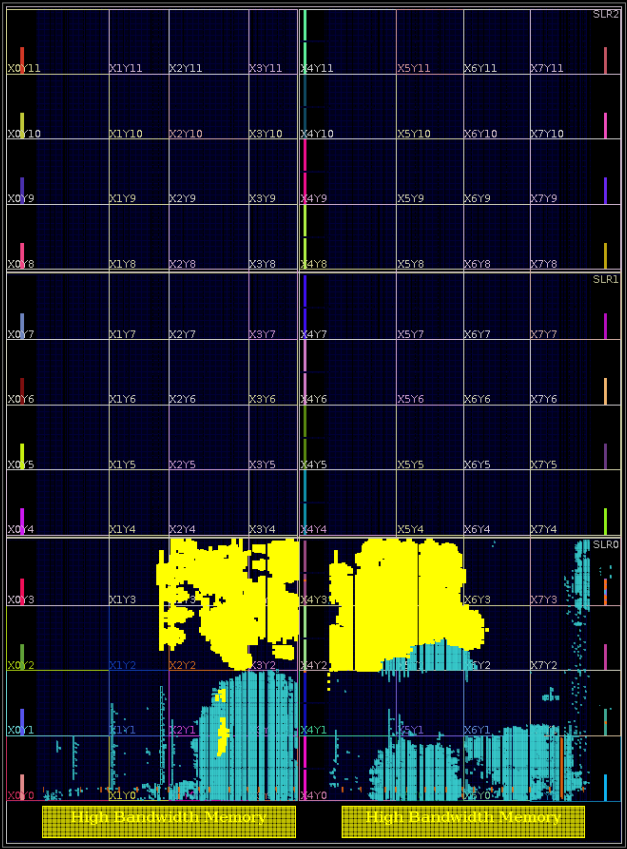}
         \caption{ucode}
         \label{fig:fpga-x1-ucode}
     \end{subfigure}
     \begin{subfigure}[b]{0.4\textwidth}
         \centering
         \includegraphics[width=\textwidth]{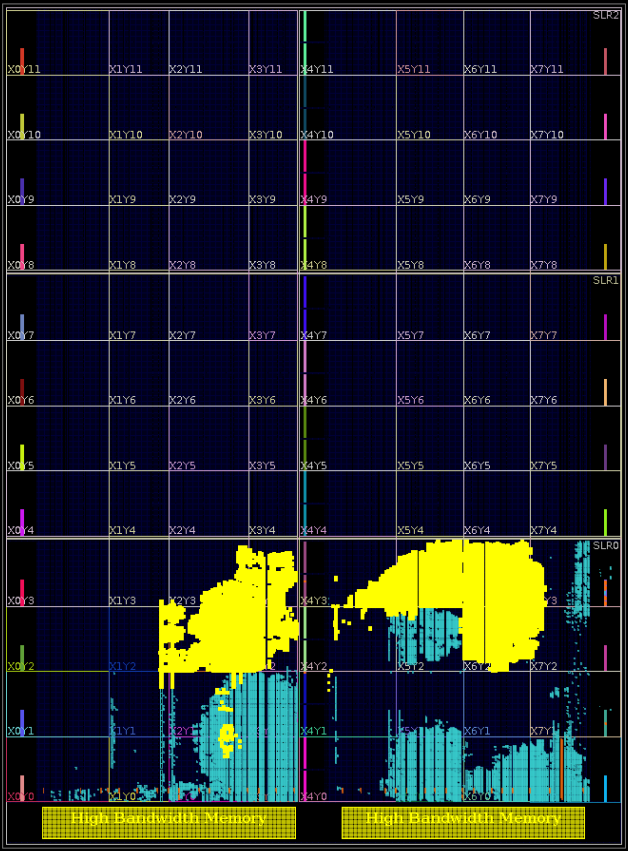}
         \caption{Hybrid FSM}
         \label{fig:fpga-x1-hybrid}
     \end{subfigure}
     \hspace{10mm}
     \begin{subfigure}[b]{0.4\textwidth}
         \centering
         \includegraphics[width=\textwidth]{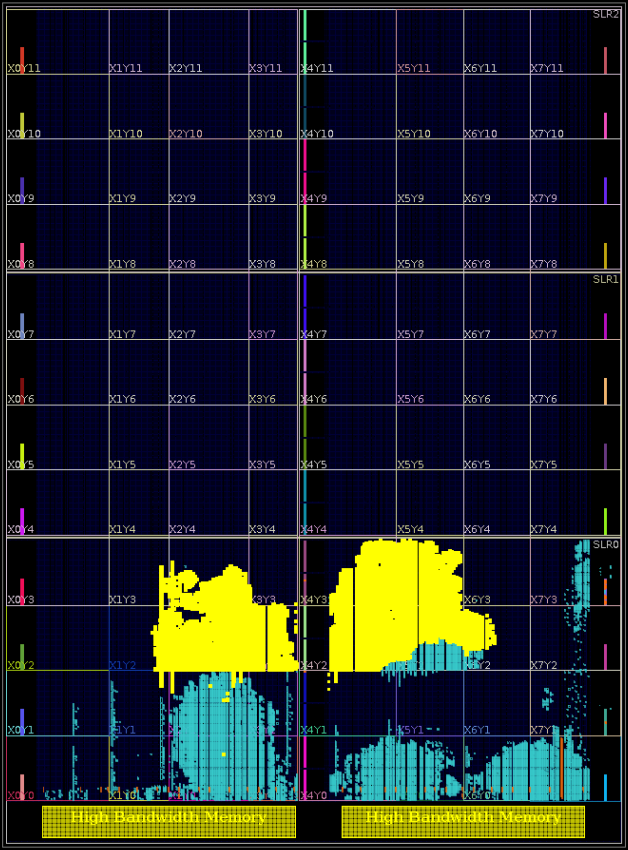}
         \caption{Hybrid}
         \label{fig:fpga-x1-hybrid-prog}
     \end{subfigure}
    \caption{\bpbedrock{} FPGA Layout - 1 core}
    \label{fig:fpga-layout-x1}
\end{figure}

\clearpage

\begin{figure}
     \centering
     \begin{subfigure}[b]{0.4\textwidth}
         \centering
         \includegraphics[width=\textwidth]{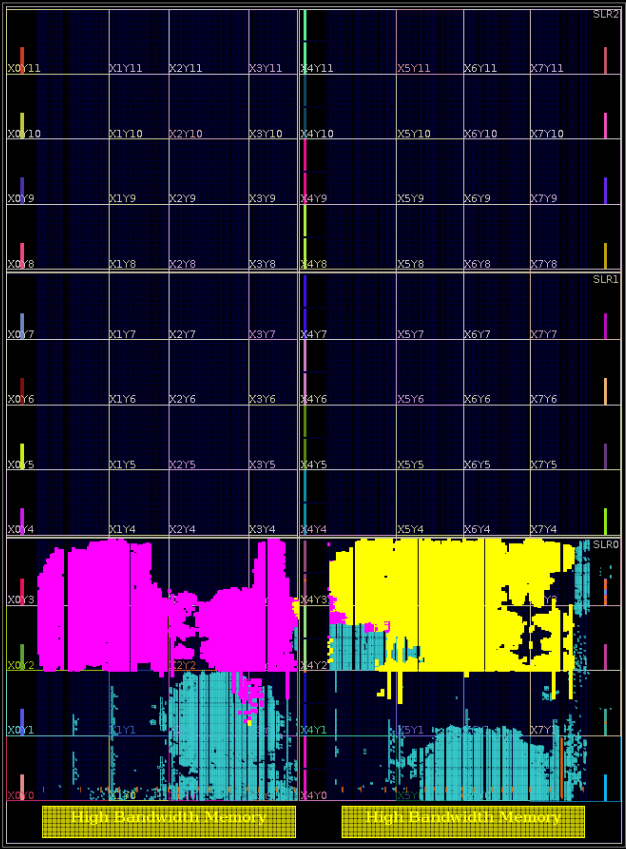}
         \caption{FSM}
         \label{fig:fpga-x2-fsm}
     \end{subfigure}
     \hspace{10mm}
     \begin{subfigure}[b]{0.4\textwidth}
         \centering
         \includegraphics[width=\textwidth]{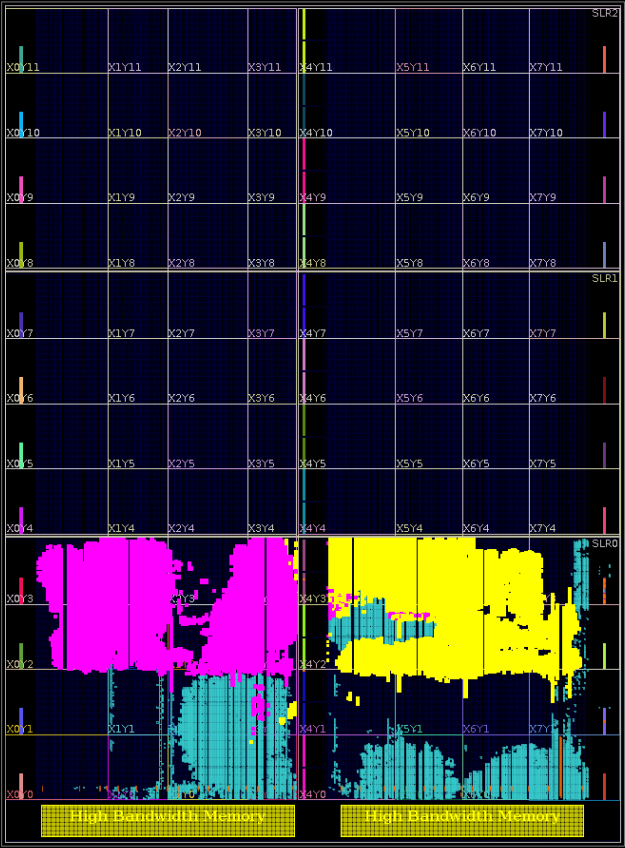}
         \caption{ucode}
         \label{fig:fpga-x2-ucode}
     \end{subfigure}
     \begin{subfigure}[b]{0.4\textwidth}
         \centering
         \includegraphics[width=\textwidth]{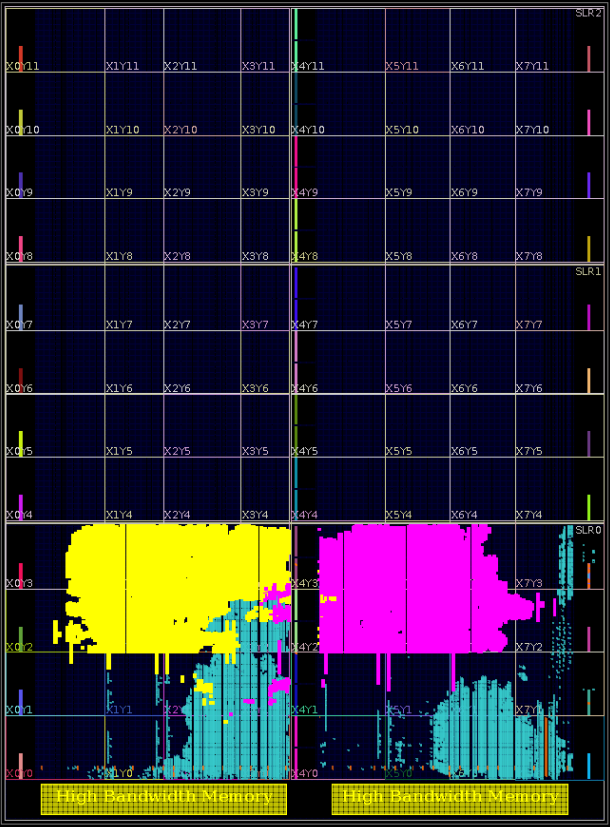}
         \caption{Hybrid FSM}
         \label{fig:fpga-x2-hybrid}
     \end{subfigure}
     \hspace{10mm}
     \begin{subfigure}[b]{0.4\textwidth}
         \centering
         \includegraphics[width=\textwidth]{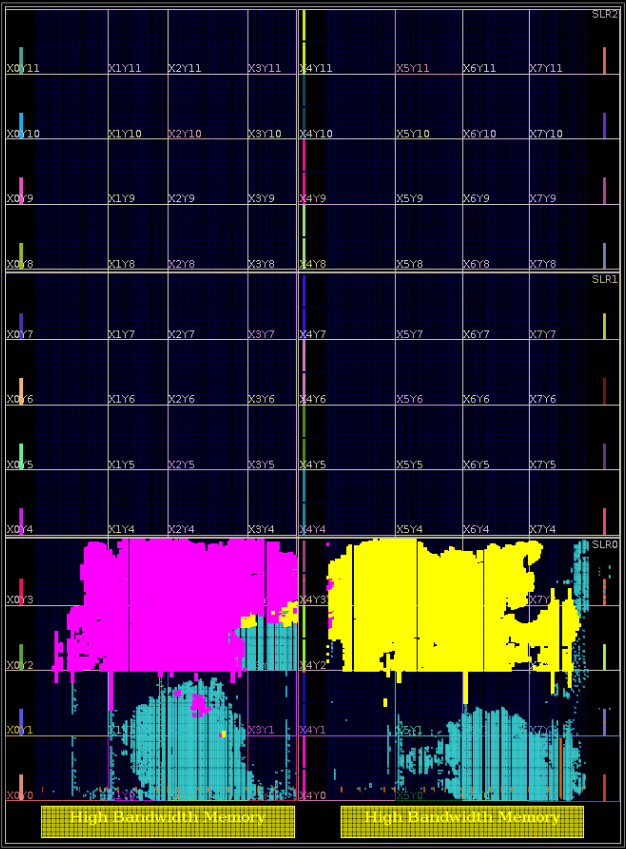}
         \caption{Hybrid}
         \label{fig:fpga-x2-hybrid-prog}
     \end{subfigure}
    \caption{\bpbedrock{} FPGA Layout - 2 core}
    \label{fig:fpga-layout-x2}
\end{figure}

\clearpage

\begin{figure}
     \centering
     \begin{subfigure}[b]{0.4\textwidth}
         \centering
         \includegraphics[width=\textwidth]{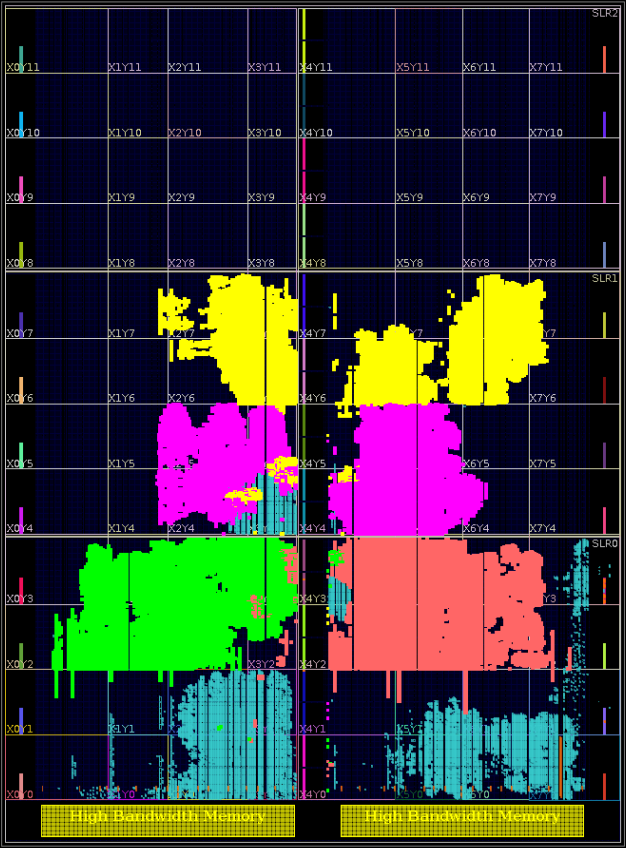}
         \caption{FSM}
         \label{fig:fpga-x4-fsm}
     \end{subfigure}
     \hspace{10mm}
     \begin{subfigure}[b]{0.4\textwidth}
         \centering
         \includegraphics[width=\textwidth]{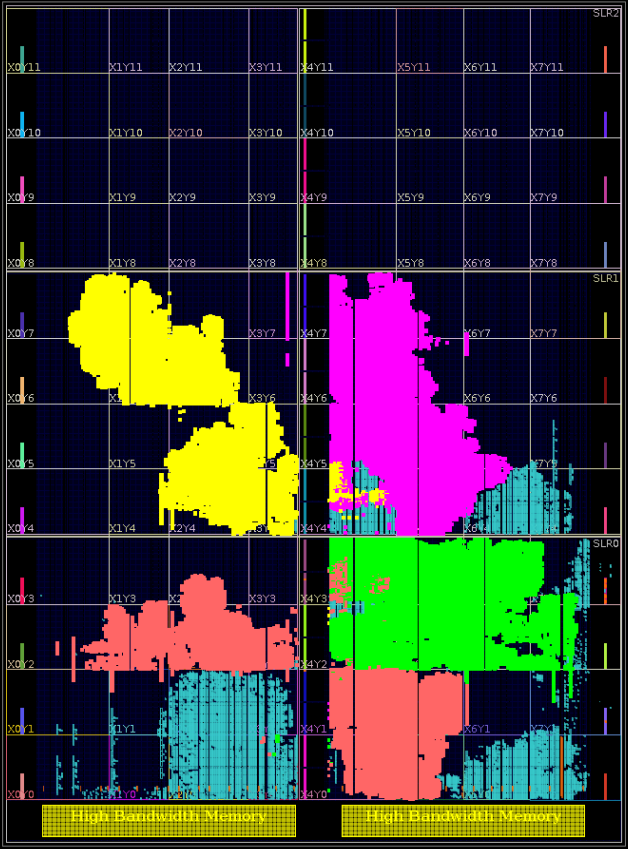}
         \caption{ucode}
         \label{fig:fpga-x4-ucode}
     \end{subfigure}
     \begin{subfigure}[b]{0.4\textwidth}
         \centering
         \includegraphics[width=\textwidth]{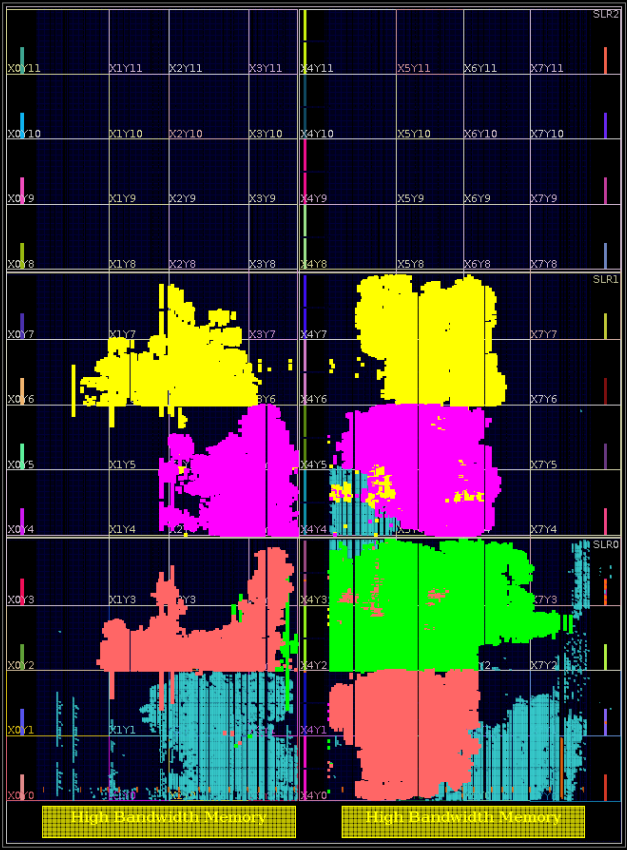}
         \caption{Hybrid FSM}
         \label{fig:fpga-x4-hybrid}
     \end{subfigure}
     \hspace{10mm}
     \begin{subfigure}[b]{0.4\textwidth}
         \centering
         \includegraphics[width=\textwidth]{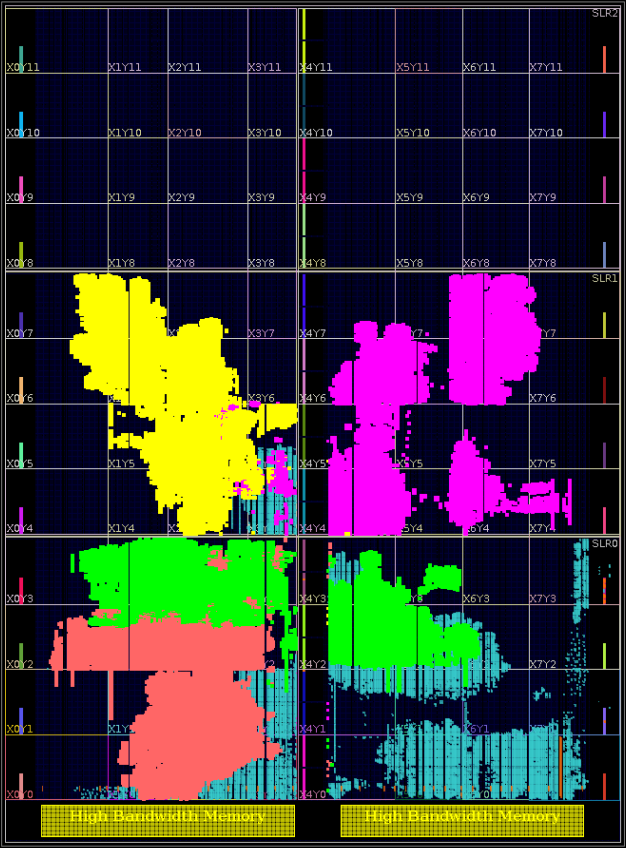}
         \caption{Hybrid}
         \label{fig:fpga-x4-hybrid-prog}
     \end{subfigure}
    \caption{\bpbedrock{} FPGA Layout - 4 core}
    \label{fig:fpga-layout-x4}
\end{figure}

\clearpage

\begin{figure}
     \centering
     \begin{subfigure}[b]{0.4\textwidth}
         \centering
         \includegraphics[width=\textwidth]{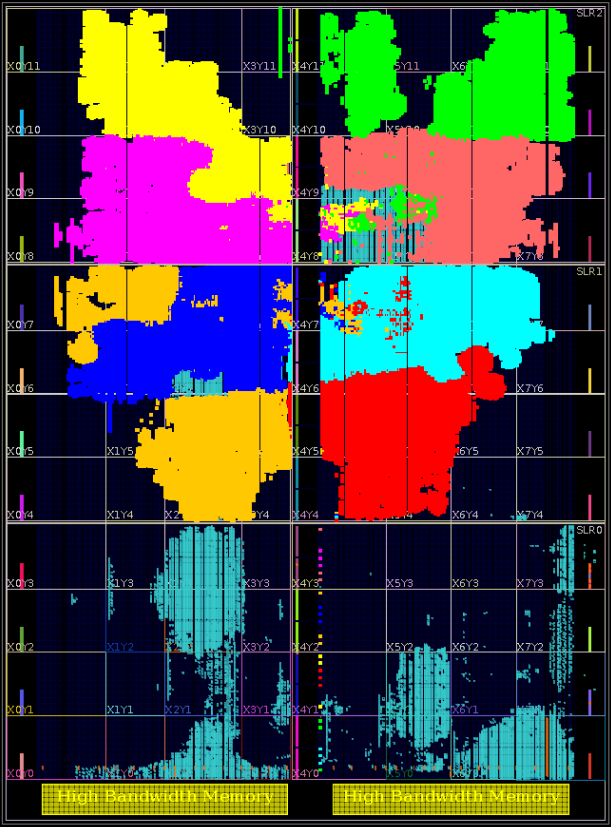}
         \caption{FSM}
         \label{fig:fpga-x8-fsm}
     \end{subfigure}
     \hspace{10mm}
     \begin{subfigure}[b]{0.4\textwidth}
         \centering
         \includegraphics[width=\textwidth]{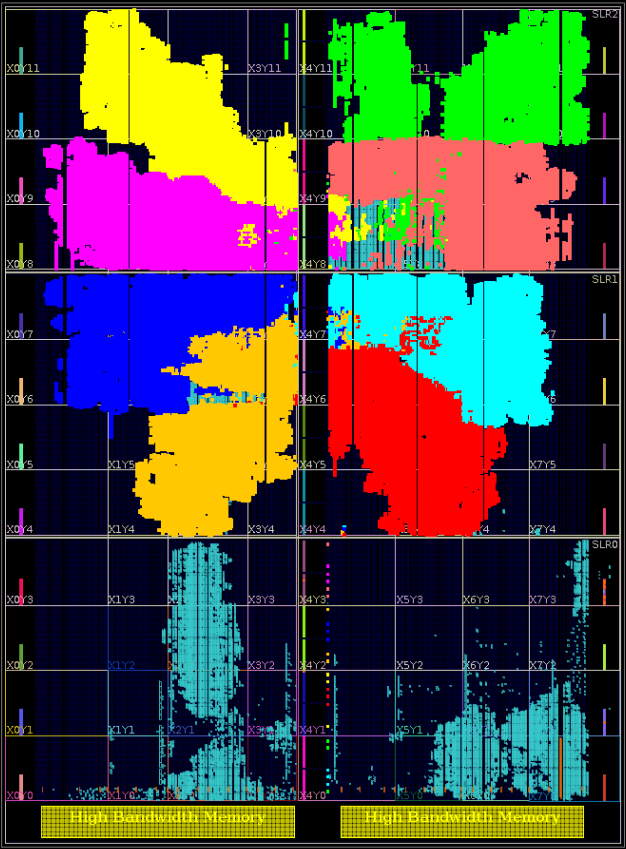}
         \caption{ucode}
         \label{fig:fpga-x8-ucode}
     \end{subfigure}
     \begin{subfigure}[b]{0.4\textwidth}
         \centering
         \includegraphics[width=\textwidth]{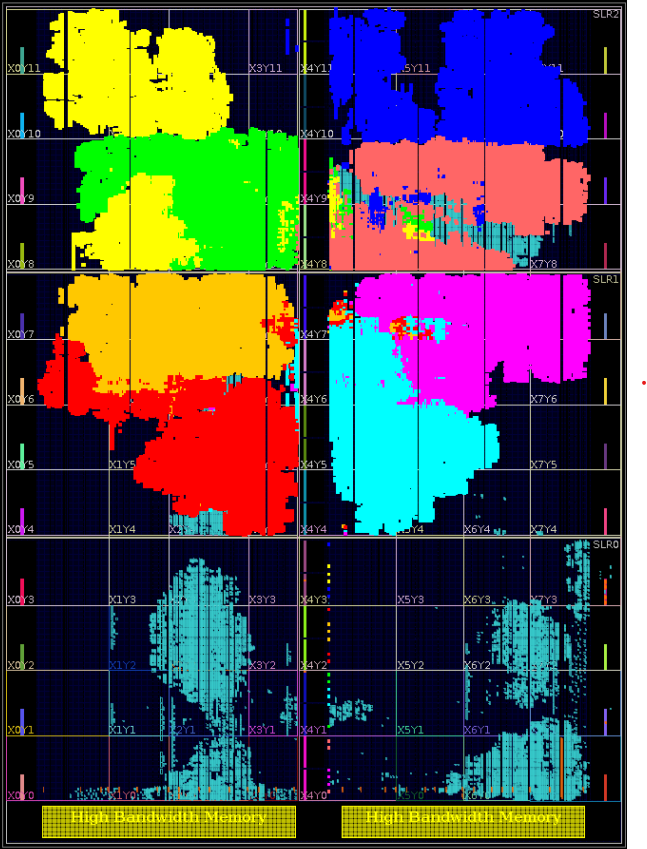}
         \caption{Hybrid FSM}
         \label{fig:fpga-x8-hybrid}
     \end{subfigure}
     \hspace{10mm}
     \begin{subfigure}[b]{0.4\textwidth}
         \centering
         \includegraphics[width=\textwidth]{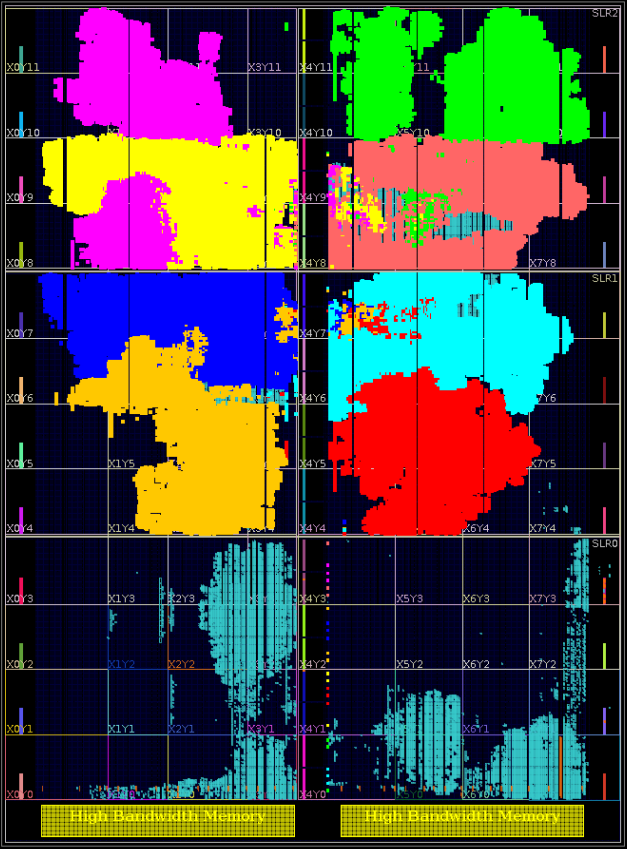}
         \caption{Hybrid}
         \label{fig:fpga-x8-hybrid-prog}
     \end{subfigure}
    \caption{\bpbedrock{} FPGA Layout - 8 core}
    \label{fig:fpga-layout-x8}
\end{figure}

\clearpage

\begin{figure}
     \centering
     \begin{subfigure}[b]{0.4\textwidth}
         \centering
         \includegraphics[width=\textwidth]{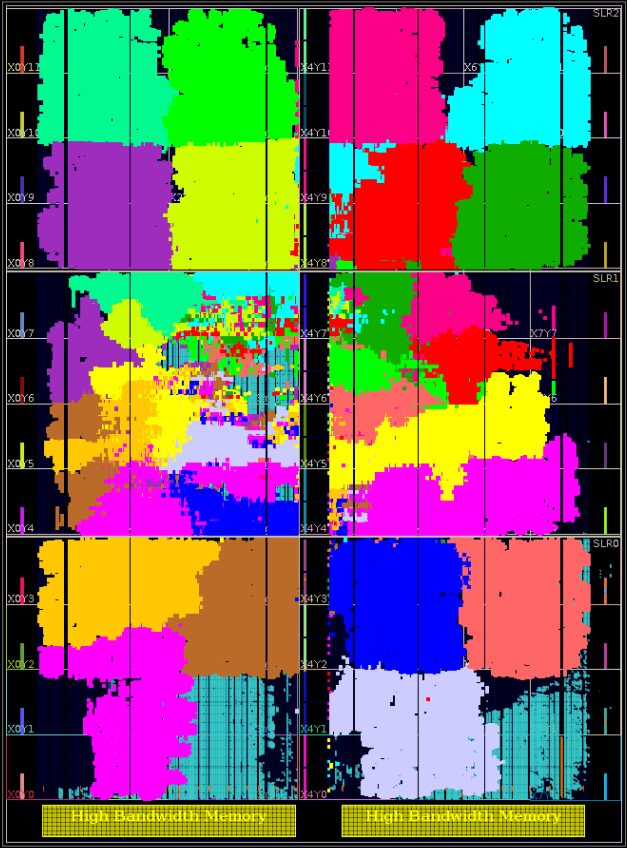}
         \caption{FSM}
         \label{fig:fpga-x16-fsm}
     \end{subfigure}
     \hspace{10mm}
     \begin{subfigure}[b]{0.4\textwidth}
         \centering
         \includegraphics[width=\textwidth]{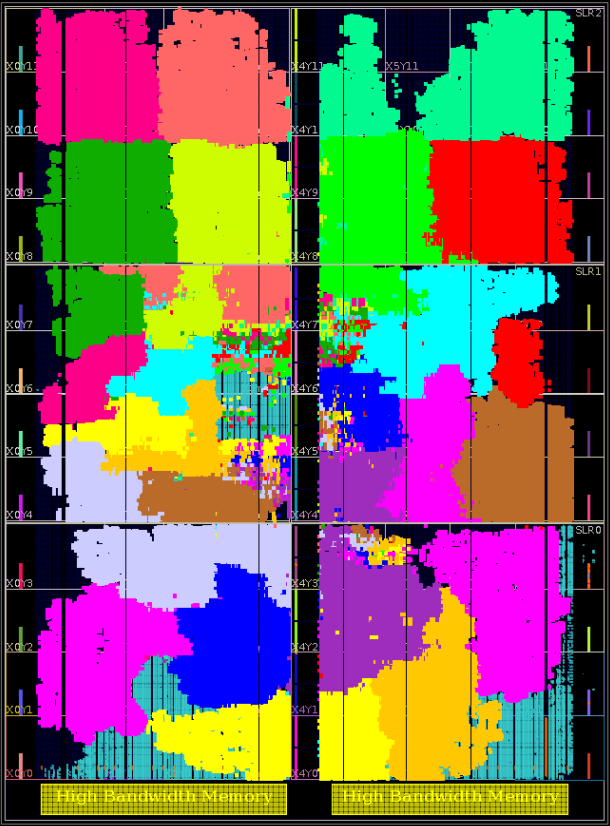}
         \caption{ucode}
         \label{fig:fpga-x16-ucode}
     \end{subfigure}
     \begin{subfigure}[b]{0.4\textwidth}
         \centering
         \includegraphics[width=\textwidth]{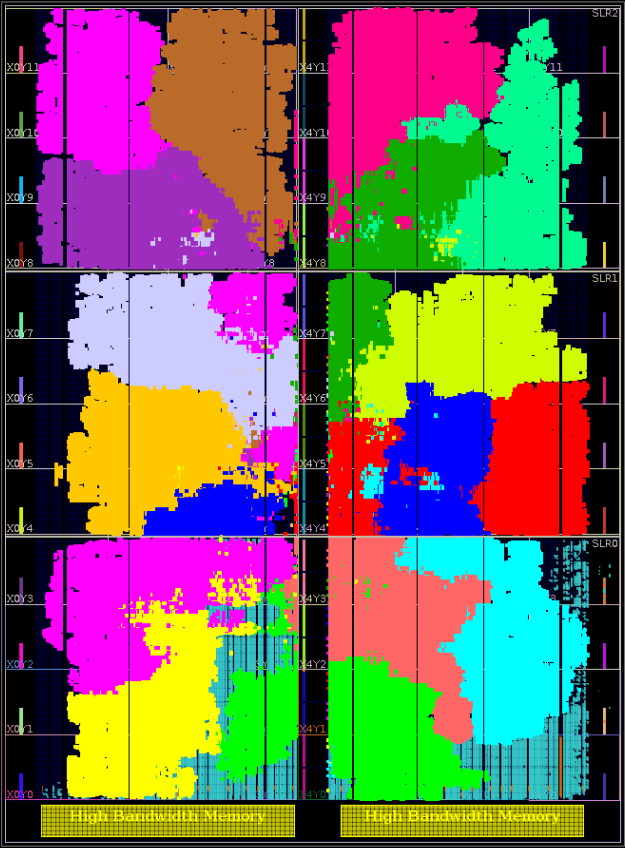}
         \caption{Hybrid FSM}
         \label{fig:fpga-x16-hybrid}
     \end{subfigure}
     \hspace{10mm}
     \begin{subfigure}[b]{0.4\textwidth}
         \centering
         \includegraphics[width=\textwidth]{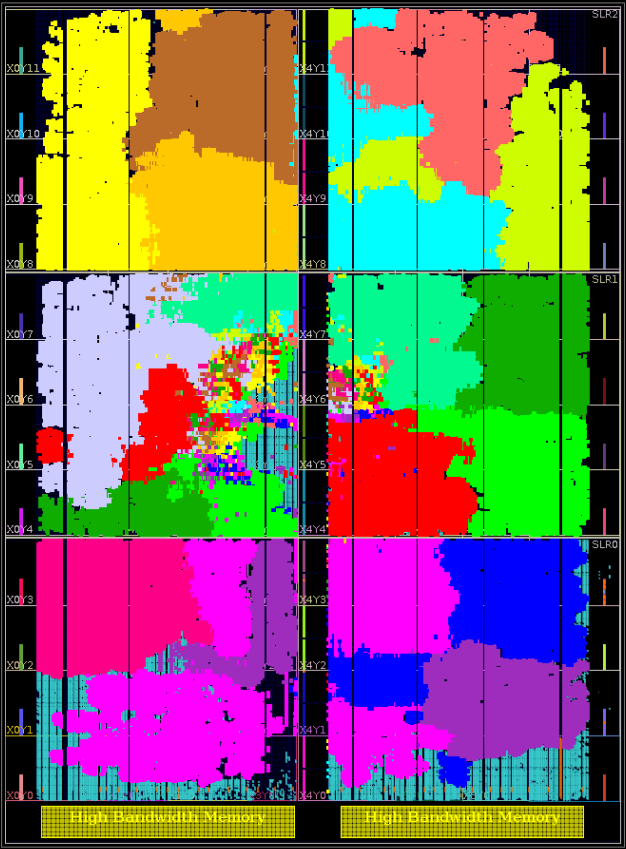}
         \caption{Hybrid}
         \label{fig:fpga-x16-hybrid-prog}
     \end{subfigure}
    \caption{\bpbedrock{} FPGA Layout - 16 core}
    \label{fig:fpga-layout-x16}
\end{figure}
% Bibliography
\printbibliography[heading=bibintoc, title={References}]

\end{document}